# Coding Concepts

# and

# Reed-Solomon Codes

**A.J. Han Vinck**

*University Duisburg-Essen, Germany*

Coding Concepts and Reed Solomon Codes

# Dedication

This book is dedicated to my wife Martina, with whom I shared almost all of my life. Her support and love made my work possible and also enjoyable.

I am grateful to my PhD supervisor Piet Schalkwijk for introducing me to the field of Information theory at the University of Eindhoven, the Netherlands. The work in his group was the basis for my scientific career.

I also want to thank the numerous students that participated in my research during a period of over forty years. During my stay in the Institute for Experimental Mathematics, also many foreign guests participated in the research and over 100 papers were published in journals. Part of these results are presented in this book.

My colleges Hendrik Ferreira, Yuan Luo, Hiro Morita and Kees Schouhamer Immink, joined me in many research projects and organization of workshops and conferences. These activities contributed to the very stimulating research atmosphere in my working group in the Institute for Experimental Mathematics in Essen.

# Contents

# Preface

The material in this book is presented to graduate students in Information and Communication theory. The idea is that we give an introduction to particular applications of information theory and coding in digital communications. The goal is to bring understanding of the underlying concepts, both in theory as well as in practice. We mainly concentrate on our own research results. After showing obtainable performance, we give a specific implementation using Reed-Solomon (RS) codes. The reason for using RS codes is that they can be seen as optimal codes with maximum obtainable minimum distance. Furthermore, the structure of RS codes enables specific applications that fit perfectly into the developed concepts. We do not intend to develop the theory of error correcting codes. We summarize our idea in Figure 0.1.

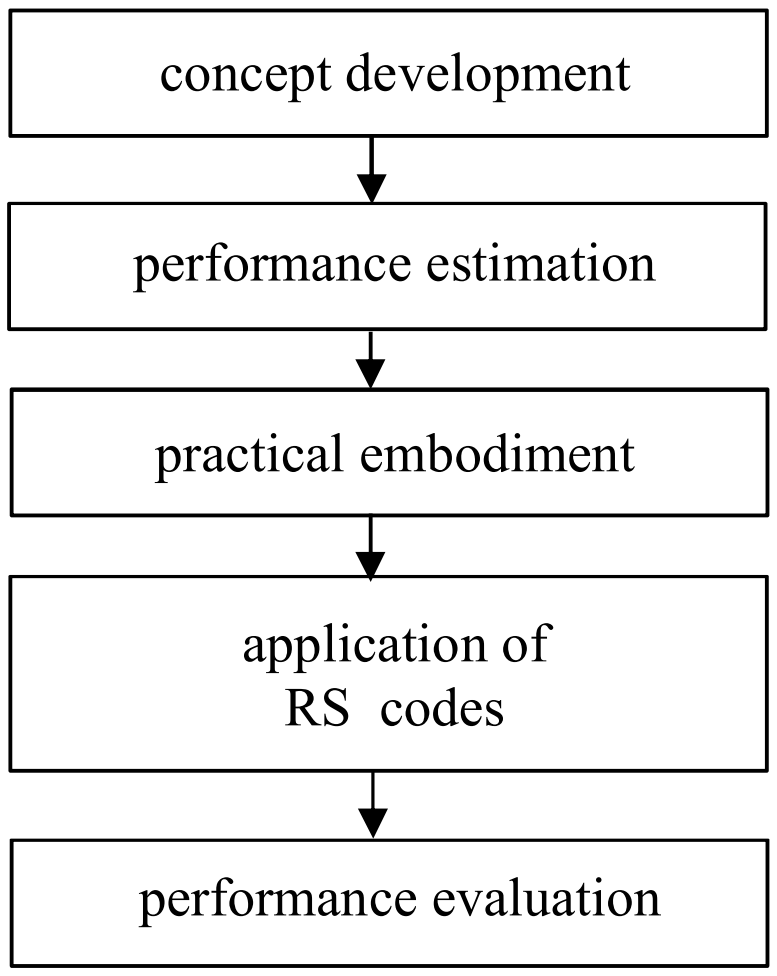

**Figure 0.1** Concept of the book

Chapter 1 considers networks where information is transmitted in packets. Reliable packet transmission can be realized via networking protocols using feedback and error control coding. When the data in the packets are code words of an RS code and in addition the packets have an identification (ID) number, provided by the network protocol in use, we can use error control to improve the throughput efficiency especially in the situation where packets get lost. Lost packets can be recovered by erasure decoding.

The random access situation, where users transmit packets without knowledge about other users is solved using the Aloha protocol. We discuss different random access strategies. The first one is the classical Aloha protocol, where feedback is given to the users about successful or not successful transmission (collision). The theoretical throughput of the Aloha system $\eta = Ge^{-G}$, where G is the total offered load by the users involved in the random access protocol, in packets per time slot, see Figure 2.

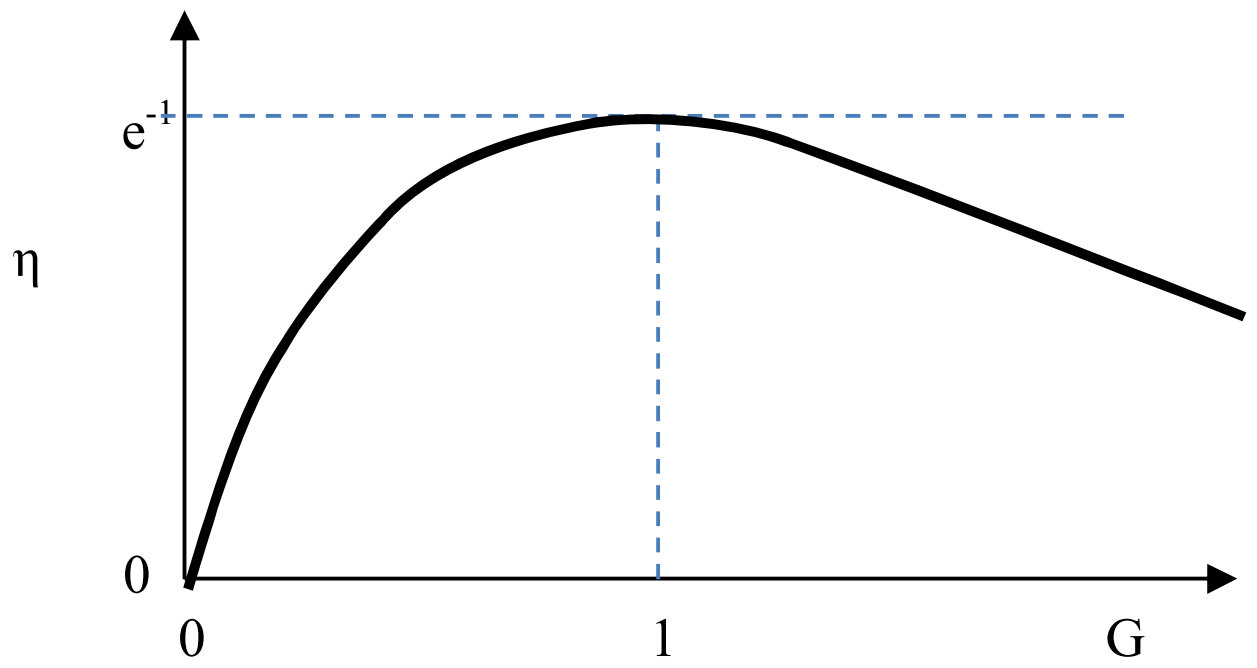

**Figure 0.2** The throughput for the slotted Aloha

We discuss a randomized scheme using RS codes, where there is no feedback, but collision detection. We also present a simple and efficient error correction scheme for array-like data structures. This scheme is used for packet transmission using the Aloha protocol without feedback and without collision detection.

In Chapter 2 we develop the concepts of random access communications, where we are especially interested in the situation where users cannot use knowledge of the participants in the access situation. Information theory considers the capacity or maximum mutual information of multi access channels (MAC) for users that are permanently sending information. In random access the users are not permanently active, and in addition cannot communicate with each other. We discuss coding in a multi user concept where the set of active users is small compared to the total amount of potential users. In this case, time or frequency division is an inefficient way of communication. We introduce three models for "non-trivial" access. For these models we calculate the maximum throughput and the way how to obtain this maximum throughput. We consider transmission using signatures. For a simple construction we can calculate the performance and compare the results with random signatures. We give a class of codes that can be used to uniquely determine the set of active users from the composite signature at the channel output.

Interesting applications for RS codes appear in Chapter 3, for channels that are not typical. One of the non-typical channels is the power line communications channel. In this channel we have all different types of noise and disturbances. RS codes give interesting optimal solutions to some of the problems.

One of the dominant parameters in power line communications is that of attenuation. Attenuation can vary between 10-100 dB per kilometer. In power line communications, attenuation prevents signals to propagate over distances that are longer than say 500 meters. Investigations to overcome the high attenuation are therefore highly important. We discuss the efficiency of communication over links that are in tandem. This situation occurs when users in a power line communication network are connected to the same line and the power lines can be seen as "bus systems", where all connected users can listen to the signals present on the bus. There is an interesting connection between links in tandem and the classical problem of broadcasting in information theory. We relate the tandem channel with the degraded broadcast channel and give the information theoretical capacities.

A popular topic in coding is the application of soft decision decoding for additive white Gaussian noise channels. The reason for this is the performance improvements in the decoding error rate that can be obtained. It is well know that, due to complexity reasons, RS codes cannot be decoded using soft decision information. In Chapter 4, we first explain the idea of soft decision and then give examples of the applications. We combine RS codes with single parity check codes and show that a 3 dB gain can be attained at low complexity.

In the second part of the chapter we consider a special class of codes, permutation codes. These codes are effective against different kinds of noise, like impulse and narrowband noise. Extended RS codes with low rate (2/n) can be seen as an optimum class of permutation codes. We develop the concept of same-weight codes to be used for channels with permanent disturbances.

In Chapter 5, we describe the classical problem of data transmission in the presence of a wiretapper listening to the communication between two parties. The encoder's objective is to maximize the amount of information transmitted to the legal receiver, while keeping the wiretapper as uninformed as possible. We consider a noiseless and a noisy main channel, see Figure 0.1 and Figure 0.2, respectively.

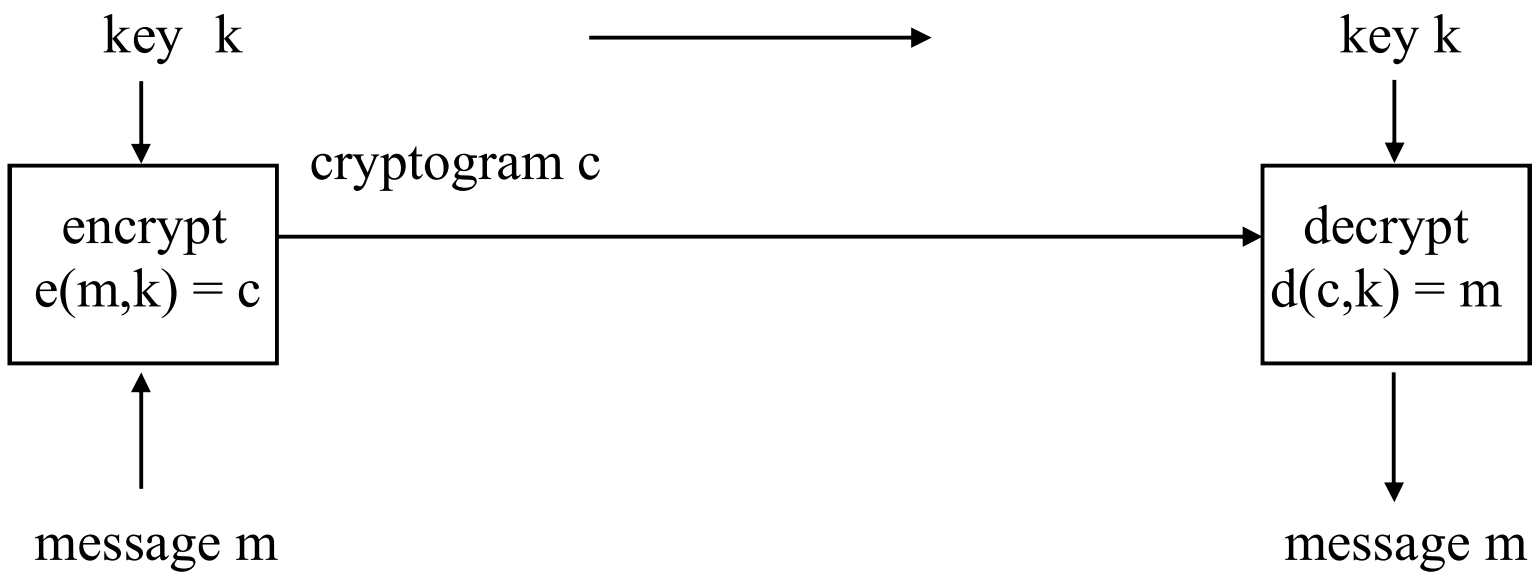

**Figure 0.1** Schematic representation of the noiseless Shannon cipher model.

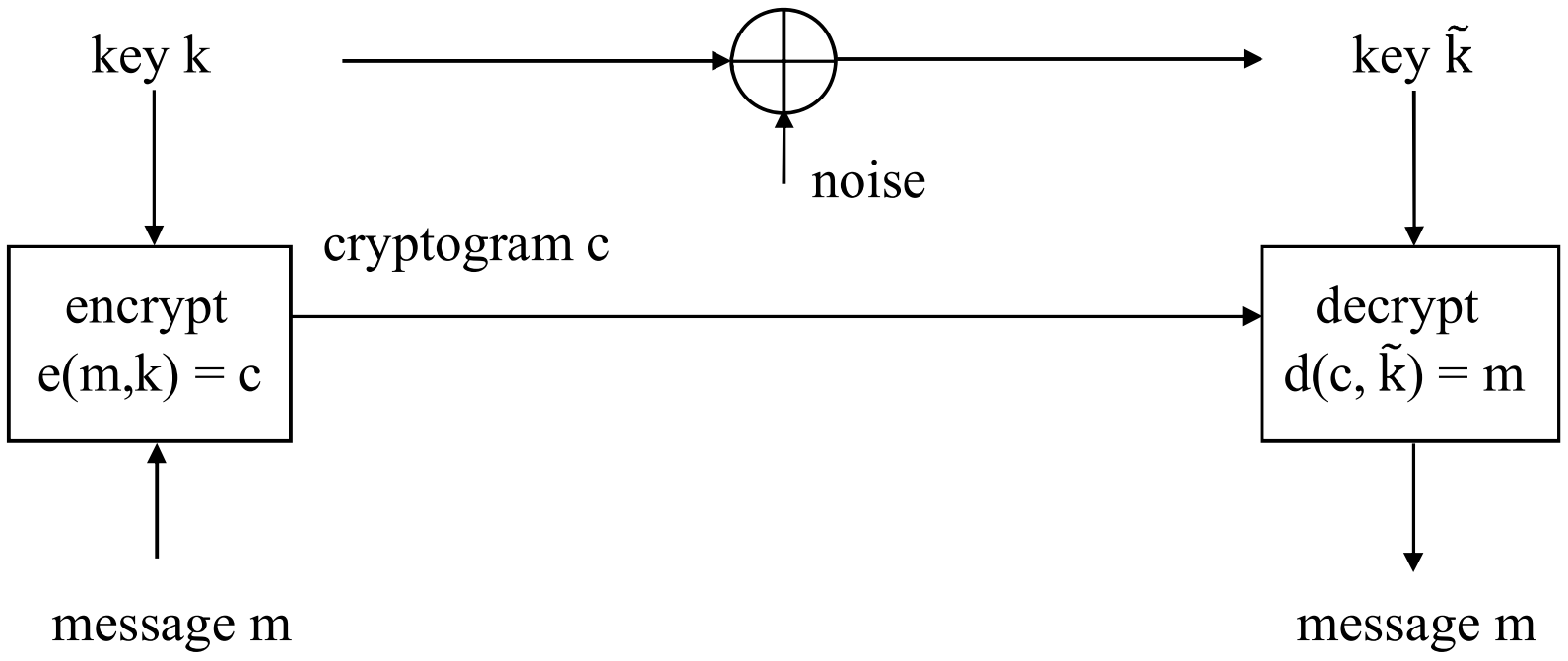

**Figure 0.2** Schematic representation of the noisy cipher system

For both situations, the use of a special form of RS encoding gives optimum results.

In Chapter 6, we show that one of the interesting consequences of the developed theory is that there is a close connection with the problem of secure biometric authentication and verification. The biometric verification scheme as developed by Juels-Wattenberg can be seen as a noisy cipher system. The biometric properties act as a key at encryption, and as a noisy key at decryption.

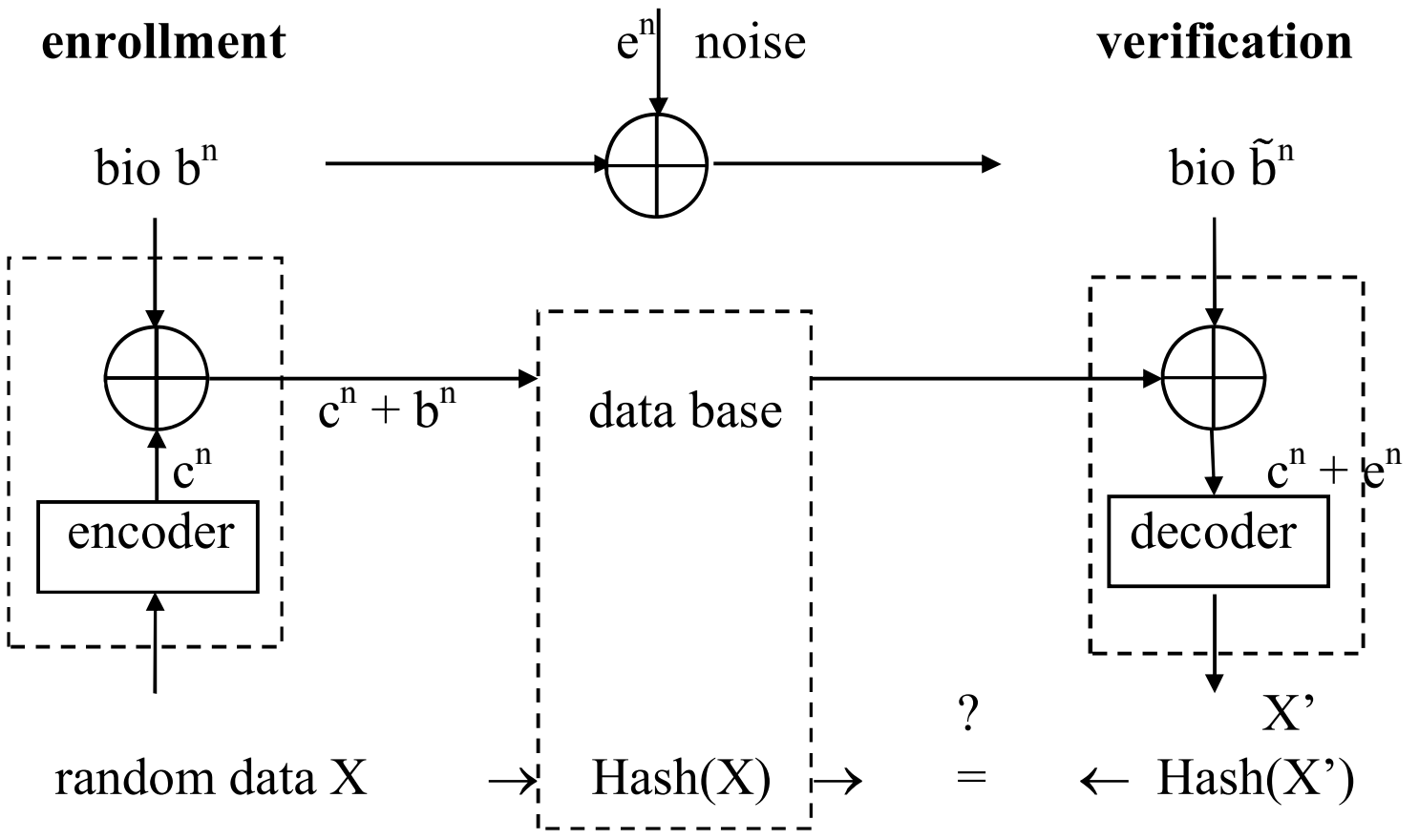

**Figure 0.2** Schematic representation of the Juels-Wattenberg scheme

We include the Juels-Sudan scheme and show that there is an interesting implementation of both schemes using RS codes. The chapter includes performance bounds for the False Acceptance Rate and the False Rejection Rate.

Chapter 7 contains topics that can be seen as constrained coding. The first topic is the avoidance of certain symbols to occur in the output of an RS encoder. The second topic is the combination of RS and run-length constrained coding with application in bandwidth limited channels. In communication systems, coding is a fixed part of the design. To change a code is impossible due to standards and already implemented decoding algorithms. In cognitive systems, we want to adapt the efficiency of a transmission scheme to the actual circumstances. For this we have to be able to change the modulation and also the error correcting or error detecting code. Some concepts are well known, such as puncturing or shortening/lengthening an encoding matrix of a linear code. In this way, the efficiency of the code is changed by changing the length n of the code words. We choose another method: row extension/deletion of the encoding matrix for an (n,k) code, and thus change the parameter k.

An important part of a computing system is the memory. Due to improvements in process technology and clever circuit design, we can produce large memory systems on a chip with a high packing density. A high packing density has its limits and may cause errors in the memory cell structure. A popular error model is that of defects. A defect always produces the same output when being read, irrespective of the input. The model for defective memory cells is given in Figure 0.3.

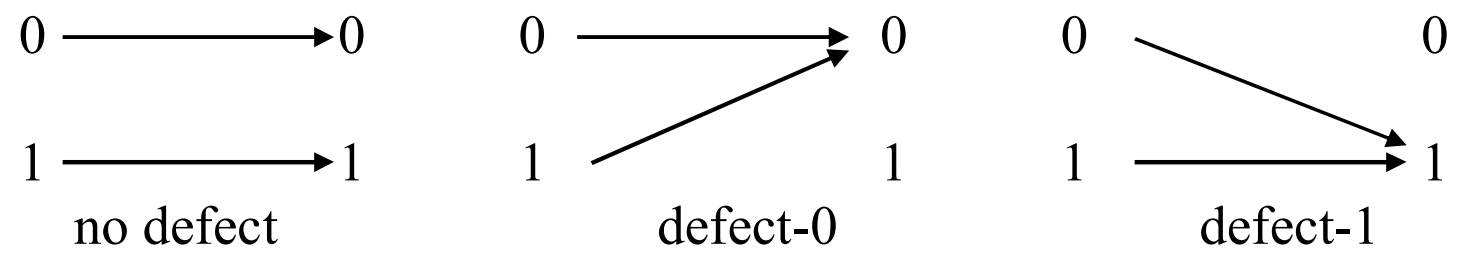

**Figure 0.3** Model for the defects in a memory

Kuznetsov and Tsybakov showed that, when the writer knows and the reader does not know the value and the position of the defect, the writing capacity is (1-p) bits/memory cell. In Chapter 8, we show how to use RS codes as optimum symbol defect matching codes.

The appendix contains the background information about RS codes as needed for the understanding of the rest of the chapters where we discuss the different applications. We also give some basic properties of the concepts of entropy in information theory. We give a brief "engineering" summary of the theory for Galois fields as far as we need this. We include the necessary properties of the Vandermonde determinant and we complement the description of RS codes with a possible decoding procedure. In the last parts we explain the idea of water-filling for two and more parallel channels.

# Chapter 1

# Packet Transmission

In networks, data is transmitted in the form of packets. A packet consists of a block of data bits plus control information that helps in its proper routing and reassembly into the original form at the receiving end. We assume that packets have an identification number (ID) provided by the network protocol in use. Reliable packet transmission can be realized via networking protocols using feedback and error control coding. Error control coding can improve the throughput or transmission efficiency especially in the situation where packets get lost.

We discuss different protocols depending on the actual networking situation:

- the classical Aloha protocol [1], where feedback is given to the users about successful or not successful transmission;
- a protocol with randomized packet transmission using Reed-Solomon (RS) codes, when there is no feedback but there is collision detection;
- a simple and efficient error correction scheme for array-like data structures. This scheme is used for packet transmission using the Aloha protocol without feedback and without collision detection.

## 1.1 Introduction

RS codes play an important role in networking. We will highlight some of the "simple" applications. We first recall the main properties of linear error correcting codes that are needed for the protocol descriptions.

**Property 1.1** The code words of a linear (n,k) error correcting code are linear combinations of the rows of a k × n encoding matrix $G_{k,n}$ of rank k, i.e. $c^n = x^k G_{k,n}$, where $x^k$ and $c^n$ are of length k and n, respectively.

We may distinguish between bit and symbol oriented codes. For symbol oriented codes over GF($2^m$), the components of $x^k$, $c^n$ and $G_{k,n}$ are m bits wide. The minimum number of differences (measured in symbols) between any two code words is denoted as minimum distance $d_{min}$.

**Property 1.2** For an (n,k) linear error correcting code, the minimum distance $d_{min}$ is upper bounded by $d_{min} \leq n - k + 1$. RS codes achieve the upper bound with equality.

**Property 1.3** An (n,k) error correcting code with minimum distance $d_{min}$ is able to detect at least t symbol errors for $d_{min} \geq t + 1$.

**Property 1.4** An (n,k) error correcting code with minimum distance $d_{min}$ is able to correct at least t symbol errors for $d_{min} \geq 2t + 1$.

If the position of an error in a code word is known, without knowing the value of the error, we call this an erasure.

**Property 1.5** Any k × (n - ($d_{min}$ -1)) sub-matrix of the encoding matrix $G_{k,n}$ has rank k.

A non-zero input gives a non-zero output, even if we delete $d_{min}$ - 1 columns of the encoder matrix. This property plays an important role in the recovery of erased or missing packets.

**Property 1.6** For an (n,k) linear code with minimum distance $d_{min}$, $d_{min}$-1 erasures can be corrected.

Error-detection is a powerful tool to decrease the decoding error rate. The idea is:

- use a linear (n,k) block code to transmit information;
- if the received word is not equal to one of the possible transmitted code words, errors are detected.

Misdetection occurs iff (if and only if) the error pattern changes a transmitted code word into another valid code word. For linear codes, the undetected error probability is determined by the probability that we do receive a code word with $d_{min}$ differences from the transmitted code word. In [4] we discuss the undetected error probability for linear block codes on channels with memory.

For narrow band applications, packet transmission combined with Automatic Repeat reQuest (ARQ) is a very robust way of communicating in a point-to-point connection. For this, data is encoded with an (n,k) linear error correcting code with code efficiency $R = k/n < 1$. Packets detected to be in error, are requested to be retransmitted. In general, for a packet error rate p and code efficiency R, the overall transmission efficiency for the basic ARQ is

$$\eta_{ARQ} = (1-p)\, R.$$

Since long packets are certainly received in error, the packet length needs to be adapted to the channel error behavior. This requires accurate modeling.

A problem we do not touch is that of packet synchronization. It can be shown that there is a synchronization problem when there is a gap in time between the transmitted packets, see [5,6].

## 1.2 Transmission Using an Encoding Matrix

### 1.2.1 Packets as symbols of an RS code

We consider the transmission of n packets, where every packet contains a symbol (typically 8-16 bits) from an RS code word. Since the minimum distance of the RS code is $n – k + 1$, we can accept $n - k$ erasures or we need k correctly received packets to be able to reconstruct (decode) the encoded information.

### 1.2.2 Binary packet combination

An alternative option is to encode k binary data words ($P_1, P_2, \cdots, P_k$) with a <u>binary</u> encoding matrix $G_{k,n}$ for a code with minimum distance $d_{min}$. The $i^{th}$ data word $Q_i$, $1 \le i \le n$, is encoded as

$$Q_i = g_{1,i} P_1 \oplus g_{2,i} P_2 \cdots \oplus g_{k,i} P_k,$$

where $g_{j,i}$ is the binary component from $G_{k,n}$ at row j and column i.

Using Property 1.5, any sub-matrix with dimension $k \times (n - (d_{min} - 1))$ has rank k. Thus, if at the receiver we do receive at least $d_{min} - 1$ correct packets, we are able to decode the transmitted k data words by using the ID of the lost (erased) packets.

**Example** We use a (7,3) encoding matrix for a code with minimum distance 4, i.e.

$$G_{3,7} = \begin{bmatrix} 1\,0\,0\,0\,1\,1\,1 \\ 0\,1\,0\,1\,0\,1\,1 \\ 0\,0\,1\,1\,1\,0\,1 \end{bmatrix} \tag{1.1}$$

Any sub-matrix with dimension $k \times (n - (d_{min} - 1)) = 3 \times 4$ has rank 3 and can thus be used to reconstruct the three encoded data words when not more than 3 transmitted packets out of 7 are lost (erased).

Suppose that we use matrix (1.1) for the encoding of data words that are RS code words. The RS codes are good error detection codes and thus after transmission, the receiver knows whether a data word contains errors or not. The data words detected in error can be declared to be erased or missing. The same decoding principle applies as before.

**Example** We encode three RS code words over $GF(2^m)$ as

$$(RS_1, RS_2, RS_3)\, G_{3,7} = (Q_1, Q_2, \cdots, Q_7).$$

For this encoding scheme, $Q_7 = (RS_1 \oplus RS_2 \oplus RS_3)$, where $\oplus$ denotes the exclusive-OR (XOR) of the code words $RS_1$, $RS_2$ and $RS_3$. By linearity, $Q_7$ is also a code word and can thus be treated as a code word from the RS code. If $Q_3$, $Q_4$ and $Q_5$ are lost, we can use the columns 1, 2, 6 and 7 to reconstruct $(RS_1, RS_2, RS_3)$. The part of the encoding matrix

$$\widetilde{G}_{3,4} = \begin{bmatrix} 1\,0\,1\,1 \\ 0\,1\,1\,1 \\ 0\,0\,0\,1 \end{bmatrix},$$

has rank three and the $3 \times 3$ matrix formed by columns 1, 2 and 4 has an inverse. Multiplying the received vector of data words $(Q_1, Q_2, Q_7)$ with $\widetilde{G}^{inv}$, we find back $RS_1$, $RS_2$ and $RS_3$, i.e.

$$(Q_1, Q_2, Q_7) \cdot \begin{bmatrix} 1\ 0\ 1 \\ 0\ 1\ 1 \\ 0\ 0\ 1 \end{bmatrix} = (RS_1, RS_2, RS_3).$$

Another option for the receiver is to decode the received RS code words. After successful decoding of $d_{min}$ - 1 received code words, we can do the inverse operation and reconstruct the three RS code words.

### 1.2.3 Non-binary packet combination

Suppose that we have k data words, each with symbols from $GF(2^m)$. We combine these data words using an encoding matrix for an RS code, again with symbols from $GF(2^m)$. Since the data words are multiplied with elements from $GF(2^m)$, the linear combinations are again data words with symbols from $GF(2^m)$. If the combining matrix has minimum distance $d_{min}$, we can reconstruct a maximum of $d_{min}$ - 1 erased (or missing) data words.

For an erasure rate of pn packets, (n - k) = pn redundant packets allow for correct decoding of the erased (missing) data words. The transmission efficiency of this scheme is thus

$$k/n = \eta_{coded} = (1- p),$$

which is roughly the same as for ARQ. The disadvantage is that we have to wait for k correctly received packets before decoding can start. The advantage is that we do not need the feedback.

**Note** This method can also be seen as a kind of parallel processing. Instead of working on packets, we could also work on a symbol level. If we consider only the $i^{th}$ symbol from each packet, we have the normal RS encoding, see Section 1.2.1. In the present situation, the "side-information" is available for all symbols in a packet and not for each symbol individually.

**Example** Suppose that we have k = n - 2 data words. These words are encoded with the RS encoding matrix $G_{n-2,n}$ as

$$(P_1, P_2, \cdots, P_{n-2}) \cdot \begin{pmatrix} 1\,0\,0 & & & 0 & \alpha^{n-2} & 1 \\ 0\,1\,0 & & & 0 & \alpha^{n-1} & 1 \\ \cdots & & & \cdots & & \\ 0\,0 & \cdots & 0 & 1\;0 & \alpha^{2} & 1 \\ 0\,0 & \cdots & & 0\;1 & \alpha & 1 \end{pmatrix} = (C_1, C_2, \cdots, C_n).$$

The RS code has minimum distance 3 and thus 2 erasures or missing data words can be corrected. Using the syndrome former

$$H^T = \begin{bmatrix} \alpha^{n-2} & 1 \\ \alpha^{n-1} & 1 \\ \cdots & \\ \alpha & 1 \\ 1 & 0 \\ 0 & 1 \end{bmatrix},$$

one can also correct one data word that is in error. Suppose that we have one data word in error and that the error word is E. Then,

$$(C_1, C_2, \cdots, C_i \oplus E, C_{i+1}, \cdots, C_n)\, H^T = (S_1, S_2).$$

The syndrome $(S_1, S_2)$ can have the following output:

| | |
|---|---|
| (0, E) | error E in $C_n$; |
| (E, 0) | error E in $C_{n-1}$; |
| $(\alpha^i E, E)$ | error E in position $n - i - 1$, $1 \le i \le n - 2$. |

From the correct $(C_1, C_2, \cdots, C_n)$ the words $(P_1, P_2, \cdots, P_{n-2})$ can be reconstructed.

**Note** We can use RS code words as input words and since then the output words are also RS code words, these can be decoded first. A *decoding* error can then be corrected with the above method in a following step.

## 1.3 Feedback and Collision Detection

One of the famous random access transmission protocols is the Aloha protocol [1]. We consider only a simplified version, namely the slotted Aloha protocol. Transmitters have an input buffer, where incoming packets are stored. In the simplified slotted Aloha protocol with T active users, we assume the following steps:

1. an active user takes a packet from the input buffer with probability p. A packet is transmitted in the next available time slot. If no packet is taken from the buffer, with probability (1-p), we do not use the next time slot and repeat the procedure;
2. if a collision (two or more users use the same time slot) is detected and reported back to the transmitter via a noiseless feedback, the transmitter repeats the last packet in the next time slot with probability p until no collision is detected;
3. if no collision is detected, we go to step 1.

The transmission efficiency for this Aloha protocol is given by the fraction of successful packet transmissions (no collisions), i.e.

$$\eta = pT(1 - p)^{T-1}, \qquad (1.2)$$

which approaches 1/e packet per time slot for T = 1/p. For our model, we can define G = pT as the total average traffic offered to the protocol by the users. For small p we then have

$$\eta \sim Ge^{-G}, \qquad (1.3)$$

which is illustrated in Figure 1.1.

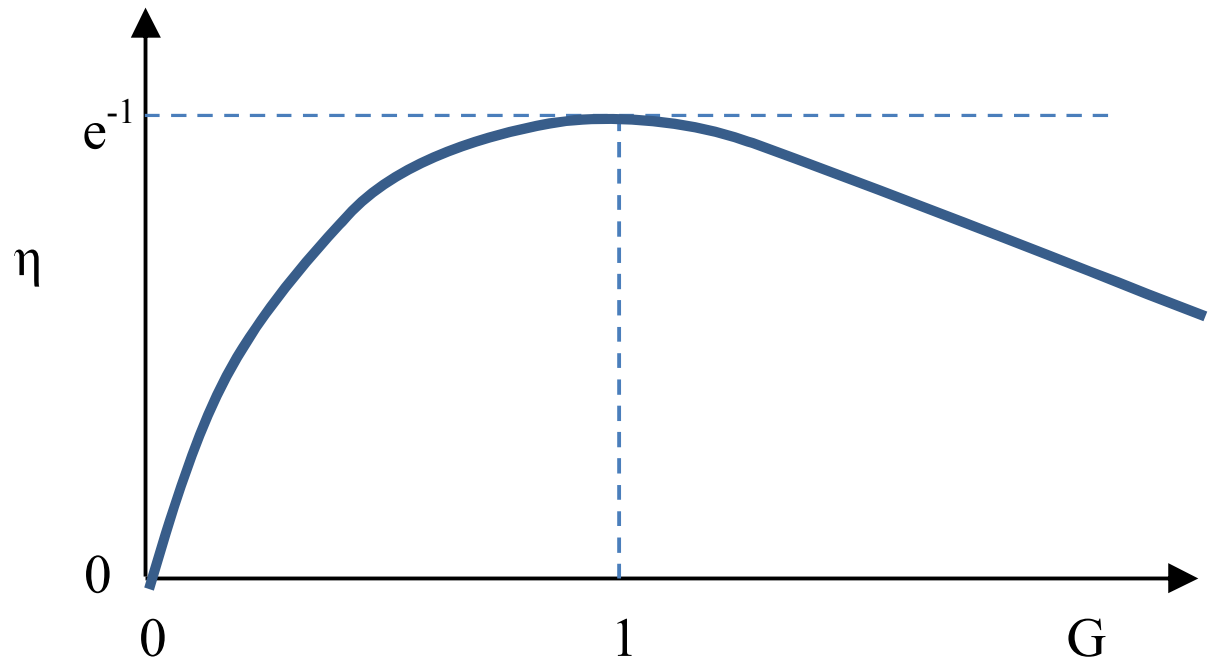

**Figure 1.1** The transmission efficiency for the slotted Aloha

Complicated problems occur in the calculations, when the model is not as ideal as in our example.

## 1.4 No Feedback, Collision Detection

RS codes can be used in a collision detection protocol. It works as follows. Suppose that $T \le Z$ active users from a large population transmit symbols from an RS code of length n using a transmission array with Z rows and n columns. Every user has a particular "random" signature of length n that indicates in which row and column a symbol can be transmitted. We denote $nP_e$ as the number of expected collisions. For a particular user, a collision occurs if at least one of the remaining T-1 users uses the same slot, i.e. the number of expected collisions

$$nP_e = n(1-(1-\frac{1}{Z})^{T-1}). \tag{1.4}$$

For $nP_e \le n-k$, we expect correct decoding at the receiver. From (1.4) we have the condition that for correct decoding

$$\frac{k}{n} \le (1-\frac{1}{Z})^{T-1} < e^{-\frac{T-1}{Z}}.$$

The overall transmission efficiency of the protocol for T users each sending k information symbols in an array with dimension $Z \times n$, is then

$$\eta = \frac{Tk}{Zn} < \frac{T}{Z} e^{-\frac{T-1}{Z}}. \tag{1.5}$$

For large Z, The maximum approaches $e^{-1}$ for $T/Z \to 1$, which is the same as for the slotted Aloha with collision feedback.

**Example** In Figures 1.2, 1.3 and 1.4 we give the code array for four users with code words of length 6 symbols, the "spreading" code array with 6 rows and 6 columns, and the received "de-spread" code array, respectively.

| | | | | | |
|---|---|---|---|---|---|
| $c_{11}$ | $c_{12}$ | $c_{13}$ | $c_{14}$ | $c_{15}$ | $c_{16}$ |
| $c_{21}$ | $c_{22}$ | $c_{23}$ | $c_{24}$ | $c_{25}$ | $c_{26}$ |
| $c_{31}$ | $c_{32}$ | $c_{33}$ | $c_{34}$ | $c_{35}$ | $c_{36}$ |
| $c_{41}$ | $c_{42}$ | $c_{43}$ | $c_{44}$ | $c_{45}$ | $c_{46}$ |

**Figure 1.2** Code array for 4 users, code words are horizontal

| | | | | | |
|---|---|---|---|---|---|
| $c_{11}$ | | | $c_{24}$ | | $c_{46}$ |
| | $c_{22}$ | | | | |
| | $c_{12}$ | | $c_{34}$ | $c_{35}$ | |
| $c_{21}$ | | x | | x | x |
| $c_{41}$ | $c_{32}$ | | x | | $c_{16}$ |
| $c_{31}$ | $c_{42}$ | $c_{43}$ | | $c_{25}$ | $c_{26}$ |

**Figure 1.3** "Spreading" array, where x indicates collision

| | | | | | |
|---|---|---|---|---|---|
| $c_{11}$ | $c_{12}$ | x | x | x | $c_{16}$ |
| $c_{21}$ | $c_{22}$ | x | $c_{24}$ | $c_{25}$ | $c_{26}$ |
| $c_{31}$ | $c_{32}$ | x | $c_{34}$ | $c_{35}$ | x |
| $c_{41}$ | $c_{42}$ | $c_{43}$ | x | x | $c_{46}$ |

**Figure 1.4** "De-spread" code array for the active users

## 1.5 No Feedback, no Collision Detection

### 1.5.1 Error correction for array-like data structures

Before we describe the protocol, we have to introduce error correction for array-like data structures, see also [2,3,7].

A packet array $P_{k,N}$ consists of N columns and k rows, as shown in Figure 1.5a. Every column in this array belongs to a specific information word to be encoded. The elements of this array are symbols of GF($2^m$).

$P_{k,N}$ =

| $P_1$ | $P_2$ | ••• | | $P_N$ | |
|---|---|---|---|---|---|
| $p_{11}$ | $p_{12}$ | ••• | | $p_{1N}$ | row 1 |
| $p_{21}$ | $p_{22}$ | ••• | | $p_{2N}$ | row 2 |
| ••• | ••• | | | | ••• |
| | | | | | |
| $p_{k1}$ | $p_{k2}$ | ••• | | $p_{kN}$ | row k |

**Figure 1.5a** Packet array $P_{k,N}$ with k rows and N columns

A code array $C_{n,N}$ with n rows and N columns is given in Figure 1.5b. The elements of this array are symbols of GF($2^m$). Every column contains the n symbols of an RS code word.

$C_{n,N}$ =

| $C_1$ | $C_2$ | ••• | | $C_N$ | |
|---|---|---|---|---|---|
| $c_{11}$ | $c_{12}$ | ••• | | $c_{1N}$ | row 1 |
| $c_{21}$ | $c_{22}$ | ••• | | $c_{2N}$ | row 2 |
| ••• | ••• | | | | ••• |
| | | | | | |
| $c_{n1}$ | $c_{n2}$ | ••• | | $c_{nN}$ | row n |

**Figure 1.5b** Code array $C_{n,N}$ with n rows and N columns

The encoding of an array of information is given in Figure 1.6. The information array, consisting of N columns of length k is multiplied by the encoding matrix $G^T$ leading to a code array of dimensions $n \times N$. The arrangement of code words as columns of an array can be interpreted as block-interleaving.

Transmission of the array $C_{n,N}$ is done row wise, symbol by symbol. We furthermore assume that channel output is binary (hard decision). The errors inserted by the channel can be described by an additive error array $E_{n,N}$, where $R_{n,N} = C_{n,N} \oplus E_{n,N}$. The channel is assumed:

- to behave such that each row of a received array is either error-free or corrupted by many symbol errors;
- the errors which disturb the rows of the array $C_{n,N}$ are such that the rows of $E_{n,N}$ are linearly independent;
- the number of erroneous rows is t.

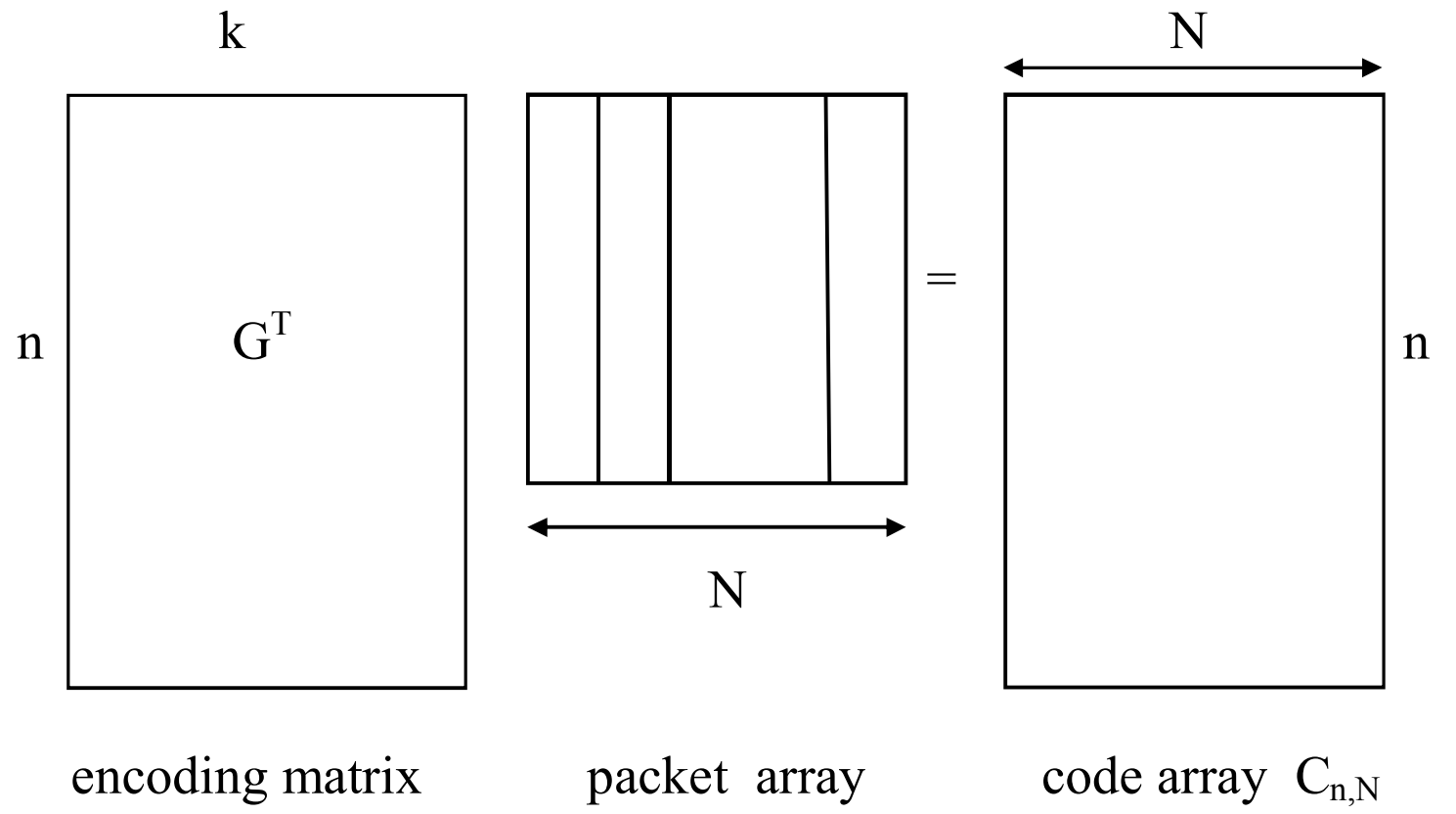

**Figure 1.6** Encoding operations on a block of information

At the receiver side, we can use the property of the check matrix $H_{n-k,n}$, and calculate the syndrome array

$$
\begin{aligned}
S_{n-k,N} &= H_{n-k,n} R_{n,N} \\
&= H_{n-k,n} (C_{n,N} \oplus E_{n,N}) \\
&= H_{n-k,n} E_{n,N}.
\end{aligned}
$$

We visualize the process as given in Figure 1.7

In order to use the properties of $S_{n-k,N}$, we will take a closer look at the syndrome array. The t rows that are in error determine the sub-array $E_{sub}$ and the collection of row numbers form the set E. Consequently, the numbers in E define t columns in $H_{n-k,n}$, called the sub-array $H_{sub}$ and clearly,

$S_{n-k,N} = H_{sub}E_{sub}$.

We now will consider the performance of the scheme. The assumption is that t independent error words of length N corrupt the code array $C_{n,N}$. For $t \leq (n - k)$, t columns of $H_{n-k,n}$ are linearly independent and hence $\text{rank}(H_{sub}) = t$. Furthermore, $\text{rank}(E_{sub}) = t$ because linearly independent error words are assumed. It thus follows that

$$\text{rank}(S_{n-k,N}) = \text{rank}(H_{sub} \cdot E_{sub}) = t.$$

For $t < (n-k)$ the syndrome array $S_{n-k,N}$ does not have maximum rank. Consequently, (n-k-t) all-0 rows can be constructed from $S_{n-k,N}$ by performing elementary row operations according to Gaussian elimination. Let the resulting array with (n-k-t) all-0 rows be denoted by $S^0$. Performing on $H_{n-k,n}$ the same row operations that were performed on $S_{n-k,N}$ we get the matrix $H^0$ where $H^0 R_{n,N} = S^0$.

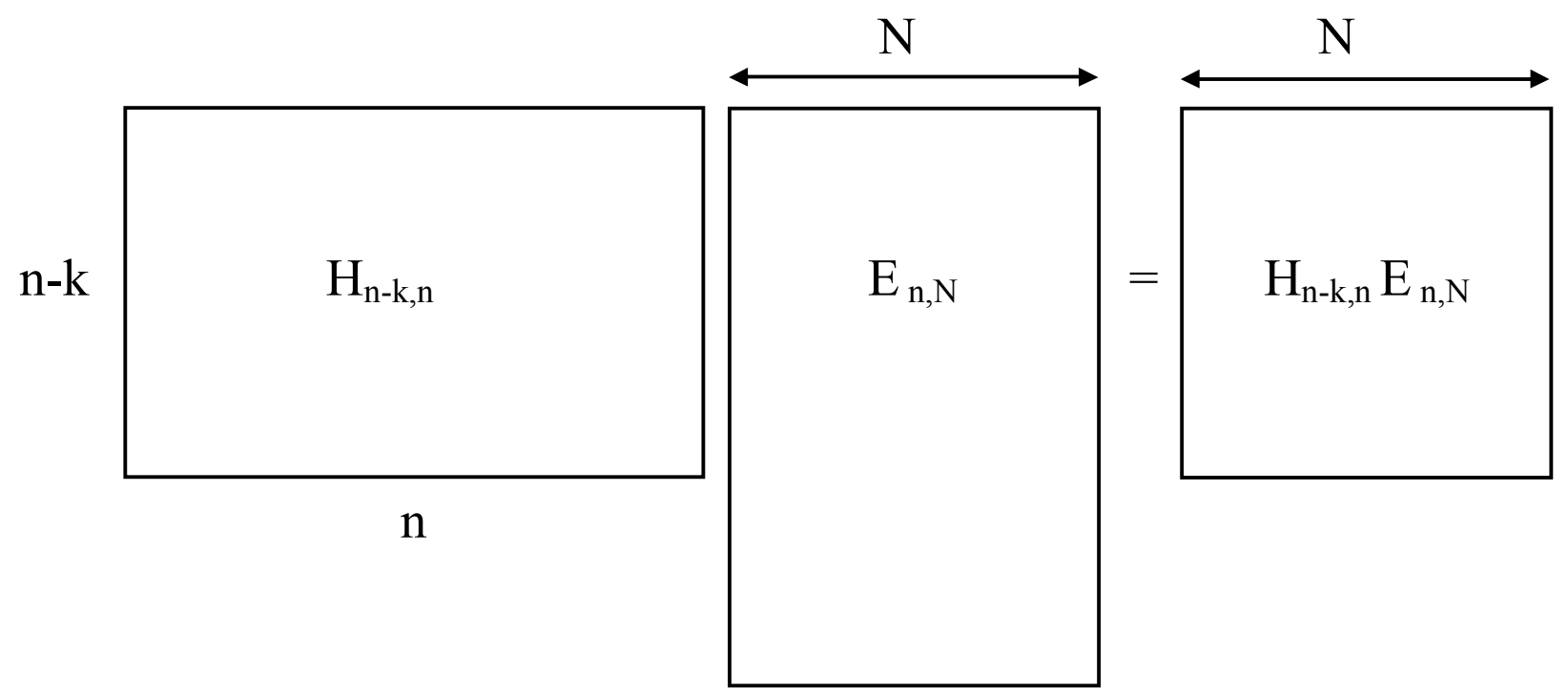

**Figure 1.7** Array syndrome forming

The reconstruction of the original information follows immediately. We use the following observations:

- we need k correct positions in a code word to be able to reconstruct the original code word;
- an RS code with minimum code word weight k+1 is generated by $H_{n-k,n}$;

- the product $H^0 \times E_{n,N} = 0_{n-k-t, N}$, can only be true if the non-zero symbols of $H^0$ correspond to error free rows of $E_{n,N}$.

Hence, $H^0$ indicate at least k+1 error free positions in $C_{n,N}$ and thus the original encoded information can be reconstructed. For RS codes, we need only one row from $H^0$. For general linear codes, we refer to [2].

**Remark** We want to state that the main ideas of the presented algorithm were presented in a publication of J. J. Metzner and E. J. Kapturowski [7].

### 1.5.2 The access algorithm

The Aloha system is a simple random access scheme, but it requires collision detection and feedback from the receiver or channel to the transmitters. We will now use the proposed decoding algorithm to construct a random access scheme that requires no feedback. No collision detection is required because our decoding algorithm performs this collision detection inherently by searching for error-free rows. Consequently, we do not need redundancy in the rows for error detection. In conclusion, collisions are not detected as such, but appear as errors. Therefore, normal erasure decoding is not possible.

| C1 | C2 | C3 | C4 | ••• | |
|---|---|---|---|---|---|
| $c_{11}$ | $c_{12}$ | | | | row 1 |
| $c_{21}$ | $c_{22}$ | | | | row 2 |
| $c_{31}$ | $c_{32}$ | | | | |
| $c_{41}$ | $c_{42}$ | | | | |
| $c_{51}$ | $c_{52}$ | | | | row 5 |

⟵ N ⟶

**Figure 1.8** Code array for a particular user

Assume that we have T independent users that have access to Z parallel and independent transmission channels. Each user encodes his data as described in Figure 1.8. Let C be an RS code with code symbols from $GF(2^m)$. In addition, every user possesses a signature sequence (s(1), s(2), ···, s(n)), of integers s(i), with $1 \le s(i) \le Z$ . They are chosen with

Probability(s(i) = j, $1 \le j \le Z$) = 1/Z.

The signature connected with each user is known at the receiver side. At time interval i the transmitter selects channel s(i) for the transmission of the ith code row. Furthermore, the users are assumed to be row-synchronized. If two or more users use the same channel at the same time, we assume that the received row is erroneous. Hence, we have exactly the same situation as is required for the proposed decoding algorithm.

| | | | | |
|---|---|---|---|---|
| $r_{11}$ | | | $r_{42}$ | |
| | $r_{22}$ | | | |
| | $r_{21}$ | | $r_{43}$ | $r_{53}$ |
| $r_{12}$ | | x | | x |
| $r_{14}$ | $r_{23}$ | | x | |
| $r_{13}$ | $r_{24}$ | $r_{34}$ | | $r_{52}$ |

**Figure 1.9** Randomized transmission (6 rows, 5 columns), where x indicates collision, T = 4 active users, Z = 6 parallel channels

| C1 | C2 | C3 | C4 | ••• | |
|---|---|---|---|---|---|
| $c_{11}$ | $c_{12}$ | $c_{13}$ | $c_{14}$ | | row 1 |
| $c_{21}$ | $c_{22}$ | $c_{23}$ | $c_{24}$ | | row 2 |
| x | x | x | $c_{34}$ | | |
| x | $c_{42}$ | $c_{43}$ | x | | |
| x | $c_{52}$ | $c_{53}$ | x | | row 5 |

**Figure 1.10** Received code array for a particular user

A user manages to decode the received array correctly if not more than (n-k-1) code rows are corrupted by the other users and if the error vectors that result from the collisions are linearly independent. Whether the second condition is fulfilled depends on how the channel reacts in the case of a collision. We consider the case where a collision as channel output is a random sequence. Hence, all error vectors are equally likely.

The probability that a row is received as a collision is given by

$$P_e = 1 - \left(\frac{Z-1}{Z}\right)^{T-1}.$$

We can expect correct decoding with an arbitrary small decoding error probability for large block length n and constant rate k/n if

$$nP_e < (d_{min} - 2) = (n - k - 1),$$

or

$$P_e < 1 - k/n.$$

The transmission of one code array per user requires a total of Z×n×N slots. The total transmitted un-coded information is T×k×N = T×n×N×r symbols. Substituting Pe = 1- k/n, we get the following expression for the transmission efficiency of this random access scheme

$$\eta = \frac{T}{Z}\left(\frac{Z-1}{Z}\right)^{T-1}.$$

For large Z we get

$$\eta \propto Ge^{-G}, \text{with } G = \frac{T}{Z}.$$

The efficiency of the proposed system is the same as the efficiency of the Aloha protocol. The maximum transmission efficiency is equal to $e^{-1}$.

# 1.6 Concluding remarks

The classical way of handling corrupted packets in a point-to-point connection in a network is that of error detection and retransmission. We call this method ARQ. For an error rate of p packets per time slot, the transmission efficiency approaches (1-p)R, where R is the loss in code efficiency due to the needed redundancy. We can also apply ARQ for missing packets with a more complicated protocol.

Packets can be transmitted in combination with an error correcting code. The RS encoding can be used with a redundancy equal to the amount of lost, erased or collided packets. We can use the erasure option in the decoding process when missing or erroneous packets can be identified. Furthermore, feedback is not necessary.

The Aloha protocol is a basic random access protocol in networking when feedback and collision detection is available. This probabilistic repeat request protocol has a maximum transmission efficiency of $e^{-1}$ packets per packet slot. A simple interleaved RS coding scheme can obtain the same performance without feedback. We describe a protocol that corrects packet errors without feedback and collision detection. The protocol was part of a PhD thesis by Christoph Haslach[1]. We acknowledge the fact that, in principle, the protocol was described by Metzner and Kapturowski [7].

The packet reconstruction methods are summarized in Figure 1.11.

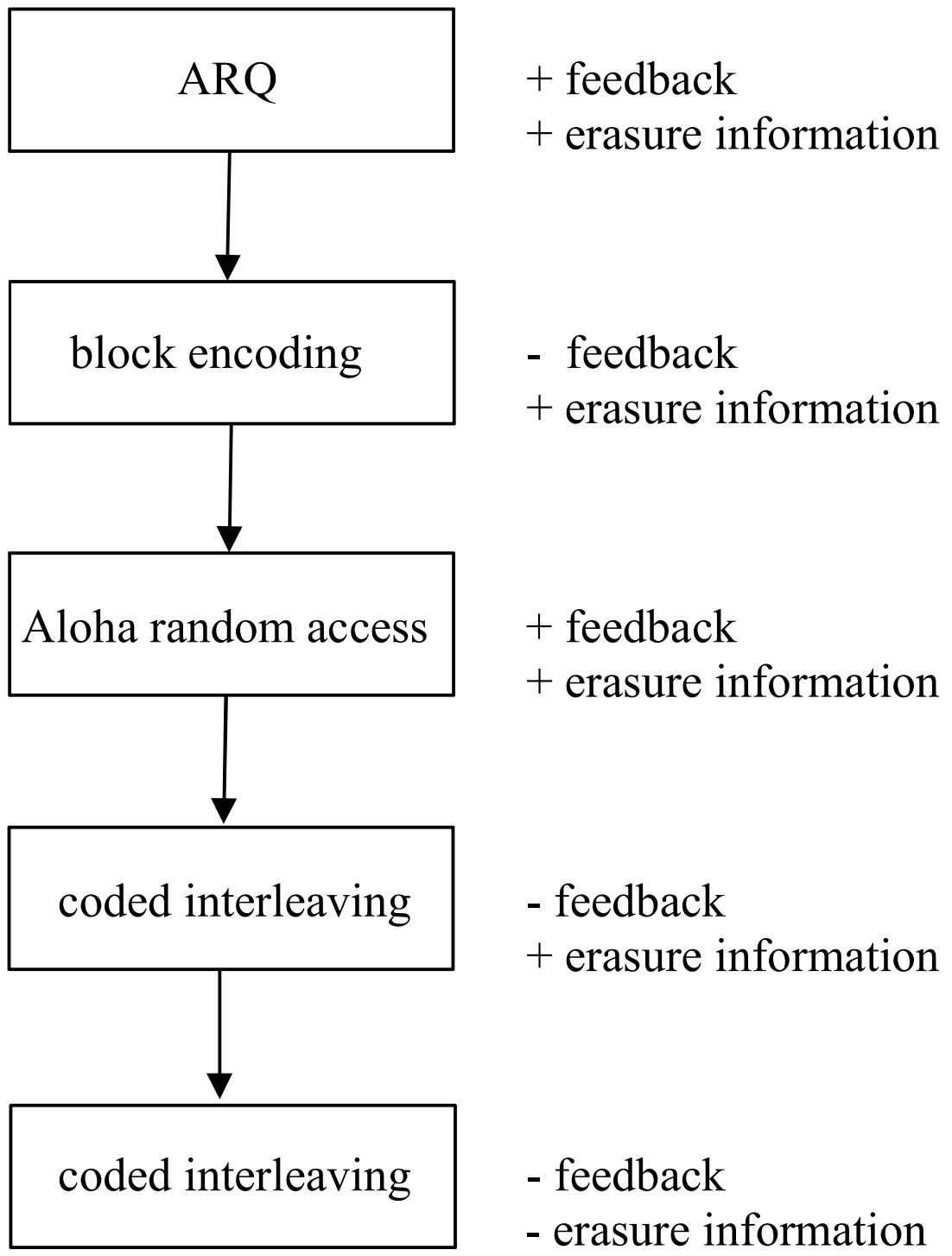

**Figure 1.11** Overview of described communication protocols

A problem, not considered here, is that of error detection. Packet transmission protocols very often depend on the availability of error detection via a Cyclic Redundancy Check (CRC). A report on this topic can be found in Kuznetsov et al.[2]

Packets usually contain a header (prefix) for synchronization purposes. Jim Massey[3] derived the conditions for the best prefix for continuous transmission of packets. However, when there is a time gap between packets, called burst mode, this prefix has to be changed. A new optimum design rule is therefore necessary[4]. This topic was part of the PhD thesis of Adriaan van Wijngaarden[5]. When markers are used in the header, these markers are not allowed as a pattern in the packet itself. This reduces the transmission efficiency and optimum markers have to be found. Results were obtained in a cooperation with Hiro Morita[6]. In [50] we consider a symbol oriented method for synchronization to be combined with RS codes.

1 Christoph Haslach, *Array-Codes auf der Basis Interleavter Block codes*, Universität Essen, 1999

2 A. Kuznetsov, F. Swarts, A.J. Han Vinck and H.C. Ferreira, "On the Undetected Error Probability of Linear Block Codes on Channels with Memory," IEEE Trans. on Information theory, Vol. 42, No. 1, pp. 303-309, Jan. 1996

3 J. L. Massey. "Optimum frame synchronization ". IEEE trans. comm., com-20(2):115-119, April 1972

4 A.J. Han Vinck and A.J. van Wijngaarden, “On Synchronization for Burst Transmission,” IEICE, pp. 2130-2135, Nov. 1997

5 Adriaan van Wijngaarden, *Frame Synchronization Techniques,* Universität Essen, 1998

6 Hiroyoshi Morita, Adriaan J. van Wijngaarden and A.J. Han Vinck, ”On the Construction of Maximal Prefix-Synchronized Codes,” IEEE Trans. on Information Theory, pp. 2158-2166, Nov. 1996

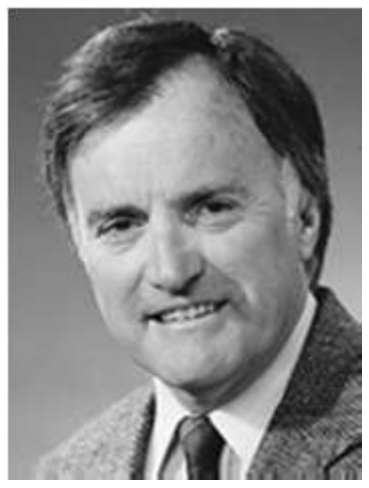
Professor James L. Massey

# Chapter 2

# Random Access

We discuss coding in a multi user concept where the set of active users is small compared to the total amount of potential users. The active transmitters use signatures to be transmitted over a common channel. We introduce the concept of q-ary superimposed codes, and we give a class of codes that can be used to uniquely determine the set of active users from the composite signature at the channel output. The so called T-user M-frequency noiseless multiple access channel without intensity information, as defined by Chang and Wolf [8], is an example of such a channel.

Information theory considers the capacity or maximum mutual information of multiple access channels (MAC) for users that are permanently sending information. In random access the users are not permanently active, and in addition do not cooperate with each other. As a reference we recommend [96].

## 2.1 Channel Capacity and Access Models

In information theory we model random access communications using simple channel models. In Figure 2.1 we give three models for “non-trivial” binary multiple access channels, the OR, the XOR and the θ channel, respectively. We assume that a user has as input the symbol 0 or 1, and the channel input is 0 in case a user is not active.

**OR channel**: $Y = X_1.OR.X_2.OR. \cdots .OR.X_T$

**XOR channel**: $Y = X_1.XOR.X_2.XOR. \cdots .XOR.X_T$

**θ channel**: $Y = 1$ if more than θ inputs are 1, otherwise $Y = 0$

**Figure 2.1** Three different binary multiple access models

**Uncoordinated access**
In uncoordinated multiple access one assumes that T individual users, out of a large community of U users, can access the communications channel without knowledge of the identity of the other users. Furthermore, the number of active users T is small compared with the total number of potential users in the network. The users are time aligned, i.e. have the same clock. The decoder for a particular user does not use the knowledge it might have about the other users and also considers the influence of the other users as interference or noise. The general model is given in Figure 2.2.

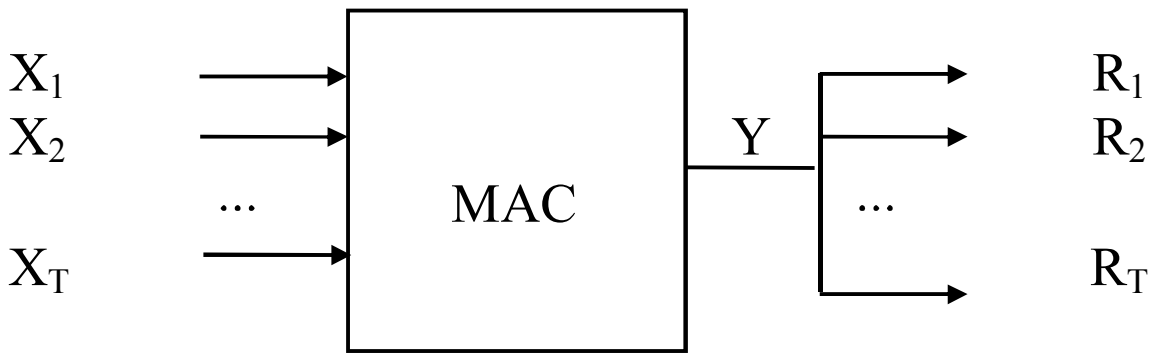

**Figure 2.2** The T-user random multiple access channel model

To calculate an achievable rate (transmission efficiency or throughput at a vanishing error probability) in a random access situation, we assume that every user has the same input probability distribution. Furthermore, we consider the problem as a "normal" channel coding problem and thus the achievable sum rate for the random access channel is equal to the sum of the individual achievable rates, i.e.

$$R_{\text{uncoordinated}} = T \max_{P(X_i)} I(X_i;Y). \qquad (2.1)$$

The maximum achievable rate is also called capacity.

**Coordinated access**

In the situation where the users are coordinated, i.e. know each other's identity, the decoders can use the information simultaneously transmitted by all users. The capacity region [22] is given by the closure of the convex hull of all rate pairs $R_1, R_2, \cdots, R_T$ satisfying

$$R_i \leq I(Y;X_i \mid \text{all } X_j, X_j \neq X_i), \qquad (2.2a)$$

$$R_1 + R_2 + \cdots + R_T \leq I(Y;X_1, X_2, \cdots, X_T), \qquad (2.2b)$$

for a probability product distribution $f(X_1, X_2, \cdots, X_T) = f(X_1)f(X_2)\cdots f(X_T)$.

In a fully coordinated situation, the number of available time slots or carrier frequencies (dimensions) determines the number of users and all dimensions are used by the participants. Examples of this method are Time Division Multiple Access (TDMA) and Frequency Division Multiple Access (FDMA). The transmission efficiency is optimal if all users do indeed use the communication channel. In all other situations, the channel contains idle or not-used dimensions. The principle drawback of dimension division is that each of the U potential users gets only 1/Uth of the dimensions. The channel contains idle or not-used dimensions when less than U users are active. Furthermore, the DMA schemes suffer from inflexibility inherent in pre-assigned dimensions. To accommodate more users than the available number of dimensions, a demand assignment protocol must be established.

## 2.1.1 The binary OR channel

The channel mostly investigated in literature is the binary OR access channel. The binary-input/binary-output relation for this channel is given by

$$Y = X_1.OR.X_2.OR.\cdots.OR.X_T. \qquad (2.3)$$

For the binary input T-user OR-channel, the output Y is 1 if at least one input is equal to 1. If all inputs are 0, the output is also 0. This model can be seen as a model for optical pulse modulation. The equivalent "AND" channel can be found in communication busses like the "Wired-AND" computer connection or the CAN (Controller Area Network) network. In Figure 2.3 we give the transmission channel as it is seen by a particular user for $p_0 = 1 - p_1 = p$.

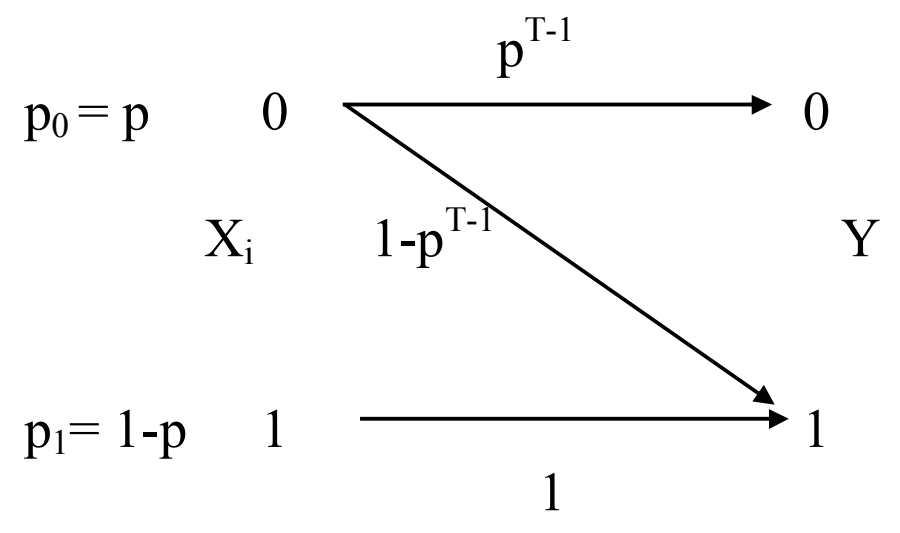

**Figure 2.3** The binary OR channel model for user i

Under the specified conditions, the achievable rate (2.1) for the uncoordinated OR channel becomes

$$R_{OR} = T \max_{P(X_i)} I(X_i; Y)$$
$$= T \max_{p} \{h(p^{T}) - ph(p^{T-1})\}, \tag{2.4}$$

where $h(p) = - p\log_2 p - (1- p)\log_2(1- p)$ is the binary entropy function. For $(1 - p) = \frac{1}{T} \times \ln 2$, we obtain that for large T the sum rate is lower bounded by

$$R_{OR,\, T\to\infty} \geq \ln 2 = 0.6931 \text{ bit/transmission.} \tag{2.5}$$

Numerically, one can show that (2.4) approaches (2.5) from above and we postulate that (2.5) is indeed the limiting capacity for the uncoordinated OR channel [9].

Evaluation of (2.2) in the coordinated case, gives a maximum sum rate of 1 bit/transmission. This rate can easily be obtained by using a time-sharing transmission strategy, where every user gets a unique assigned time slot. Hence, the rate loss due to the uncoordinated access is about 30%.

### 2.1.2 The M-ary OR channel

The M-ary OR channel has M-ary inputs and at each time instant the detector outputs which subset of symbols occurred as input to the channel at that time instant, but not how many of each occurred. For M symbols, we can thus have $2^M - 1$ different sets of outputs.

**Coordinated access**
Chang and Wolf [8] showed that capacity of the coordinated M-ary OR channel approaches (M-1) bits per transmission. This capacity can be obtained by a very simple time-sharing strategy. Assume that we divide the T active users into groups of size (M-1). Within one group the users are numbered from 1 up to M-1. User i, $1 \leq i \leq M-1$, uses as the transmitting symbols 0 and i. At the receiver, we are able to detect whether user i uses symbol 0 or i and thus, for M-1 users we transmit M-1 bits per transmission. Note that here we have a "*central*" symbol 0.

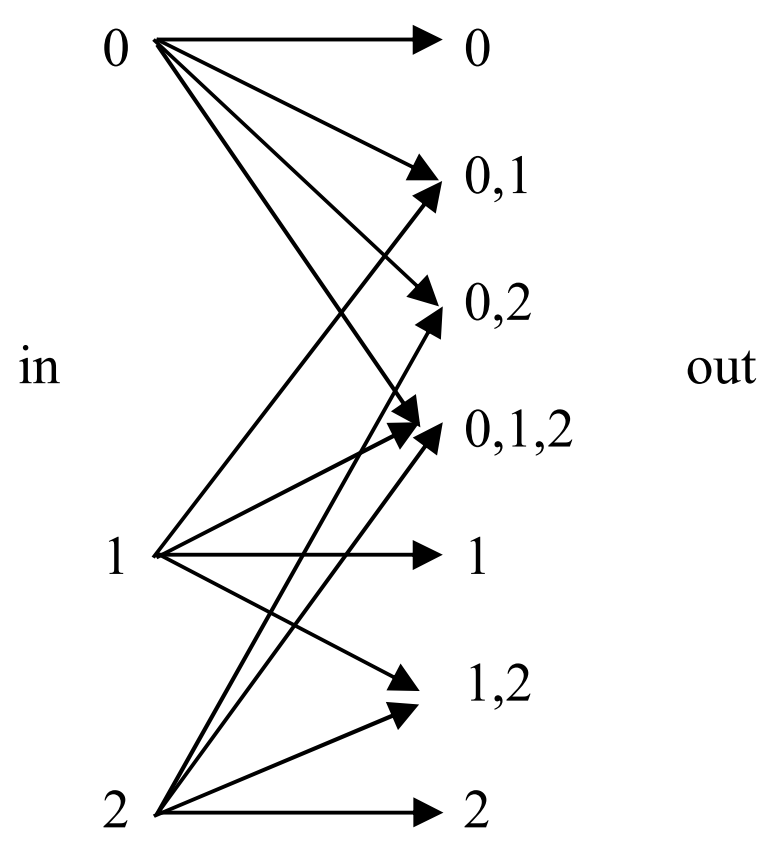

**Figure 2.4** The 3-ary OR channel model

**Example** For M = 3, T/2 users use the pair (0 or 1) as channel input, the other T/2 users use as input (0 or 2). The output of the channel is thus (0), (0,1), (0,2), or (1,2). Note that in this coordinated situation the input of the channel is uniquely decodable.

**Uncoordinated access**

Suppose that for the uncoordinated M-ary OR channel, some coordination between the T users exists. Namely, a fraction T/(M-1) of users uses the input symbol pair (0,i), $1 \leq i \leq M - 1$. According to the previous section, the transmission per group can be considered as transmission over a Z-channel, where one output contains the symbol i and the other outputs contain only symbols not equal to i. The achievable sum rate for this channel is ln2 bits per transmission. Hence, for M-1 groups we obtain a normalized sum rate of (1 - 1/M) ln2 bits per dimension.

**Example** For the example M = 3, as given in Figure 2.4, we have 7 subsets at the output: {0}, {1}, {2}, {0,1}, {0,2}, {0,1,2}, and {1,2}. The input probability distribution is chosen as: $P_\infty(0) = 1 - (2/T)\ln 2$, and $P_\infty(1) = P_\infty(2) = (\ln 2)/T$. The limiting input distribution shows that for one symbol the probability approaches 1. The asymptotic rate R(3,T) approaches 2ln2 bits per channel use, or a total of (2/3)ln2 bits per dimension.

For M > 3, we postulate a probability distribution. For this distribution we calculate the maximum mutual information which is, from an information theoretical point of view, equivalent to the achievable rate or ε-error rate. This result is thus a lower bound for the capacity, since channel capacity is the maximum mutual information over all possible input probability distributions. We give an achievable rate for the uncoordinated M-ary OR channel in Theorem 2.1.1. The achievable rate reduces from (M-1) bits per channel use in the fully coordinated multiple access situation, see Chang and Wolf [8], to (M-1)ln2 bits per channel use if we assume no coordination between users or one-to-one communication.

**Theorem 2.1.1** For the input probability distribution that puts a probability 1-((M-1)ln2)/T on one symbol and a probability (ln2)/T on each of the remaining (M-1) symbols, the total achievable rate R for T users becomes (M-1)ln2 bits per transmission or (1-1/M)ln2 bits per dimension for $T \to \infty$.

**Example** In Figure 2.5a, for M = 2, the output is ternary: 0, 1, or (0,1). The transitions are a result of the interference of the other users. Optimizing with respect to p and letting T go to infinity, the achievable rate for this channel becomes $R(2,T) \to \ln 2$ bits per channel use, or a total of 0.5ln2 bits per dimension. For each individual user, the 2-symbol input multiple access channel is, from a rate point of view, equivalent to the binary input-binary output Z-channel. Although the 2-symbol input multiple access channel has a ternary output, the achievable sum rate is the same as if we make a hard decision in case of an ambiguous reception of two symbols. The limiting input distribution for both examples puts all probability mass on one symbol. For M = 2, $P_\infty(0) = 1 - (\ln 2)/T$, $P_\infty(1) = (\ln 2)/T$, respectively.

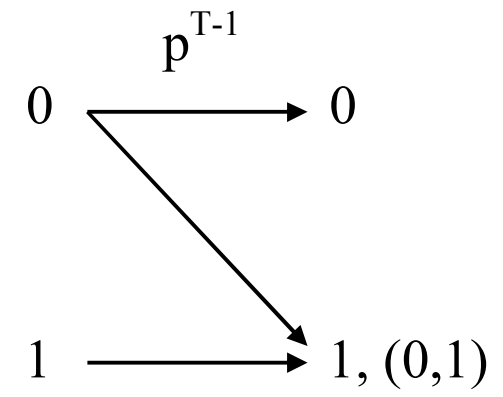

**Z-Channel**

**Figure 2.5a** Binary Z access channel model

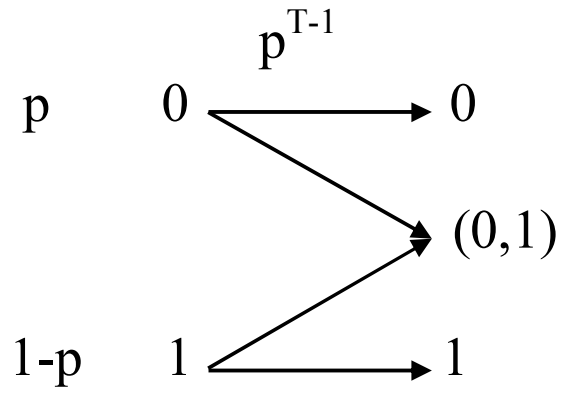

**3-ary output**

**Figure 2.5b** Binary input ternary output access model

**Remark** The 3-ary output channel as given in Figure 2.5b looks like an erasure channel, where the symbol pair (0,1) can be considered as the erasure. One normally uses these channels with symmetric input probability distribution to obtain channel capacity. However, in this case, the rate of the channel would give a value of $C = (1/2)^{T-1}$ for p = ½, which goes to zero for large T. The reason for this discrepancy is the fact that the transition probability is a function of the input probability. In information theory, capacity is a convex cap function of the input probability distribution for constant transition probabilities. In our situation, this is no longer the case and we have to search for the global optimum. The problem is to express the transition probabilities as a function of the input probabilities.

The M-ary Erasure-channel has as inputs $X_i \in \{0, 1, ..., M-1\}$, and as outputs $Y \in \{0, 1, ..., M-1, E\}$, where E occurs if two or more different symbols occur at the output. We conjecture that the M-ary Erasure-channel as given in Figure 2.6 has a capacity of ln2 bits per transmission for $T \to \infty$. This result was checked for M = 2, 3 and 4.

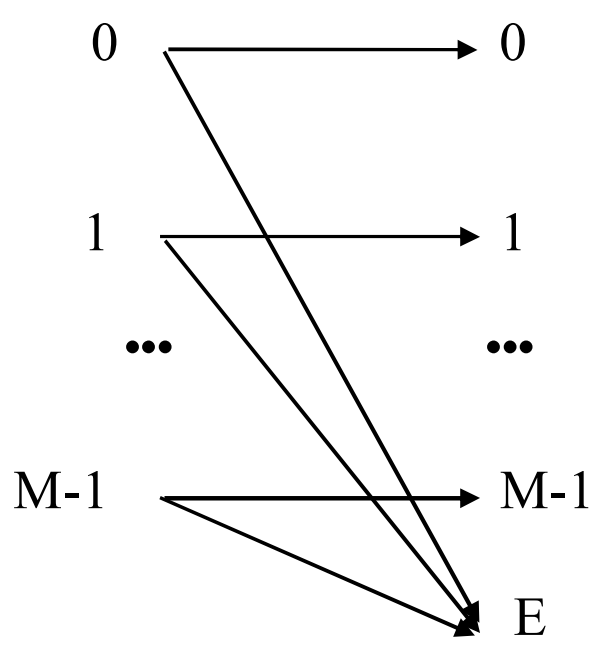

**Figure 2.6** M-ary Erasure-channel

**Example** An example of a practical M-ary OR channel is that generated by Pulse Position Modulation (PPM). Pulse position modulation divides the transmission time into time slots of length $\tau$ and further sub-divides every time slot into M sub-slots of time $\tau/M$. An M-ary symbol is modulated as a pulse in one of the M available sub-slots.

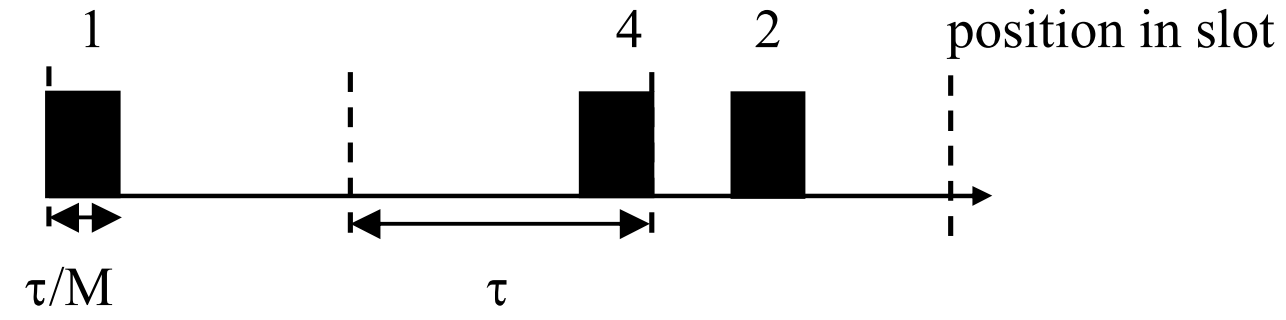

**Figure 2.7** Pulse Position Modulation for one user, M = 4

If we consider the M positions in a slot as M independent parallel channels, the <u>uncoordinated</u> sum rate for T users is $R_{unc}(M,T)$ = ln2 bits per position. For M positions we thus have a sum rate of Mln2 bits per slot. Note that the number of possible different combinations in a slot is equal to $2^M$.

If we consider a time slot as a channel where we put a pulse in a particular position, we have a total of M different inputs and the channel is equivalent to the M-ary OR channel. The maximum number of different combinations in a slot is now equal to $2^M$-1, since users have to give a pulse as input. However, instead of using one out of $2^M$ combinations of M symbols, the same rate can be achieved by using only a single symbol from M, which is the advantage of the PPM modulation. The total achievable rate R for T users becomes (M-1)ln2 bits per transmission, or (1-1/M)ln2 bits per dimension for $T \to \infty$. A vanishing loss compared with the unrestricted case for large values of M. To obtain this rate, we generate a pulse with probability $(1 - \frac{M-1}{T}\ln 2)$ for one specific position in a time slot, whereas all other positions have a probability (ln2)/T of having a pulse.

For the transmission scheme of Figure 2.7, we may consider transmitting 1 pulse of duration τ/M in τ seconds using a theoretical "bandwidth" of 2B = M/τ Hz. For M-ary Frequency Shift Keying (MFSK) we transmit one frequency out of M possible frequencies with a duration of τ seconds, using a theoretical bandwidth of 2B = M/τ Hz. Hence, from a "bandwidth" point of view PPM and MFSK are equivalent, and the rate results apply to both systems. Instead of using time, we can thus also use the frequency domain.

## 2.2 Signaling with Random Signatures

Access schemes for PPM using M-ary signatures can be found in [10,11]. In Figure 2.8 we give the M-ary scheme as described by Cohen, Heller and Viterbi [10], where an active user is allowed to transmit an M-ary signature in L slots. Users are assumed to be frame aligned, i.e. every user knows the starting time for the L slots. In general, the receivers look for a particular signature connected with a message that belongs to their respective transmitter. A decoder makes an error if it detects a valid signature that is the result of the "OR" of the other users. Codes are to be designed to minimize the decoding error probability for a given number of active users.

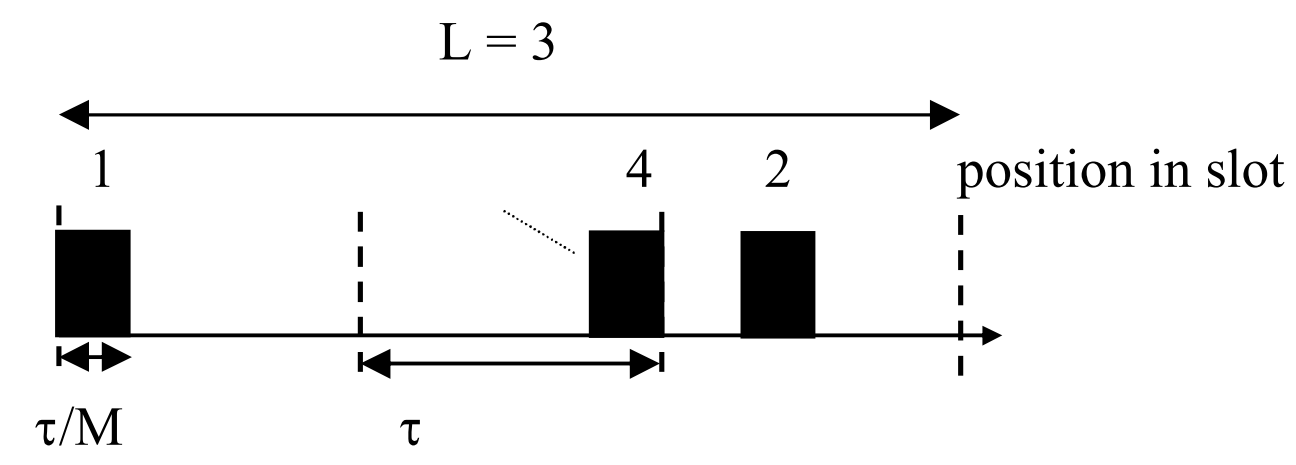

**Figure 2.8** Access model with signature (1,4,2)

### 2.2.1 Signatures with a symmetric input distribution

We first describe a transmission scheme using random signatures. We assume that T users are connected with their T respective receivers. The starting and end point of transmission is known to all transmitters and receivers, and they are time aligned. The parameters of the transmission scheme are:

- every user has its own set of M-ary signatures of length L. The cardinality of the set is M;
- every symbol in a signature is chosen with probability 1/M and corresponds with a pulse in the respective time sub-slot position.

The decoder for a particular user has to detect a valid signature that belongs to the user. Suppose that a user transmits a particular signature. The decoding error probability $P_e$ for this particular user is upper bounded by the probability that the T-1 other users together generate one of the M-1 other valid signatures in L slots, i.e.

$$P_e \leq (M-1)\times(1-(1-\frac{1}{M})^{T-1})^{L}. \tag{2.6}$$

For $M = 2^{(1-\varepsilon)L}$ and $T = M\cdot\ln 2$, the error probability

$$P_e \to 2^{-\varepsilon L}, \varepsilon > 0,$$

and the normalized rate

$$\eta = T\frac{\log_2 M}{LM} \to (1-\varepsilon)\ln 2 \qquad \text{bit/transmission.}$$

In order to decrease the symbol error probability, we have to increase L and thus, also M.

One sees from (2.6), that the average error probability for T users simultaneously, $TP_e$, does not approach zero for increasing L. If we change the variables such that $MT = 2^{(1-\varepsilon)L}$ and $T = M\cdot\ln 2$, then the average error probability for T users is given by

$$TP_e \rightarrow 2^{-\varepsilon L}, \varepsilon > 0, \tag{2.7}$$

but the normalized rate

$$\eta = T\frac{\log_2 M}{LM} \rightarrow \frac{1-\varepsilon}{2}\ln 2 . \qquad \text{bit/transmission.}$$

This is roughly 50% of the maximum achievable rate.

### 2.2.2 Signatures with an asymmetric input distribution

We can improve the rate by using a different probability assignment and signature scheme. For this, the signatures for a particular user consist of (M-1) signatures with symbols not equal to 1. We reserve the all-1 signature for message m = 1. In Figure 2.9a, we transmit the position 1 in three subsequent time intervals. In Figure 2.9b we transmit a "*spread*" sequence of three positions.

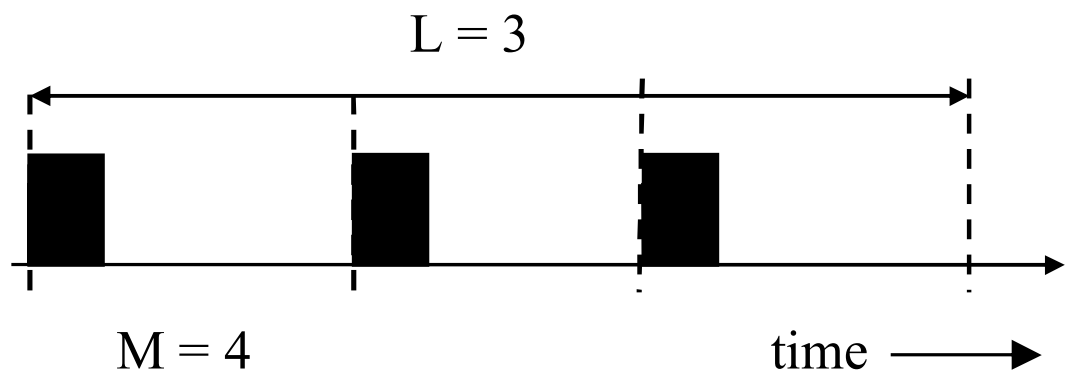

**Figure 2.9a** Cohen, Heller and Viterbi model, the message is m = 1

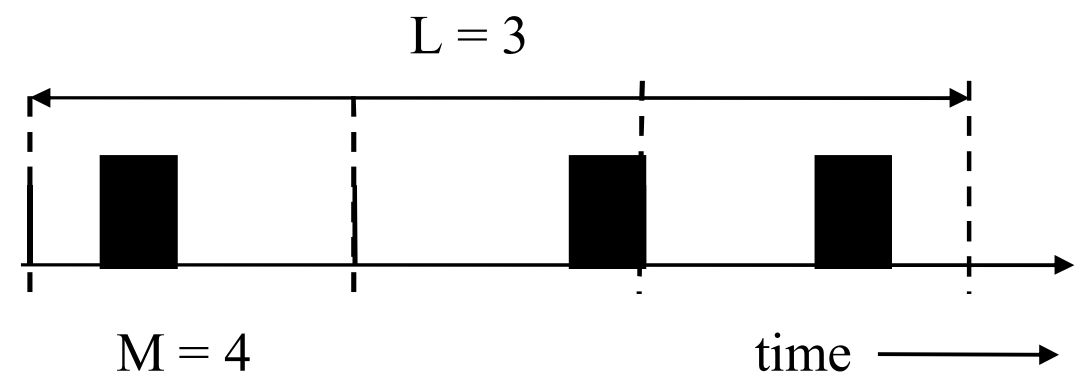

**Figure 2.9b** Signature (2,4,3)

For a message probability

$\Pr(m = 1) = 1 - \frac{M-1}{T}\ln 2 = 1 - p,$

$\Pr(m = k) = \frac{\ln 2}{T}, 2 \leq k \leq M,$

the amount of information that is transmitted per user is $R = h(p) + p\log_2(M-1)$, where h(p) is the binary entropy function. The decoder estimates $m' = k$ if the valid signature for message k is detected in L slots. The decoder is assumed to always decode message 1, unless another valid signature is detected. The decoding error probability $P_e$ for a particular user is then upper bounded by the probability that the other users together generate one of the other valid signatures in L slots, i.e.

$$P_e \leq (M-1)\times(1-(1-\frac{1}{M-1})^{pT-1})^L. \quad (2.8)$$

We can approximate (2.8) as

$$P_e \approx M\times(1-e^{-pT/M})^L. \quad (2.9)$$

For $p = \frac{M-1}{T}\ln 2$, $M = 2^{\varepsilon L}$ and $T = 2^{(1-2\varepsilon)L}\ln 2$, the error probability (2.8) vanishes as

$$P_e \rightarrow 2^{-(1-\varepsilon)L}, \varepsilon > 0. \quad (2.10)$$

From (2.10) it also follows that the average error probability for T users simultaneously goes to zero for increasing L, since

$$TP_e \rightarrow 2^{-\varepsilon L}\ln 2 \rightarrow 0, \varepsilon > 0. \quad (2.11)$$

At the same time the normalized rate

$$\eta = T\times\frac{h(p) + p\log_2(M-1)}{LM} \rightarrow (1 - 2\varepsilon)\ln 2 \text{ bit/transmission} \quad (2.12)$$

The signature scheme with an asymmetric distribution clearly performs better than the signature scheme with a symmetric distribution. As before, one position is selected as a common position for all users.

**Remark** The asymmetric input probability distribution scheme has $M = 2^{\varepsilon L}$. We can keep the value of M constant by choosing $\varepsilon = O(1/L)$ leading to a vanishing probability of error and also a constant M.

## 2.3 Signaling with Constructed Signatures

### 2.3.1 The Titlebaum construction

The first signature construction we discuss is the Titlebaum [13] construction. Suppose that we extend the first two rows of an RS encoding matrix as

$$G_{2,M} = \begin{bmatrix} 1 & 1 & 1 & 1 & \dots & 1 \\ 0 & 1 & \alpha & \alpha^2 & \dots & \alpha^{M-2} \end{bmatrix} . \qquad (2.13)$$

It is easy to see that the minimum distance of the code is M-1, where M is the length of the extended code. The number of different non-zero code words formed with the second row only is M-1. These code words are used as signatures for M-1 different users. We assume that the users are fully synchronized. In addition, every user may send one of M messages by adding a multiple of the first row to his signature, and create a modulated signature. At the receiver side, we assume error free detection of the transmitted symbols. For the encoding matrix as given in (2.13), we have the following properties:

- the Hamming distance between the modulated signatures for a particular user is M;
- the Hamming distance between signatures for two different users is at least M-1.

Since the $d_{min} \geq M-1$, the maximum number of agreements between two signatures from different users is equal to 1. Consequently, M-2 or less active users can never produce a valid signature for another particular user, since at least one symbol is left for the unique identification.

**Example** For M = 5, we can form the 4 signatures: $C_1$ = (0,1,2,3,4), $C_2$ = (0,2,4,1,3), $C_3$ = (0,3,1,4,2) and $C_4$ = (0,4,3,2,1). These 4 signatures have a minimum distance of 4 and are used as the signatures for 4 different users. From these signatures we can easily form the set of 5 possible transmitted modulated signatures for a particular user. For user 1, with signature (0,1,2,3,4) these are

$$C = \{(0,1,2,3,4), (1,2,3,4,0), (2,3,4,0,1), (3,4,0,1,2), (4,0,1,2,3)\}.$$

**Remark** If we shorten the signatures from length M to length L, the minimum distance within a set of signatures is reduced from M to L and the minimum distance between different sets of signatures is reduced from M-1 to L-1. Shortening can be used to influence the efficiency of the modulation scheme. Therefore, we can give the normalized efficiency or rate of a transmission scheme using the Titlebaum construction for T active users as

$$\eta = T\frac{\log_2 M}{L \cdot M}.$$

The receiver consists of a bank of M energy detectors matched to the M chips of the M-ary signaling scheme. The M energy detectors provide M·L outputs per received word. The decoder for a particular user compares the demodulator output with its M possible transmitted signatures. It outputs the signature for which the maximum number of agreements with the symbols at the demodulator output occurs (minimum distance decoding).

### 2.3.2 Decoding error probability for the Titlebaum construction

We consider the situation when only user interference occurs. Furthermore, we assume that the demodulator for a particular user always outputs the transmitted symbols. These assumptions make the analysis easier. Due to the structure of the demodulation process, decoding errors may occur if and only if the other users together cause the occurrence of a modulated signature for a particular user.

Since the minimum distance between signatures is L-1, a signature for a particular user interferes in exactly one position with a signature from another user. This enables us to calculate the error probability. For T active users, an error may occur if at least L out of T-1 users together generate an additional valid signature for the remaining active user. Assuming random selection of the messages, the error probability is given by

$$P_e \le (M-1)[(\tfrac{1}{M})^L \cdot \binom{T-1}{L} \cdot L!]. \qquad (2.14)$$

This follows from the fact that, for a particular set of L users, L! permutations can give a particular signature with probability $(1/M)^L$. Furthermore, we can choose L out of T active users that will produce the particular signature.

In Figure 2.10, we plot (2.14) for several values of M, T and L.

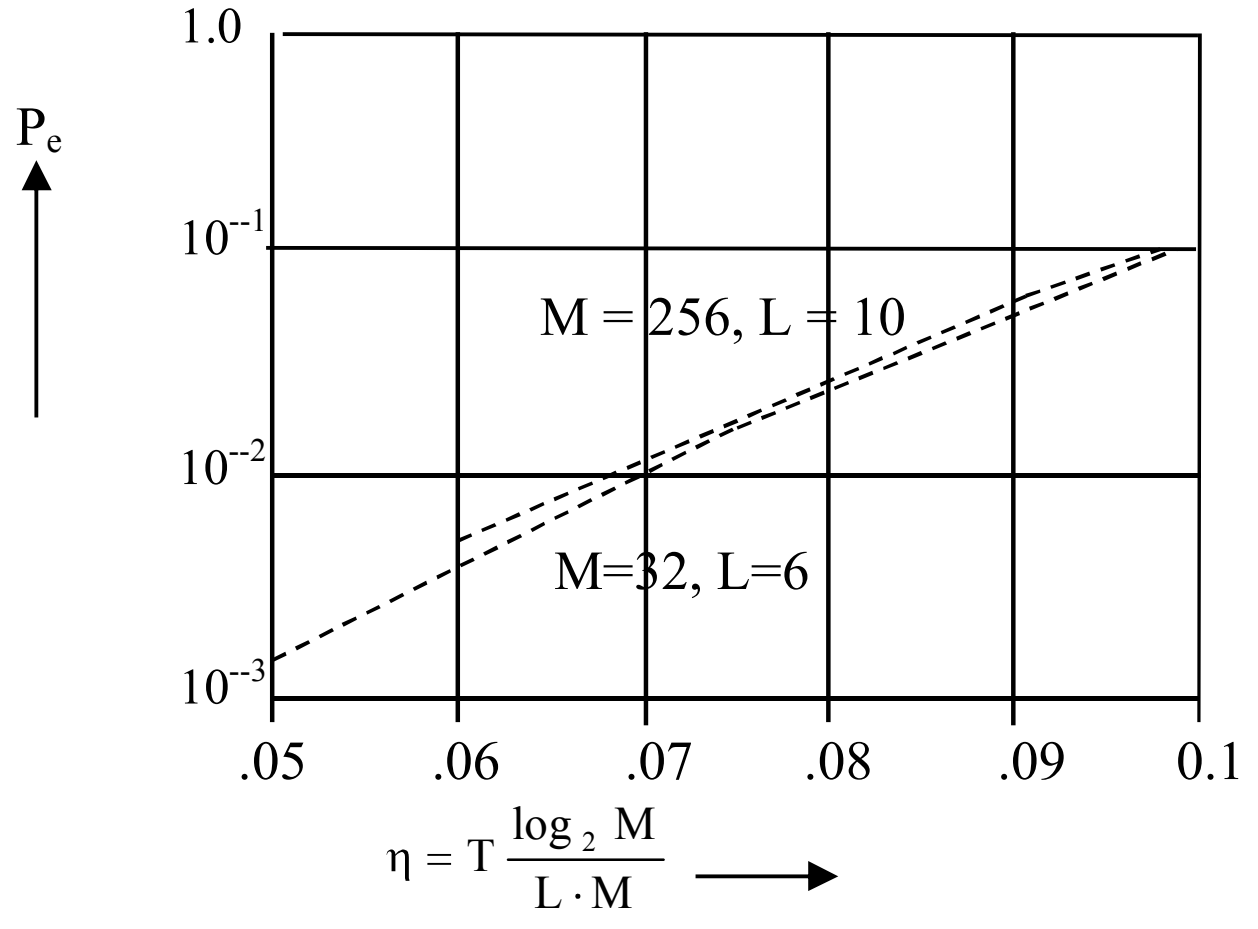

**Figure 2.10** $P_e$ as a function of the efficiency $\eta$

Note that for large T and small L, (2.14) can be approximated as

$$P_e \approx M(\tfrac{T}{M})^L. \qquad (2.15)$$

The asymptotic behavior of (2.15) is the same as that of (2.6) where we use random signatures. The above transmission scheme assumes that the number of active users T < M. Extension to T > M is straightforward.

### 2.3.3 Superimposed Codes

Kautz-Singleton (KS) [11] introduced the class of binary superimposed codes (SIC) to construct signatures. A binary superimposed code SIC(U, n, 2, T) consists of U binary code words of length n, with the property that the Boolean sum (OR) of any T-subset does not cover any word not in the T-subset.

We extend the definition of binary SICs to the situation where code words have q-ary symbols and the channel output is a symbol which identifies which subset of symbols occurred as input to the channel (no how many of each). We first have to give some additional definitions.

**Definition** The union of a set {a,b, ⋯,c} denoted as $\cup$(a,b, ⋯,c) is defined as the set of different symbols of the argument (a,b, ⋯,c).

**Example** $\cup(1,2,3,3,2) = \{1,2,3\}$.

**Example** $\cup(0,1,1,0,0) = \{0,1\}$.

Let $V \subset \{ 0, 1, \cdots, (q-1) \}^n$, $|V| = U$, be a code book with U q-ary code words.

**Definition** The union of T code words in V, $\cup^n(r^n, s^n, \cdots, t^n)$ is defined as the component wise $\cup$ of the symbols.

**Example** $\cup^4(1223,1321,1111) = (\{1\},\{1,2,3\},\{1,2\},\{1,3\})$.

**Definition** The union $\cup^n$ of T code words $\cup^n(r^n, s^n, \cdots, t^n)$ covers a code word $v^n$ if

$$\cup^n(r^n, s^n, \cdots, t^n) = \cup^n(r^n, s^n, \cdots, t^n, v^n).$$

**Example** The word ({1},{1,2,0},{1,0}) covers the code word $v^n = (1,0,0)$.

**Definition** A q-ary superimposed code with parameters U, n, q, T contains U q-ary code words of length n with the property that the $\cup^n$ of any set S containing T or less code words does not cover any code word not in S.

The following proposition then follows.

**Proposition**

$$(2^q - 1)^n \geq \sum_{i=1}^{T} \binom{U}{i}. \tag{2.16}$$

Another lower bound on n can be obtained as follows. We first consider (for simplicity) the situation where n = T·s.

**n = T·s**

We partition the code words into T parts of length s. Every code word must have a part different from the corresponding part of all other code words. This part contains at least one symbol, called special element, that can be used to distinguish a code word from the union of any set S of T or less code words. If the number of special elements in a particular part is exactly $q^s$, we have $U = q^s$. We must therefore assume that every part contains at most $q^s$-1 special elements. The maximum number of different parts is an upper bound for the number of code words in the code, and thus $U \leq T(q^s-1)$. Since n = Ts we obtain

$$n \geq \frac{T}{\log_2 q} \log_2 \frac{U}{T}, \tag{2.17}$$

which improves on (2.16).

We can also give a lower bound for n using a probabilistic approach. Consider a U · n matrix containing q-ary elements that are uniformly and independently distributed. An error occurs if a selection of T or less rows covers a row not in the selection. For a q-ary SIC the probability that this happens can be bounded by

$$P_e(\text{q-ary SIC}) \leq (U - T)\binom{U}{T}\left[1 - \left(\frac{q-1}{q}\right)^T\right]^n. \tag{2.18}$$

For $T = q\ln 2$, $n(1+\varepsilon) > T\log_2 U$, the probability $P_e$(q-ary SIC) → 0 for increasing n.

**T > n**

For T > n, every code word must have a special element in at least one of its positions. Again, if one of the positions for all code words contains exactly q special elements, then U = q. Therefore, every position must contain no more than q-1 special elements. Hence, we obtain as a lower bound for n,

$$U \le n(q-1).$$

**Example** The following example gives a q-ary SIC(U = 15, n = 5, q = 4, T), where $n \le T < n(q-1)$ and U = n(q-1). The SIC(15, 5, 4, T) contains the following code words

$$\mathrm{SIC}(15, 5, 4, T) = \begin{Bmatrix} 10000 & 01000 & 00100 & 00010 & 00001 \\ 20000 & 02000 & 00200 & 00020 & 00002 \\ 30000 & 03000 & 00300 & 00030 & 00003 \end{Bmatrix} .$$

The example can easily be generalized to other values of n and q.

We now consider the construction of q-ary SICs using RS codes due to Kautz-Singleton [11]. The parameters for these codes are:

- RS "outer" codes over GF(q) with length n = q-1, dimension k, minimum distance $d_{min}$ = n-k+1;
- a mapping of every RS code symbol to a binary word of length q and weight 1.

Note that the number of disagreements between two code words is larger or equal to $d_{min}$ and thus, the number of agreements is less than or equal to n - $d_{min}$ = k – 1. As a consequence, the Boolean sum of T code words agrees with an arbitrary code word in at most T(k-1) positions. If T(k-1) < n, we have the following theorem.

**Theorem 2.3.1** There exists a

$$\mathrm{SIC}\,(U = q^k, n_{KS} = q(q-1), q, T < \frac{n}{k-1}).$$

We compare the KS codes in rate with TDMA for T active users. We give every user a set of $q^i$ code words out of $q^k$. The number of potential users is thus $q^{k-i}$. If a SIC exists, the normalized rate is

$$\eta_{KS} = \frac{T\log_2 q^i}{q(q-1)}.$$

For TDMA every user gets a slot where a symbol can be sent. The rate is

$$\eta_{TDMA} = \frac{T}{q^{k-i}}.$$

We can see that, for $k > i + 1$, the KS codes are superior in rate.

**Example** Let $k = 4$, $i = 2$, $q = 2^5$. For these parameters:

- $T = 7 < n/(k-1)$;
- The number of possible users is $q^4/q^2 = 2^{10}$;
- The rate $\eta_{KS} \approx 0{,}08$; $\eta_{TDMA} \approx 0.007$.

**Theorem 2.3.2** The extended RS code with parameters ($n = q^s$, $k = q^{s-1}$, $d_{min} = n–k+1$) code over $GF(q^s)$, where q is any prime and $T \le q$, defines a superimposed code $SIC(q^{sk}, q^s, q^s, T)$.

**Proof** It is easy to check that for $T \le q$, the condition of Theorem 2.3.1 is fulfilled.

**Remark** The condition on T in Theorem 2.3.1 is sufficient but not necessary for the existence of a q-ary SIC. This follows from the next example.

**Example** Let $q = 3$, $T = 2$ and $n = 4$. The following code has minimum distance 2. The corresponding q-ary SIC

$$SIC(12, 4, 3, 2) = \begin{bmatrix} 0\ 0\ 0\ 0 & 0\ 1\ 1\ 0 & 0\ 2\ 2\ 1 & 1\ 1\ 2\ 2 \\ 1\ 2\ 0\ 1 & 1\ 0\ 1\ 0 & 2\ 2\ 1\ 1 & 2\ 0\ 2\ 1 \\ 2\ 1\ 0\ 1 & 2\ 2\ 2\ 0 & 0\ 0\ 1\ 2 & 2\ 2\ 0\ 2 \end{bmatrix}$$

does not satisfy the condition on T in Theorem 2.3.1.

The following theorems construct longer codes from shorter ones.

**Theorem 2.3.3** If there exists a SIC($U_0$, $n_0$, $q_0$, T) and a SIC($U_1$, $n_1$, $q_1$, T), where $q_1 \le U_0$, then there also exists a SIC($U_1$, $n_0$ $n_1$, $q_0$, T).

**Proof** Assign to each symbol {0, 1, ···, $q_1$-1} a different code word from SIC($U_0$, $n_0$, $q_0$, T). Replace the symbols in SIC($U_1$, $n_1$, $q_1$, T) by these code words. Since we replaced all $q_1$-ary elements by different code words from SIC($U_0$, $n_0$, $q_0$, T) we thus obtain a SIC($U_1$, $n_0$ $n_1$, $q_0$, T), see also Figure 2.11.

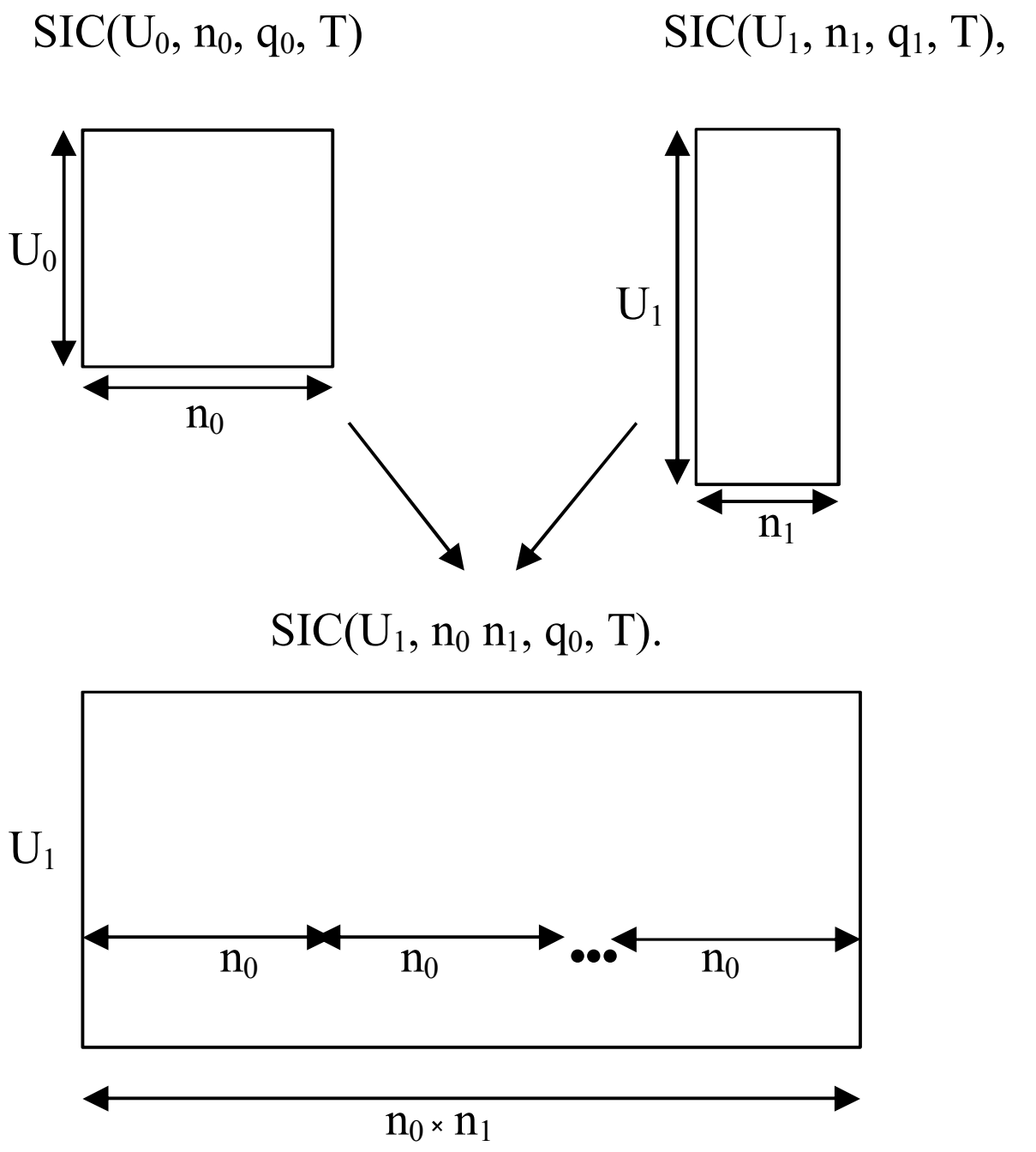

**Figure 2.11** Construction of a SIC out of two others

**Corollary** If there exists a SIC($U_0$, $n_0$, 2, T) and a SIC($U_1$, $n_1$, $q_1$, T), where $q_1 \le U_0$, then there also exists a SIC($U_1$, $n_0$ $n_1$, 2, T).

The codes constructed in Theorem 2.3.3 can be seen as a generalization of the Kautz-Singleton codes [11].

**Example**

1. Suppose that we have the following starting SIC(4, 3, 2, T = 2) with the 4 code words

$$( 1\ 0\ 0, 0\ 1\ 0, 0\ 0\ 1, 1\ 1\ 1 ) \equiv ( 0, 1, a, b ).$$

2. The second code to be used is an RS code over GF($2^2$) with parameters (n, k, $d_{min}$) = (3, 2, 2). For this code, T < n/(k-1). Hence, we can construct a SIC(16, 3, $2^2$, 2). We replace every element with a code word from the first code and obtain a SIC(16, 9, 2, 2) with 16 code words, i.e.

$$\text{SIC}(16, 9, 2, 2) = \left\{ \begin{array}{l} 000,\ 01a,\ 0ab,\ 0b1,\ 1a0,\ ab0,\ b10,\ a01 \\ b0a,\ 10b,\ 1ba,\ a1b,\ ba1,\ 111,\ aaa,\ bbb \end{array} \right\}.$$

3. As a third code we construct an RS code over GF($2^4$) with parameters (n = 15, k = 8, $d_{min}$ = 8), where T < n/(k-1). From this code we obtain a SIC($2^{32}$, 15, $2^4$, T = 2). For the combination with the second code we obtain a SIC($2^{32}$, 9×15 = 135, 2, 2).

**Example** Let q = 4 and T = 3.

1. The first code we use is an RS code over GF($2^2$) with parameters (n = 4, k = 2, $d_{min}$ = 3). Since T < n/(k-1), we obtain a SIC($2^4$, 4, $2^2$, 3).
2. The second code we choose is a shortened RS code over GF($2^4$) with parameters (n = 13, k = 5, $d_{min}$ = 9). Since T < 13/4, we obtain a SIC($2^{20}$, 13, 16, 3).
3. Combining both codes, we obtain SIC(U = $2^{20}$, 4×13 = 42, q = 4, T = 3).

These examples show that we can construct a series of codes, see also [12].

## 2.4 Signatures for the XOR Channel

We consider the situation where T active users from a population of size U transmit signatures over the XOR multiple access channel. The symbols in a signature are elements from GF($q = 2^m$). We furthermore assume that the users are block synchronized.

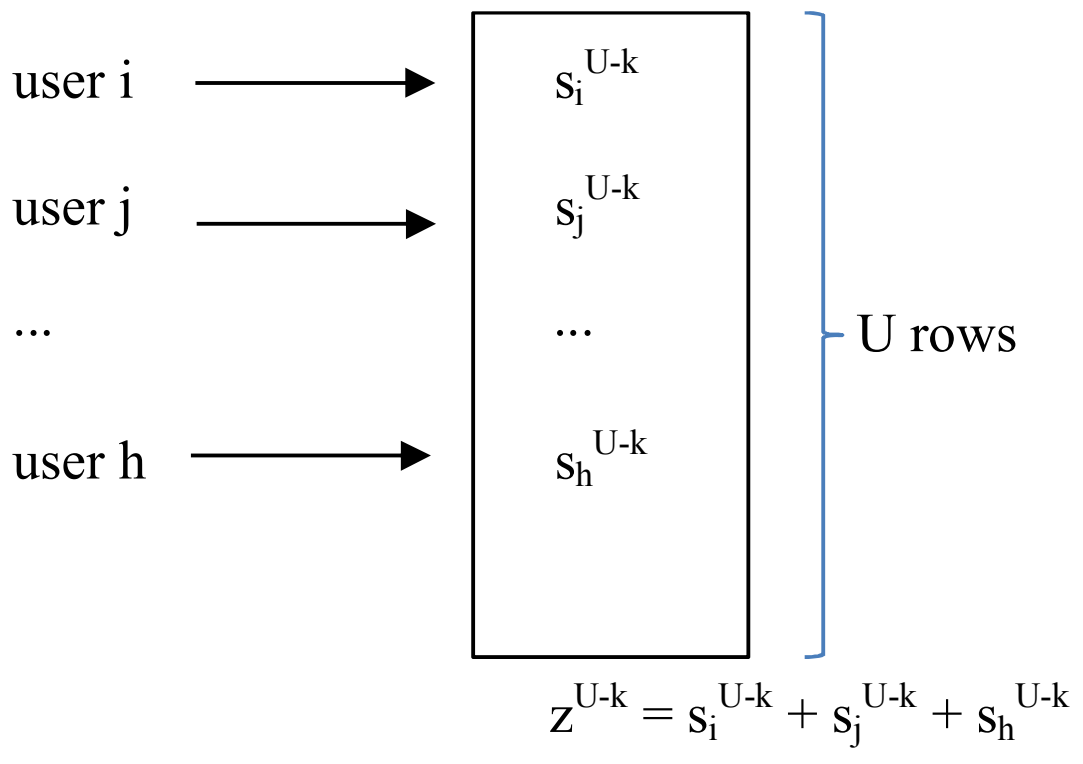

**Figure 2.12** RS syndrome former used for random access

Every user has a unique q-ary signature $s_i^{U-k}$, $1 \le i \le U$, that corresponds to a particular row of the parity check matrix $H^T$ of a (U,k) RS code over GF($2^m$) with (U - k) columns and U rows. We use the property that for a minimum distance $d_{min} = (U - k + 1)$ any (U – k) rows of $H^T$ are linearly independent.

Suppose that we have a set S of T = pU active users, each transmitting their signature multiplied with the information $t_i$, $t_i \in$ GF($2^m$), $t_i \neq 0$. At the receiver we have

$$z^{U-k} = \sum_{i \in S} t_i s_i^{U-k},$$

where the addition is in GF($2^m$). For $T \le (U-k)/2$ active users, the receiver uses the RS decoding algorithm, and decodes the information (noise) transmitted by the active users. Hence, the maximum sum rate becomes

$$\eta = \frac{(U-k)\log_2(q-1)}{2(U-k)m} < \frac{1}{2} \text{bit/transmission.}$$

In case the decoder knows the active users, the transmission efficiency $\eta = 1$ bit/transmission, since every user can transmit $\log_2 q$ bit/transmission. This is the maximum possible over this binary output channel. We visualize the access scheme in Figure 2.12.

We calculated the achievable rate for this channel in the same way as we did for the OR channel. For a particular user, we have a binary symmetric channel with crossover probability

$$P_e = ½ (1 - (1 - 2p)^{T-1}),$$

where p is the probability of an input symbol equal to 0. Note that for p = ½, $P_e$ = 1/2 and the normal channel capacity is zero. We have to assume an asymmetric distribution for p to improve the result. The following achievable rate was obtained.

$$R_{XOR} = \lim_{T\to\infty} T \max_{P(X_i)} I(X_i;Y)$$

$$= ce^{-2c}\log_2\left(\frac{1+e^{-2c}}{1-e^{-2c}}\right) \geq 0.3138 \text{ bit/transmission.}$$

for the input probability distribution

$$p = P(0) = c/T;\ \ P(1) = 1 - c/T;\ c = 0.1999.$$

Observe that a positive rate remains for $T \to \infty$. For more information about the applications and background of the XOR channel we refer to [80,81,82,83].

# 2.5 Concluding Remarks

In this chapter, for which we summarize the concept in Figure 2.13, we start with calculating the maximum achievable transmission efficiency (throughput) in a random access situation. The considered problem is different from the classical information theory problem, in the sense that users do not know each other and thus cannot cooperate. Furthermore, in a multiple user situation, the interference depends on the channel input probability distributions and thus classical maximization methods using concavity arguments to obtain channel capacity do not work.

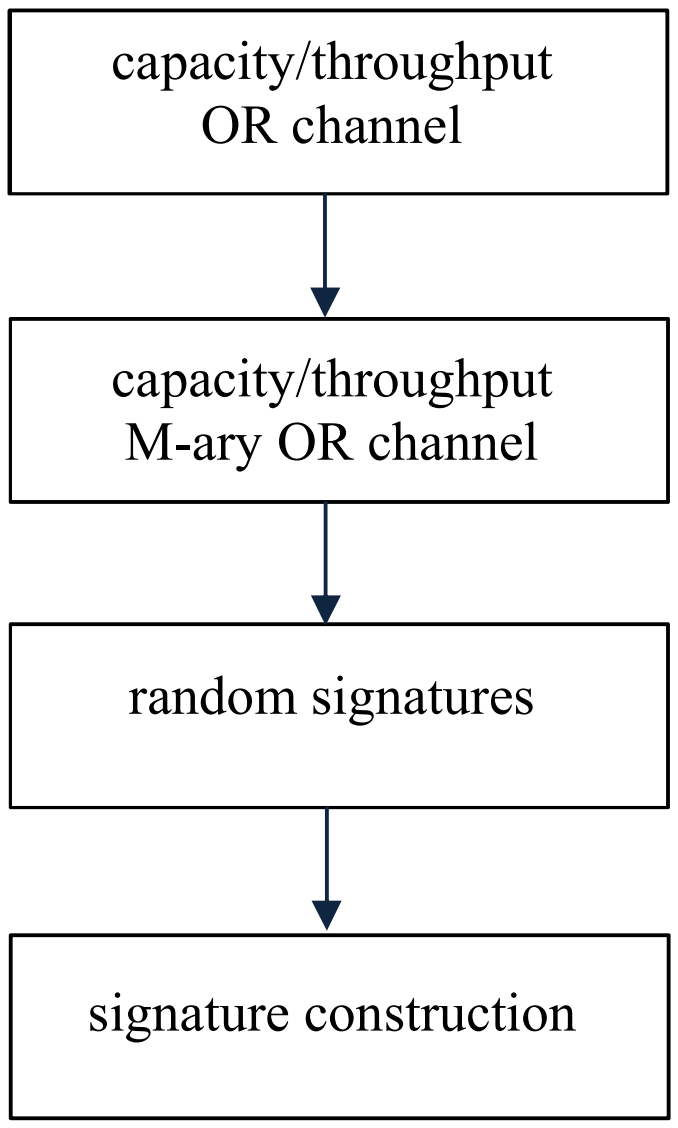

**Figure 2.13** Concept of the chapter

For the binary OR channel and for the M-ary OR channel, the throughput approaches ln2 bits per channel use. This throughput has to be shared by the number of active users. The loss in efficiency, compared with fully coordinated access, is about 30%. Remark that the M-ary OR channel is equivalent to the Pulse Position Modulation scheme. By using only one pulse in M time slot, we obtain the same throughput as if we are using M parallel binary OR channels. These topics were considered by Jeroen Keuning[2], a student form the University of Eindhoven, the Netherlands, and Peter Gober[3] a PhD student from the University of Duisburg-Essen.

An interesting approach was taken by Young-Gil Kim in [85], where the OR channel model was used to improve a collision arbitration algorithm in RFID (Radio Frequency Identification) applications.

For the M-ary OR channel, we develop and analyze a transmission scheme using signatures. For random signatures, we show that an asymmetric input distribution improves the throughput. This implies that a particular user almost always transmits a pulse in a "central" position. Sometimes one of the remaining time positions is used. We also indicate the problem of error probability for a set of active users simultaneously. For the random signaling scheme, a throughput of (½ln2) bits per channel time unit can be achieved. We conclude with two constructions for signature sequences. One is called the "Titlebaum" construction, based on an extended RS encoding matrix. Together with Samwel Martirossian[4], who was a guest in the Institute for Experimental Mathematics, we further developed the concept of "superimposed codes." From this concept a series of codes can be derived.

We conclude the chapter with the construction of signatures for the XOR channel. This research was initiated by T. Ericson and V. Levenshtein [80]. V. Levensthein was a visitor of the Institute for Experimental Mathematics and we published a paper on peak-shift correcting codes [84].

In information theory, network users are assumed to be able to cooperate. In certain network situations this is indeed possible. The main problems considered are the achievable rate regions and their converses. Examples in our research can be found in [90,91].

The reported research was inspired by lectures from Jack Wolf[1] at the 1981 Nato Advanced Institute in Norwich, England.

[1] J.K. Skwirzynski, New Concepts in Multi-User Communication, Norwich, UK, August 4-16, 1980

[2] A.J. Han Vinck and Jeroen Keuning, „On the Capacity of the Asynchronous T-user M-frequency Noiseless Multiple Access Channel without Intensity Information," IEEE Transactions on Information theory, Nov. 1996

[3] P. Gober. *Aspekte des unkoordinierten Mehrfachzugriffs,* Verlag dissertation.de, 2000.

[4] A. J. Han Vinck, Samvel Martirossian and Peter Gober, "On q-ary Superimposed Codes," 1999 IEEE ITW, Kruger National Park, South Africa, June 20 – 25, 1999

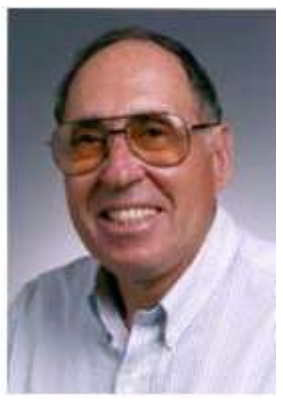
Professor Jack Wolf (1935-2011)

# Chapter 3

# Cooperative Coding

## 3.1 Introduction

One of the dominant parameters in power line communications is that of attenuation. Attenuation can vary between 10-100 dB per kilometer. In power line communications, attenuation prevents signals to propagate over distances that are longer than say 500 meters. Investigations to overcome the high attenuation are therefore highly important, see [14,15,16].

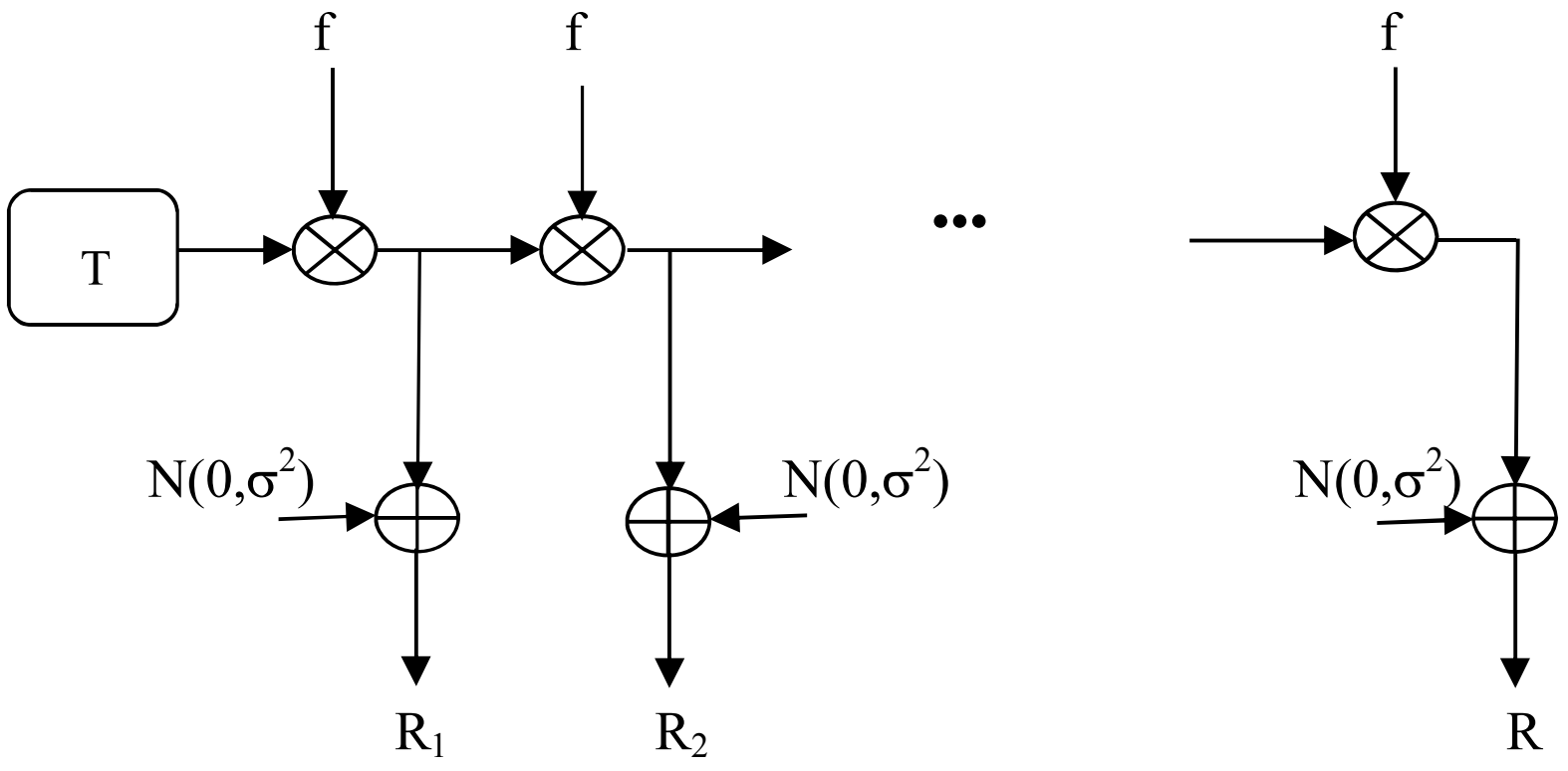

**Figure. 3.1** L links connecting T and R

We discuss the efficiency of communication over links that are in tandem [17]. This situation occurs when users in a power line communication network are connected to the same line and the power lines can be seen as "bus systems", where all connected users can listen to the signals present on the bus. An injected signal can be observed by all connected users.

Figure 3.1 gives a model for L links connecting T and R using the power line. We assume for simplicity that all links have the same length, same attenuation as well as the same noise variance. The attenuation factor f is a function of the length of a link, i.e. f = f(d), where d = D/L and D is the distance between transmitter T and user R.

To illustrate the problem, consider an AWGN channel with signal attenuation factor f and binary modulation with $E_b$ the energy per information bit and $N_0/2$ the noise variance. The signal-to-noise ratio (SNR) in dB is defined as SNR = $10\log_{10}(E_b/N_0)$. The detection error probability at the receiver

$$P_1 \propto Q\left(\sqrt{\frac{E_b}{N_0}f^2}\right) \approx \exp(-\frac{E_b}{2N_0}f^2), \tag{3.1}$$

where the Q-function is the tail probability of the standard normal distribution. If we have L links in tandem, then the attenuation is $f^L$ and the detection error probability after L links becomes

$$P_2 \propto Q\left(\sqrt{\frac{E_b}{N_0}f^{2L}}\right) \approx \exp(-\frac{E_b}{2N_0}f^{2L}) \approx P_1^{f^{2(L-1)}}. \tag{3.2}$$

We can see a dramatic degradation in performance for f < 1, even with only two links. To overcome this degradation, we can use the fact that every user can observe the signal and thus can possibly act as a repeater or relay. We investigate several possibilities in Section 3.2.

## 3.2 Communication Strategies

In the following, we discuss possible communication strategies to efficiently use and operate repeaters under the condition that not more than one user can use the bus simultaneously. The first two strategies fall into the category of conventional multi-hop, where messages are forwarded by a repeater, while the third makes use of incremental redundancy.

**Detect-and-forward**

In the connection between the transmitter T and receiver R, the intermediate receivers can act as detect-and-forward repeaters. After taking a hard decision on the received signal, repeater i sends the re-modulated signal to receiver i + 1. Therefore, if at the first repeater the detection error probability is $P_e$, then the error probability at the Lth receiver is approximately $LP_e$. While this linear increase of error probability with distance is far better than the exponential growth experienced without repeaters, the use of repetitions reduces the overall transmission efficiency by a factor of L. Since $P_e$ is determined by the attenuation factor f(d), and d = D/L, the number and location of repeaters can be optimized such that L is minimized subject to an upper bound on the error rate $LP_e$ at the destination.

**Decode-and-forward**

As in point-to-point connections, we can also use coding to improve the communication efficiency. Given the model described above, we have the capacity $C(d) = 1 - h_2(P_e)$ for a link of distance d, where $h_2(\cdot)$ is the binary entropy function. Assuming that C(d) is a measure for the achievable rate with coding, which is monotonically decreasing with increasing d, and considering that L hops reduce the overall achievable rate to

$$C_{df} = C(d)/L, \tag{3.3}$$

we again have an optimization problem for the number and thus location of repeaters.

**Cooperative coding**

Since we have a bus structure, we can use the fact that all users connected to the bus are able to hear the communication and cooperate to improve the communication efficiency, see also [19]. The achievable rate for the communication between transmitter T and the first receiver is $k/n_1 = C(d)$, i.e. after $n_1$ transmissions the first receiver can decode the transmitted k bits of information. Simultaneously, the amount of information received at the second receiver is $n_1C(2d)$. To assist the second receiver in decoding the transmitted message, repeater 1 sends $n_2$ encoded symbols, such that $k = n_1C(2d) + n_2C(d)$. For the third transmission, repeater 2 transmits $n_3$ encoded symbols such that $k = n_1C(3d) + n_2C(2d) + n_3C(d)$ and the third receiver can decode (in theory) the message. For the destination, we have

$$k = \sum_{i=1}^{L} n_i C((L-i+1)d). \tag{3.4}$$

In summary, given k and the link capacities C(id), the lengths $n_i$, i = 1, 2, ···, L, are obtained from

$$\begin{bmatrix} C(d) & 0 & 0 & \\ C(2d) & C(d) & 0 & \\ \cdots & & & \\ C(Ld) & C((L-1)d) & \cdots & C(d) \end{bmatrix} \begin{bmatrix} n_1 \\ n_2 \\ \cdots \\ n_L \end{bmatrix} = \begin{bmatrix} k \\ k \\ \cdots \\ k \end{bmatrix}. \tag{3.5}$$

The overall achievable rate is

$$C_{coop} = \frac{k}{\sum_{i=1}^{L} n_i}. \tag{3.6}$$

If we do not use the cooperative coding, we have from (3.3),

$$C_{df} = k/(Ln_1).$$

**Example** Let us consider L = 2. With decode-and-forward the rate is $C_{df}$ = ½C(d) = $k/(2n_1)$. With cooperative coding we obtain

$$C_{coop} = \frac{k}{n_1 + n_2} = \frac{C(d)}{2 - \frac{C(2d)}{C(d)}} > \frac{1}{2} C(d). \tag{3.7}$$

Clearly, for f = 1, C(2d) = C(d) and thus $n_2 = 0$ and the repeater is not used. However, for f < 1, C(2d) < C(d), and cooperative coding has an advantage.

**Example** If for example, $P_e = 10^{-3}$ and f = −10 dB, then C(d) = 0.99 bit/transmission and C(2d) = 0.36 bit/transmission. Hence, cooperative coding increases the rate from 0.36 bit/transmission using no repetition or 0.49 bit/transmission using decode-and-forward to C(d)/(2 − C(2d)/C(d)) = 0.60 bit/transmission. To outline the advantages of cooperative coding, consider a binary modulation scheme with a transmitter side SNR of 13 dB, and a T-R distance of D = 1km. The attenuation for a link of length d = D/L is given by $f^2 = 10^{-\delta d/10}$, where we choose $\delta$ = 40, 60, 100 dB/km.

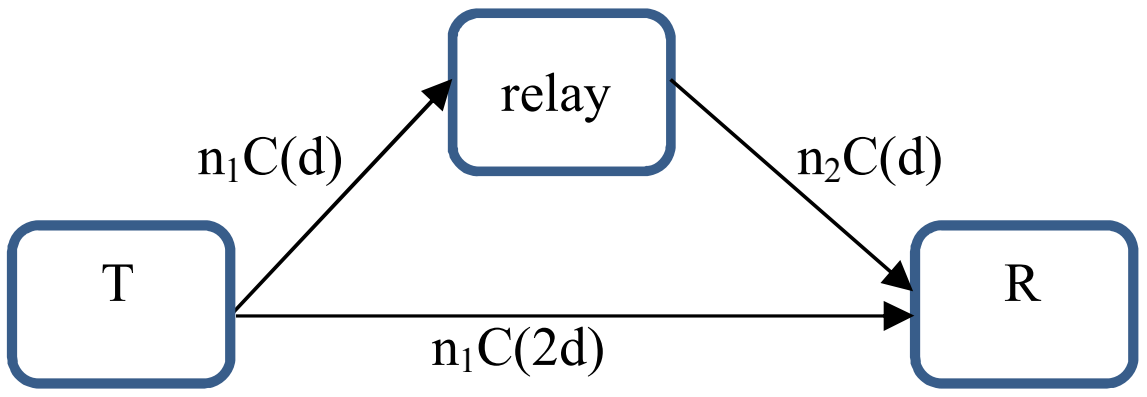

**Figure 3.2** The two link connection as a relay channel.

The error probability for direct transmission over i links follows from (3.2), and the corresponding link capacity as $C(id) = 1 - h_2(P_i)$. We then can compute the achievable data rate for decode-and-forward and cooperative coding from (3.3) and (3.6), respectively. Figure 3.3 shows the achievable rate as function of the number of repeaters N.

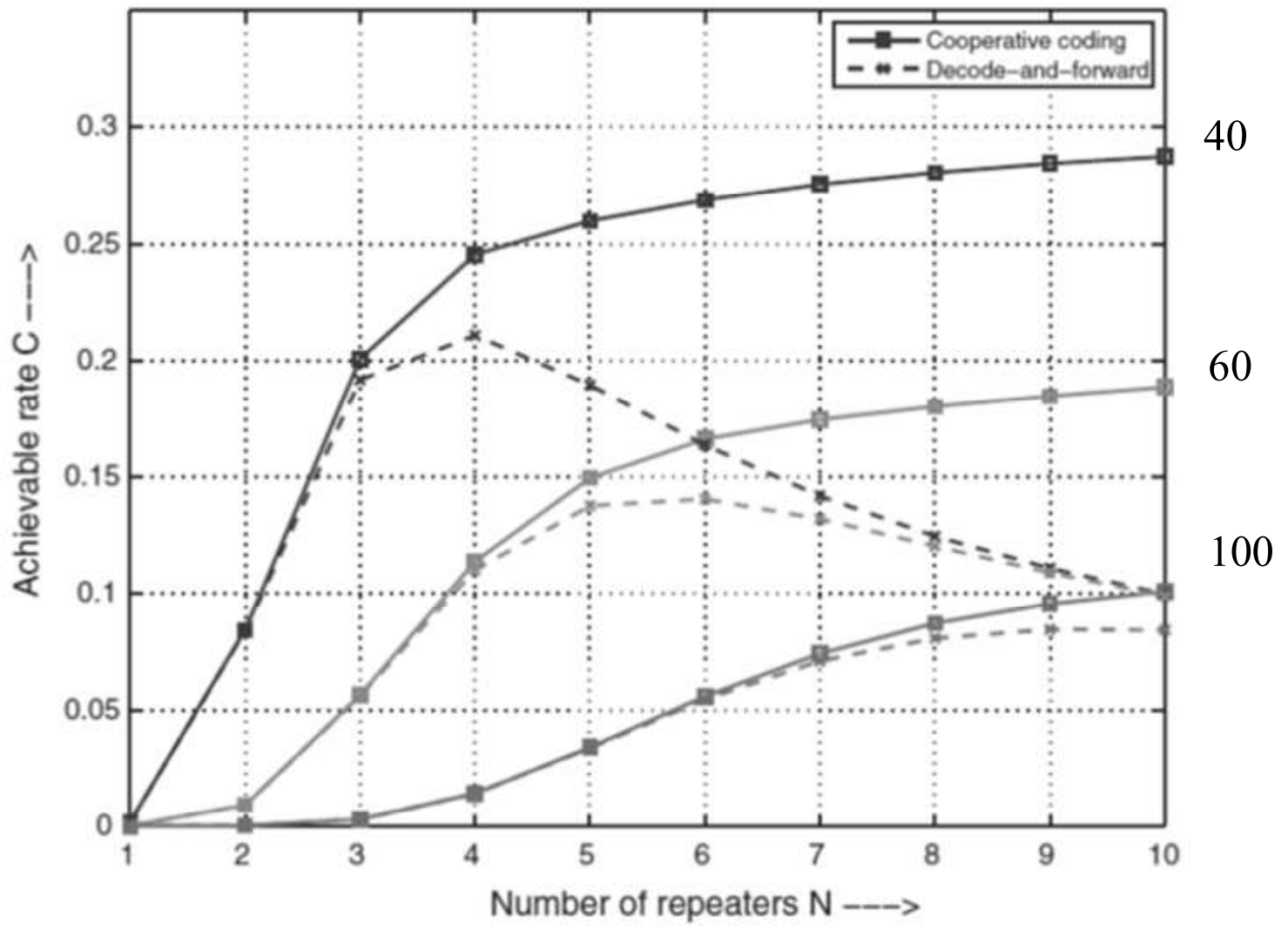

**Figure 3.3** Achievable rate for cooperative coding and decode-and-forward

We observe that cooperative coding achieves a higher rate for a given number of repeaters. In particular, conventional multi-hop reaches a maximum rate at a certain repeater distance d = D/N, which decreases when repeater distance decreases through the addition of repeaters. Cooperative decoding, however, experiences a continuing increase in achievable rate for increasing N and thus decreasing repeater distance d.

**Example** We use the (7,4) Hamming code to illustrate an implementation of cooperative coding with a fixed-rate code. Suppose that we have a bus with only L = 2 links. The length of the links is such that we expect a single error in a received code word per link. For the transmission using two links in tandem via detect-and-forward we expect two errors to occur, which cannot be corrected using the Hamming code. Applying decode-and-forward would allow error free transmission. But in both cases, the rate is only 4/14 bits per transmission. For cooperative coding, we use the Hamming code generated by the encoding matrix

$$G = \begin{bmatrix} 1 & 0 & 0 & 1 & 1 & 0 & 0 \\ 0 & 1 & 0 & 1 & 0 & 1 & 0 \\ 0 & 0 & 1 & 0 & 1 & 1 & 0 \\ 1 & 1 & 1 & 1 & 1 & 1 & 1 \end{bmatrix}. \qquad (3.8)$$

The reason for this choice of encoding matrix will be made clear in the following:

1. the first receiver can decode the transmitted code word if it has no more than a single error. The second receiver also receives a corrupted code word, but it cannot reliably decode the transmitted code word since the number of errors might be above the correcting capacity;
2. the first receiver (repeater), after decoding, encodes the first three information bits with a minimum distance 3 shortened Hamming code of length 6, generated by the code words [100110], [010101], [001011] (see (3.8));
3. the second receiver can decode the received code word, since it contains no more than a single error. After decoding, the second receiver can subtract the code word connected to these three information bits from the first reception and decodes the fourth information bit using the minimum distance 7 code generated by [1111111] which is able to correct three errors. The overall efficiency is 4/(7 + 6) > 4/14.

The example shows the importance of the encoding matrix representation. Following this strategy, the optimum distance profile (ODP) of linear block codes is defined in [18]. Using the ODP, for particular codes, we are able to specify the maximum minimum distance of a sub-code. The ODP depends on the initial encoding matrix, as in the example.

The extension to RS codes can be explained using the k/n RS encoding matrix

$$G = \begin{bmatrix} 1 & 1 & 1 & \ldots & 1 \\ 1 & \alpha & \alpha^2 & \ldots & \alpha^{n-1} \\ 1 & \alpha^2 & \alpha^4 & \ldots & \alpha^{2(n-1)} \\ & \ldots & & & \\ 1 & \alpha^{k-1} & \alpha^{2(k-1)} & \ldots & \alpha^{(k-1)(n-1)} \end{bmatrix}, \qquad (3.9)$$

where the $\alpha$ is a primitive element from the $GF(2^m)$ and $n \leq 2^m - 1$. Suppose again that $L = 2$ and that we start with the $k \times n$ encoding matrix G at T. The repeater has an encoding matrix generating the $(m_2,k_2)$ RS code consisting of the top $k_2$ rows of length $m_2$ of G, where $m_2 < n$. T encodes k symbols and forwards the code word to the repeater. The repeater can decode the received word from T, whereas $R_2$ can listen but not decode since the reception is too noisy. The repeater then encodes the first $k_2$ symbols, and forwards the code word that corresponds to the $(m_2,k_2)$ RS code. $R_2$ is assumed to be able to decode these $k_2$ symbols and subtracts the influence from the first received code word. Since the minimum distance of the remaining code is increased, $R_2$ can decode the whole code word. The performance of this method depends on the knowledge of the respective channel parameters and can be optimized for particular values of the channel parameters.

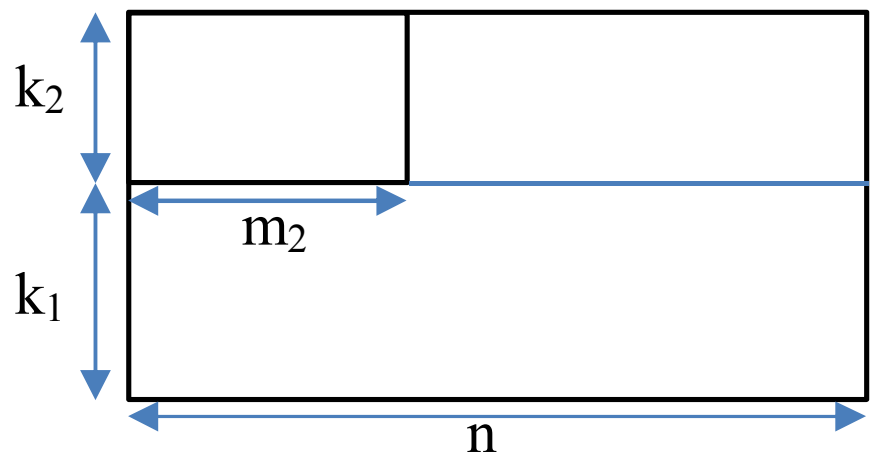

**Figure 3.4** Dimensions for the RS encoding

Suppose that the link error rate is upper bounded by $t_1$ and the two link error rate is upper bounded by $t_2$, where ½ > $t_2 > t_1$. The dimensions of the RS canonical encoding matrix that we are going to use are given in Figure 3.4. The relations between the dimensions and error rate are as follows:

$$n - (k_1 + k_2) = 2t_1 n, \tag{3.10a}$$
$$n - k_1 = 2t_2 n, \tag{3.10b}$$
$$m_2 - k_2 = 2t_1 m_2. \tag{3.10c}$$

The encoded information to be transmitted to user $R_2$ is given by $k = (k_1 + k_2)$ symbols. The communication uses three steps.

1. In the first step we encode k information symbols and transmit the n code symbols to $R_1$. Using (3.10a), $R_1$ can decode. Receiver $R_2$ receives a noisy version of the code word with a maximum of $t_2 \times n$ errors, which is assumed to be beyond the error correcting capability of the code with the given parameters.
2. In the second step, receiver $R_1$ forwards the code word, that corresponds to $k_2$ information symbols, of length $m_2$. Using the RS encoding matrix properties, this is again an RS code word.
3. From (3.10c), it follows that $R_2$ is able to decode the $k_2$ information symbols. Since $R_2$ can subtract the influence of the decoded information from the code word in the first step, $R_2$ is able to decode the remaining $k_1$ information symbols using property (3.10b).

We can evaluate the transmission efficiency for simple repeating and RS encoded retransmission as

$$R_{REP}(T \to R_2) = \frac{k}{2n},$$
$$R_{RS}(T \to R_2) = \frac{k}{n + n\dfrac{2t_2 - 2t_1}{1 - 2t_1}}. \tag{3.11}$$

Since ½ > $t_2 > t_1$, we can see the improvement that depends on the error rates $t_2$ and $t_1$. For $t_2$ close to $t_1$ the efficiency is close to the situation without repeater. For $t_2$ close to ½, the un-coded repeating efficiency appears.

**Remark** For error rates larger than $t_2$ or $t_1$, we accept decoding errors.

## 3.3. Multi User Tandem Communication

In this section, we first look at the communication from one transmitter to two receivers (see Figure 3.5a). We give the efficiency for time-sharing as a reference value for further comparisons and we use the water-filling argument [22] to find the optimum frequency-sharing power allocation. We use principles from information theory to translate the tandem channel into a degraded broadcast channel and calculate its performance [20]. We show that this translation improves the performance compared with the time- and frequency-sharing results.

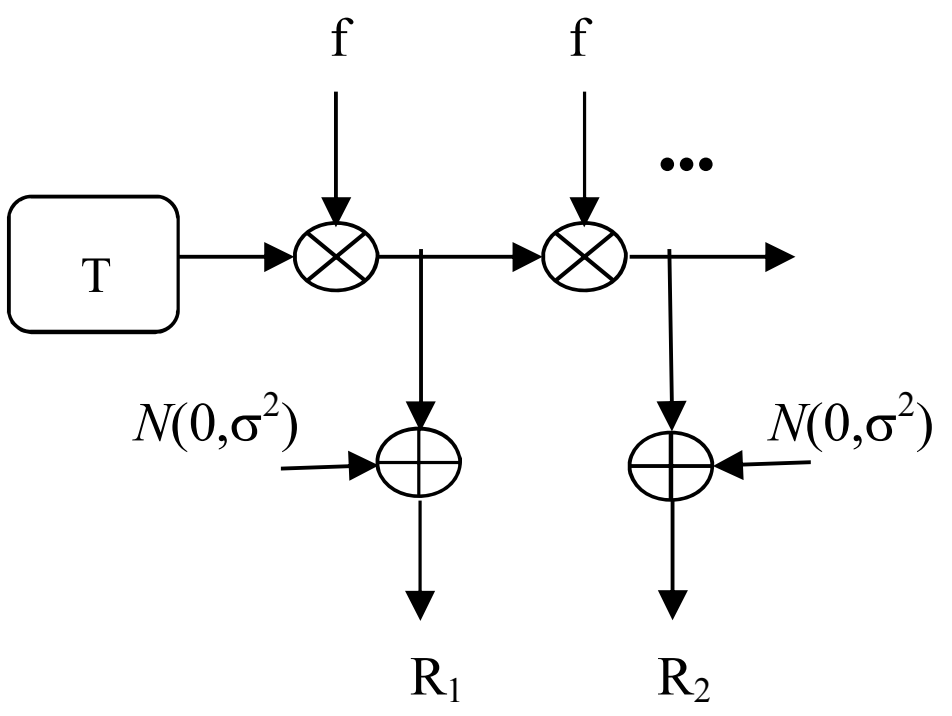

**Figure 3.5a** One transmitter to two receivers.

In Section 3.2 we describe the use of a repeater and its consequences for the transmission efficiency. It is interesting to compare the repeater strategy with the broadcast results and determine the condition for which it is not beneficial to use a repeater. Since the design of practical coding scheme that achieves the capacity region for broadcast channels is not a solved problem yet, we conclude that for the tandem channel without repeater, time-sharing is the best, low complexity, option.

Finally, we introduce a multiple access model for the communication between two transmitters and one receiver using the tandem channel model, see Figure 3.5b. Information theoretical principles are used to show equivalence between the broadcast and the multiple access models.

### 3.3.1 Point-to-point communication

In this section we consider only two links and study several communication topologies. For simplicity, we assume that the attenuation factor f, $f < 1$, is the same for both links. The noise is considered to be additive white Gaussian noise (AWGN) with power spectral density $\sigma^2$ W/Hz and average value 0. The bandwidth and transmit power are denoted by B and P, respectively.

The Gaussian channel capacity $C_1$ for a single link and $C_2$ for 2 links in tandem, are given by

$$C_1 = B\ln\left(1+\frac{Pf^2}{2\sigma^2 B}\right) \text{ nat/s,}$$

$$C_2 = B\ln\left(1+\frac{Pf^4}{2\sigma^2 B}\right) \text{ nat/s,} \tag{3.12}$$

respectively. All further derived transmission efficiencies (capacities) are expressed in a similar way.

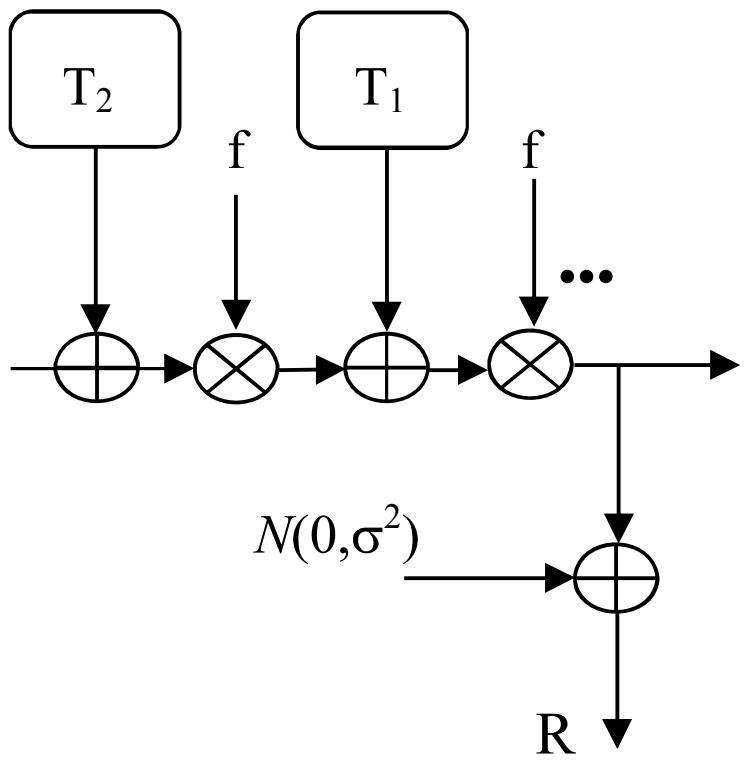

**Figure 3.5b** Two transmitters to one receiver.

### 3.3.2. Communication strategies

**A. Time-sharing.**

If we use time-sharing (TS) with time-sharing parameter $\alpha$ from transmitter to both the receivers, then the achievable communication rate or transmission efficiency is given by

$$R_{TS}(T \rightarrow R_1) = \alpha B \ln\left(1 + \frac{Pf^2}{2\sigma^2 B}\right) \text{ nat/s,}$$

$$R_{TS}(T \rightarrow R_2) = \bar{\alpha} B \ln\left(1 + \frac{Pf^4}{2\sigma^2 B}\right) \text{ nat/s,} \tag{3.13}$$

$$0 \leq \alpha \leq 1; \bar{\alpha} = 1 - \alpha.$$

**B. Frequency-sharing.**

For frequency-sharing, we use the water-filling (see Appendix 8.8) concept from information theory to obtain the power allocation that maximizes the efficiencies. We divide the frequency band into the part $\alpha$ and $\beta$, $\alpha + \beta = 1$, and find the achievable rates

$$R_{FS}(T \rightarrow R_1) = \alpha B \ln\left(1 + \frac{P_1 f^2}{2\alpha\sigma^2 B}\right) \text{ nat/s,}$$

$$R_{FS}(T \rightarrow R_2) = \beta B \ln\left(1 + \frac{P_2 f^4}{2\beta\sigma^2 2B}\right) \text{ nat/s,} \tag{3.14}$$

where $P = P_1 + P_2$. To obtain the rate region for the frequency-sharing using the water-filling argument, we have to maximize (3.14) for all different values of $\alpha$ and $\beta$, $\alpha + \beta = 1$ under the sum power constraint. This problem was solved in [99]. The key point in the optimization is the assumption that (3.14) crosses the time-sharing line in two points. The conclusion is that the envelope of all solutions dominates the time sharing performance. The general behavior is given by the dashed line in Figure 3.8. Analyzing the results from A and B, we conclude that frequency-sharing is always better than time-sharing. Note that both efficiency regions have the same boundary points. For $\alpha + \beta = 1$, we can use the approach from Appendix A.8.3. Furthermore, for $P_1 = \alpha P$ and $P_2 = (1 - \alpha)P$, we obtain the "naïve" time-sharing result as given in (3.13).

**C. Receiver $R_1$ acts as relay.**

If receiver $R_1$ can act as an intelligent relay, a simple decoding and re-encoding strategy would give a transmission rate region

$$\begin{aligned} R_{REP}(T \to R_1) &= \alpha B \ln\left(1 + \frac{Pf^2}{2\sigma^2 B}\right) \\ &= \alpha C_1 \text{ nat/s}, \\ R_{REP}(T \to R_2) &= \frac{\bar{\alpha}}{2} B \ln\left(1 + \frac{Pf^2}{2\sigma^2 B}\right) \\ &= \frac{\bar{\alpha}}{2} C_1 \text{ nat/s}, \end{aligned} \tag{3.15}$$

where $\alpha$ acts as a time-sharing parameter. Comparing the limiting points, i.e. for $\alpha = 0$ or for $\alpha = 1$, we can conclude that repeating is worse than time-sharing for

$$\frac{P}{\sigma^2 B} > \frac{1 - 2f^2}{f^6}.$$

We see that in a situation where power is large and bandwidth is small, time-sharing is better than repeating. For small f, repeating is preferable.

**Example:** For $\sigma^2 = 10^{-8}$ W/Hz and $B = 10^5$ Hz, $P = 25$ W, an attenuation factor $f > 0.2$ satisfies the condition.

We show in [17] and (3.7), that the performance for the repetition strategy can be improved to

$$R^*_{REP}(T \to R_2) = (1 - \alpha)\frac{C_1}{2 - \frac{C_2}{C_1}} \text{ nat/s.} \tag{3.16}$$

In general, this improved repetition strategy needs a complex encoding and decoding strategy. In a practical situation, we can use RS codes.

**D. The Gaussian Broadcast channel.**

The two link model can be transformed into a broadcast model, i.e.

$$y_1 = x\,f + n_1, \qquad y_2 = x\,f^2 + n_2,$$

where $n_1$ and $n_2$ represent the Gaussian noise with variance $\sigma^2$ and $\eta^2$, respectively. The input average power constraint is given by $E(X^2) \le P$. The capacity region for the Gaussian broadcast channel has been determined [22]

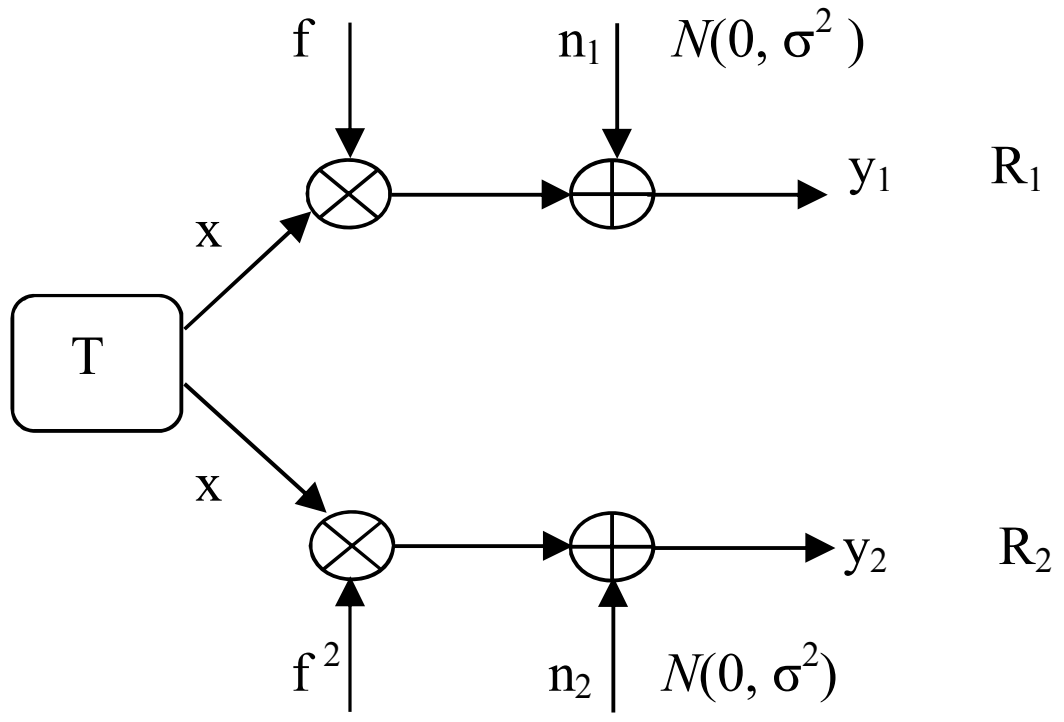

**Figure 3.6** The 2-link broadcast model.

and is given by

$$R_{BC}(T \to R_1) = B\ln\left(1 + \frac{\gamma P f^2}{2\sigma^2 B}\right) \text{ nat/s,}$$

$$R_{BC}(T \to R_2) = B\ln\left(1 + \frac{\bar{\gamma} P f^4}{2\sigma^2 B + \gamma P f^4}\right) \text{ nat/s.} \tag{3.17}$$

Note that for $(\gamma,\alpha) = (1,1)$ or $(0,0)$ we have the same result for the broadcast efficiency as for time- or frequency-sharing. However, simple calculations show that the capacity region for the broadcast channel is strictly larger than the frequency- and time-sharing result for all other values of $\gamma$. The rate regions approach each other for $B \to \infty$. This conclusion can also be found in [21].

**Example** For $\sigma^2 = 10^{-8}$ W/Hz, $B = 10^5$ Hz, $P = 25$ W, $f = 0.3$, $\alpha = 1/2$ and $\gamma = 0.02$ we obtain $R_{TS}(T\to R_1) = R_{BC}(T\to R_1)$ but $R_{BC}(T\to R_2) \approx 2\, R_{TS}(T\to R_2)$.

Hence, we obtain a clear improvement of the broadcast efficiency over time-sharing. The challenging problem is to design practical coding strategies that achieve this gain in efficiency. This is still an open problem.

In [20] a general tandem channel model with more than 2 receivers is given. In this case the broadcasting capacity is given by

$$R_{BC}(T \to R_1) = B\ln\left(1+\frac{\alpha_1 P f^2}{2\sigma^2 B}\right) \text{ nat/s,}$$

$$R_{BC}(T \to R_2) = B\ln\left(1+\frac{\alpha_2 P f^4}{2\sigma^2 B+\alpha_1 P f^4}\right) \text{ nat/s.} \tag{3.18}$$

$$R_{BC}(T \to R_3) = B\ln\left(1+\frac{\alpha_3 P f^6}{2\sigma^2 B+\alpha_1 P f^6+\alpha_2 P f^6}\right) \text{ nat/s.}$$

$$\cdots,$$

where $\alpha_1 + \alpha_2 + \cdots + \alpha_N = 1$. Calculation of the capacity then becomes an optimization problem.

**E. The multiple access channel**

The two transmitter one receiver concept of Figure 3.5b, can be transformed into a multiple access channel as shown in Figure 3.7. The output y at the receiver can be written as

$$y = f\,x_1 + f^2\,x_2 + n,$$

where n is the additive white Gaussian noise with variance $\sigma^2$. The average power is $P = P_1 + P_2$, where $P_1 = \delta P$ and $P_2 = (1-\delta)P$. The capacity region for

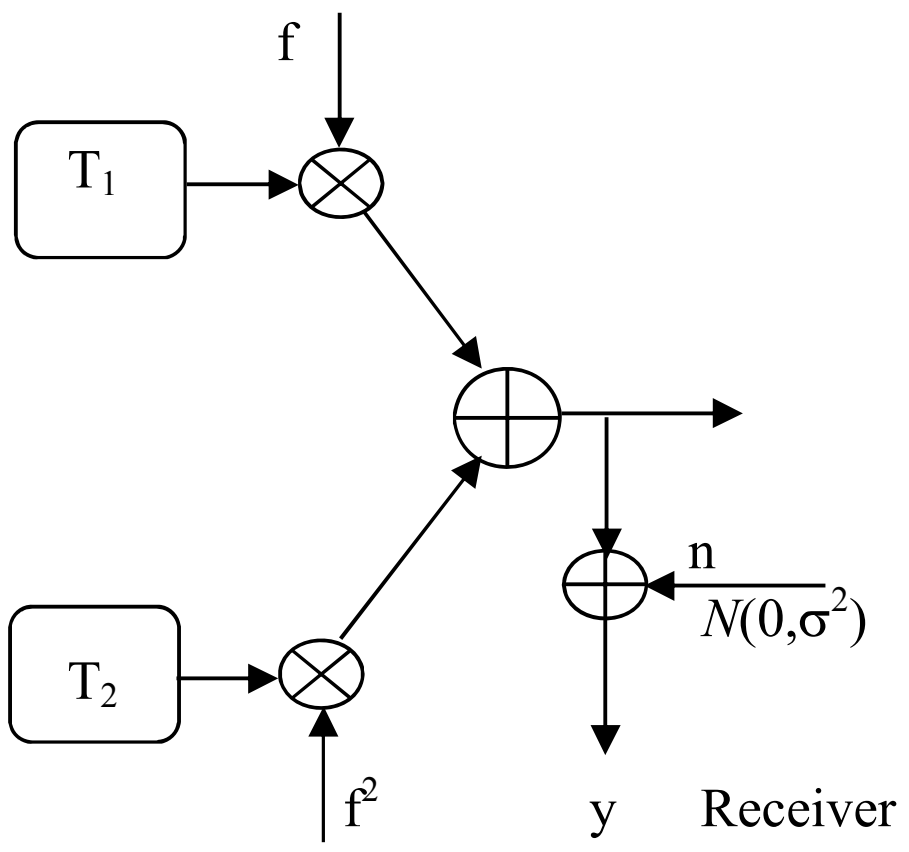

**Figure 3.7** The 2-link multiple access model.

the two access channel [22] is given by

$$C_{MAC}(T_1 \rightarrow R) = B\ln\left(1 + \frac{\delta P f^2}{2\sigma^2 B}\right) \text{ nat/s,}$$
$$C_{MAC}(T_2 \rightarrow R) = B\ln\left(1 + \frac{\bar{\delta} P f^4}{2\sigma^2 B + \delta P f^4}\right) \text{ nat/s.} \tag{3.19}$$

To achieve the capacity, the receiver first decodes $x_2$, then subtracts $x_2$ from y and decodes $x_1$.

As was observed in [23], the multiple access capacity region and the broadcast capacity region can be shown to be equivalent when the sum of the average power in the two transmitters is the same as the sum power in the broadcast channel model. As a consequence, any point in the capacity region for the broadcast channel can also be achieved in the multiple access channel capacity region. The proof is based on the transformation of power from the access to the broadcast domain.

For the time- and frequency-sharing, we refer to the first part of the section. The results are the same, since for this situation, the links in tandem can be considered to be point-to-point connections.

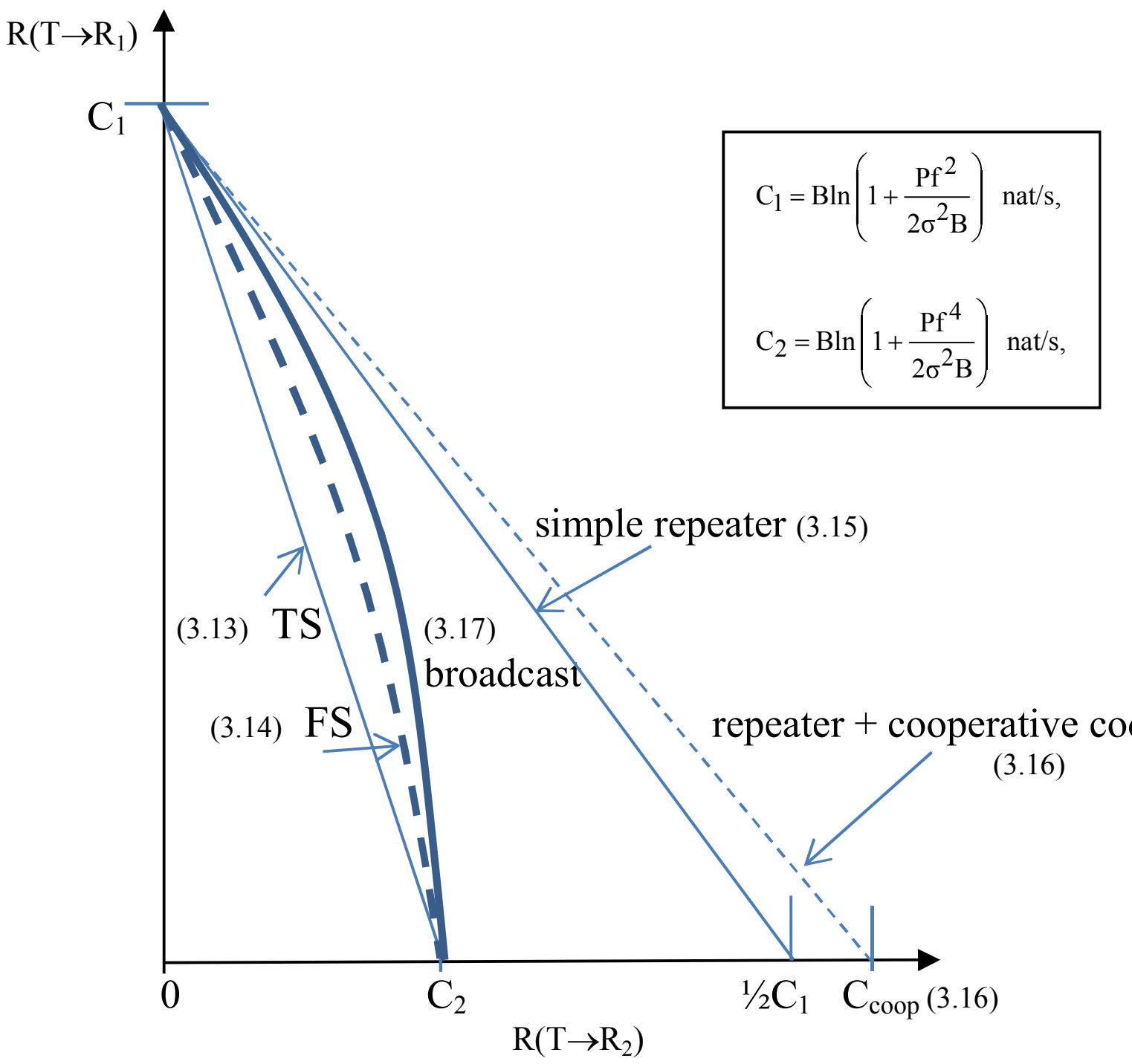

**Figure 3.8** Overview of transmission efficiencies

In Figure 3.8 we present the results for the different strategies in a general way. FS and TS are the frequency- and the time-sharing regions, respectively. We also indicate the performance for the degraded broadcast channel together with the repeating strategies. Of course, the different points have to be evaluated for particular link channel parameters and conditions. We show that frequency-sharing is preferable over time-sharing. Time-sharing also outperforms repeating for specific values of the link attenuation factor f. From the presented material, it follows that there is an interesting connection between multi-user information theory and real communications, such as communication over links in tandem.

## 3.4 Concluding remarks

This chapter originated from actual problems in power line communications. The concept of the chapter is summarized in Figure 3.9.

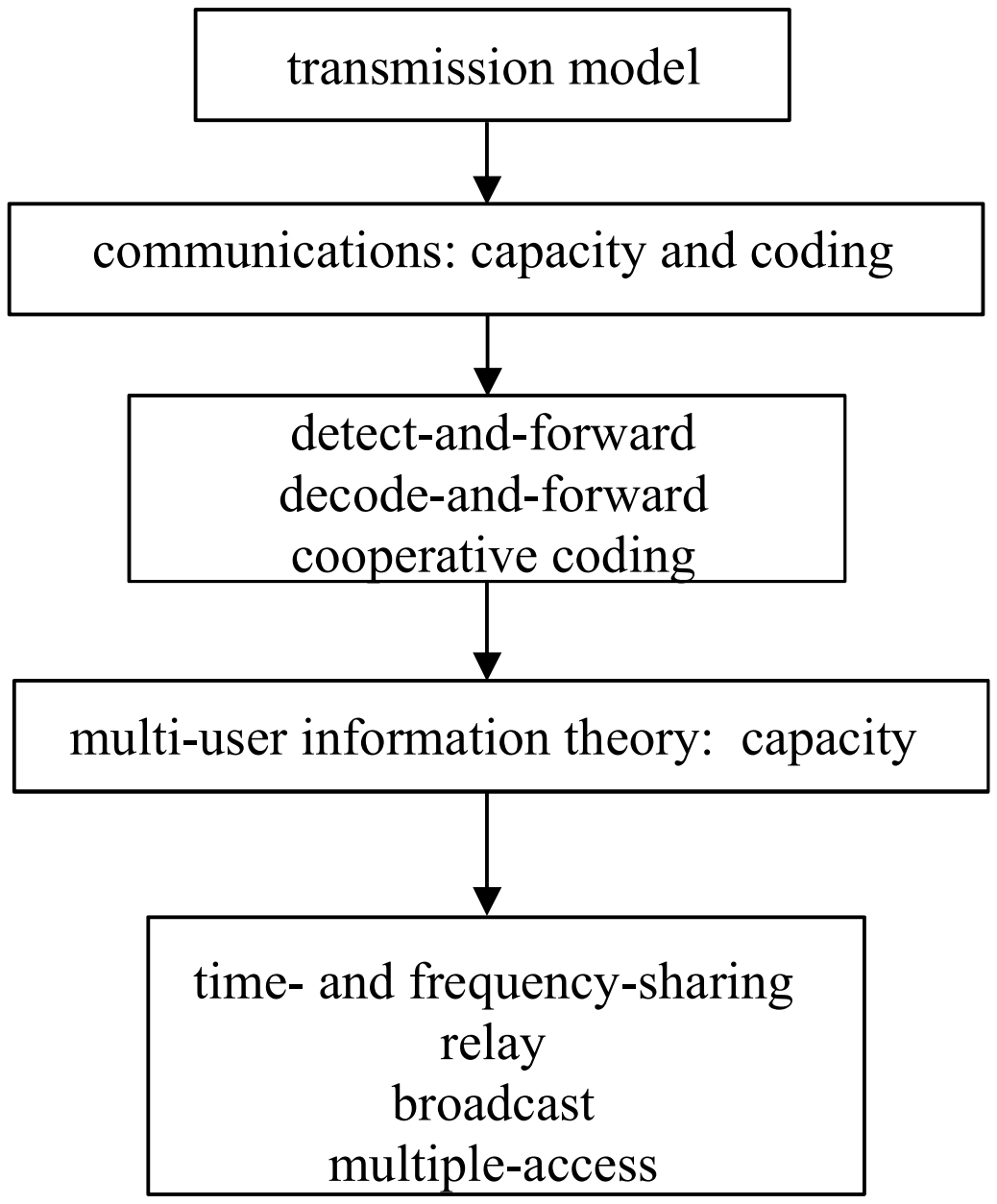

**Figure 3.9** Structure of the material presented in Chapter 3

High attenuation hinders reliable transmission using power lines beyond 500 meters. Improving this number is the target of many investigations. Research in this area started in 1994 with Olaf Hooijen[1], a PhD student in the Institute for Experimental Mathematics. We developed several models and described the power line channel parameters. Measurements were carried out to confirm the theory.

A possible solution to the attenuation problem is the application of repeaters. We consider this possibility and investigate "intelligent" repeating: detect-and-forward, decode-and-forward, and cooperative repeating. We calculate the channel capacity for an artificial connection consisting of several links.

For cooperative repeating, we use RS codes for an easy implementation and optimal efficiency results.

In the Section 3.3 we translate the bus- or link communication concept to a multi-user information theory situation. The concepts of "degraded broadcast" and "multi-access" are shown to be suited for the description of two-link connections.

[1] Olaf Hooijen, Aspects of residential power line communications. Dissertation, University Duisburg-Essen, 1998, ISBN 3-8265-3429-8

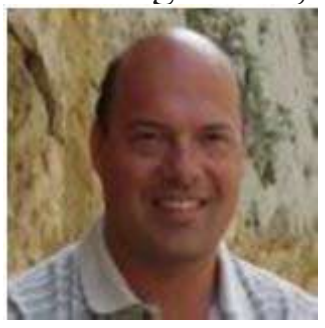

# Chapter 4

# Modulation and Detection

One of the popular topics in coding is the application of soft decision decoding for additive white Gaussian noise channels. The reason for this is the performance improvements in the decoding error rate that can be obtained. We explain the idea of soft decision and then give examples of the applications in combination with RS codes.

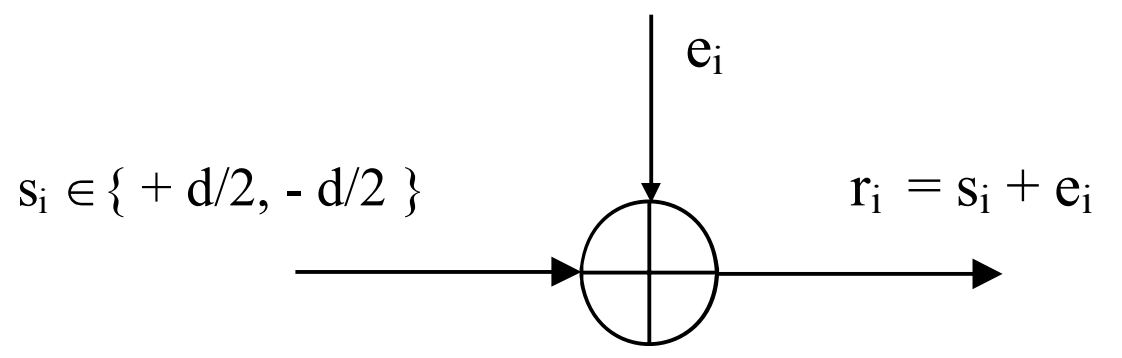

**Figure 4.1** The additive white Gaussian noise channel

The second part of this chapter is concerned with non-Gaussian noise channels. As an example, we assume that in addition to Gaussian noise, the channel can have impulse noise, narrowband disturbances and fading. We give a specific sub-class of RS codes that can be used to avoid degradation due to these atypical disturbances.

## 4.1 The Gaussian Channel

The channel model that we use is the Gaussian channel as given in Figure 4.1. The noise is additive white Gaussian noise (AWGN) with the following properties:

- the average $\overline{e_i} = 0$;
- the probability density function is given by

$$p_{e_i}(\alpha) = \frac{1}{\sqrt{2\pi\sigma^2}} e^{-\frac{\alpha^2}{2\sigma^2}}, \tag{4.1.1}$$

$$\sigma^2 = E(e_i^2).$$

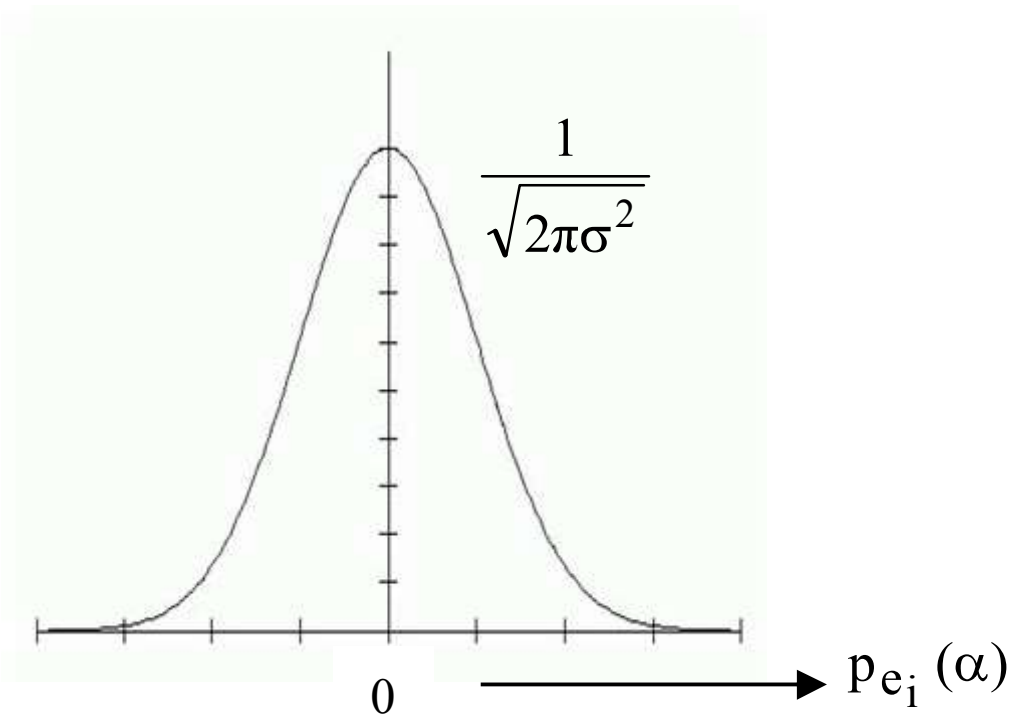

**Figure 4.2** The Gaussian distribution

We assume that the transmitted message is binary, with components represented as $s_i = +d/2$ or $s_i = -d/2$, respectively. We express the energy per information bit $E_b$ as $(d/2)^2$. This is important if we want to compare uncoded transmission with coded transmission.

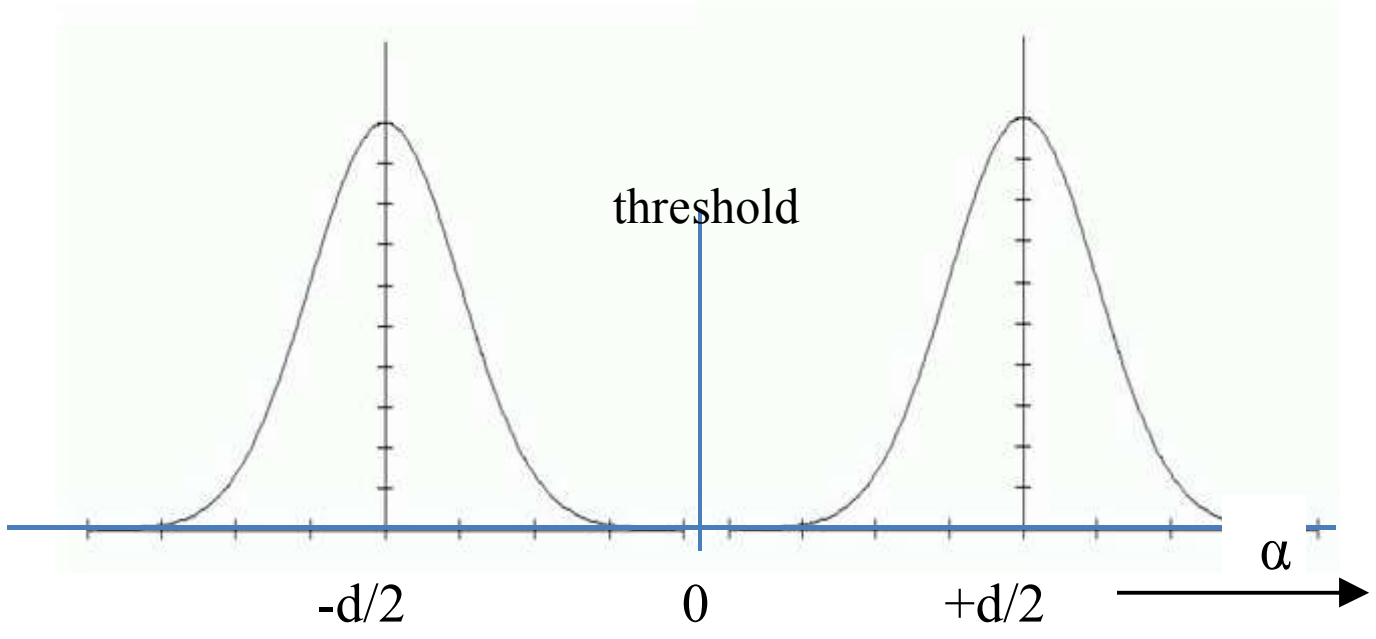

**Figure 4.3** Conditional probability distribution

The hard decision detector decides for + or – and makes an error if for $s_i = +d/2$ the received $r_i = s_i + e_i < 0$. The probability of error is thus given by

$$P_{BSC} := P(\text{ error } | s_i = +d/2) = \int_0^{-\infty} \frac{1}{\sqrt{2\pi\sigma^2}} e^{-(-\frac{d}{2}+\alpha)^2/2\sigma^2} d\alpha$$

$$= \int_{-\frac{d}{2}}^{-\infty} \frac{1}{\sqrt{2\pi\sigma^2}} e^{-\beta^2/2\sigma^2} d\beta$$

$$= Q\left(\sqrt{\frac{d^2}{4\sigma^2}}\right)$$

$$\approx e^{-\frac{d^2}{8\sigma^2}} = e^{-\frac{E_b}{2\sigma^2}},$$

where we used the approximation $Q(a) \approx \exp(-a^2/2)$.

### 4.1.1 Error correction

For coded transmission, the code words are transmitted as $s^n = (s_1, s_2, ..., s_n)$, where $s_i \in \{+ d'/2, - d'/2\}$. The energy per transmitted symbol is then defined as $E_s = (d'/2)^2$. For a fair comparison, the amount of energy for the n coded symbols has to be equal to the amount of energy for k un-coded symbols, i.e. $nE_s = kE_b$ or $E_s := RE_b$

A code with minimum distance $d_{min}$ can correct a maximum of t errors where $d_{min} > 2t$. An indication of the behavior of the decoding error probability is then given by

$$P_{e,BSC} = (P_{BSC})^{t+1}$$

$$\approx e^{-\frac{d'^2}{8\sigma^2}(t+1)} = e^{-\frac{E_s}{2\sigma^2}(t+1)} = e^{-\frac{RE_b}{2\sigma^2}(t+1)}.$$

We can observe, that the product R(t+1) determines the performance of the coded system. The practical coding problem is to maximize R(t+1).

### 4.1.2 Error and erasure decoding

An (n,k) error correcting code with minimum distance $d_{min}$ is able to decode correctly if for the number of erasures E and the number of errors t, the following relation holds:

$$d_{min} \geq E + 2t + 1.$$

The decision process is illustrated in Figure 4.4. The area where we declare the detected symbols as erasure is denoted as E1 and E2.

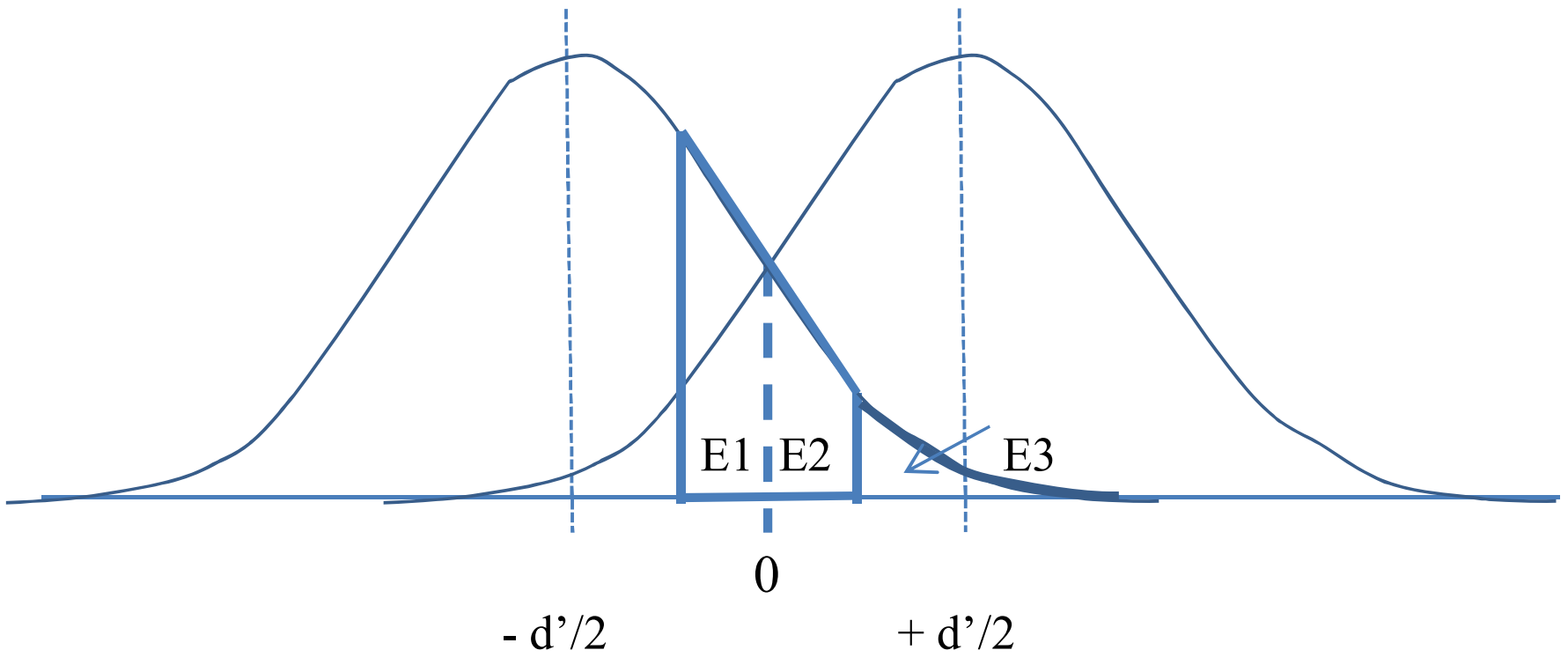

**Figure 4.4** Decision regions for erasure and error decoding

- The probability of an erasure is thus given by the probability that the received symbol is in the region E1 or E2. This probability is determined by the probability density function of the noise given the transmitted symbol.
- The probability that an error occurs is equal to the probability that we are in region E3 given that we transmit $s_i = - d'/2$. Hence, using Figure 4.4, we can say that for reliable decoding we have the condition that

$$d_{min} > n\,(P(E1 + E2) + 2P(E3)).$$

For hard decision decoding, we make an error if the received signal is positive, given that we transmit $s_i = - d'/2$. The condition on correct decoding then becomes

$$d_{min} > n\,(2P(E2) + 2P(E3)).$$

Since P(E1 + E2) > 2P(E2), we conclude that hard decision is the better option for additive white Gaussian noise.

### 4.1.3 Soft decoding

The Maximum Likelihood (ML) detector searches for the code word that maximizes $L := p(r^n \mid s^n)$, where

$$L = p(r^n \mid s^n) = \frac{1}{\sqrt{(2\pi\sigma^2)^n}} e^{-\sum_{i=1}^{n}(s_i - r_i)^2/2\sigma^2}.$$

This is equivalent to minimizing the ***squared Euclidean*** distance between $r^n$ and $s^n$, which is defined as

$$d^2(s^n, r^n) = \sum_{i=1}^{n}(s_i - r_i)^2.$$

If $\hat{s}^n$ is a code word of length n, then the squared Euclidean distance between $c^n$ and $\hat{s}^n$ is defined as

$$d^2(s^n, \hat{s}^n) = \sum_{j \in D} (s_j - \hat{s}_j)^2 = |D| d'^2 ,$$

where D is the set of positions where $s_j \neq \hat{s}_j$.

The probability of error, given $s^n$ is transmitted and $r^n = s^n + e^n$ is received, is given by:

$$P\{ p(r^n | s^n) < p(r^n | \hat{s}^n ) \}$$

$$= P\{ p(e^n = r^n - s^n ) < p(\hat{e}^n = r^n - \hat{s}^n ) \},$$

where,

$$p(e^n) = \frac{1}{\sqrt{(2\pi\sigma^2)^{|D|}}} e^{-\sum_D (s_i - r_i)^2 / 2\sigma^2} ,$$

$$p(\hat{e}^n ) = \frac{1}{\sqrt{(2\pi\sigma^2)^{|D|}}} e^{-\sum_D (\hat{s}_i - r_i)^2 / 2\sigma^2} .$$

From $p(e^n) < p(\hat{e}^n)$ it follows that a decision error occurs when

$$e^{\sum_D 2r_i * (\hat{s}_i - s_i) / 2\sigma^2} > 1 ,$$

which is equivalent to

$$\sum_D (s_i + e_i)(s_i - \hat{s}_i) < 0.$$

For linear codes the distance spectrum (number of code words at a certain distance) is the same for all code words. Therefore, we can calculate the probability of error for the all-0 code word where all symbols are transmitted as +d'/2. From this it follows that for $s_i = - \hat{s}_i$ a decision error occurs when

$$-|D|\frac{d'}{2} > \sum_D e_i \ .$$

Now we use two properties of independent zero-mean Gaussian random variables (GRV):

1. the sum of, zero-mean, independent GRV's is again a GRV with mean zero;
2. the variance of the sum is the sum of the variances.

From which,

$$E((\sum_D e_i)^2) = |D|\sigma^2.$$

The probability of error is thus given by

$$P_e = Q\left(\frac{|D|\frac{d'}{2}}{\sqrt{|D|\sigma^2}}\right) = Q\left(\sqrt{\frac{D(d')^2}{4\sigma^2}}\right).$$

For $kE_b$ = $nE_s$ and $E_s$ = $\sqrt{d'/2}$ we then have the un-coded bit error probability

$$P_{un\text{-}coded} = Q\left(\sqrt{\frac{E_b}{\sigma^2}}\right), \qquad (4.1.2)$$

and the probability that we decode a specific code word at distance $d_{min}$ from the transmitted code word

$$P_{coded} = Q\left(\sqrt{\frac{|D|\frac{k}{n}E_b}{\sigma^2}}\right) \le Q\left(\sqrt{\frac{d_{min}\frac{k}{n}E_b}{\sigma^2}}\right) \approx e^{-\frac{RE_b}{2\sigma^2}d_{min}} . \qquad (4.1.3)$$

One can observe a gain in the decoding error probability when the product $Rd_{min}$ is larger than 1. If $d_{min}$ is the minimum value of |D| for any pair of code words, the union bound on the error probability for a binary (n,k) code is given by

$$P_{union} \le (2^k) \cdot Q\left(\sqrt{\frac{d_{min}\frac{k}{n}E_b}{\sigma^2}}\right). \qquad (4.1.4)$$

We define the coding gain $G_c$ in dB as

$$G_c := 10\log_{10} d_{min}\frac{k}{n} \quad \text{dB}. \qquad (4.1.5)$$

For practical codes, the product of the minimum distance and efficiency R must be maximized to get the highest coding gain. The problem is to find codes for which the net gain is positive.

For binary or hard quantized decoding of an error correcting code, the asymptotic decoding error probability is upper bounded by the probability that we have more than $d_{min}/2$ errors for every code word, i.e.

$$P_{hard} \le 2^k \times Q\left(\sqrt{\frac{\frac{k}{n}E_b}{\sigma^2}}\right)^{d_{min}/2} \approx 2^k \times e^{-\frac{RE_b}{2\sigma^2}(d_{min}/2)}. \qquad (4.1.6)$$

For soft decision decoding

$$P_{soft} \le 2^k \times Q\left(\sqrt{\frac{d_{min}\frac{k}{n}E_b}{\sigma^2}}\right) \approx 2^k \times e^{-\frac{RE_b}{2\sigma^2}d_{min}}. \qquad (4.1.7)$$

From this, we see that soft decision brings a coding gain of 3 dB [30] (factor of 2 in energy).

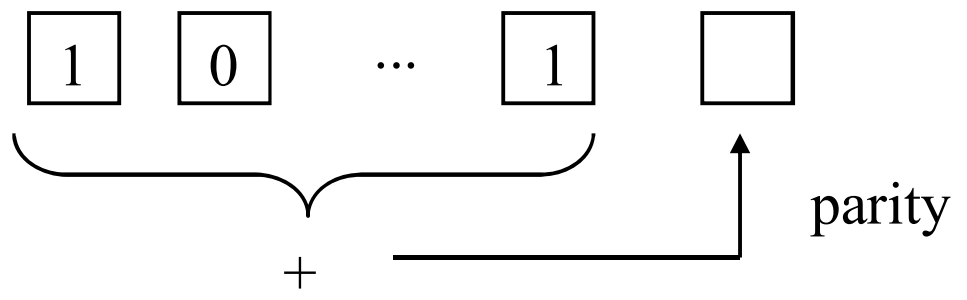

**Figure 4.5** The Single Parity Check (SPC) code

**Example** The simplest code that we can find is the single parity check code (SPC). For this code all code words have even Hamming weight and thus the minimum distance is 2. Adding a parity bit to k = (n-1) information bits gives the code word of length n and thus a code efficiency of (n-1)/n.

The asymptotic coding gain over the un-coded case is $G_c = 10\log_{10}2 = 3$ dB. In practice, the gain is less because we have to take the number of code words into account. From simulation results, it follows that for n = 7 or 8, a net coding gain of about 2 dB can be realized. If we fix the rate and increase the code word length, the number of code words increases exponentially, and the coding gain disappears.

**Decoding for the SPC**

To find the code word $c^n$ at minimum squared Euclidean distance from the received word $r^n$ we act as follows:

1) take a hard (binary) decision on the received values. If the word is a code word with even parity, stop and output the word as correct;

2) if the hard decision word is not a code word (odd parity), search for the position with the smallest received absolute value, say $r_{min}$. Invert the decision at this position.

Inversion of the hard decision value increases the squared Euclidean distance with the smallest possible amount and the resulting hard decision word is a code word.

These two steps are the basis for soft decision algorithms to be discussed later. The SPC decoding is characterized by its extremely low complexity since the decoder has simply to invert the least reliable bit (corresponding to the smallest absolute demodulator output) whenever a parity error occurs in the binary quantized received word.

**Example** Using the previous AWGN assumptions:

| | | | | | | |
|---|---|---|---|---|---|---|
| **encode** | +5 | +5 | -5 | +5 | -5 | +5 |
| **receive** | +4 | +3 | -4 | -1 | -3 | +5 |
| **hard decision** | 1 | 1 | 0 | 0 | 0 | 1 |
| **correct** | 1 | 1 | 0 | 1 | 0 | 1 |

We change the decision at position four, since the hard decision at this position is most likely to be wrong.

**Remark** If we compare the bit error rate for the coded situation using soft decision and the un-coded situation using hard decision, we can expect a gain of 3 dB. In theory, we can have a gain of 2 dB in channel capacity for low signal to noise ratios (bad channels). Remark that the gain in channel capacity is 2 dB at a decoding error probability approaching zero.

## 4.2 The Combination RS and SPC

We give several attempts to include “soft decision” decoding into the decoding of RS codes. We do this by combining the SPC code with the RS code, called concatenation. The SPC code has a minimum distance 2, and thus the combination can in principle double the distance of the RS code. This gain is the maximum that we can expect to achieve when using the RS code only and performing “soft decision” RS decoding, which is considered to be impossible due to its complexity and algebraic decoding structure. In our combination, the 3dB gain is obtained by the SPC decoding at the cost of a small rate loss.

### 4.2.1 RS symbols extended with a parity bit

We en- and decode in the following way:

**Encoding** Every symbol of the RS code is extended with a parity bit.

**Decoding** The decoding steps are:
1. apply soft decision decoding to the symbols;
2. decode the symbols with the RS decoder.

Performance can be predicted from the results from SPC decoding.

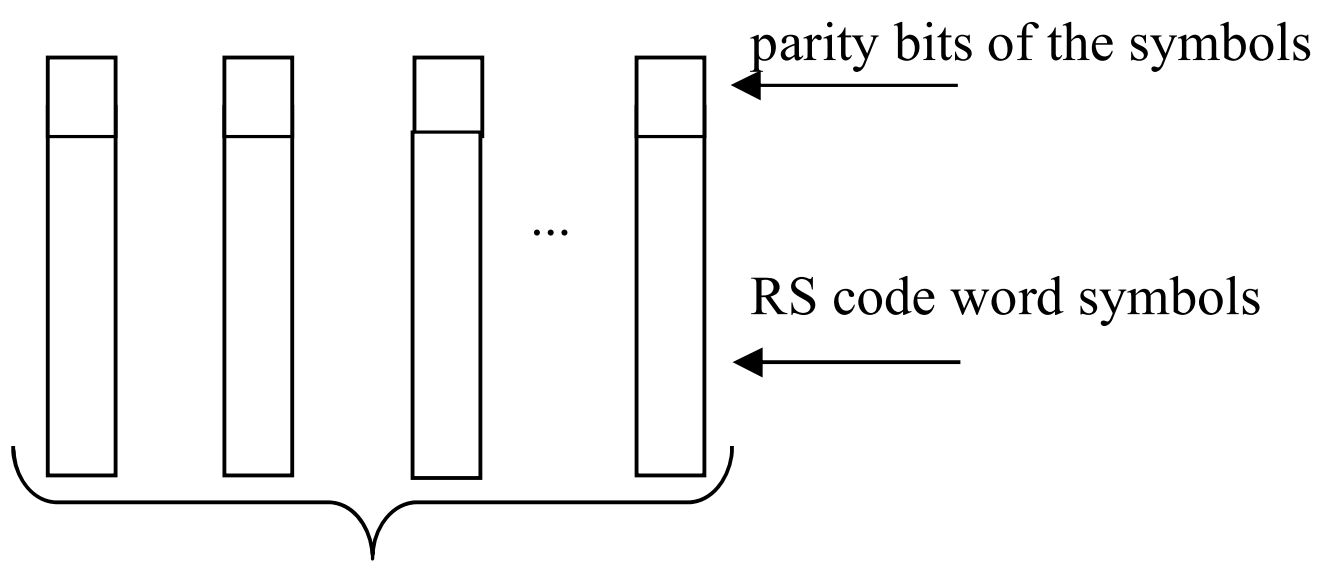

**Figure 4.6** RS code word combined with the SPC

### 4.2.2 Block wise RS-SPC coding

In this scheme, we encode and decode in the following way:

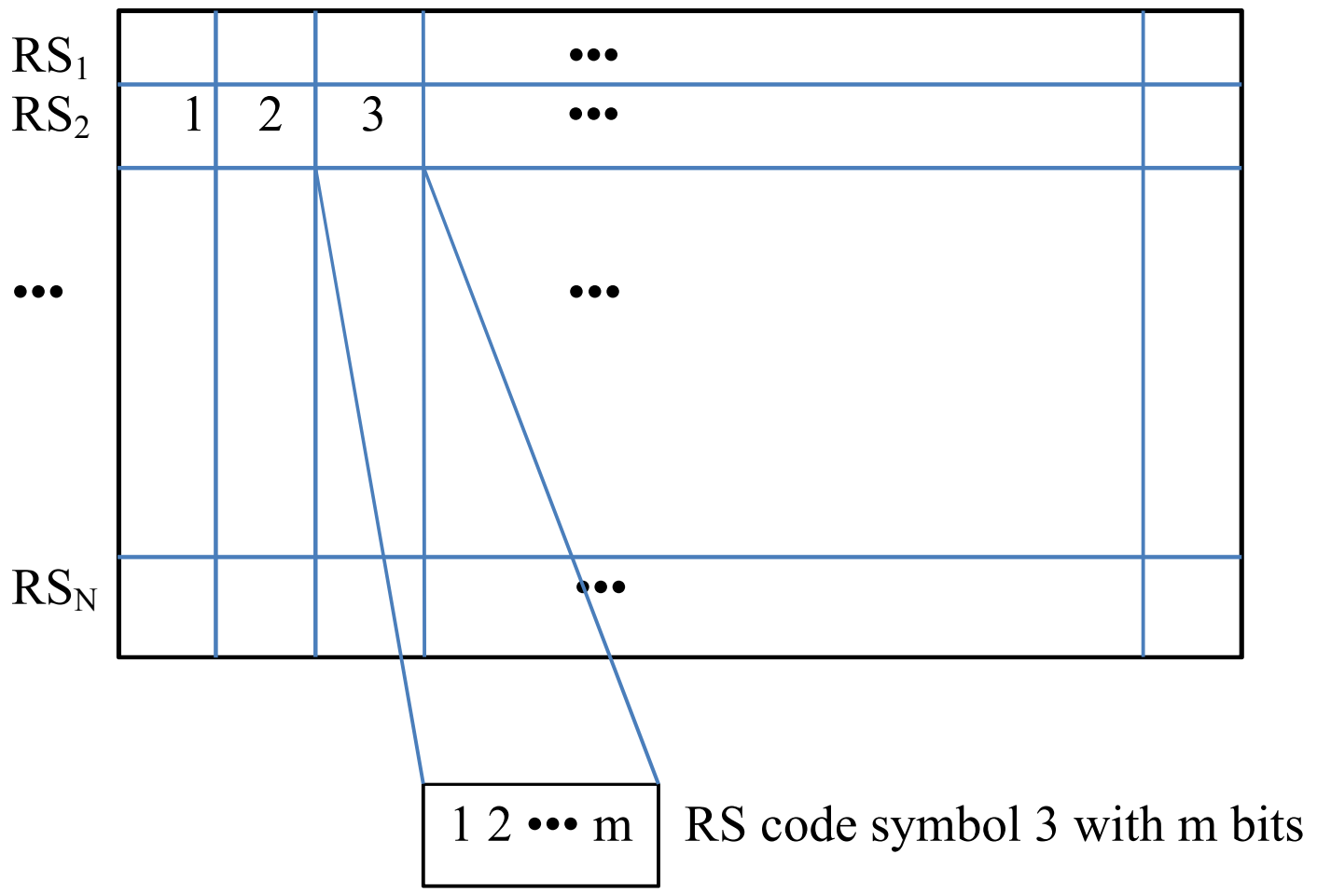

**Figure 4.7** Structure of the concatenated RS-SPC code

**Encoding** We assume transmission of (N-1) RS code words from an (n,k) code over GF($2^m$) followed by a word that is the bit wise modulo-2 sum of the previous words. By linearity, this is also an RS code word.

**Decoding** For each received block of transmitted RS code words, the decoder performs the following three major steps:

1. apply soft decision for each individual column (each column is an SPC code word);
2. apply RS decoding for each row. If a detected, but uncorrectable error pattern occurs, mark the row with a flag;
3. after receiving a block of N words $R_i$, i = 1, ···, N, we calculate the word $E^n$ as the modulo-2 sum of the N rows, i.e.

$$E^n := \sum_{\substack{i=1 \\ R_i \text{ not flagged}}}^{N} R_i \quad .$$

Initially, $E^n = 0^n$.

Four different situations can occur:

- no flag, $E^n = 0^n$. No erasures and no detected errors. Continue with the next block;
- no flag, $E^n \neq 0^n$. In this case we assume that there is only one RS code word in error.
- one flag. Hence, one erasure occurs which can be solved by using the fact that the modulo-2 sum of all the code words is $0^n$ and thus the erased RS word is equal to modulo-2 sum of the non-erased words;

  Suppose that we define the initial block likelihood (equivalent to Euclidean distance) as

$$L_0 = \prod_{k=1}^{N} p(R_k \mid RS_k) \,.$$

  If only one code word is assumed to be in error, we have to add to one of the decoded words the error word $E^n$. The best choice is the one that maximizes the likelihood of the modified block. Hence, the decoder determines the position that maximizes the corresponding likelihood, i.e.

$$\max = \arg \max_{i=1,\cdots,N} p(R_i \mid RS_i \oplus E^n) \times \prod_{\substack{k=1 \\ k \neq i}}^{N} p(R_k \mid RS_k), \qquad (4.2.1)$$

and gives as output $RS_{max} + E^n$ instead of $RS_{max}$ for the RS code word according to equation (4.2.1). The decoder thus calculates

$$L_{max} = p(R_{max} \mid RS_{max} \oplus E^n) \times \prod_{\substack{k=1 \\ k \neq max}}^{N} p(R_k \mid RS_k) \ . \qquad (4.2.2)$$

Note that in (4.2.1) we only have to consider the locations in $RS_k$ where $E^n$ has nonzero components;

- more than one flag. More than one erasure occurs. Since the sum of all RS code words is equal to $0^n$, we can calculate the modulo-2 sum of the erased RS code words. For this block, given $E^n$, we can start again to do the soft column decoding, followed by RS-row decoding. To avoid looping, we put the condition that for the new block, the number of erased RS code words is less than before. If not, decoding is stopped.

### 4.2.3 Performance

In this section, we discuss the performance of the different coding schemes.

A. We first consider RS-SPC (Section 4.2.1) decoding for an AWGN channel with noise variance $N_0/2$. The symbol error rate can be approximated using (4.1.3) as

$$_A P_s = \binom{m+1}{2} Q\left(\sqrt{\frac{2RE_b}{N_0/2}}\right),$$

where $R = \frac{k}{n} \times \frac{m}{m+1}$ denotes the overall code rate, k/n is the code rate of the (n,k) RS code over $GF(2^m)$ and $E_b$ is the transmitted energy per information bit.

B. For the concatenated RS-SPC coding scheme (Section 4.2.2) the symbol error rate for high signal-to-noise ratios is approximately

$$ {}_{B}P_s = m(N-1)Q\left(\sqrt{\frac{4RE_b}{N_0}}\right), $$

with $R = \frac{k}{n} \times \frac{N-1}{N}$. Since the symbol error rate for hard decision RS decoding can be approximated by

$$ P_s = mQ\left(\sqrt{\frac{2\frac{k}{n}E_b}{N_0}}\right), $$

an asymptotic gain due to soft decision decoding of $10\log_{10}(2m/m+1))$ dB (structure A) and $10\log_{10}(2(N-1)/N)$ dB (structure B) is expected for n = 31 (m=5), that is about 2.2 dB.

Since the minimum distance of the concatenated code (structure B) is twice the minimum distance of the (n,k) RS code, an additional gain corresponding to distance doubling can be expected. For exploiting this potential gain erasure RS decoding as given by structure B is employed in addition to soft decision SPC inner decoding.

The simulation results, for the different coding schemes that are based on the use of antipodal signaling with coherent detection in a channel with additive white Gaussian noise, are given in Figure 4.17. The RS code employed is the short (31,21) code. For block structure B, N = 9 is chosen in the simulations; in addition the maximum number of iterations is limited to 3. As can be seen, symbol wise soft decision SPC decoding (curve 3) results in a gain over the hard decision RS decoding of to about 1.5 dB at a bit error rate (BER) of $10^{-5}$: this corresponds to an overall coding gain of about 4 dB. For code structure B the best performance (curve 4) for BER < $10^{-3}$ is obtained. Compared with the symbol wise soft decision SPC decoding, the gain is about ½ dB at a BER = $10^{-5}$.

**Remarks** We applied the same technique to the concatenation of convolutional codes and a SPC code. We consider a concatenated coding scheme that is characterized by a set of N multiplexed convolutional codes forming the inner code and an outer (N,N-1) SPC code. The inner convolutional codes are maximum-likelihood (ML) decoded via N parallel, independently operating Viterbi decoders. The decoded overall parity check sequence is used to correct decoding errors resulting from Viterbi decoding.

This outer soft decision SPC decoding can be realized exploiting the analogue demodulator output used for Viterbi decoding before, with only a small amount of additional hardware [31].

## 4.3 Non-Gaussian Disturbances

Soft decision decoding is based on the fact that the channel noise is Gaussian distributed and additive. Furthermore, the channel parameters must be known to the receiver, otherwise a mismatch will occur. In practice, channel noise can be non-Gaussian and/or non-additive. Examples are narrowband and impulse noise. In addition, channel characteristics might be unknown to the receiver, see [16,24]. We consider MFSK modulation combined with coding as an efficient way to handle these disturbances.

### 4.3.1 MFSK modulation

MFSK modulation schemes [26,27] modulate symbols as one of the sinusoidal waves described by

$$s_i(t) = \sqrt{\frac{2E_s}{T_s}}\cos(2\pi f_i t);\;\; 0 \le t \le T_s\,, \tag{4.3.1}$$

where i = 1, 2, ···, M. The signal energy per modulated symbol is represented by $E_s$ and

$$f_i = f_0 + \frac{i-1}{T_s}\,,\; 1 \le i \le M.$$

The signals are orthogonal and for non-coherent reception the frequencies are spaced by $1/T_s$ Hz. Every $T_s$ second we transmit a new symbol. We restrict ourselves to the ideal MFSK modulation signaling as given in (4.3.1). The bandwidth B required is approximately

$$B = \frac{M}{T_s}.$$

For an information rate of b bits per second, we have a symbol duration time

$$T_S = \frac{\log_2 M}{b},$$

and thus a bandwidth efficiency of

$$\frac{b}{B} = \frac{\log_2 M}{M} \text{ bit/second/Hz.}$$

For large M, MFSK is spectrally inefficient.

MFSK modulation can provide for:

- a constant envelope modulation signal, to enable a transmitter's power amplifier to operate at or near saturation levels;
- frequency spreading to avoid bad parts of the frequency spectrum;
- time spreading to facilitate correction of frequency disturbances and impulse noise simultaneously.

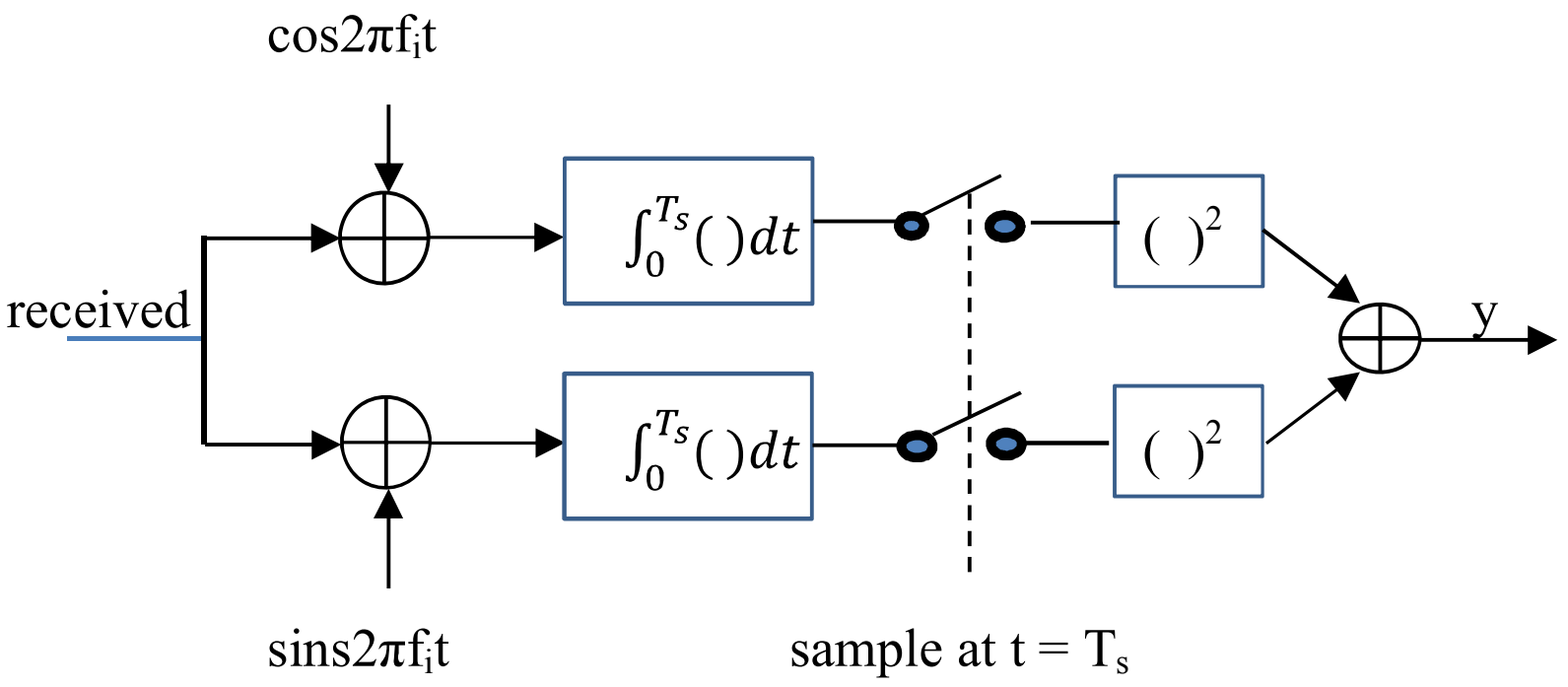

**Figure 4.8** Basic envelope detector for frequency $f_i$

### 4.3.2 Detection

**Non-coherent MFSK detection** is implemented by using a bank of 2M correlators, with a quadrature pair for each frequency. For each quadrature pair the output is added together using the square law to produce a metric for the corresponding frequency candidate, see Figure 4.8. Non-coherent MFSK detection measures received energy for M possible frequencies used. The M detected envelopes can be used in the decoding process in several ways.

**Soft "envelope" detection**
Normally, one chooses from a set of frequencies the one with the highest energy present at a sampling instance $T_s$, assuming that all the M frequencies are transmitted with energy $E_s$. For a fair comparison with un-coded transmission, we set $E_s/\log_2 M = E_b$, where $E_b$ is the energy per information bit. Furthermore, the Signal-to-Noise Ratio (SNR) for such a system is defined as $E_b/N_0$ (refer to [26], p. 258).

The *symbol* error probability for transmission over an additive white Gaussian noise (AWGN) channel with single sided noise power spectral density $N_0$, at high values of $E_s/N_0$, can be approximated as

$$P_{s\text{-uncoded}} \propto M e^{-\frac{E_s}{2N_0}}$$
$$= e^{\log_2 M\,[\ln 2 - \frac{E_b}{2N_0}]}, \qquad (4.3.2)$$

where the energy per bit $E_b = E_s/\log_2 M$. Note that we transmit $\log_2 M$ bits per MFSK symbol. For AWGN channels the probability of a symbol error can thus be made arbitrarily small by increasing M, provided that $E_b/2N_0 > \ln 2$. The cost for increasing M is the bandwidth required to transmit the signals.

In a coded situation, for an (n,k) code where $k = \log_2|C|$, (4.3.2) becomes

$$P_{s\text{-coded}} \propto \log|C|\, e^{-d_{min}\cdot\frac{\log_2|C|}{n}\cdot\frac{E_b}{2N_0}}. \qquad (4.3.3)$$

**Example** Suppose that we have an RS code with M-ary symbols, length M, $|C| = M^k$ and $d_{min} = M - k + 1$. Then, for large M, (4.3.3) is

$$P_{s\text{-coded}} \propto e^{MR\log_2 M[\ln 2 - (1-R)\cdot\frac{E_b}{2N_0}]},$$

where $k/M = R$. For $(1 - R)\cdot E_b/2N_0 > \ln 2$, the error rate goes to zero for increasing M.

**A sub-optimum detector**

As indicated before, spectral noise disturbs the demodulation process of M-ary FSK. To overcome this problem, we give a possible structure for a sub-optimum demodulator using a threshold after every envelope detector, under the assumption of equal energy - and noise distribution for all M frequencies.

We modify the demodulator for our modulation/coding scheme as follows. After every envelope detector we introduce a threshold T. For practical reasons the value of T is set equal to ½$\sqrt{E_S}$ , where $\sqrt{E_S}$ is the envelope detector output for a transmitted frequency when no noise is present. For noisy channels, we have to add a positive constant, which incorporates the variance of the noise, to the threshold [24]. Envelopes that are larger than the respective thresholds are marked with a symbol 1, otherwise we assign the symbol 0. Thus, for a particular time instant, the output of the envelope detector is a binary vector of length M.

We assume that we transmit M-ary code words of length n. These code words can be represented in a binary matrix of dimension M × n, where every column contains exactly one single symbol 1. In a hard decision detector (detecting presence of a frequency), we therefore also put the binary outputs in an M × n decoding matrix. We output the message corresponding to the code word that has the maximum number of agreements with the demodulator output.

The influence of the channel disturbances on the demodulator output can be summarized as:

- <u>narrowband</u> noise causes large demodulator envelopes for particular frequencies hit by the narrowband noise and thus may cause an envelope detector output to be equal to 1 during many time instants. Narrowband noise sources are emergency services, radio amateurs, etc. The disturbance may be intentionally, or caused by an unknown source of interference;

- <u>impulse</u> noise may put energy in larger parts of the spectrum in use for a short period. Impulse noise can also have a periodic character. Impulse noise results mainly from switching transients and is characterized by its duration, amplitude and the inter-arrival time. We assume that impulse noise present at a certain time interval causes all M outputs of the demodulator output column vector to be equal to 1;

- the additive white Gaussian noise, also called the "background" noise, introduces insertion or deletion of a symbol 1;

- in frequency selective fading, some frequencies may be in deep fade, or experience high attenuation. As a consequence, the particular detector output will be 0 for a long time (assuming this is the only disturbance). In this case, an entire row of the decoding matrix is set to 0.

The effect of the different kinds of noise on the multi valued detector output can be seen from Figure 4.9. We assume that M = 4 and transmit the code word (1, 2, 3, 4) as frequencies ($f_1$, $f_2$, $f_3$, $f_4$).

| | | |
|---|---|---|
| 1 0 0 0 | 1 0 1 0 | 1 0 0 0 |
| 0 1 0 0 | 0 1 0 0 | 0 0 0 0 |
| 0 0 1 0 | 0 0 1 0 | 0 0 1 0 |
| 0 0 0 1 | 0 0 0 1 | 0 0 0 1 |
| no noise | background noise | |
| 1 1 1 1 | 1 0 0 1 | 1 0 0 0 |
| 0 1 0 0 | 0 1 0 1 | 0 0 0 0 |
| 0 0 1 0 | 0 0 1 1 | 0 0 1 0 |
| 0 0 0 1 | 0 0 0 1 | 0 0 0 1 |
| narrowband | impulse | fading |

**Figure 4.9** Several types of noise in the channel

As one can see from Figure 4.9, row errors can be catastrophic for RS codes. This can be explained as follows. Let $E_N$ denote the energy detected at the demodulator due to a narrowband noise source. If $E_N$ exceeds the threshold T, the demodulator will have a symbol that is "always on". This poses a problem, since the all-c vector (that is, the vector with all its symbols equal to c, where $c \in GF(2^m)$) is a valid code word for the RS code. Therefore, the presence of narrowband noise disturbances will result in undetected errors. We will discuss further consequences in the next section.

## 4.4 Combined Modulation and Coding

### 4.4.1 Permutation codes

Consider again an M-ary FSK signal set that consists of an orthogonal set of M frequency-shifted signals that is used to encode information. We may assume non-coherent demodulation and thus the bandwidth required is given by $B = M/T_s$. For a code of cardinality |C| and code words of length n, the number of information bits transmitted per second is given by

$$b = \frac{\log_2 |C|}{nT_s} \text{ bits/sec.},$$

where $T_s$ is the signal duration time. The transmission efficiency per second per Herz is given by

$$\frac{b}{B} = \frac{\log_2 |C|}{nM} \text{ bits/sec/Hz.}$$

We consider a special class of codes, permutation codes, that is very well suited for the noise problems mentioned before.

**Definition** A permutation code C consists of |C| code words of length M, where every code word contains M different symbols.

**Example** M = 4, |C| = 4.

| message | transmit |
|---|---|
| 1 | ( 1, 2, 3, 4 ) |
| 2 | ( 2, 1, 4, 3 ) |
| 3 | ( 3, 4, 1, 2 ) |
| 4 | ( 4, 3, 2, 1 ) |

As an example, message 3 is transmitted in time as the series of frequencies ($f_3$, $f_4$, $f_1$, $f_2$). Note that the code C has 4 words with the property that any two words always differ in 4 positions.

**TABLE 1**

Code book for M = 4, $d_{min}$ = 3

| Message | Code word | Message | Code word |
|---|---|---|---|
| 1 | (1, 2, 3, 4) | 7 | (4, 2, 1, 3) |
| 2 | (1, 3, 4, 2) | 8 | (4, 3, 2, 1) |
| 3 | (2, 1, 4, 3) | 9 | (1, 4, 2, 3) |
| 4 | (2, 4, 3, 1) | 10 | (2, 3, 1, 4) |
| 5 | (3, 1, 2, 4) | 11 | (3, 2, 4, 1) |
| 6 | (3, 4, 1, 2) | 12 | (4, 1, 3, 2) |

**TABLE 2**

Two code books for M = 3

| $d_{min}$ = 2 | $d_{min}$ = 3 |
|---|---|
| (1, 2, 3) | (1, 2, 3) |
| (1, 3, 2) | |
| (2, 1, 3) | (2, 3, 1) |
| (2, 3, 1) | |
| (3, 1, 2) | (3, 1, 2) |
| (3, 2, 1) | |

**TABLE 3**

Code book sizes for M = 2, 3, 4, 5

| M | $d_{min}$ = | | | |
|---|---|---|---|---|
| | 2 | 3 | 4 | 5 |
| 2 | 2 | | | |
| 3 | 6 | 3 | | |
| 4 | 24 | 12 | 4 | |
| 5 | 120 | 60 | 20 | 5 |

The code as given in Table 1 has 12 words, each with M = 4 different symbols and a minimum difference between any two words or *minimum Hamming distance* $d_{min}$ of 3. For M = 3 we have the code books as given in

Table 2. An interesting problem is the design of codes and the effect of coding on the transmission efficiency. In Table 3 we give the code construction results for $M < 6$. To maximize the efficiency, we have to find the largest $|C|$ for a given M and $d_{min}$. It is easy to see that for a code with code words of length M each having M different numbers $d_{min} \geq 2$. The cardinality $|C|$ of this code is M!. Hence, the bandwidth efficiency can be defined as

$$\frac{b}{B} = \frac{\log_2 M!}{M^2} \approx \frac{\log_2 M}{M},$$

for large M. This is the same efficiency as un-coded M-ary FSK. The next theorem gives an upper bound on the number of code words in a permutation code.

**Theorem** For a permutation code of length M with M different code symbols in every code word and minimum Hamming distance $d_{min}$, the cardinality of the code is upper bounded by

$$|C| \leq \frac{M!}{(d_{min} - 1)!}.$$

**Proof** Choose $(M - d_{min})$ different code symbols and put them at $(M - d_{min})$ different positions. Then, with the remaining $d_{min}$ symbols at the remaining $d_{min}$ positions, we can construct a maximum number of $d_{min}$ code words at a distance $d_{min}$ apart. For every set of positions we can distribute the chosen $(M - d_{min})$ different code symbols in $(M-d_{min})!$ different ways. We can choose

$$\binom{M}{M - d_{min}}$$

different sets of positions. Hence, the overall product gives the maximum number of code words with a minimum Hamming distance $d_{min}$,

$$|C| \leq d_{min} \binom{M}{M - d_{min}} (M - d_{min})!$$
$$= \frac{M!}{(d_{min} - 1)!}.$$

It can be shown mathematically, that for M = 6 and $d_{min} = 5$ the upper bound cannot be met with equality [97]. For $d_{min} = 2$, we always have equality for any M. Blake [28] uses the concept of sharply k-transitive groups to define permutation codes with distance M – k + 1. The structure of all sharply k-transitive groups is known for k ≥ 2. In [28] it is also shown that the group of even permutations on M letters is a permutation code with |C| = M!/2 code words and minimum Hamming distance 3. To find good codes in general appears to be quite difficult [29,74,98].

### 4.4.2 Permutation codes derived from RS codes

We put a constraint on the code words of an RS code over $GF(2^m)$, namely that all symbols in a word must be different. This code will be a permutation code. To construct a simple permutation code we use the first two rows of the extended RS encoding matrix given by

$$G_T = \begin{bmatrix} 1 & 1 & 1 & 1 & \dots & 1 \\ 0 & 1 & \alpha & \alpha^2 & \dots & \alpha^{M-2} \end{bmatrix}.$$

The minimum distance of the code is (M-1), where $M = 2^m$ is the length of the extended code. If only the second row is used to generate code words, the minimum distance is (M-1). Including the first row, the last M-1 columns generate a code with minimum distance (M-2). Since the symbol at the first position is non-zero, the overall minimum distance is (M-1).

**Property** Suppose that for the M-ary information pair (x,y) the information symbol y ≠ 0. Then, the code words contain every symbol from the set $\{0, 1, \alpha, \alpha^2, \cdots, \alpha^{M-2}\}$ only once. Every code word is a permutation of another code word.

This property follows from that fact that multiplication of the second row with a nonzero element of $GF(2^m)$ realizes a permutation of the last (M-1) symbols. Adding the same constant (first row) to all symbols realizes a permutation over all M symbols.

The construction thus generates a permutation code with minimum distance (M-1), and M(M-1) code words, which meets the upper bound with equality. It furthermore allows the decoding using the regular RS decoding algorithms. More constructions can for instance be found in [28,74].

### 4.4.3 Performance for permutation codes

A message is encoded as a code word of length M with the integers 1, 2, ···, M as symbols. The symbols of a code word are transmitted in time as the corresponding frequencies.

**Example** For M = 4 and $d_{min} = 4$, a permanent disturbance (narrowband noise) present at the sub-channel for frequency $f_4$ and transmission of code word (3, 4, 1, 2) could lead to a demodulator output {(3,4), (4), (1,4), (2,4)}.

The decoder compares the demodulator output with all possible transmitted code words. It outputs the code word for which the maximum number of agreements with the symbols at the demodulator output occurs. For the example, all symbols corresponding to code word 3 are present and thus correct decoding follows. Since code words are different in at least $d_{min}$ positions, $d_{min} - 1$ errors of this type still allow correct decoding. The example code has $d_{min} = 4$ and hence, we can tolerate the presence of 3 permanent disturbances in the demodulator output.

**Example** Suppose that we transmit the example code word {3, 4, 1, 2}. If an impulse noise causes all envelopes to be present at three symbol transmissions, then we may have as a demodulator output {(1,2,3,4), (1,2,3,4), (1,2,3,4), (2)}.

Comparing this output with the possible transmitted code words gives a difference (distance) of zero to the correct code word and one to all other code words. Thus, even if three of these multi-valued outputs occur, we are still able to find back the correct code word since there is always a remaining symbol that gives a difference of one to the incorrect code words.

**Example** Background noise degrades performance by introducing unwanted (called insertions) demodulator outputs or by causing the absence (called deletion) of a transmitted frequency in the demodulator output.

For this type of *"threshold"* demodulation, the decoding is still correct for $d_{min} - 1$ errors of the insertion/deletion type, since:

- the absence of a frequency in the demodulator output always reduces the number of agreements between a transmitted code word and the received code word by one. The same is true for the other code words having the same symbol at the same position. If the symbols are different, the number of agreements does not change;

- the appearance of every unwanted output symbol may increase the number of agreements between a wrong code word and the received code word by one. It does not decrease the number of agreements between a transmitted code word and the received code word.

**Example** In frequency selective fading the symbol energy disappears and thus the symbol 0 will be generated by the hard decision detector. The symbol disappears for the correct code word as well as for the competing code words. Correct decoding is possible if less than $d_{min}$ symbols are set to 0.

In conclusion, we can say that the introduced thresholds in the modified demodulator in combination with a code allow the correction of $d_{min} - 1$ incorrect demodulator outputs caused by narrowband, impulse, back-ground noise or frequency selective fading.

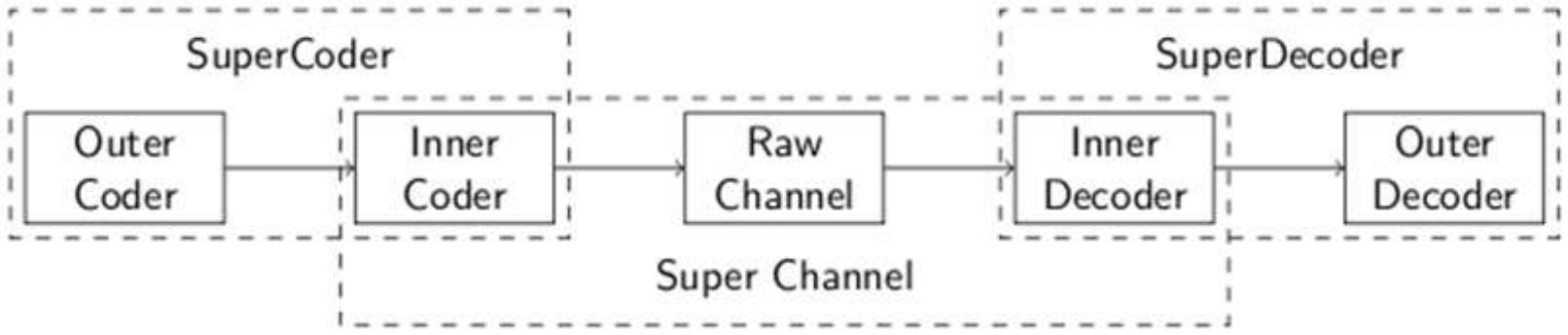

**Figure 4.10** General concatenation from Forney [33]

**Conclusion** We describe a modulation/coding scheme that is capable of handling frequency disturbances like narrowband noise or impulse noise. For this we use the concept of MFSK modulation combined with permutation codes. We modify the "simple" non-coherent demodulator in such a way that the demodulator output becomes multi-valued. The demodulator output is used in the decoding of the permutation code. The described modulation/coding scheme leads to the implementation of an overall robust system. Applications of permutation codes can be found in power line communications [88].

## 4.5 Other Code Combinations

The original idea of code concatenation can be found in the work of David Forney [33]. The raw channel together with an inner decoder behaves like a super channel, see Figure 4.10. It turns the raw channel into another channel, the super-channel, that is suited for the decoding process of the outer code. The goal of concatenation is thus to make the coding schemes suitable for the particular channel conditions.

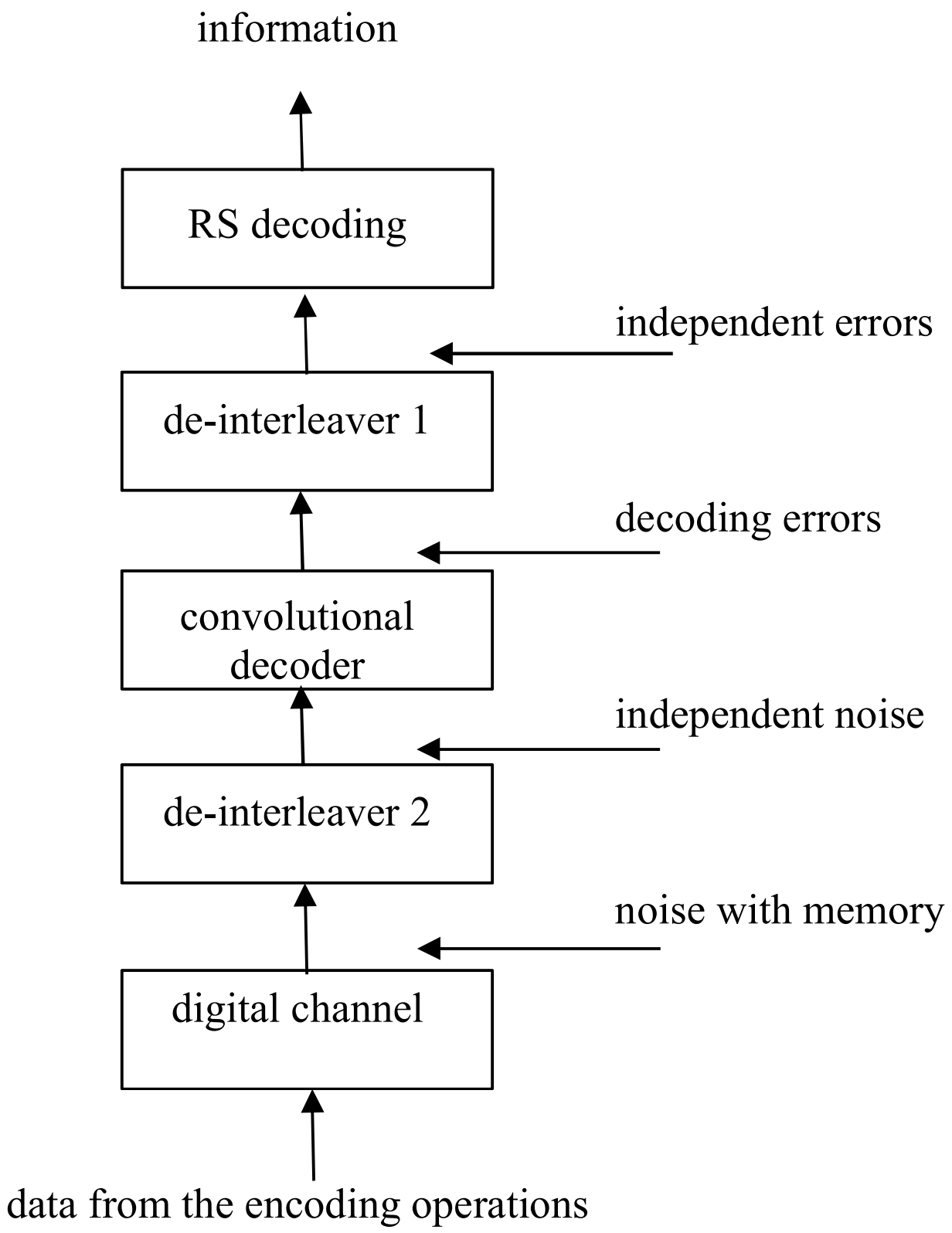

**Figure 4.11** General concatenated coding scheme

### 4.5.1 RS-Convolutional code combination

In Figure 4.11 we give a concatenation scheme with an RS outer code and a convolutional inner code. The de-interleaving is supposed to change an arbitrary channel noise distribution into a distribution that looks completely "random" to the decoder. In this way, only average error probabilities are of importance, not the distribution itself. As a consequence, complicated channel modeling can be avoided. What remains is the design of the interleaving procedure.

The digital channel delivers data for the convolutional decoder. The memory in the data can be removed by the de-interleaver 2, such that the convolutional decoder receives independent symbols. Convolutional decoders are sensitive to dependencies between noise symbols. The convolutional decoder, in general, produces decoding errors that have a burst nature. We collect the output in symbols to be processed by the RS decoding. To be able to decode long bursts of errors, we spread the symbols over different RS decoders. In this way, simple low complexity parallel RS decoders, each decoding a small number of errors, can handle bursts of symbol errors.

If the channel has AWGN, we do not need de-interleaver 2. The convolutional decoder does the soft decision decoding, whereas the RS decoding takes care of the burst decoding errors.

### 4.5.1 RS-constrained code

In recording, we have to generate constrained sequences that can be stored into or written onto a medium. Here, RS codes also play an important role. For disk applications, very often an RS code over GF($2^8$) is used, because the 8-bit symbols can be translated in an efficient way into, for instance, Run Length Limited (RLL) blocks, see [32]. An RLL block puts constraints on the runs of bits. There can be a minimum and a maximum run of the same type of bits for synchronization or control purposes. An encoding scheme is given in Figure 4.12. The idea is that at reading, errors in the RLL sequence produce symbol errors after mapping back to the 8-bit symbols. The RS code can also be combined with other constrained codes, like equal weight codes.

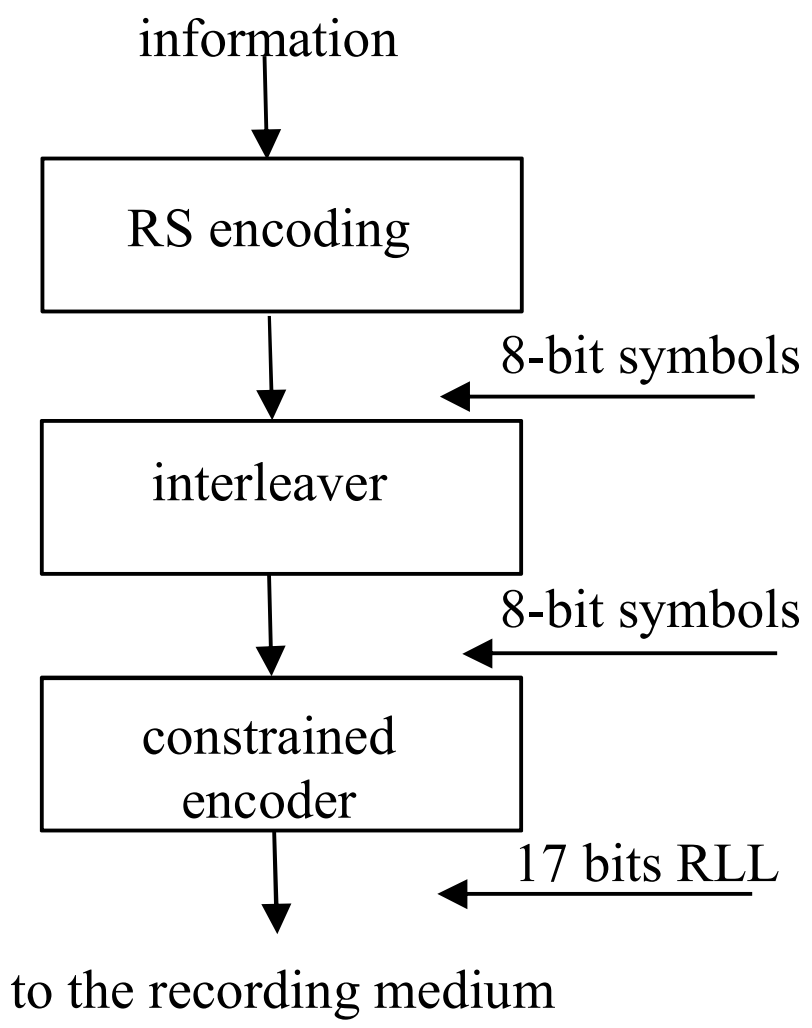

**Figure 4.12** Concatenation of RS with RLL

### 4.5.2 RS Product code

A classical combination of codes is that of a product code. The idea, generated by Peter Elias [34], of a product code is given in Figure 4.13, where code words are formed by n rows and N columns. A column of $k_r$ information symbols is converted into a column of n code symbols. Then, every row with $k_c$ symbols is converted into a row with N symbols. We end with a $n \times N$ matrix. The row and/or the column code can be an RS code.

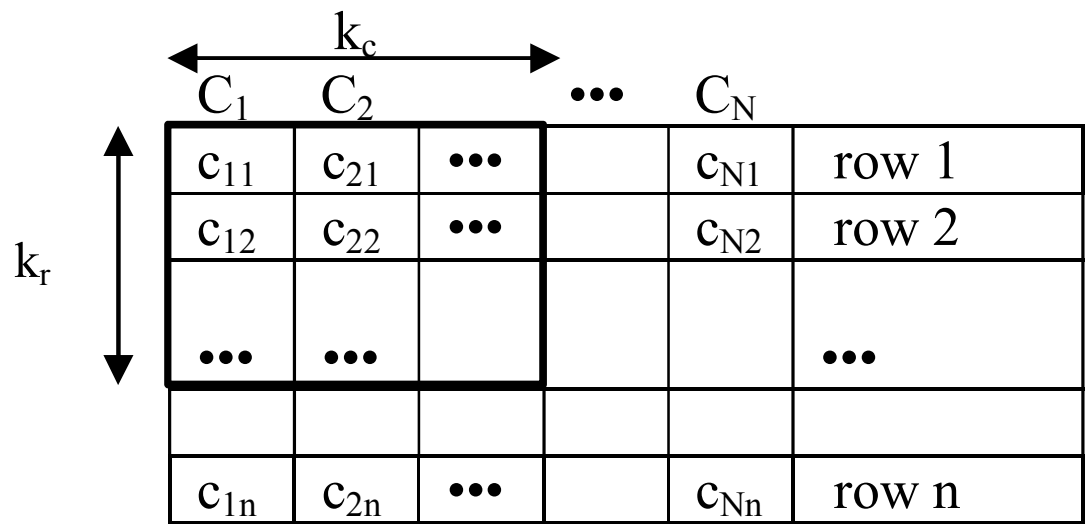

**Figure 4.13** A product code with n rows and N columns

The minimum distance of the product code is larger than or equal to the product of the distances of the individual codes. This can be seen as follows. Any symbol of a nonzero column word of minimum weight $d_c$ gives rise to a row code word with minimum weight $d_r$ and thus the minimum weight of product code word is at least $d_c \times d_r$. Since low weight column code words do not necessarily give low weight row code words, we have a lower bound on the minimum distance. We can see this product code as a generalization of the combination RS-SPC discussed earlier.

**Remark** A row decoding error occurs if at least $d_r/2$ column errors occur. A column error occurs if at least $d_c/2$ errors occur in a column. Hence, we can guarantee correct decoding up to $d_c \times d_r/4$ errors.

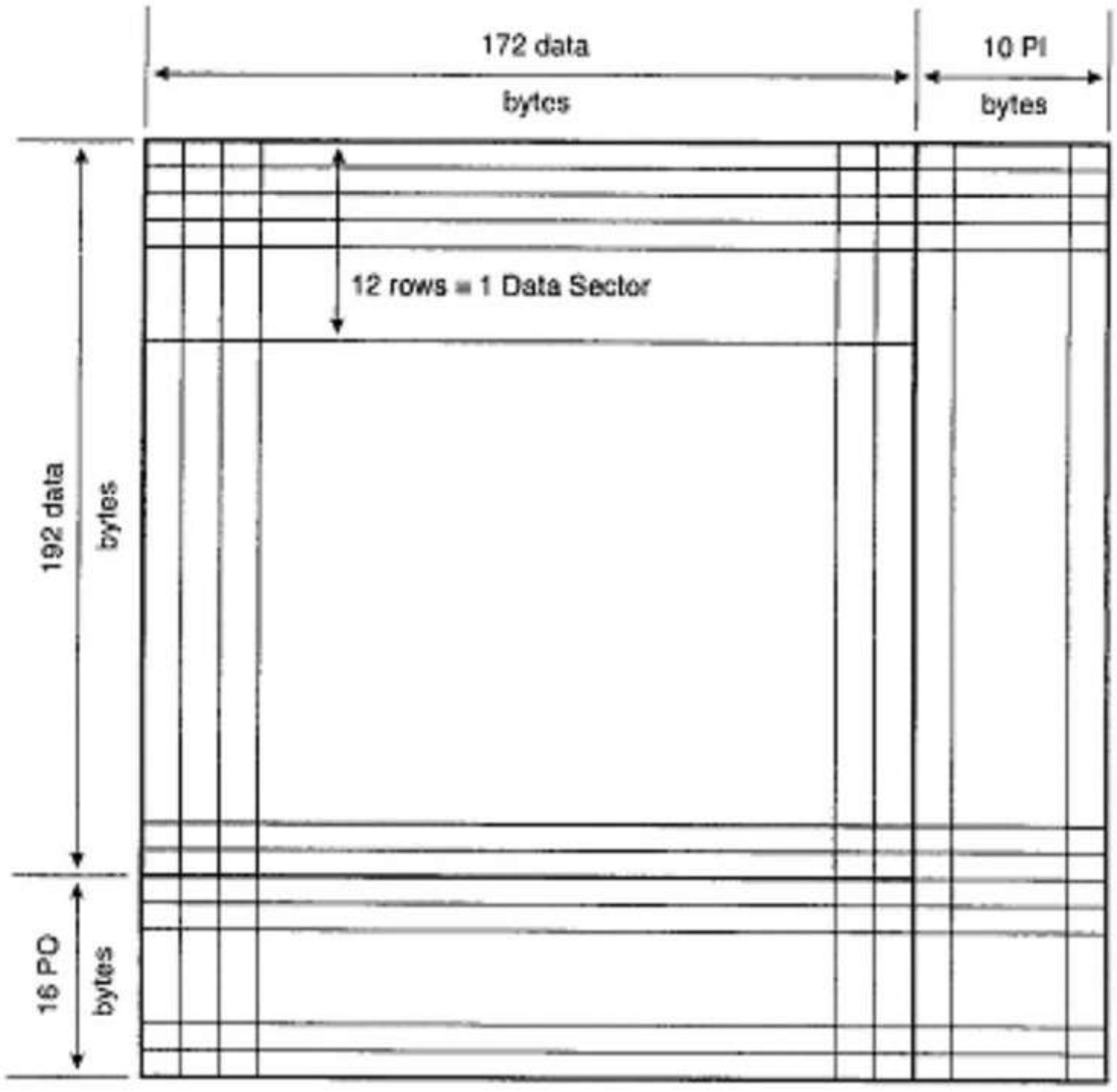

**Figure 4.14** RS product code for DvD applications

In Figure 4.14 we give the RS product code used in DvD. The data symbols are contained in 172 columns and 192 rows. Each of the 172 columns is extended with 16 parity symbols which give a total of 208 rows. Every row is extended with 10 parity symbols which give a total of 182 columns. This procedure gives an RS product code with 208 rows and 182 columns.

If we do a simple row and column decoding, the combination can correct 8 column errors and 5 row errors. By using iterative decoding where row and column decoding is repeated, performance improvements can be obtained. In Figure 4.15 we give a general concept for such an approach (source: Scholarpedia, G.D. Forney)

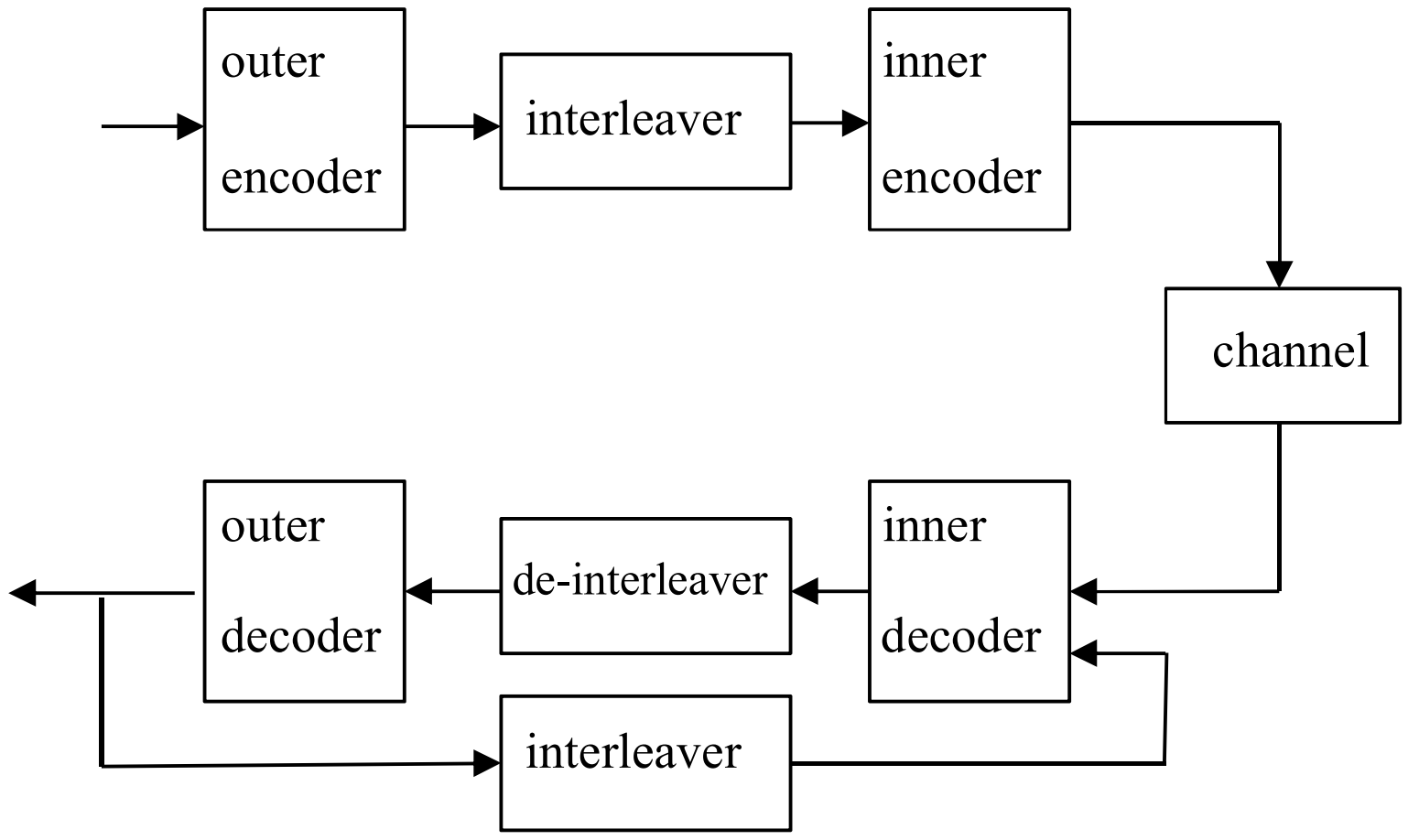

**Figure 4.15** Concept of iterative decoding of concatenated codes

# 4.6 Concluding remarks

Soft decision decoding based on additive white Gaussian noise promises an asymptotic 3 dB gain in the decoding error probability. This is equivalent to a distance doubling using the regular hard decision decoding. Therefore, it is interesting to see how this gain can be realized. The gain for a simple R = 7/8 SPC code is about 2 dB[1]. About the same additional gain can be expected for a concatenated SPC-RS coding scheme.

[1] Peter Foerster, Soft Decision Decoding, PhD University of Darmstadt

The problem of the soft decision decoding is the knowledge of the noise source. If the noise is pure AWGN, then the soft decision metrics are correct. However, in practical systems this knowledge is often not available and thus the advantages of soft decision decoding disappear or soft decision decoding even makes the performance worse.

RS codes are symbol oriented codes and thus, in principle the transmission over an AWGN channel with independent transmissions does not exploit the power of an RS code. Therefore, concatenation with a short high rate inner code that performs the soft decision decoding or that decodes single errors is preferred. The RS code can act as an outer code that corrects the symbol errors caused by decoding errors from the inner code or caused by channel symbol errors.

In the second part of the chapter we consider non-Gaussian noise. Practical communication systems using wireless or power line communications are disturbed by impulse or narrowband noise. We chose diversity in time and frequency to facilitate correct decoding. As a modulation scheme we use MFSK for frequency spreading. The permutation code performs the time spreading. We show that several types of noise can be corrected using a simple detection scheme. We give a simple (n-1,n) RS code that can be seen as a permutation code with $d_{min}$ = n-1. Further constructions using trellis codes can be found in [95].

We summarize the content of this chapter in Figure 4.16.

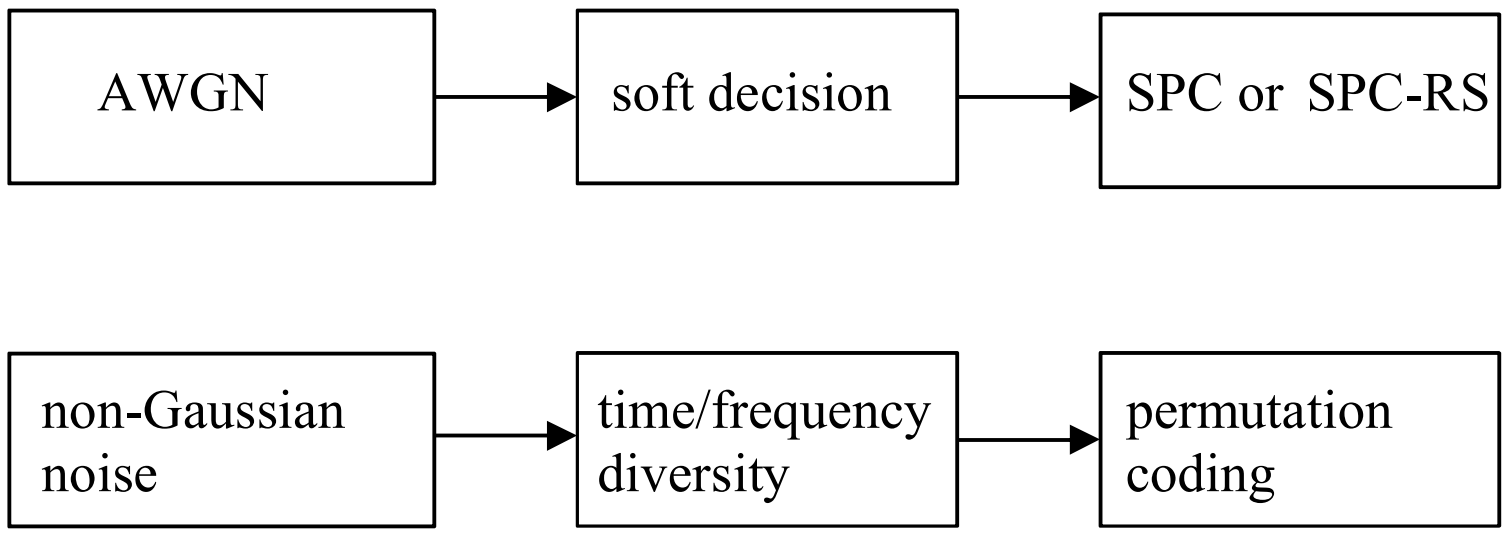

**Figure 4.16** Overview of the content of Chapter 4

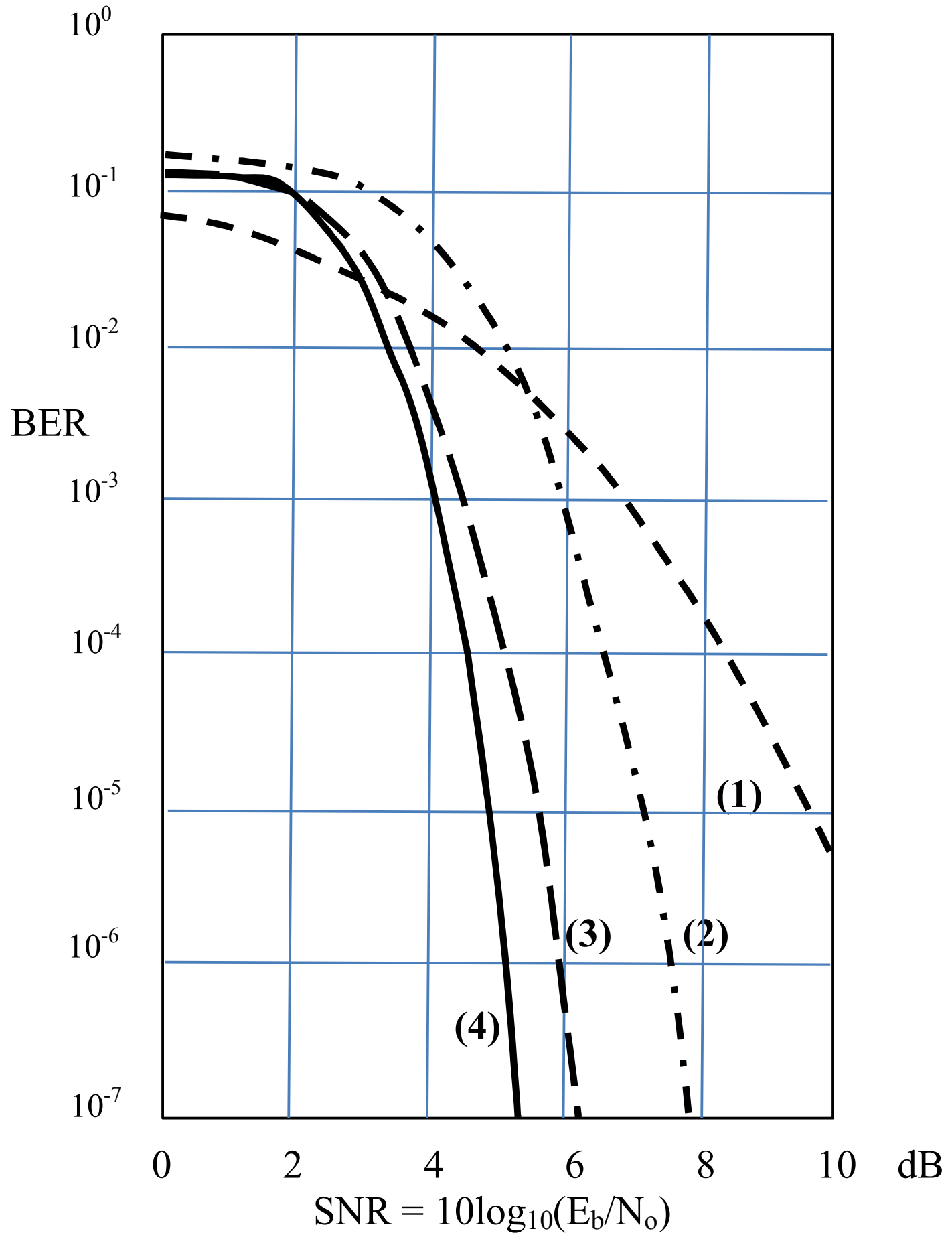

**Figure 4.17** Bit error rate (BER) versus signal-to-noise ratio for antipodal signaling in AWGN channels: (1) un-coded; (2) hard decision (31,21) - RS decoding; (3) concatenated(31,21) - RS - (6,4) SPC soft decoding as in structure A, R = 0.56; (4) concatenated (9,8) SPC, (31,21) - RS decoding, R = 0.6

# Chapter 5

# The Wiretap Channel

A classical problem in data transmission is that of wiretapping. The wiretap channel was introduced by Wyner [35], see Figure 5.0. The idea is that a wiretapper can listen to the normal communication between two parties. The encoder's objective is to maximize the amount of information transmitted to the legal receiver, while keeping the wiretapper as uninformed as possible [39,40]. Another option is the transmission of a secret to the legal receiver without giving any information to the wiretapper. One of the interesting consequences of the developed theory is that there is a close connection between wiretap communication and the problem of secure biometric authentication and verification schemes, see also Chapter 6.

In the wiretap communication model, the message $s^k$ is encoded as $x^n$, transmitted via a noisy channel and received as $y^n$ by the legal receiver. The wiretapper also receives a disturbed word $z^n$.

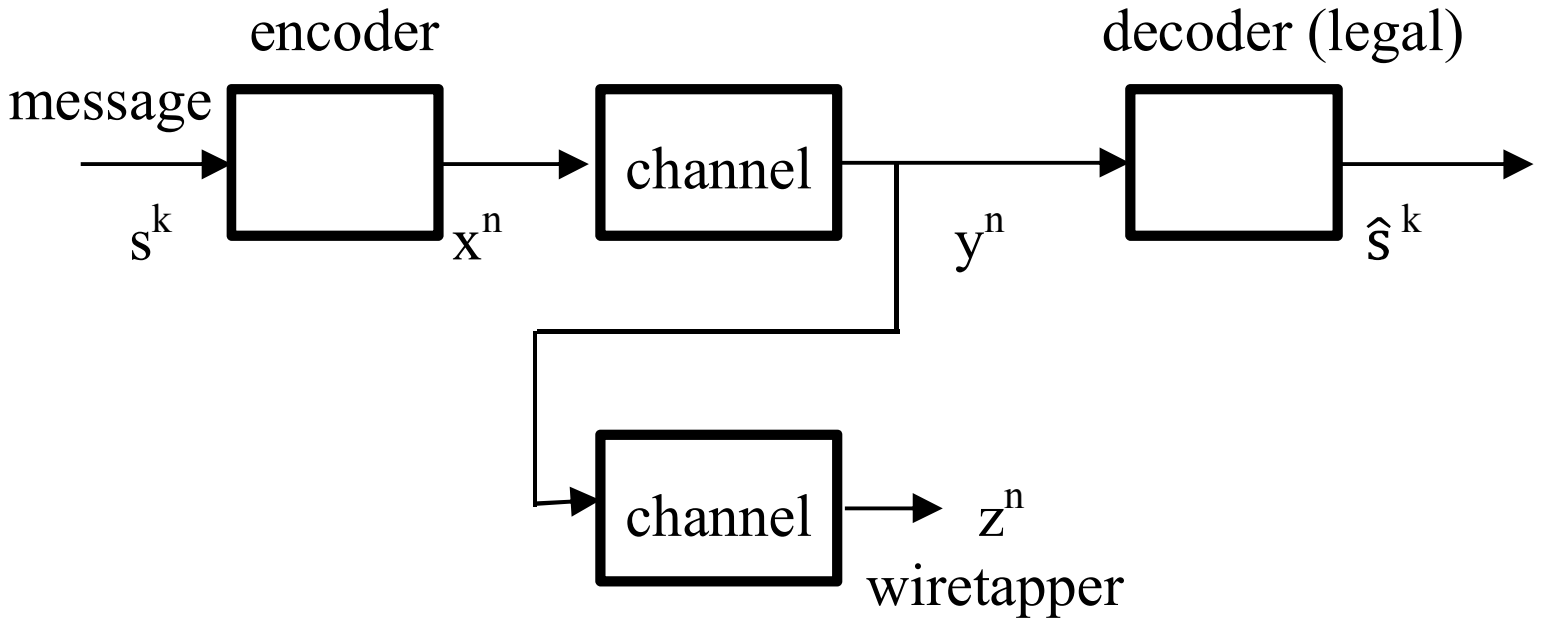

**Figure 5.0** Classical wiretap channel model

Practical encoding schemes are to be designed that achieve the goal of privacy and transmission efficiency.

## 5.1 Wiretap Channel Models

We consider two basic communication models each with a different position for the wiretapper, see Figure 5.1 and Figure 5.2. We only deal with discrete wiretap channels, and we assume that the transmission is binary, using Binary Symmetric Channels (BSC) with cross over or transition probabilities p, q' and q, respectively. For $p < q'$, $p < ½$, the configuration B can be converted into configuration A by substituting $q = (q'- p)/(1 - 2p)$, see Figure 5.3.

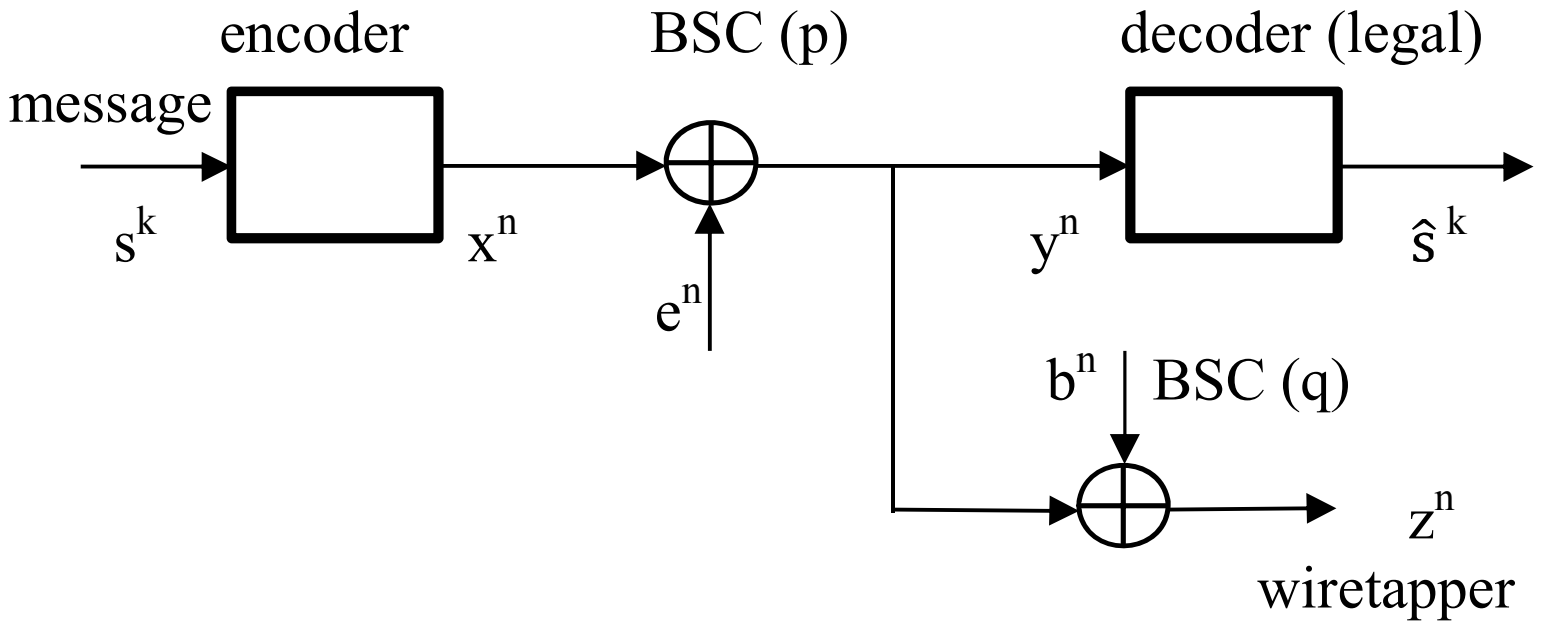

**Figure 5.1** Binary wiretap channel model A

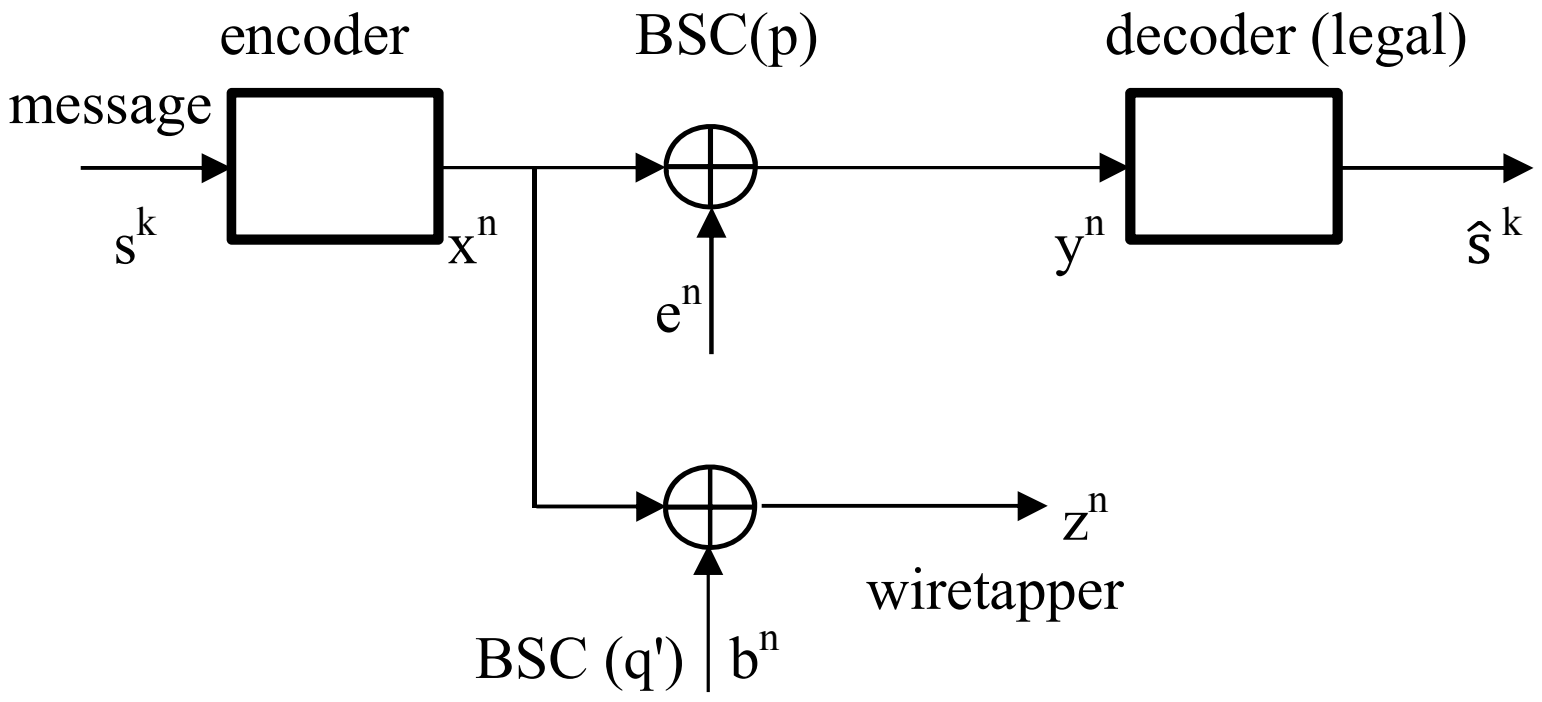

**Figure 5.2** Binary wiretap channel model B

A message $s^k$ of length k bits is transmitted as a code word $x^n$ of length n. The wiretapper, receiving $z^n$, tries to estimate the selected input message $s^k$.

The first problem is the maximization of the amount of transmitted information to the receiver, $I(S^k;Y^n)$ over the main channel from X to Y, while at the same time minimizing the amount of information, $I(S^k;Z^n)$, leaked to the wiretapper. As usual in cryptography, the wiretapper knows the encoding scheme for the transmitter and the decoding scheme used at the receiver.

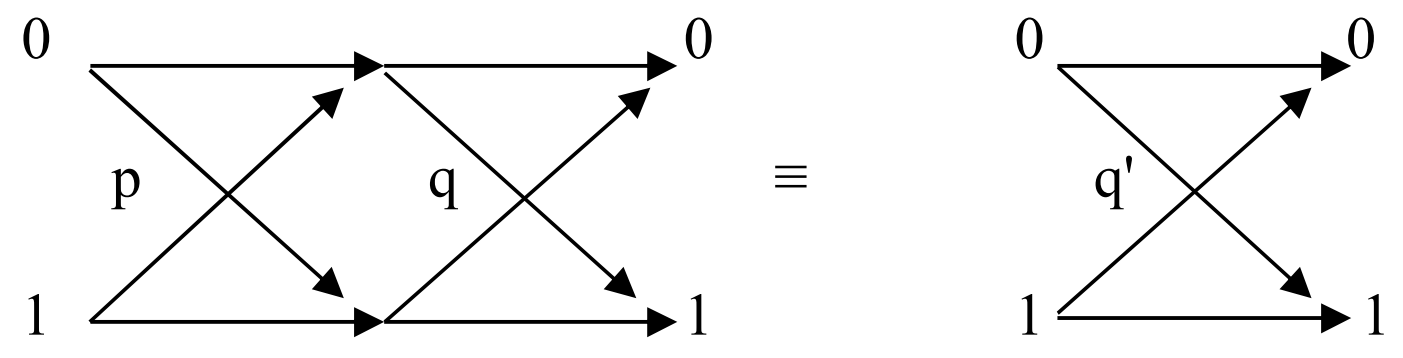

**Figure 5.3** Transformation of two "tandem" channels

In information theory, we express the uncertainty about the message $s^k$, given the received vector $z^n$, by the equivocation $H(S^k | Z^n)$, i.e.

$$H(S^k | Z^n) = H(S^k) - I(S^k;Z^n). \quad (5.1)$$

The information $s^k$ is encoded as a code word $x^n$ and we assume that this code word is transmitted to the legal receiver at an efficiency $I(X^n;Y^n)$. The mutual information $I(S^k;Z^n) = I(X^n;Z^n)$ when we assume that every $s^k$ is uniquely connected to $x^n$ (bijection). Hence, (5.1) can be written as

$$H(S^k | Z^n) = I(X^n;Y^n) - I(X^n;Z^n). \quad (5.2)$$

The goal of the transmitter is thus to maximize $H(S^k | Z^n)$. It has been shown, that for a binary symmetric main and a binary symmetric wiretap channel the maximum is given by the secrecy capacity $C_s$,

$$
\begin{aligned}
C_s &= \max_{P(X^n)} \left[ I(X^n;Y^n) - I(X^n;Z^n) \right] \\
&= \max_{P(X^n)} I(X^n;Y^n) - \max_{P(X^n)} I(X^n;Z^n) \qquad (5.3) \\
&= n\left[C_{main} - C_{wt}\right],
\end{aligned}
$$

where we assume that the main channel is less noisy than the wiretap, i.e. $I(X^n;Y^n) \geq I(X^n;Z^n)$ for all input distributions $P(X^n)$. We can reformulate (5.3) as

$$
\begin{aligned}
&R \times \Delta \leq C_{main} - C_{wt}, \\
&R := \frac{H(S^k)}{n}, \qquad (5.4) \\
&\Delta := \frac{H(S^k \mid Z^n)}{H(S^k)}.
\end{aligned}
$$

Figure 5.4 gives a graphical illustration of (5.3).

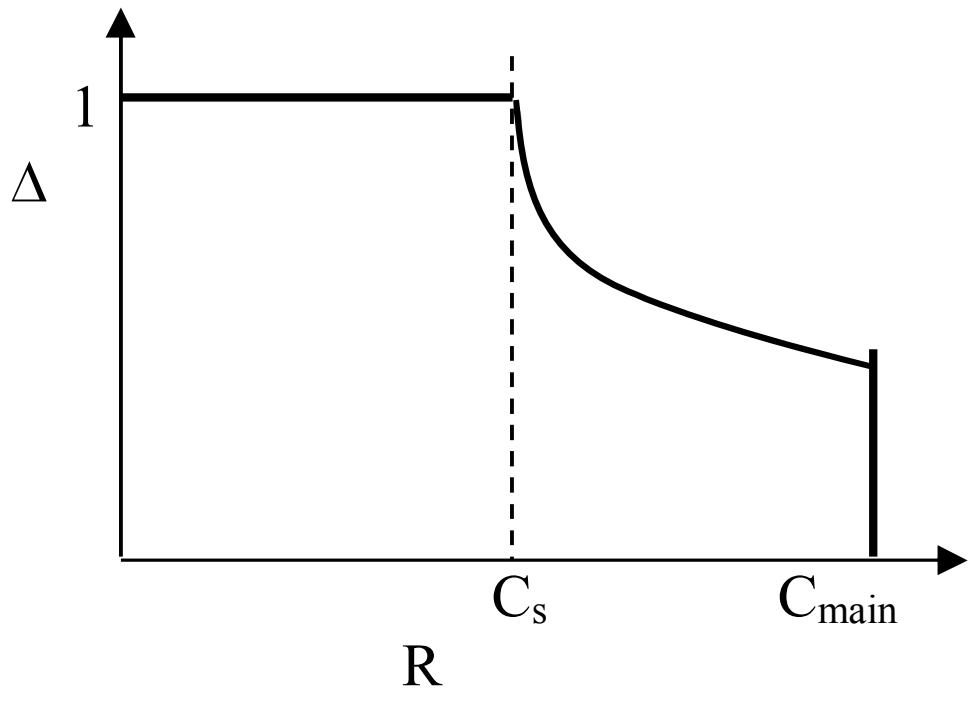

**Figure 5.4** Information rate R versus equivocation Δ

The secrecy capacity $C_s$ is thus the maximum difference between the information transmitted to the legal receiver and the information transmitted to the wiretapper.

For binary symmetric channels, the normalized capacity for the main channel with transition probability p is $C_{main} = 1 - h(p)$ bit/transmission, where h(p) is the binary entropy function. For the wiretap channel the normalized capacity is $C_{wt} = 1 – h(q)$ bit/transmission. Hence, according to the definition, the normalized secrecy capacity

$$C_s = h(q) - h(p) \text{ bit/transmission.} \tag{5.5}$$

The interpretation is, that the wiretapper, after observing a block of n received symbols, has an uncertainty of n(h(q) - h(p)) bits for the transmitted message.

In Chapter 6, we describe the concept of biometric authentication. This concept can be modeled as a wiretap channel. Performance improvements can be obtained by using the fact that the legal receiver can also observe the data received by the wiretapper. The legal receiver thus has the availability of two parallel channels: from X to Y and from X to Z, see also Figure 6.5. We therefore reformulate the secrecy capacity for the binary symmetric channels as

$$\begin{aligned} C_s^+ &:= \max_{P(X^n)} [I(X^n;Z^nY^n) - I(X^n;Z^n)] \\ &= \max_{P(X^n)} [I(X^n;Y^n) + I(X^n;Z^n \mid Y^n) - I(X^n;Z^n)] \\ &= \max_{P(X^n)} \left[ I(X^n;Y^n) - I(X^n;Z^n) + H(Z^n \mid Y^n) - H(Z^n \mid X^nY^n) \right], \end{aligned}$$

which enlarges the secrecy capacity as given in (5.3), since conditioning makes the entropy smaller. For i.i.d. inputs X, where P(x=0) = P(x=1) = ½, the secrecy capacity for the channel from Figure 5.1 is given by

$$\begin{aligned} C_s^+ &= n[H(B \oplus E) - H(E)] \\ &= n[h(p\bar{q}+q\bar{p}) - h(p)], \end{aligned}$$

where $(p\bar{q} + q\bar{p})$ is the probability that $b \oplus e = 1$.

## 5.2 Noiseless Main Channel

We first consider the noiseless main channel, where p = 0 and thus the secrecy capacity is h(q) bits per transmission.

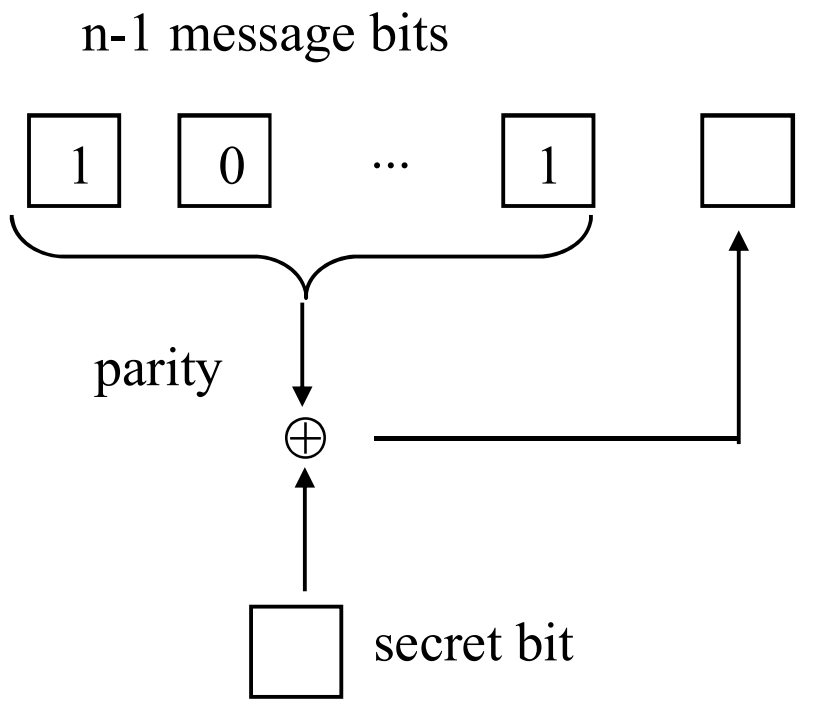

**Figure 5.5** One bit secrecy using the single parity check code

### 5.2.1 Single Parity Check code

For this channel a simple strategy that allows transmission of one secret bit to the legal or main receiver can be explained as follows. The transmitter uses a single parity check code of length n with n-1 information bits and one parity bit. The transmitter adds his secret bit modulo-2 to the parity, see Figure 5.5.

The wiretapper can try to recover the secret bit by adding up, modulo-2, the received n bits. The probability that the secret bit is incorrect is the probability that an odd number of errors occurred during the transmission of the n bits, since this changes the parity of the word, i.e.

$$P_e = 1 - ½ (1+(1-2p)^n),$$

which converges quickly to ½ for appropriate values of n. For $P_e$ = ½, the wiretapper has a probability of ½ of being correct. This is equivalent to an entropy of one bit, which is exactly the purpose of the coding.

**Note** We transmitted (n-1) message bits and one secret bit to the legal receiver. The wiretapper only has a corrupted message (errors occur with probability p) .

### 5.2.2 Hamming code

Before we discuss the application of RS codes, we give an example using the Hamming code.

For the (7,4) Hamming code, we choose the encoding matrix $G_{4,7}$ and the syndrome former $H^T$ as given by

$$G_{4,7} = \begin{pmatrix} 1\,0\,0\,0\,1\,1\,1 \\ 0\,1\,0\,0\,1\,1\,0 \\ 0\,0\,1\,0\,1\,0\,1 \\ 0\,0\,0\,1\,0\,1\,1 \end{pmatrix}, \quad H^T = \begin{bmatrix} 1\,1\,1 \\ 1\,1\,0 \\ 1\,0\,1 \\ 0\,1\,1 \\ 1\,0\,0 \\ 0\,1\,0 \\ 0\,0\,1 \end{bmatrix}.$$

A message $m^4$ is encoded with $G_{4,7}$ and a secret of three bits, $s^3$, is added to the last bits of the code words. Since the legal receiver has a noiseless channel it can recover the message and secret without any decoding errors.

The wiretapper obtains

$$r^7 = m^4 G_{4,7} + (0^4, s^3) + e^7.$$

To recover the secret, the wiretapper may calculate the syndrome

$$\begin{aligned} z^3 &= r^7 H^T \\ &= e^7 H^T + s^3 . \end{aligned}$$

After calculating the syndrome at the receiver, we estimate the secret $\hat{s}^3$ equal to the syndrome $z^3$. This is equivalent to the estimation of the most likely noise sequence $e^7 = 0^7$, the all-zero sequence. The probability of error is then equal to the probability that a noise sequence occurs that is not equal to a code word, since this would give the incorrect estimate. Hence,

$$P_e = \text{Probability}(e^7 H^T \neq 0^3)$$

$$= 1 - \sum_{C} A_i p^i (1-p)^{n-i} , \tag{5.6}$$

where $A_i$ is the number of code words of weight i. For the Hamming code the weight distribution is known and can be used to estimate the error rate. For the sixteen possible code words, we have $A_0 = 1$, $A_3 = A_4 = 7$, and $A_7 = 1$. A first approximation for $P_e$ gives $P_e \approx 1 - (1 - p)^7$.

**Note** The wiretapper can of course use another strategy that gives him a lower error rate.

In general, for linear codes, we encode the information $m^k$ and secret $s^{n-k}$ as

$$x^n = (m^k, s^{n-k}) \cdot \begin{bmatrix} I_k & T \\ 0 & I_{n-k} \end{bmatrix} ,$$

where we use the systematic encoding matrix $G_{k,n} = [I_k, T]$. We continue with RS codes using the general encoding method.

### 5.2.3 Reed-Solomon codes

We investigate the more complicated situation where we transmit symbols.

Suppose that we use a systematic symbol error correcting RS code over $GF(2^m)$ with a redundancy of (n-k) symbols. The equally probable messages $m^k$ are encoded as $m^k G_{k,n}$, where $G_{k,n}$ is the systematic encoding matrix of an RS code. The schematic representation is given in Figure 5.6. We add to the (n-k) check symbols the equally probable secrets of length (n-k) symbols, represented as $(0^k, s^{n-k})$.

During transmission, we assume that a symbol transmission error occurs with probability $p < ½$. At the receiver side we thus obtain

$$r^n = m^k G_{k,n} + (0^k, s^{n-k}) + e^n.$$

where $e^n$ and $0^k$ are the noise and the all-zero symbol sequence, respectively. As before, the wiretapper calculates the syndrome

$$z^{n-k} = r^n H^T = e^n H^T + s^{n-k}.$$

For equally probable secrets $s^{n-k}$, every syndrome is connected to a secret and thus a decodable noise sequence. The minimum error probability receiver would chose the secret equal to the syndrome, for this connects the secret with the most likely noise sequence, the all-0 sequence. The probability of a decision error can be estimated as the probability that a noise sequence occurs that is not equal to a code word, i.e.

$$P_e = 1 - \sum_C A_i p^i (1-p)^{n-i}$$

$$\geq 1 - (1-p)^n - ((2^m)^k - 1) \cdot p^{d_{min}} \cdot (1-p)^{n - d_{min}}, \tag{5.7}$$

where we lower bounded the error probability by assuming that all non-zero code words have minimum weight $d_{min}$. The bound (5.7) can be close to 1 for large values of n, but needs precise evaluation for specific values of k and n. The uncertainty that remains at the wiretapper is then equal to (n-k) symbols for equally probable messages.

## 5.3 Noisy Main Channel

For the noisy main channel a more complicated situation occurs. We modify the strategy as given in section 5.2. The goal is to transmit a message and a secret to the legal receiver over a noisy main channel, while keeping the wiretapper uninformed about the secret.

Suppose that the maximum fraction of channel symbol errors for the main channel is $p < ½$ and that for the wiretap channel the symbol error fraction is $q > p$. In the encoding procedure, we use an RS code with a redundancy $n - k = 2pn$.

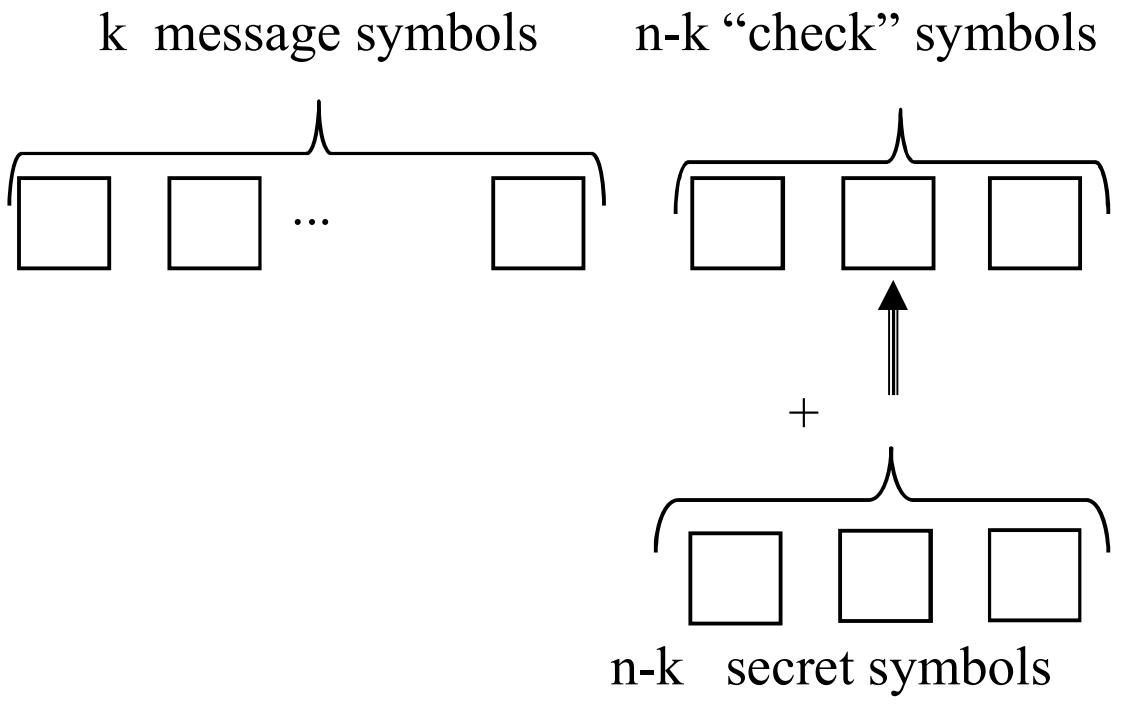

**Figure 5.6** Encoding for (n-k) secrecy symbols.

To achieve the goal, we use the following RS encoding matrix representation. The standard encoding matrix $G_{k,n}$ for an RS code as given in (A.4.1) with dimension $k \times n$, can be brought into semi-systematic form as follows.

**Step 1**. Consider the first u rows and bring this part of the matrix $G_{k,n}$ in systematic form using only the first u rows. The sub-matrix formed by the top u rows in called $G_{u,n}$. The matrix $G_{u,n}$ is an RS encoding matrix with minimum distance $(n – u + 1)$. With the systematic part we can create a zero sub-matrix in the bottom part of $G_{k,n}$. The resulting matrix $\tilde{G}_{k,n}$ is given by

$$
\begin{array}{cccccc}
 & \overset{u}{\longleftrightarrow} & \overset{k-u}{\longleftrightarrow} & \overset{n-k}{\longleftrightarrow} & & \\
\tilde{G}_{k,n} = & \left[\begin{array}{c} I_u \\ \\ 0 \end{array}\right. & \begin{array}{c} A \\ \\ C \end{array} & \left.\begin{array}{c} B \\ \\ D \end{array}\right] & \begin{array}{c} u \\ \\ k-u \end{array} & \begin{array}{c} \updownarrow \\ \updownarrow \end{array}
\end{array} .
$$

$\tilde{G}_{k,n}$ generates an RS code with minimum distance (n-k+1) and therefore, the bottom part must have rank equal to $v = (k - u)$. If not, the minimum distance cannot be equal to (n-k+1). Hence, by using row and column operations we arrive at step 2.

**Step 2.** The bottom v rows of the matrix can be manipulated in such a way that $G_{k,n}$ is equivalent to

$$G_{k,n} \equiv G_{s\text{-}sys} := \begin{bmatrix} G_{u,n} \\ 0 \quad G_{v,n-u} \end{bmatrix} = \begin{bmatrix} \overset{\xleftrightarrow{u}}{I_u} & \overset{\xleftrightarrow{k-u}}{T} & \overset{\xleftrightarrow{n-k}}{U} \\ 0 & I_v & W \end{bmatrix} \begin{matrix} \updownarrow u \\ \updownarrow v = k-u \end{matrix}$$

**Property** The RS codes generated by the matrix $G_{s\text{-}sys}$, $G_{u,n}$ and $G_{v,n-u}$ have minimum distance (n-k+1), (n-u+1) and (n-k+1), respectively.

The transmitted word $c^n$ for the information word $m^u$ and the secret word $s^v$ of length u and v, respectively, is given by

$$c^n = (m^u, s^v) \bullet \begin{bmatrix} I_u & T & U \\ 0 & I_v & W \end{bmatrix}$$

$$= (m^u, m^u T + s^v, m^u U + s^v W).$$

The word $c^n$ is transmitted and corrupted by the noise word $e_1^n$ in the main channel and by the noise word $e_2^n$ in the wiretap channel. When the number of errors in the main channel is less than $\lfloor (n-k)/2 \rfloor$, where k = u + v, these errors are correctable and the legal decoder can recover $m^u$ and $s^v$.

The wiretapper has a channel that is worse than the main channel, and thus is assumed not to be able to decode the message. However, the wiretapper may try to imitate the legal receiver and calculate a syndrome, using the syndrome former $H^T$ for $G_{u,n}$, as

$$z^{n-u} = (c^n + e_2^n) \bullet H^T$$

$$= [(m^u, m^u T + s^v, m^u U + s^v W) + e_2^n] \bullet \begin{bmatrix} T & U+TW \\ I_v & W \\ 0 & I_{n-k} \end{bmatrix}$$

$= (s^v, 0^{n-k}) + e_2^n \bullet H^T$.

To minimize the error probability, the receiver would choose the secret equal to the first v symbols of the syndrome, for this connects the secret with the most likely noise sequence, the all-0 sequence. The probability of a decision error can be estimated as the probability that a noise sequence occurs that is not equal to a code word and is given in (5.7). For $P_e \to 1$, the uncertainty that remains at the wiretapper is then equal to $v = k - u = n(2q - 2p)$ symbols.

The wiretapper may try to profit from the fact that the number of correctable error patterns is less than the number of possible syndromes. However, for code word symbols from $GF(2^m)$, code word length $n = 2^m - 1$, and a redundancy $2qn = n - k + v = n - u$, the number of correctable error patterns can be estimated as

$$| e_2^n | = \sum_{t=0}^{qn} (2^m)^t \binom{n}{t} \approx (n+1)^{2qn} = (n+1)^{n-u}. \quad (5.8)$$

Hence, every possible syndrome pattern corresponds with a correctable error patterns. In fact, for every different $s^v$ a different decodable error pattern can be given. Given $s^v$ and the error pattern, a unique $m^u$ can be determined. Since the $s^v$ word is of length v, we say that the wiretapper has v symbols of uncertainty, where $v = k - u = n(2q - 2p)$. This is the same result as given in the Gelfand-Pinsker [36] bound for the wiretap channel.

$$G = \begin{bmatrix} 1 & 1 & 1 & 1 & 1 & 1 & 1 \\ 1 & \alpha & \alpha^2 & \alpha^3 & \alpha^4 & \alpha^5 & \alpha^6 \\ 1 & \alpha^2 & \alpha^4 & \alpha^6 & \alpha & \alpha^3 & \alpha^5 \end{bmatrix}$$

**Figure 5.10a** RS encoding matrix for n = 7, k = 3

**Example** We conclude with two simple examples of codes, based on the minimum distance 5, two symbol error correcting RS code for $GF(2^3)$ generated by the minimal polynomial $1 + X^2 + X^3$, see appendix A2 and Figures 5.10a, 5.10b, and 5.10c.

$$G = \begin{bmatrix} 1 & 0 & \alpha & \alpha^6 & \alpha^5 & \alpha^2 & \alpha^4 \\ 0 & 1 & \alpha^5 & \alpha^4 & \alpha & \alpha^3 & \alpha^6 \\ 0 & 0 & 1 & \alpha^4 & 1 & \alpha^6 & \alpha^4 \end{bmatrix}$$

**Figure 5.10b** Reed-Solomon encoding matrix for one symbol security

$$G = \begin{bmatrix} 1 & 1 & 1 & 1 & 1 & 1 & 1 \\ 0 & 1 & 0 & \alpha^5 & 1 & \alpha & \alpha \\ 0 & 0 & 1 & \alpha^4 & 1 & \alpha^6 & \alpha^4 \end{bmatrix}$$

**Figure 5.10c** RS encoding matrix for two symbol security

## 5.4 Noiseless Main Channel and Partly Observed Message

Suppose that code words are transmitted over a wiretap channel of type II [36], where the main channel is noiseless, $x^n = y^n$, and the wiretapper can examine a subset $\tau$ of size-$\mu$ of the code word symbols of his choice. Ozarow and Wyner [36] examined this situation and defined the equivocation for the wiretapper as

$$\Delta = \min_{\tau:\mu} H(S^k \mid Z^\mu).$$

The concept of equivocation is equivalent to the average number of source symbols that cannot be found given $\mu$ code word symbols. The minimization in the definition of the equivocation reflects the best possible situation for the wiretapper. A system designer tries to find a coding scheme that provides an optimal tradeoff between the rate of transmission and the equivocation.

We will discuss the properties for the case when we transmit symbols from an RS code over GF($q = 2^m$). To explain the coding strategy, we make use of a systematic RS code with dimensions $k \times n$. The four encoding/decoding steps are:

1. randomly generate a code word;
2. add the secret of length n-k symbol wise to the last n-k code symbols;
3. the legal receiver immediately retrieves the secret by subtracting the code word, identified by the first k symbols, from the received word;
4. the wiretapper tries to find the secret based on a subset $\tau$ of size-$\mu$ of the code word symbols.

Performance can be analyzed using the following observation: Let i and j, $0 \leq i+j = \mu \leq n$, be the number of symbols in $y^n$ known to the wiretapper from the first k and the last n-k symbols (check part), respectively. We distinguish the following two cases.

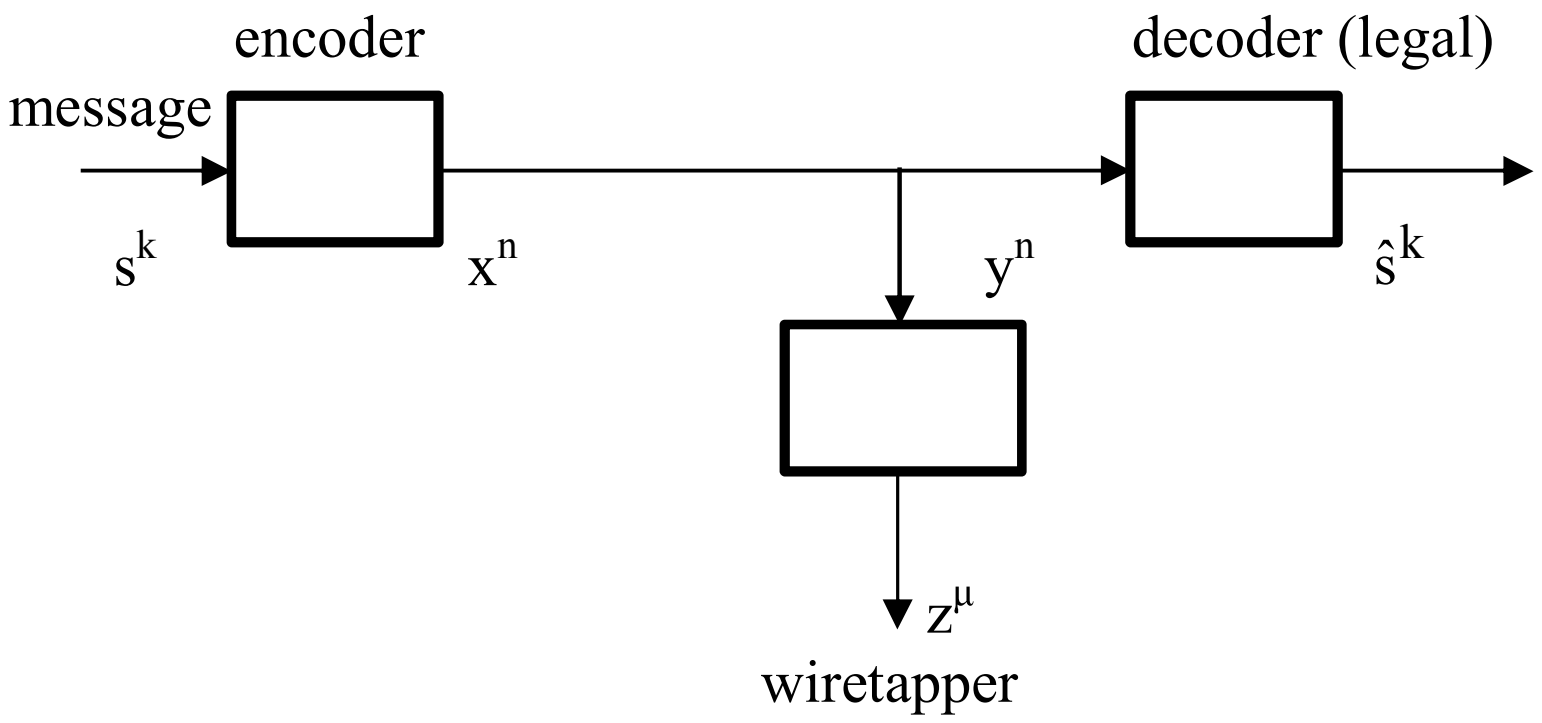

**Figure 5.7** The general wiretap channel II

**Case 1: $\mu \leq k$ (Figure 5.8a)**

For the RS encoding matrix the rank of any k × k sub-matrix is equal to k, and thus there are $q^j$ possible different segments at the j positions, specified by the wiretapper in the check part. Every particular segment, together with the observation, corresponds to j secret symbols. Since there are $q^j$ possible different segments, the wiretapper does not gain any information about the transmitted secret by observing the j symbols. For equally probable code words the total equivocation, or logarithm (base q) of the number of possible segments, is thus $\Delta_1$ = (n-k) symbols. We visualize case 1 in Figure 5.8a.

**Case 2**: **μ > k (Figure 5.8b)**

The wire tapper chooses j positions in the check part. For $k < \mu = i + j \leq n$, we have $j > k - i$. Since k positions uniquely determine the code word there are $j - k + i$ positions with symbol values determined by the specific code word. The total number of unknown symbols, equivocation, for the wire tapper is thus $(k - i) + (n - k) - j = n - (i + j) = n - \mu$. We visualize case 2 in Figure 5.8b.

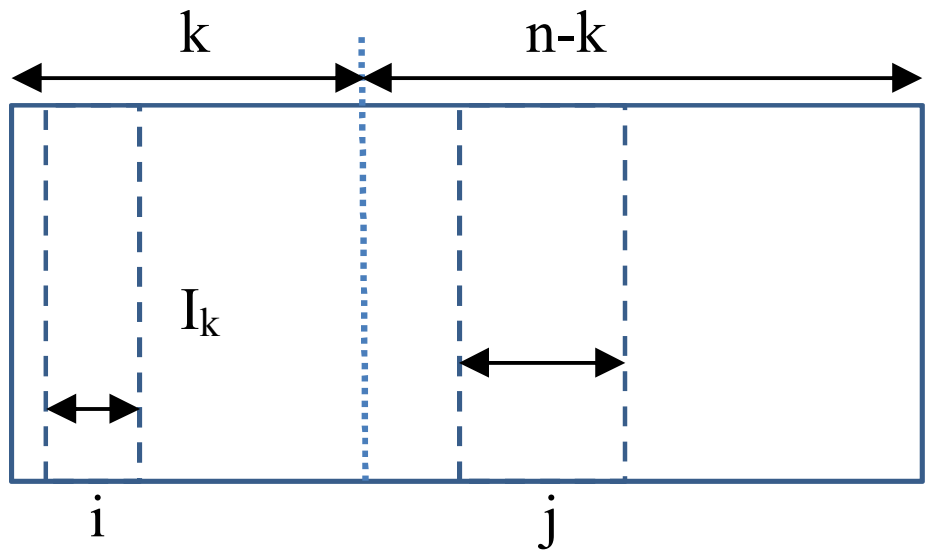

**Figure 5.8a** Selection of $i + j$ columns in G, where $i + j \leq k$

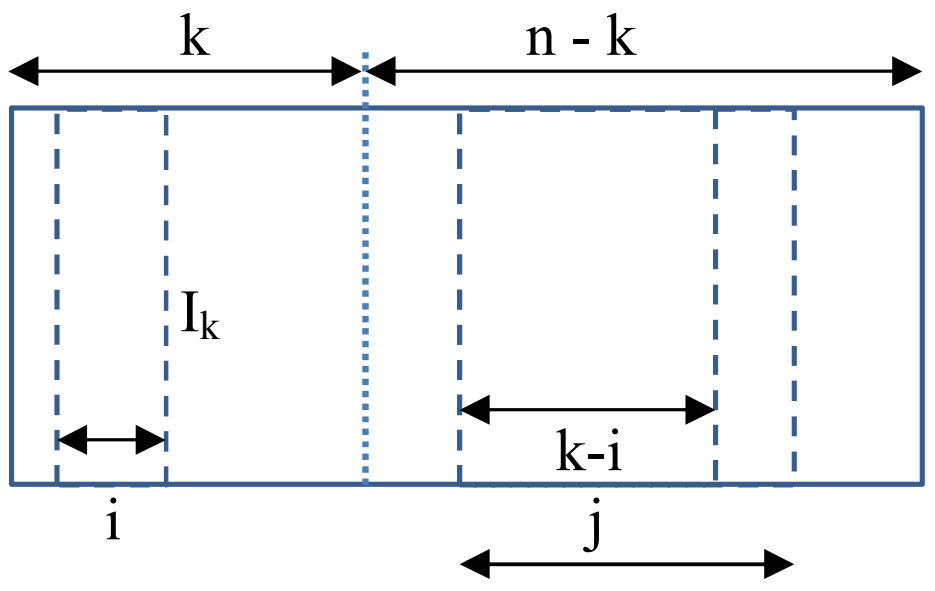

**Figure 5.8b** Selection of $i + j$ columns in G, where $k < i + j \leq n$

For $\mu \leq k$, the equivocation is (n - k) symbols. For $\mu = n$, the equivocation is zero symbols. For $\mu > k$ the equivocation is $n - \mu$. We summarize the results in the "minimum possible equivocation" region as given in Figure 5.9 which was derived by Ozarow and Wyner. We conclude that RS codes perform according to the bound!

**Example** We use the RS code of length 3 over GF($2^2$). The encoding matrix looks like

$$G_{2,3} = \begin{bmatrix} 1 & 0 & \alpha \\ 0 & 1 & \alpha^2 \end{bmatrix}.$$

The word

$$y^3 = x^2 \bullet G_{2,3} + (0, 0, s^1),$$

where $s^1$ is the secret symbol to be protected. It is easy to verify, that the possible observation of $\mu = 1$ symbol by the wiretapper does not reveal the secret symbol.

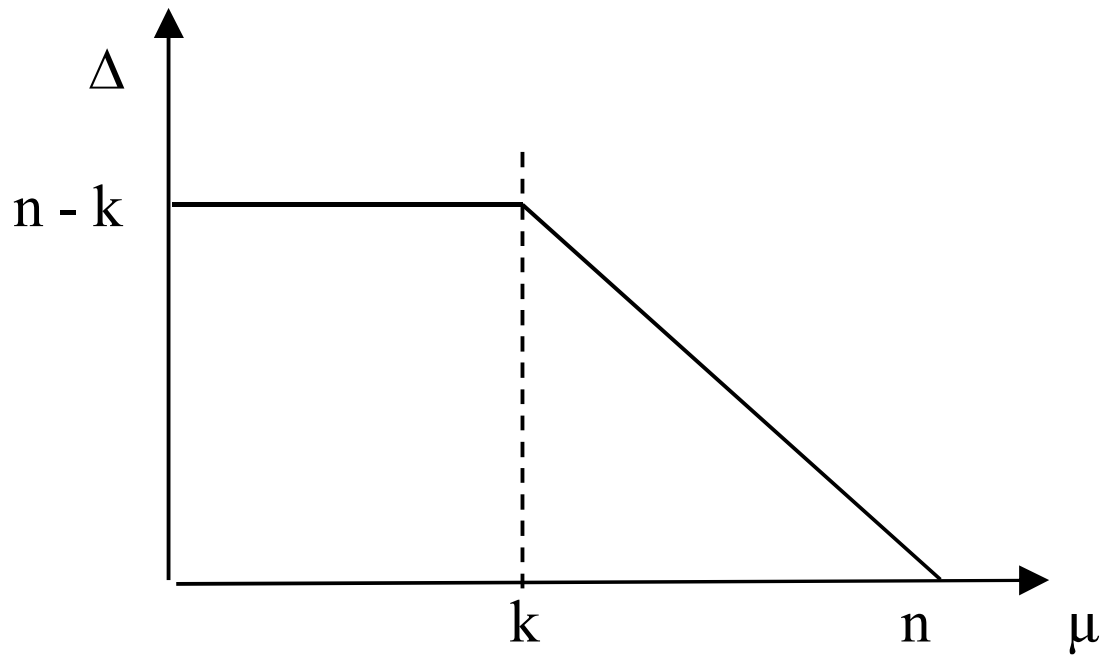

**Figure 5.9** The capacity region for μ observations by the wiretapper

For general linear block codes, the equivocation depends on the rank of the selected sub-matrix. In [37], Forney introduced the concept of Inverse Dimension/Length Profile (IDLP) of an (n,k) linear block code C, which demonstrates the dependency of the equivocation on the structure of the matrix G generating the code. The IDLP of C is defined as a sequence of length n + 1 with components

$$k_\mu(C) = \min_\tau \{ \text{rank}\,[P_\tau(C)] \}, 0 \le \mu \le n,$$

where $P_\tau(C)$ is the selected sub matrix defined by the μ positions. For our example, k(C) = {0, 1, 2}. An upper bound on IDLP is given by Forney in [37] as

$$k_\mu(C) = \begin{cases} k, & k \le \mu \le n \\ \mu, & 0 \le \mu \le k \end{cases} .$$

The RS codes are codes that meet the upper bound. In the situation where the wiretapper has additional information on the encoded data, an extension of the IDLP has been defined in [38].

## 5.5 Concluding Remarks

We describe the classical wiretap channel model with a legal receiver and a wiretapper. For this model we derive the secrecy capacity for the noiseless main channel and for the noisy main channel. We give a coding method using RS codes and compare the performance with the optimum capacity results. As a special case, we explain the idea of a wiretapper observing only a specific number of symbols from the transmitted message to the legal receiver. Here we also conclude that RS codes obtain optimum performance. The content is given in Figure 5.10.

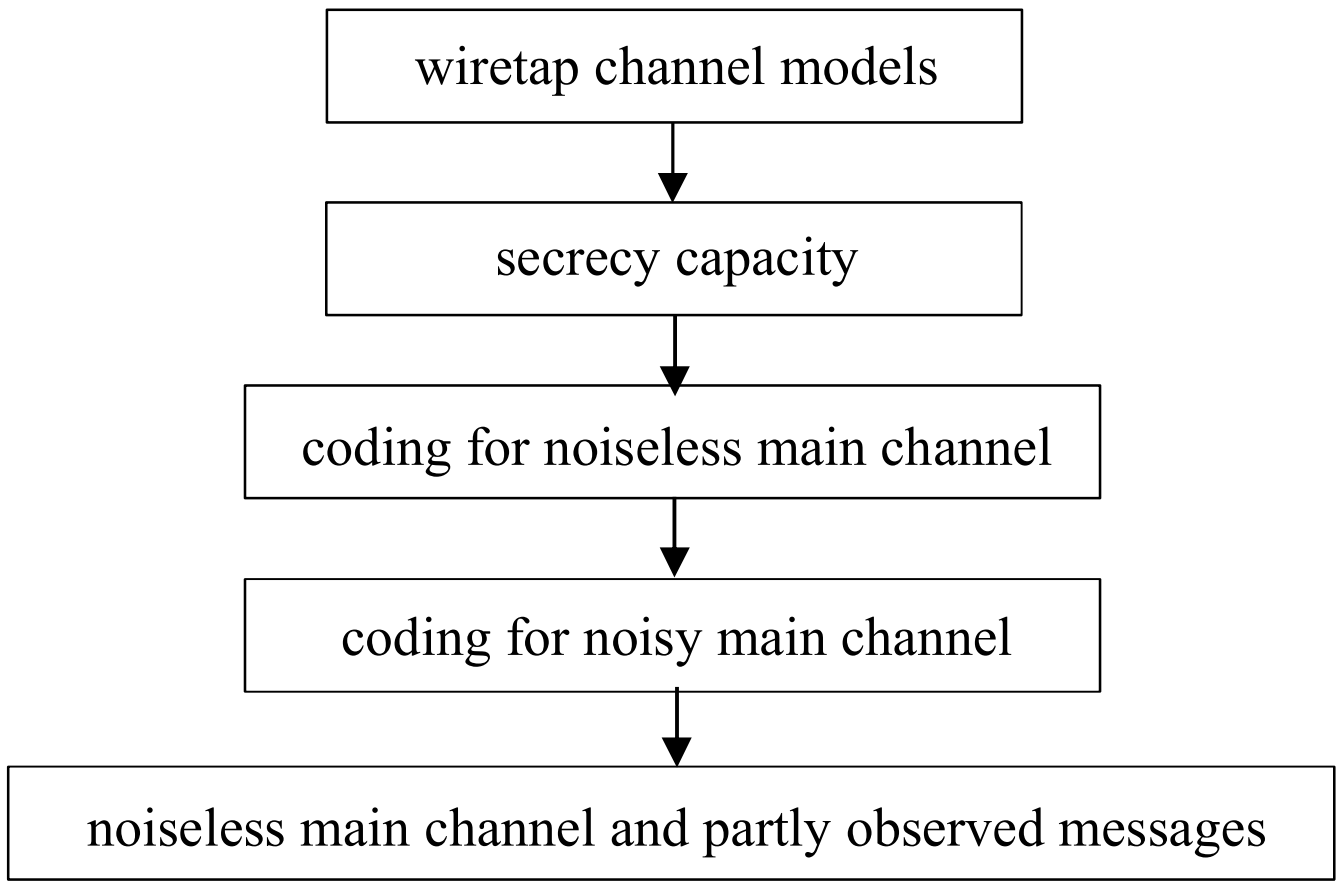

**Figure 5.10** Overview of the content of Chapter 5

An interesting result was obtained by Bin Dai in [89], where he considered the wiretap channel with noiseless feedback. Feedback can increase the secrecy capacity region! Further basic research in this area has been carried out by Chai Mitrpant[1], Yanling Chen[2], and Bin Dai[3]. They showed that side information available at the transmitter enlarges the secrecy capacity region in the Gaussian case, whilst in the discrete case the problem is still open.

[1] Chai Mitrpant, PhD, University Duisburg-Essen, Information Hiding, 2003
[2] Yanling Chen, PhD, University Duisburg-Essen, Wiretap Channel with Side Information, 2007
[3] Bin Dai, PhD, Jiao Tong University, Shanghai, 2013

# Chapter 6

# Biometrics

In Shannon's cryptography landmark paper [50], he defines a cryptosystem as depicted in Figure 6.1. A message m is encrypted with a private key into a cryptogram c. At decryption we use the same key and reconstruct the message m. A cryptosystem is said to have "perfect secrecy" when for every message m and ciphertext c (cryptogram) the probability $P(m \mid c) = P(m)$. In this case, it can be shown that the entropy $H(M) = H(M \mid C)$. In all other cases, the difference $H(M) - H(M \mid C)$ is positive and we call the difference the leakage of a system.

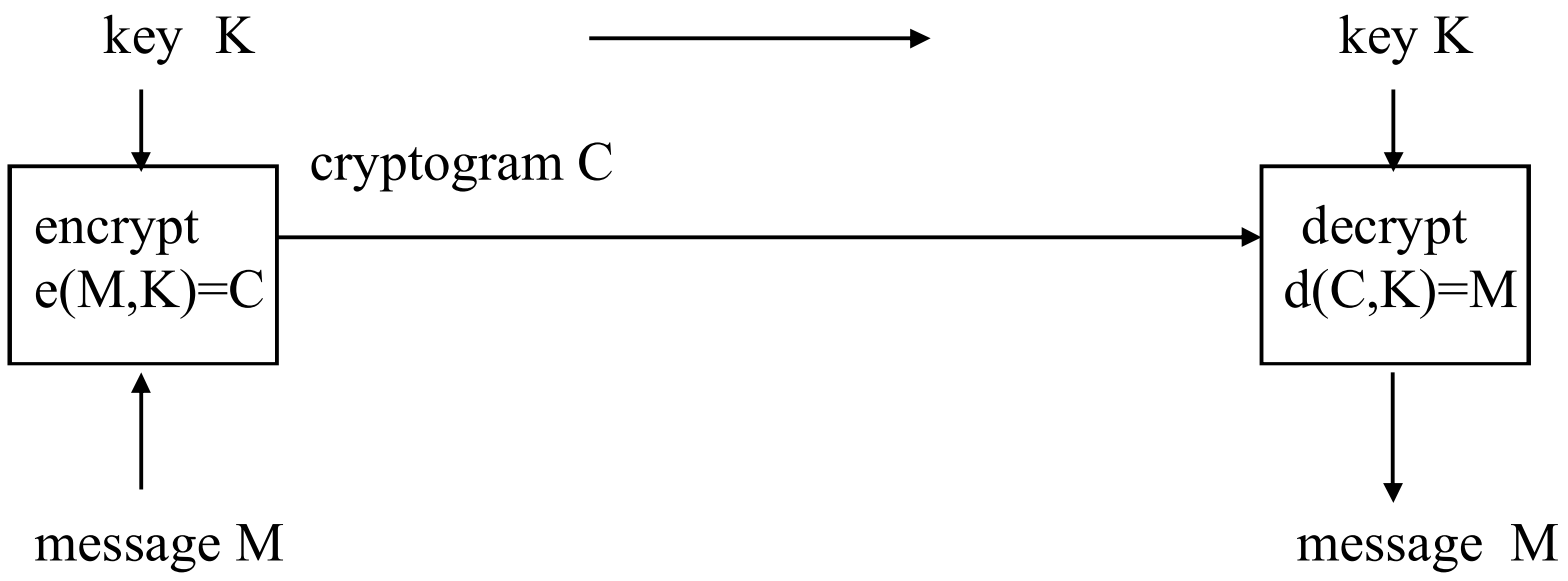

**Figure 6.1** Schematic representation of the Shannon cipher model

Without loss of too much generality, we assume that the message m and cryptogram c are connected via one unique key k. Then, the condition for "perfect security" can be given as

$$H(M) = H(M \mid C) = H(M,C \mid C) = H(K \mid C) \leq H(K).$$

The interpretation is that the minimum average description length of the message must be smaller than that for the key in order to have "perfect secrecy".

In the following situation, we assume that the receiver has a noisy version of the key, see Figure 6.2.

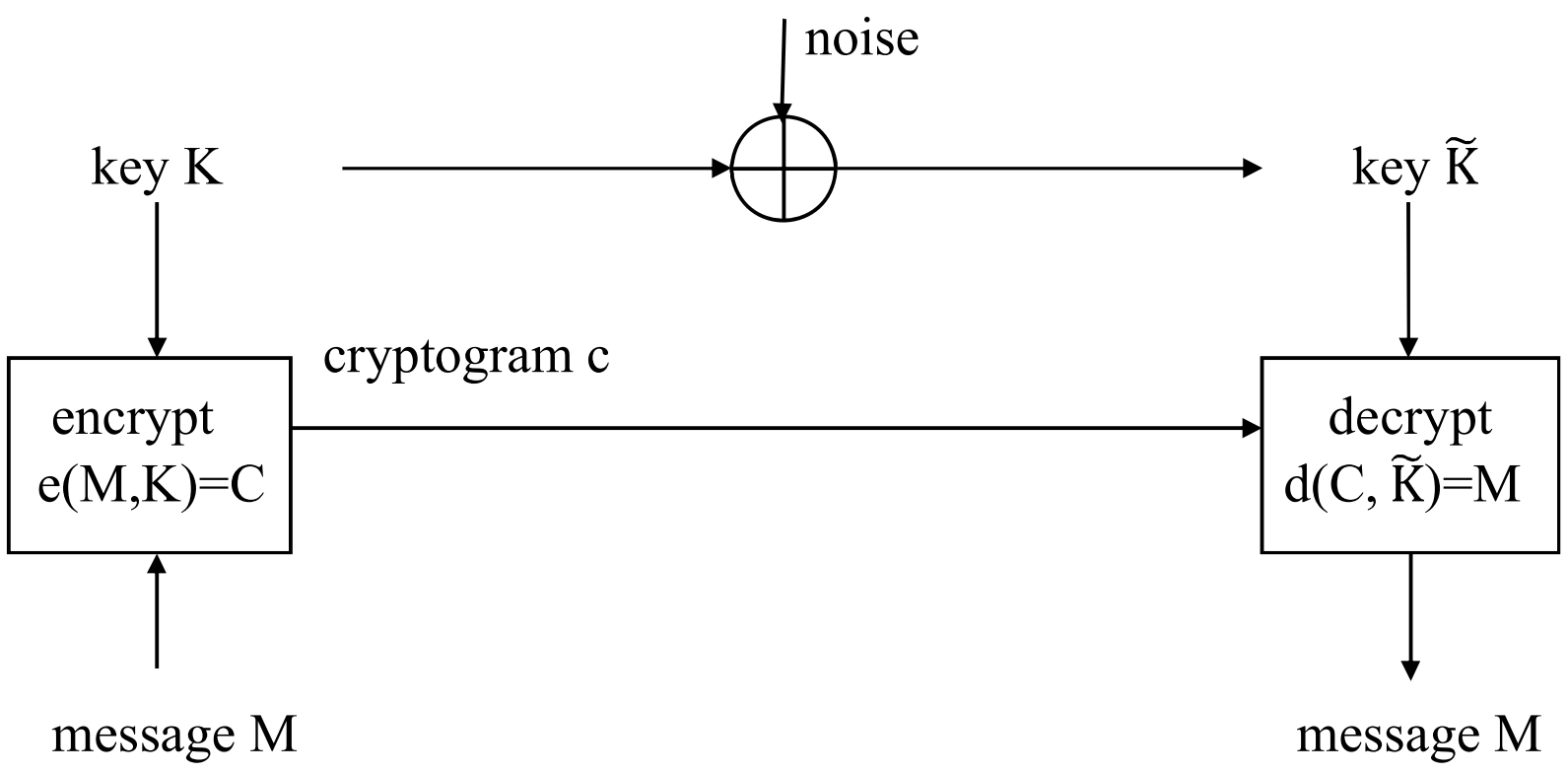

**Figure 6.2** Schematic representation of the noisy cipher system

Again, we assume that a message M and cryptogram C together also uniquely determine the key K. In addition, we expect that the noisy key $\tilde{K}$ and C uniquely determine M. Hence, there is no decryption error. For "perfect secrecy", we have the relation $H(M) = H(M \mid C)$ and thus for $(M,C) \rightarrow K$ and $(\tilde{K},C) \rightarrow M$, we have

$$
\begin{aligned}
H(M \mid C) = H(M,C \mid C) &= H(K \mid C) \\
&= H(M,\tilde{K} \mid C) - H(\tilde{K} \mid M,C) \\
&= H(\tilde{K} \mid C) + H(M \mid \tilde{K},C) - H(\tilde{K} \mid K) \\
&= H(\tilde{K} \mid C) - H(\tilde{K} \mid K) \\
&\leq H(\tilde{K}) - H(\tilde{K} \mid K)
\end{aligned}
$$

$$= H(K) - H(K \mid \tilde{K}). \quad \text{(manipulating the H function)}$$

Hence, for "perfect secrecy", we have the conditions

$$H(M) = H(M \mid C) = H(K \mid C) \leq H(K) - H(K \mid \tilde{K}),$$

where the maximum source entropy and the key entropy, given the cipher text, are reduced with the amount $H(K \mid \tilde{K})$. This is the price for having a noisy key.

## 6.1 Biometric Authentication

Biometric identifiers are the distinctive, measurable characteristics used to label and describe individuals. The biometric authentication scheme as developed by Juels-Wattenberg (JW) [41] can be seen as a noisy cipher system, see Section 6.3. The biometric $b^n$ acts as a binary key of length n at encryption and as a noisy binary key at decryption. The message $m^k$ of k bits is encoded as code word $c^n$ of length n using an error correcting code. The result $(b^n \oplus c^n)$ is stored in a data base (enrollment), where $\oplus$ is modulo-2 addition. At authentication, the code word is retrieved from the data base and added to the "noisy" biometric. The decoder gives the correct $m^k$ in case the number of errors is within the error correcting capability of the code. We will analyze the performance later.

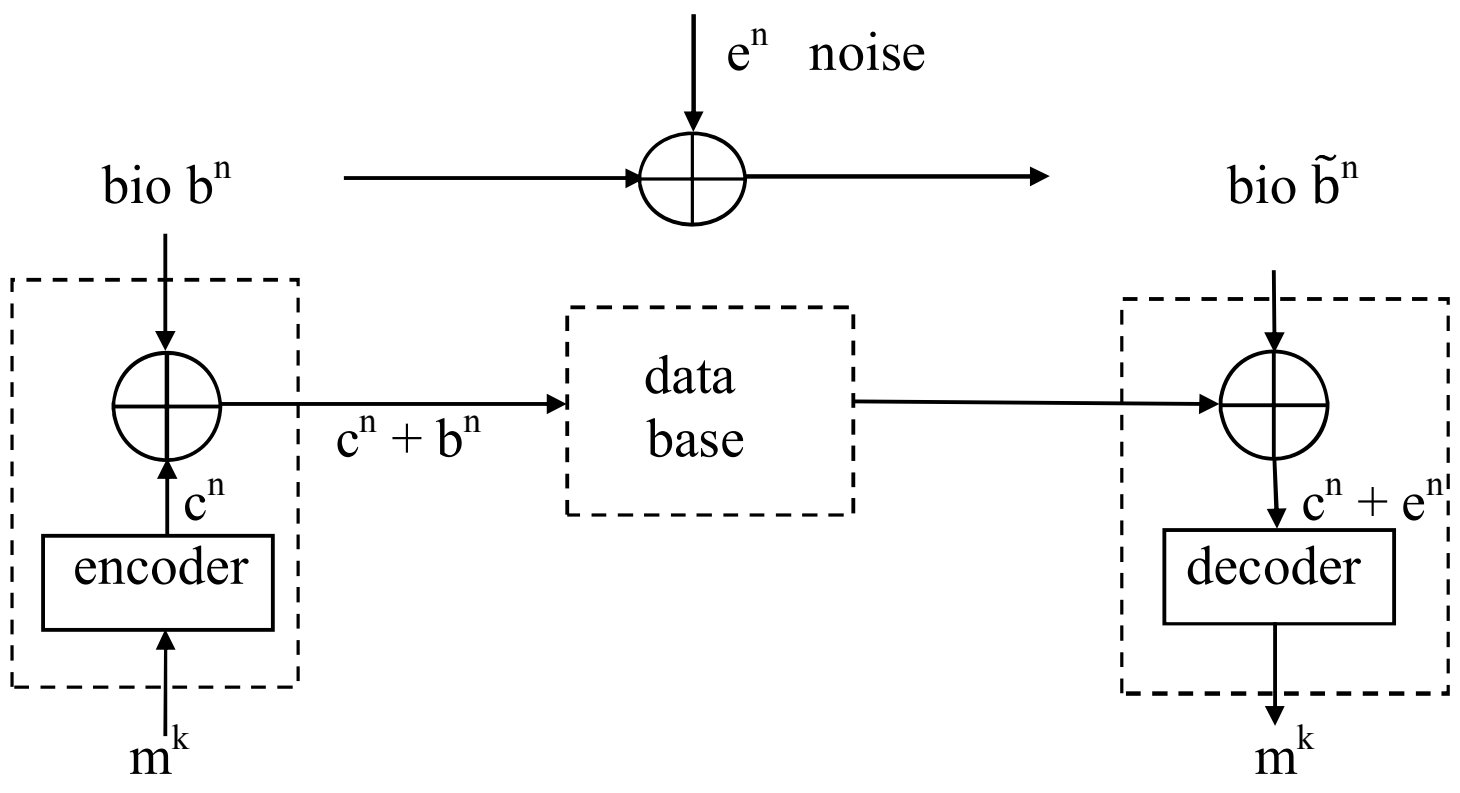

**Figure 6.3** Schematic representation of the JW scheme

The JW biometric authentication scheme can also be seen as a wiretap channel, see Figure 6.4. Encoding is done as in Figure 6.3. The legal "receiver" has the observations $z^n = c^n \oplus b^n$ and $\tilde{b}^n = b^n \oplus e^n$, where $e^n$ is an error vector (the difference between $b^n$ and $\tilde{b}^n$). This legal receiver can calculate $y^n = c^n \oplus b^n \oplus \tilde{b}^n = c^n \oplus e^n$. A data base observer has the availability of $z^n = c^n \oplus b^n$. Therefore, Figure 6.3 can be redrawn as the wiretap model of Figure 6.4. For binary inputs and binary symmetric channels, the maximum "secrecy capacity" is

$$\begin{aligned} C_S &= \max_{P(C^n)} (I(C^n;Y^n) - I(C^n;Z^n)) \\ &= n(H(B) - H(E)), \end{aligned} \tag{6.1}$$

which is the difference in biometric and noise entropy.

The observation that the legal user has the availability of the output of two parallel channels, see Figure 6.5, can also be applied to the wiretap channel to improve the secrecy capacity, see Chapter 5. In this case,

$$C_S{}^+ = n(H(B) - H(E)) + \max_{P(C^n)} \{H(Z^n \mid Y^n) - H(Z^n \mid C^n Y^n)\}, \tag{6.2}$$

which enlarges the secrecy capacity as given in (6.1). The problem that remains is the quantification of (6.2).

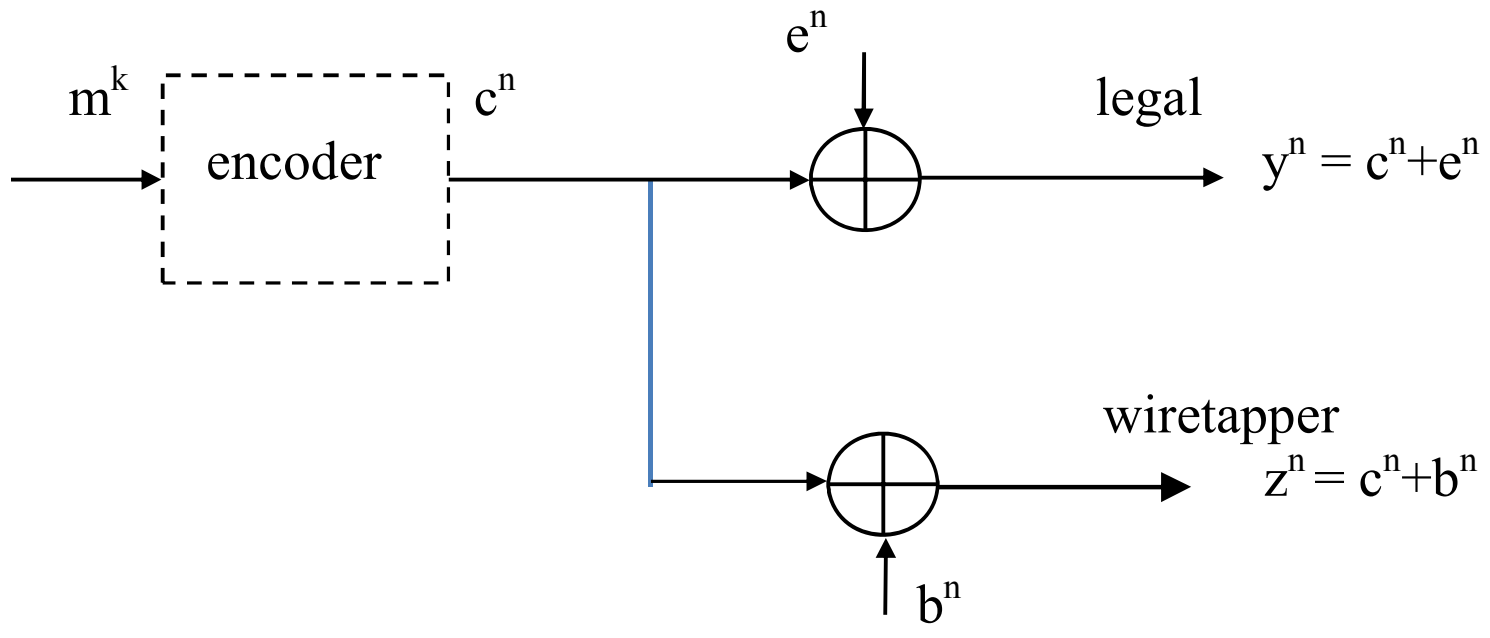

**Figure 6.4** Wiretap representation for the JW scheme

## 6.2 Biometric Reconstruction

The first problem we consider is that of reconstructing original biometric data given a noisy version of the biometric data and some related data previously stored in a data base. The condition is that it is difficult or almost impossible to guess the original biometric data from the stored data [44,45,46,49].

**Enrollment**

The reconstruction scheme uses the parity check matrix $H^T$ for a length n linear block code with k information symbols [42]. The parity check matrix $H^T$ has n rows and (n-k) columns. Hence, the inner product of a biometric vector $b^n$ with the parity check matrix gives as a result a "syndrome vector" $s^{n-k} = b^n H^T$ of reduced length (n-k), which a server stores in the data base.

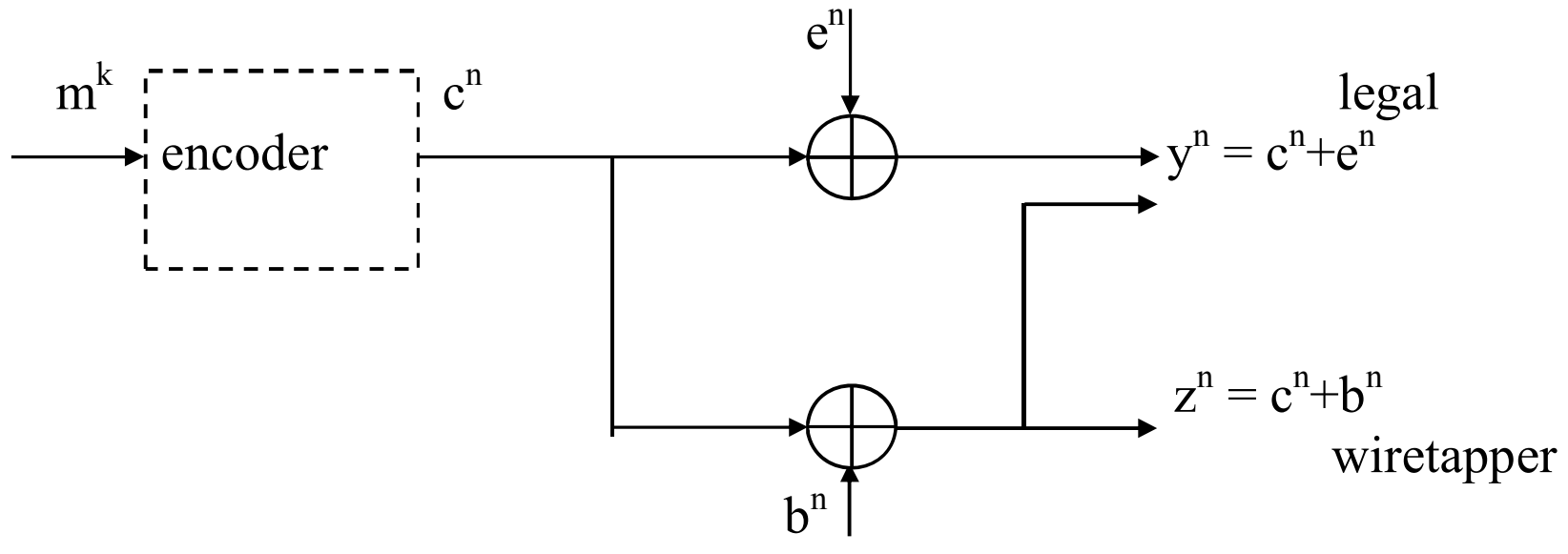

**Figure 6.5** Wiretap representation for the improved JW scheme

**Reconstruction**

We assume that, at the legal reconstruction phase, the noisy vector $\tilde{b}^n$ is offered to the server, where $\tilde{b}^n = b^n \oplus e^n$ and $e^n$ is an error vector changing $b^n$ at positions where the error vector $e^n$ has ones. To reconstruct the original biometric vector $b^n$, the server calculates $\tilde{b}^n H^T \oplus s^{n-k} = \tilde{b}^n H^T \oplus b^n H^T = e^n H^T$. A regular decoding algorithm for the corresponding error correcting code subsequently produces $e^n$ and thus reconstructs $b^n = \tilde{b}^n \oplus e^n$. Of course, knowledge of the statistical properties of $e^n$ is very important for the decoding algorithm.

The correct vector $b^n$ can be used for en- or decryption in a key-based entrance system. Incorrect decoding, caused by a $\tilde{b}^n$ with un-correctable errors, leads to the impossibility to the further use of $\tilde{b}^n$.

We use as an example an RS code over GF($q = 2^m$) of length $n = 2^m - 1$. The redundancy is n-k symbols, and p is the symbol error probability. The important performance parameters to consider are the False Rejection Rate (FRR) and the False Acceptance Rate (FAR).

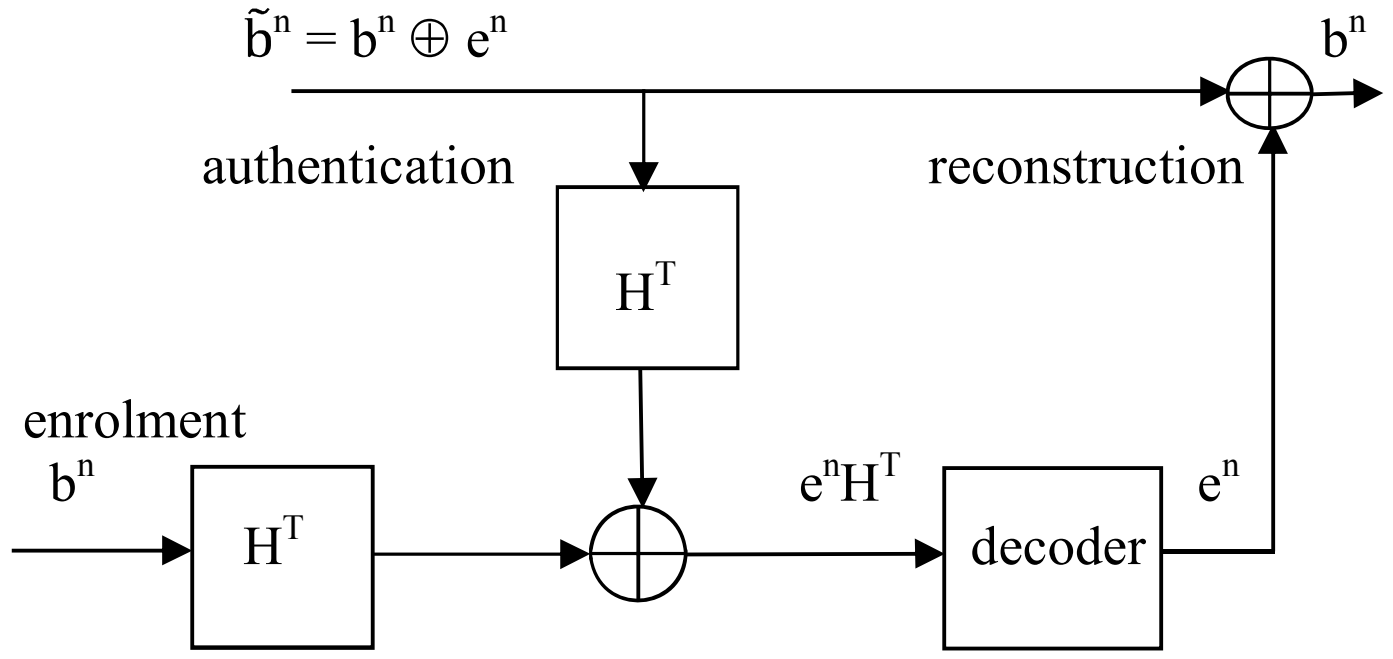

**Figure 6.6** Biometric reconstruction scheme

**False Rejection Rate**

The FRR is defined as the probability that a legal user is not accepted by the system. It can be upper bounded by the probability that more than (n-k)/2 errors occur, i.e.

$$\mathrm{FRR} \le \sum_{i=\left\lceil \frac{n-k+1}{2} \right\rceil}^{n} \binom{n}{i} p^i (1-p)^{n-i} \propto (np)^{\lceil (n-k+1)/2 \rceil} . \qquad (6.3)$$

**False Acceptance Rate**

Suppose that a decoding algorithm only accepts syndromes resulting from correctable error patterns and an illegal person is assumed to produce a random syndrome. Then, the FAR is the probability that a random syndrome is accepted as valid, i.e.

$$\mathrm{FAR} = \sum_{i=0}^{\lfloor (n-k)/2 \rfloor} \binom{n}{i} (n+1)^{-(n-k)} \propto q^{-(n-k)/2}. \qquad (6.4)$$

From (6.3) and (6.4), we see that the FRR and FAR are reduced for increasing (n-k). Calculation of the FAR and the FRR are the same as for a noisy communication channel with the same parameters.

**Correct guessing probability**

Another measure of performance is the probability that an illegal user of the system guesses the correct biometric with or without the stored syndrome from the data base. Assume that an illegal user always guesses the biometric $b^n \in B^n$, where $B^n$ is the set of possible biometrics, with the highest probability of occurrence, thus minimizing the average probability of guessing error (Maximum Aposteriori Probability, MAP). Using this principle, without data base knowledge, the correct guess probability

$$P_{guess}\,(\text{correct}) \;=\; \max_{b^n \in B^n} \; P(b^n)\,. \qquad (6.5)$$

An illegal user, with data base knowledge of $s^{n-k}$, can improve this probability by guessing the $b^n$ for which $P(b^n$ stored as $s^{n-k} | s^{n-k})$ is maximum. Let $S^{n-k}$ be the set of possible syndromes, then

$$P_{guess}\,(\text{correct} \mid s^{n-k}) = \max_{B^n : b^n \to s^{n-k}} P(b^n \mid s^{n-k})\,.$$

The average probability of correct guessing is

$$\overline{P}_{guess}(\text{correct} \mid s^{n-k}) = \sum_{s^{n-k} \in S^{n-k}} P(s^{n-k}) \max_{B^n : b^n \to s^{n-k}} P(b^n \mid s^{n-k})$$

$$= \sum_{s^{n-k} \in S^{n-k}} \max_{B^n : b^n \to s^{n-k}} P(b^n) P(s^{n-k} \mid b^n) \qquad (6.6)$$

$$\le q^{n-k} \max_{B^n} P(b^n).$$

The upper bound on the correct guessing probability in equation (6.6) is a factor $q^{n-k}$ higher than (6.5).

For every particular syndrome, there are only $q^k$ candidate biometric words and thus, the average probability that we have a correct biometric $b^n$ can be bounded as

$$q^{-k} \leq \bar{P}_{guess}(\text{correct} \mid s^{n-k}) \leq q^{n-k} \max_{B^n} P(b^n). \qquad (6.7)$$

For a small value of k, a high correct guessing probability for the illegal user can be expected, whereas a large value of k makes it more difficult to guess the correct biometric.

Using the concept of entropy with base q, the entropy $H(B^n \mid S^{n-k}) \leq k$ symbols, because there are only $q^k$ candidate biometrics for a particular syndrome vector $s^{n-k}$. The entropy $H(S^{n-k}) \leq (n-k)$, since there are $q^{n-k}$ possible syndromes. Thus, since $H(S^{n-k}) + H(B^n \mid S^{n-k}) = H(B^n) + H(S^{n-k} \mid B^n) = H(B^n)$, we obtain

$$k \geq H(B^n \mid S^{n-k}),$$

$$\geq H(B^n) - (n - k). \qquad (6.8)$$

We call (n-k) the entropy loss or leakage rate, see also (6.6), where we lose a factor $(2^m)^{n-k}$ in the guessing probability. In information theory, the concept of typicality can be used to show equivalence between (6.6) and (6.8).

## 6.3 The Juels-Wattenberg Scheme

There are several ways to describe biometrics for authentication purposes. Biometric data can be given as a collection of different symbols in an ordered or un-ordered way. In particular cases, biometric data is given by the possession of particular important properties. At authentication these properties are checked on their presence. The properties might change, disappear or other properties appear as important. When the biometrics are used as a password, one has to deal with these errors without sacrificing the security. JW designed a scheme where both aspects, are incorporated.

### 6.3.1 Juels-Wattenberg Scheme with a biometric of length n

The original JW scheme using general linear codes is described for biometrics $b^n$ of length n.

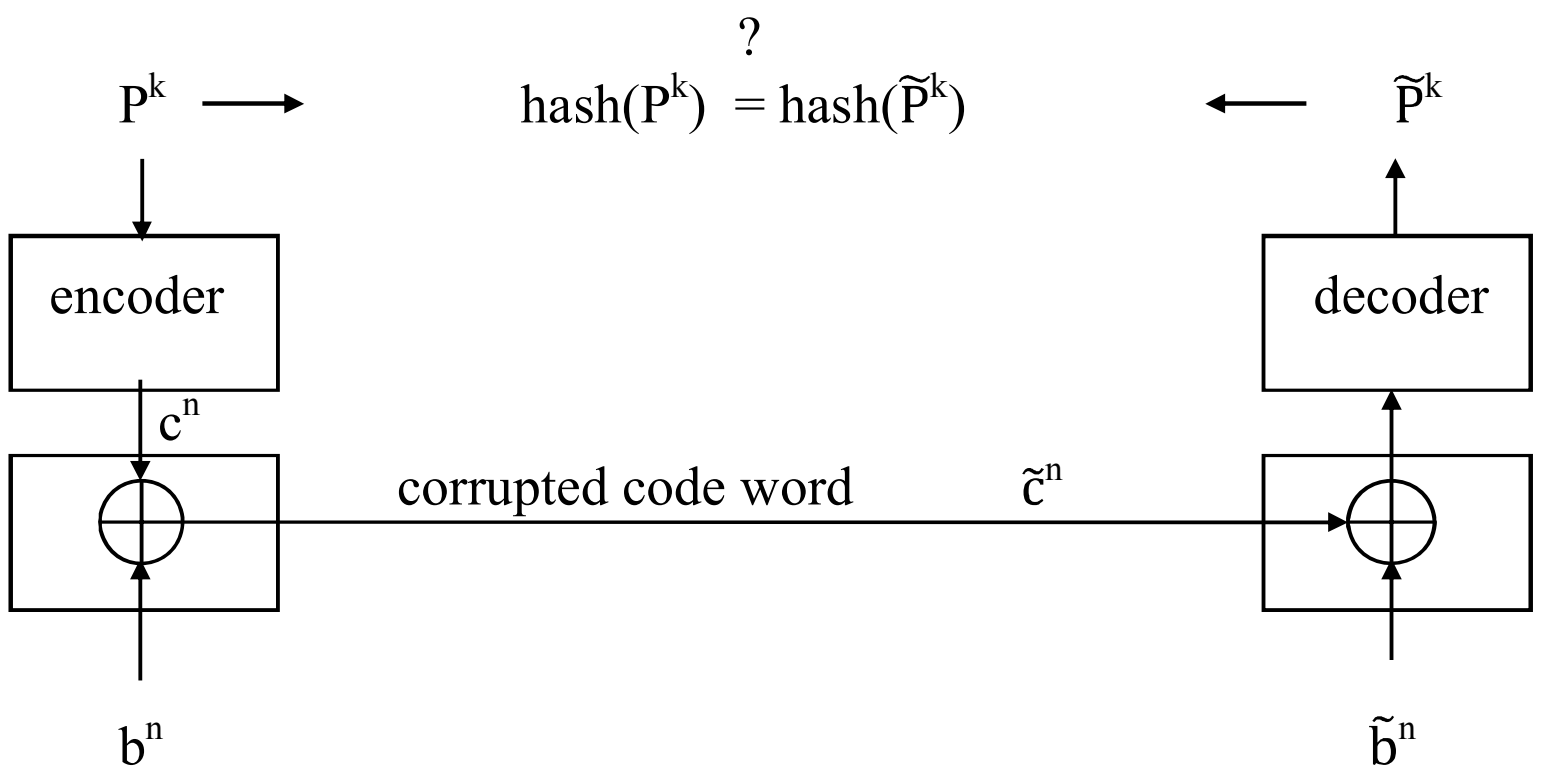

**Figure 6.7** JW scheme for linear codes

**Enrollment**

We add a biometric vector $b^n$ of length n with components from GF($2^m$) to a randomly chosen code word $c^n$ from a scheme specific RS code. The code word is generated by a random vector $P^k$. We store the sum $\tilde{c}^n = b^n \oplus c^n$ in a data base. In addition, we also store the hash($P^k$) in the data base.

**Authentication**

At authentication we add again the relevant biometric and the result is $\tilde{c}^n \oplus \tilde{b}^n = b^n \oplus c^n \oplus \tilde{b}^n = e^n \oplus c^n$, see Figure 6.7. From this, we can retrieve the original code word $c^n$ and thus $P^k$ if the number of errors is within the decoding region. The hash($P^k$) can be used to check for correct decoding.

**False Rejection Rate**

The FRR is upper bounded by the probability that we have more than $d_{min}/2$ errors in the biometric $\tilde{b}^n$. For RS codes over GF($2^m$) with minimum distance $d_{min} = n\text{-}k\text{+}1$,

$$\text{FRR} \le \sum_{i=\left\lfloor \frac{n-k}{2} \right\rfloor+1}^{n} \binom{n}{i} p^{i}(1-p)^{n-i} \propto (np)^{\left\lfloor \frac{n-k}{2} \right\rfloor+1},$$

.

which is the same as (6.3)

**False Acceptance Rate**

For the FAR, we calculate the probability that a random vector $\tilde{b}^n$ is such that $b^n \oplus \tilde{b}^n = e^n$ is within the decoding radius of the code word $c^n$, i.e.

$$\text{FAR} = \sum_{i=0}^{\left\lfloor \frac{n-k}{2} \right\rfloor} \binom{n}{i} (n+1)^{-n} \propto q^{-k} \times q^{-\frac{n-k}{2}}.$$

This FAR is a factor $q^{-k}$ better than (6.4). The calculation is based on a random biometric $\tilde{b}$, whereas it can be expected that the vector $\tilde{b}^n$ probably is not a random vector, but drawn from the same set of biometrics as the original one.

**Correct guessing probability**

The correct guessing probability is the same as given in Section 6.2.

### 6.3.2 Juels-Wattenberg scheme with a biometric of fixed length t < n

We describe a particular implementation that uses an RS code as a basic component. Let the RS code be over GF($q = 2^m$), i.e. length $n = 2^m - 1$. The biometric is characterized by t different values less than $2^m$ and is specified by $b^t = (b_1, b_2, \cdots, b_t)$, $b_i \in GF(2^m)$, $b_i \neq b_j$, The scheme works as follows.

**Enrollment**

1. generate a random word $P^k$ of k symbols from GF($2^m$), and store the corresponding hash value, hash($P^k$), in the data base. The hash($P^k$) produces a value from which we cannot reconstruct $P^k$;
2. encode $P^k$ with an RS encoding matrix as $c^n$;
3. the t different values from $b^t$ are used as indicators for the positions in $c^n$ where the code word symbols are fixed. In all other (n-t) positions, the symbols are changed in a random way from their original value. A code word $\tilde{c}^n$ is created with n - t errors;

4. the word $\tilde{c}^n$ is stored in the data base.

**Authentication**

We assume that the biometric has r error symbols. Hence, the biometric points at t – r correct positions and r incorrect positions. For the decoding the remaining positions are considered to be erasures. Since the code has dimensions (n,k), minimum distance n-k+1, the correct code word $c^n$ and thus $P^k$ can be decoded for

$$(n - k + 1) - (n - t) \geq 2r + 1.$$

For $(t - k) \geq 2r$ no decoding error can occur. After decoding $P^k$, the hash($P^k$) can be calculated and compared with the stored value from the data base at enrollment.

**Note** We can also let $b^t$ determine t incorrect positions. Correct decoding is then guaranteed for $(n - k + 1) - t \geq 2r + 1$, or $(n - t - k) \geq 2r$. A schematic representation is given in Figure 6.8.

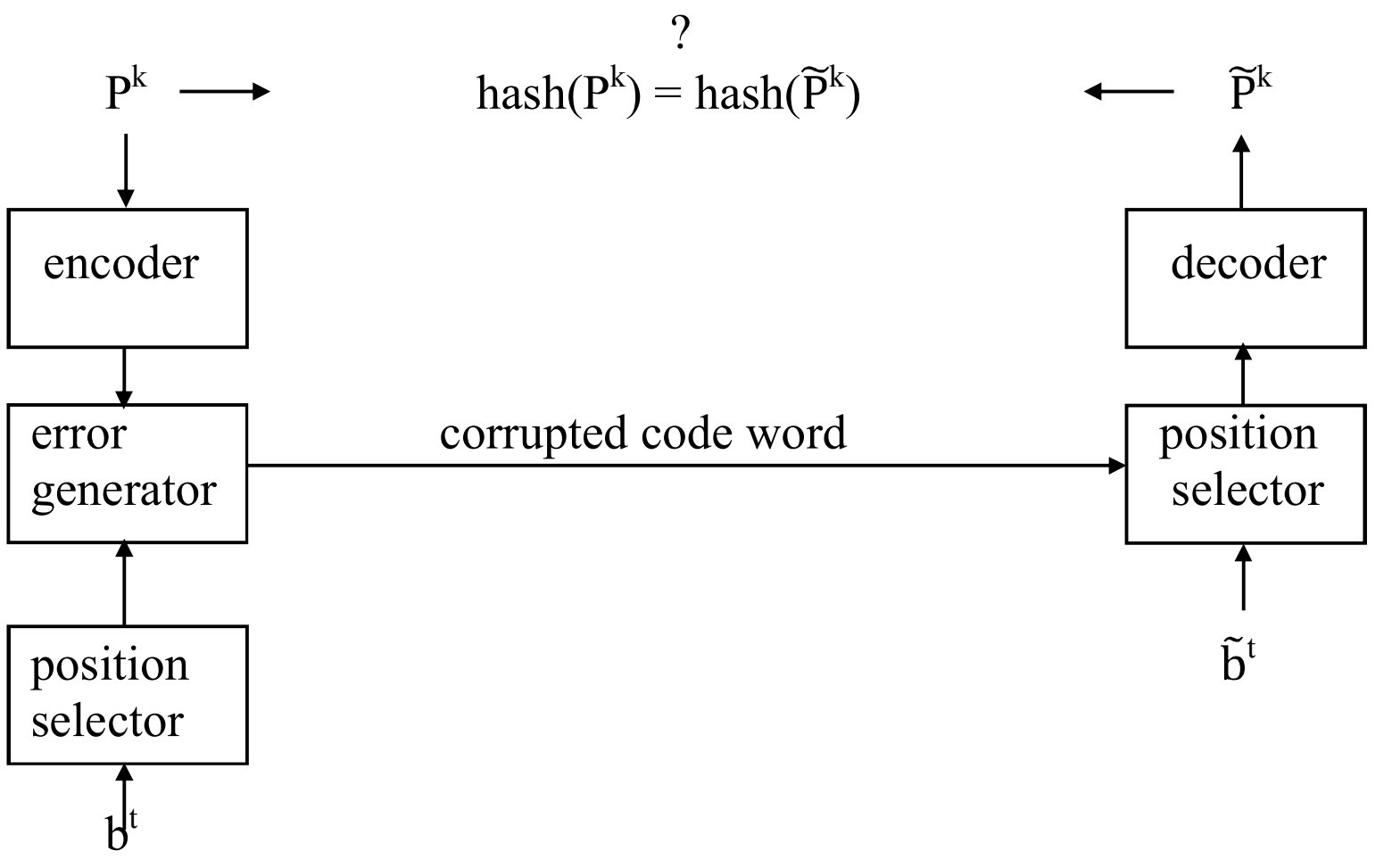

**Figure 6.8** Representation of the JW scheme using RS codes

We use this description of the JW scheme, because it fits the use of RS codes very well and it relates the JW scheme with the Juels-Sudan (JS) hereafter.

## 6.4 The Juels-Sudan (JS) Scheme

The JS scheme [47] is a biometric authentication scheme also based on RS codes. We describe a particular version of the JS scheme.

Code words of an RS code over GF($q = 2^m$) are generated by evaluating a random information polynomial P(X) of degree k-1 (equivalent to a vector $P^k$ of length k with elements from GF($2^m$)), for $X = \alpha^i$, $i = 0,1, \cdots, n-1$; $\alpha$ a primitive element of GF($2^m$); $n = 2^m - 1$. The k symbols from P(X) can be hashed, as for the JW scheme [41] and stored as hash(P(X)) in the data base for further use.

**Enrollment**

We assume that $b^t = (b_1, b_2, \cdots, b_t)$, $b_i \in GF(2^m)$, $b_i \neq b_j$, represents the biometric at enrollment. Then:

1 choose random P(X) of degree k-1;

2 store : $c^n = (c_0, c_1, \cdots, c_{n-1})$, where

$c_i = P(\alpha^i)$ for $i \in b^t$,

$c_i \neq P(\alpha^i)$ in $(n - t)$ other positions.

As mentioned before, after enrollment the word $b^t$ can change due to errors like measurement imprecision, aging, etc. We assume that for a particular user at authentication we offer a word $\tilde{b}^t$ from t biometric properties. The procedure at authentication is described as follows.

**Authentication**

Given $\tilde{b}^t = (\tilde{b}_1, \tilde{b}_2, \cdots, \tilde{b}_t)$, $\tilde{b}_i \in GF(2^m)$, $\tilde{b}_i \neq \tilde{b}_j$.

1 evaluate $P(\alpha^i)$, for $i \in \tilde{b}^t$;

2 declare the remaining code word positions as erasure;

3 decode $\tilde{P}(X)$;

4 give Reject when hash(P(X)) unequal hash($\tilde{P}$(X)), otherwise Accept.

Note, that $b^t$ determines t positions in $c^n$, whereas the values at these positions are uniquely determined by the polynomial P(X). The values at the (n - t) other positions differ from the evaluation of P(X).

**False Rejection Rate**

The legal user, having a biometric $\tilde{b}^t$, considers t positions of which possibly a maximum of r positions are wrong. The remaining (n-t) positions are considered as erasures. Correct decoding (the hash values comparison gives Accept) is guaranteed if $(n-t+2r+1) \le (n-k+1)$ or $2r \le (t-k)$. The probability that there are more than (t-k)/2 symbol errors in t positions, for a symbol error rate p, thus gives an approximation of the FRR

$$\text{FRR} \approx (tp)^{1+(t-k)/2}.$$

Making k larger will reduce the error correcting capability of the decoder and thus the FRR will increase.

**False Acceptance Rate**

For the FAR we assume that an illegal person produces a random vector of t positions. The FAR is then given by the probability that a random vector at authentication has at least k correct positions from $c^n$ that lead to the reconstruction of P(X). Hence, under this condition,

$$\text{FAR} \le \binom{t}{k}\binom{n-k}{t-k}\Big/\binom{n}{t} = \binom{t}{k}^2 \Big/ \binom{n}{k} \approx \binom{t}{k} \times \binom{t}{k} q^{-k}. \tag{6.9}$$

For t = k, the FAR is roughly equal to $q^{-k}$, the same as for a direct guess without using the biometric.

**Correct guessing probability**

Since the data is stored in a public data base, an illegal user of the data base, trying to decode on the basis of the observed data, has n-t errors in n symbols and thus correct decoding is "not" possible for $2(n-t)+1 > n-k+1$ or $2t < n+k$. One possibility is to guess k correct positions out of n and try to decode $\tilde{P}(X)$ with (n-k) erasures. Note that t positions out of n are correct. Hence, the probability of success

$$P(\text{successful guess}) = \binom{t}{k}\Big/\binom{n}{k} \approx \binom{t}{k} q^{-k}. \tag{6.10}$$

Without biometric, the probability of a correct guess is $q^{-k}$. Hence, we have a loss due to the fact that the biometric $b^t$ has t properties. An illegal user could also guess P(X) given $c^n$ using the Maximum Aposteriori Probability (MAP) principle and thus:

$$\bar{P}_{guess}(\text{correct} \mid c^n) = \sum_C \max_P \Pr(P^k \mid c^n)\Pr(c^n)$$
$$= \sum_C \max_P \Pr(P^k)\Pr(c^n \mid P^k)$$
$$= q^{-k}(q-1)^{-(n-t)} \sum_C \max_{b^t:(P^k, c^n)\to b^t} \Pr(b^t)$$
$$\le \left(\frac{q}{q-1}\right)^n q^{t-k} \max_{b^t} \Pr(b^t).$$

As in (6.7) we have

$$q^{-k} \le \bar{P}_{guess}(\text{correct} \mid c^n) \le \left(\frac{q}{q-1}\right)^n q^{t-k} \max_{b^t} \Pr(b^t). \qquad (6.11)$$

Comparing (6.11) with (6.7) we see an exponent (t-k) instead of (n-k).

## 6.5 An Improved Version of the Juels-Sudan Scheme

Dodis et al. [43] published an improved version of JS. In this version only t, instead of n symbols are stored in the data base. Performance can be shown to be the same as for the original scheme, when n symbols are stored. It can be described as follows:

**Enrollment:**
Given the biometric $b^t = (b_1, b_2, \cdots, b_t)$, $b_i \in GF(2^m)$, $b_i \neq b_j$.
1. choose a random secret P(X) of degree k-1;
2. calculate $Q(X) = P(X) + (X-b_1)(X-b_2)\cdots(X-b_t)$;
3. store Q(X) of degree (t-1).

**Authentication:**
Given $\tilde{b}^t = (\tilde{b}_1, \tilde{b}_2, \cdots, \tilde{b}_t)$, $\tilde{b}_i \in GF(2^m)$, $\tilde{b}_i \neq \tilde{b}_j$.
1. evaluate $Q(\tilde{b}_i)$, $i = 1, 2, \cdots, t$. $Q(\tilde{b}_i) = P(b_i)$ for $\tilde{b}_i = b_i$;
2. decode P(X).

For $\tilde{b}^t$, with at least k correct values, we have correct decoding of P(X). If we look at the vector $Q^t$ that corresponds to Q(X), we conclude that $Q^t$ plays exactly the same role as the vector $c^n$ in the JS scheme. One of the problems with the JS scheme is the choice of biometrics in the application. A fingerprint-based fuzzy vault implementation can be found in Nandakumar et al. [48].

**Note** The JS scheme is close to the JW scheme when we consider RS codes. In the JW scheme, we store $c^n + b^n$ in the data base, where $c^n$ is generated by $P^k$. At authentication, we add $\tilde{b}^n$ to the stored result and thus we obtain $c^n + e^n$. If we assume not more that t errors, we can decode the original code word. The difference with the JS scheme is that $b^n$ can change the whole code word $c^n$, whereas in the JS scheme only a particular set of values is changed. If $b^n$ contains exactly t positions, the two schemes are exactly the same.

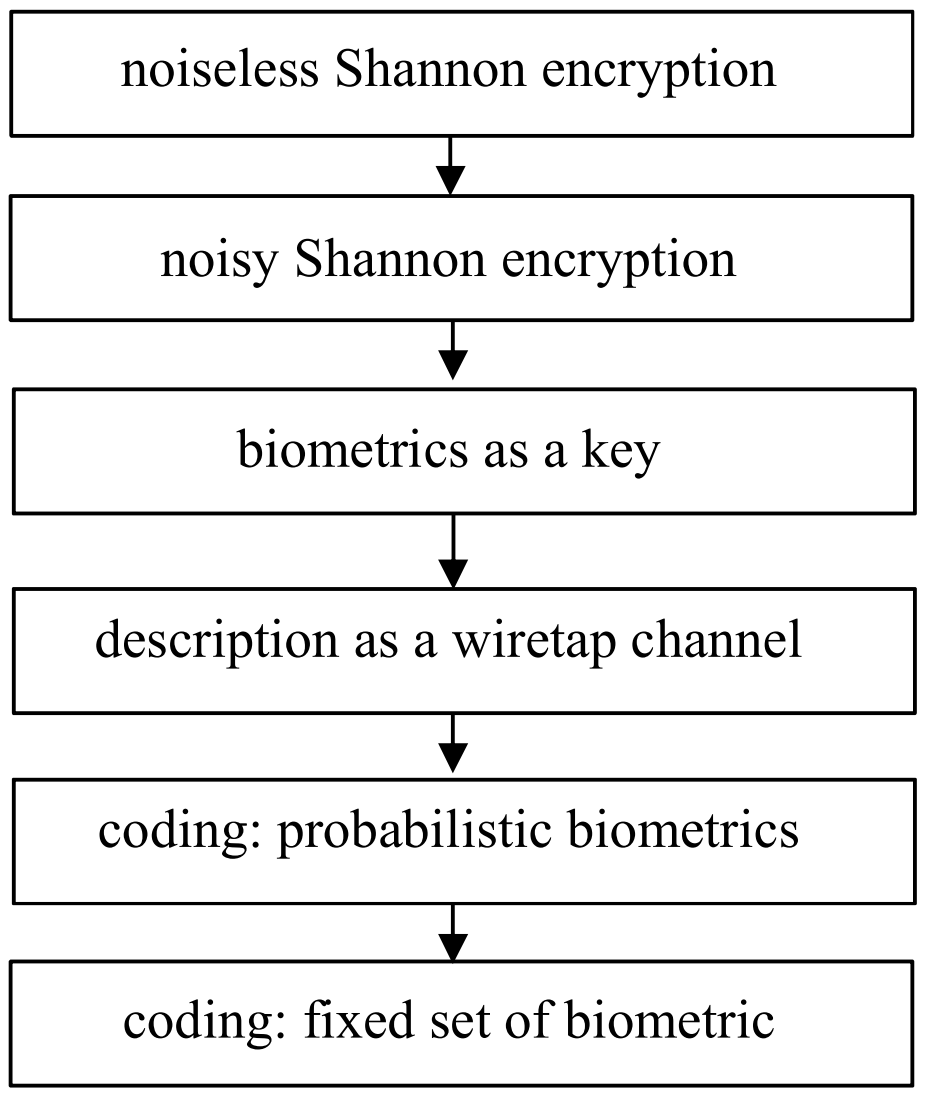

**Figure 6.9** Schematic form of the content of Chapter 6

# 6.6 Concluding Remarks

This chapter gives an interesting application for the wiretap channel. We start with a noisy crypto scheme as defined by Shannon and relate this scheme with the noisy fingerprint authentication as developed by Juels and Wattenberg. The scheme uses a probabilistic description of the biometric. Therefore, we can describe this biometric authentication principle using a wiretap channel model. In [79], an algorithm is described that estimates the entropy of finite length biometric data.

We conclude with an authentication scheme derived by Juels and Sudan. This scheme is based on a maximum number of properties from the biometric. It is interesting to see the connection between the Juels-Sudan scheme and the Juels-Wattenberg scheme. We derive and compare the performance parameters False Acceptance Rate, False Rejection Rate and probability of a successful attack. In schematic form, the content of Chapter 6 is given in Figure 6.9.

# Chapter 7

# RS Codes with Constraints

This chapter contains four topics that can be seen as coding with restrictions, also called constrained coding. The first topic is the avoidance of certain symbols to occur in the output of an RS encoder. The second topic is the combination of RS and Run-length constrained coding with application in bandwidth limited channels. The third topic introduces the distance profile of a code and we conclude with RS codes for which the maximum number of symbols that are the same is bounded.

## 7.1 Symbol Avoidance

The first problem is to generate RS code words over GF(q) in such a way, that particular symbols do not occur in the code word. To achieve this, we use a method described in a paper by G. Solomon [57]. The method is described for the avoidance of one symbol, but can be extended to exclude a particular set A of symbols with cardinality |A|.

$$G_{k,n} = \underbrace{\begin{bmatrix} I_\kappa & 0 & P \\ 0 & I_r & Q \end{bmatrix}}_{n} \Big\updownarrow k$$

**Figure 7.1** Parameters for a systematic RS encoding matrix

For instance, if we want to use an RS code over GF(11) with code words of length 10 and code word symbols from {0, 1, ···, 9}, we have to exclude the symbol 10 in the encoder output.

In Figure 7.1 we give the parameters of the RS code in systematic form that are of importance for the description. Note that, $\kappa + r = k$. We encode a message as

$$(m^{\kappa},s^{r})\, G_{k,n} = c^{n},$$

where $m^{\kappa}$ is the information to be encoded and $s^{r}$ a control word of length r.

**Property 7.1** A control word $s^{r}$ can be used to create any symbol in a particular position in the check part.

The property follows from the fact that the minimum distance for this systematic RS code is $d_{min} = (n - k + 1)$, and thus the last (n - k) symbols of any row are non-zero. Furthermore, since every element in the Galois field has an inverse, we can create any symbol in a particular position in the check part by using a particular control word $s^{r}$.

The $\kappa$ q-ary information symbols are pre-coded such that they do not contain a particular set A of symbols. The r control symbols are used to manipulate the check part (last n-k symbols of a code word) such that symbols from A do not occur in the check part. From the constraint it follows, that the maximum number of control words we can choose is $(q - |A|)^{r}$.

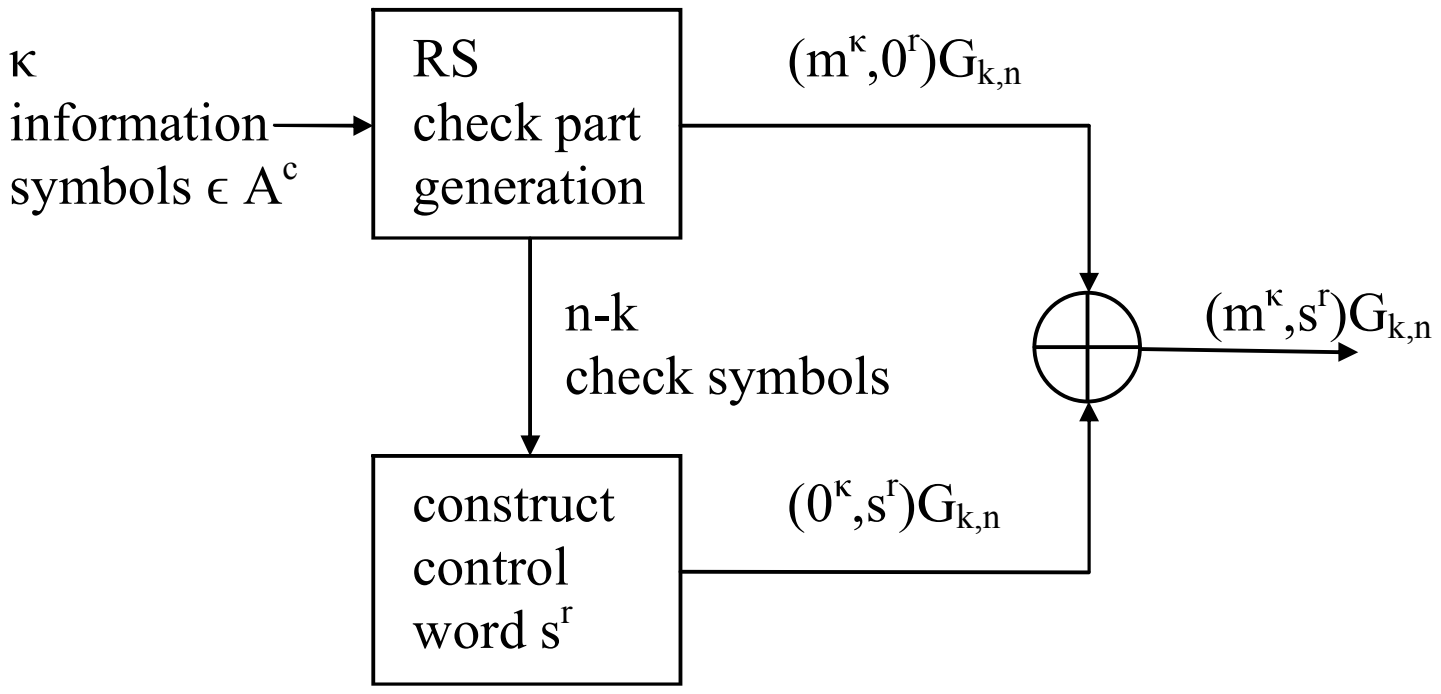

**Figure 7.2** Concept of the control word generation

**Encoding**

Let the first κ pre-coded information symbols (no symbol from A) generate a check part. A particular set of control words has to be selected that generate check parts such that the symbols from A do not occur in the summation of the check parts generated by the information and control input, see also Figure 7.2. Using property 7.1, it is easy to see that every symbol in the check part generated by the information, eliminates a maximum of $|A|(q-|A|)^{r-1}$ control words that cannot be used. If the total number of eliminated control words is less than $(q - |A|)^r$, we can always find a particular choice for the control input that realizes a check part without the undesired symbols. Thus, this condition is fulfilled when

$$|A|(n-(\kappa + r))\,(q-|A|)^{r-1} < (q-|A|)^r$$

or

$$n-(\kappa+r) < \left(\frac{q-|A|}{|A|}\right) \quad . \tag{7.1}$$

**Example** For $|A| = 1$, $r = 1$, $q = n+1$ (RS), the condition is always fulfilled.

**Example** For $|A| = 2$, $r = 1$, the redundancy $n-(\kappa+1) < (n-1)/2$, or $\kappa > (n-1)/2$.

**Example** Take an ($n = 7$, $k = \kappa + r = 3$) RS code over $GF(2^3)$. The Galois field $GF(2^3)$ is defined by a primitive polynomial, $P(X) = X^3 + X + 1$.
The generator matrix

$$G_{3,7} = \begin{pmatrix} 1\,0\,0 & 6\,1\,6\,7 \\ 0\,1\,0 & 4\,1\,5\,5 \\ 0\,0\,1 & 3\,1\,2\,3 \end{pmatrix} .$$

By choosing $r = 1$, we have $\kappa = 2$. Let $A = \{7\}$. From $G_{3,7}$, we have the following control words we can use when we encounter an information check part containing the symbol 7,

[ 0 0 0  0 0 0 0], [ 0 0 1  3 1 2 3], [ 0 0 2  6 2 4 6],
[ 0 0 3  5 3 6 5], [ 0 0 4  7 4 3 7], [ 0 0 5  4 5 1 4],
[ 0 0 6  1 6 7 1].

The word [ 0 0 7 2 7 5 2 ] cannot be used to eliminate the symbol 7 since it has 7 in its systematic part. To illustrate the elimination of symbol 7, we pick one code word with the symbol 7 (in the check part) from the list of code words produced by $G_{3,7}$ using the two information digits (0,3). The specific code word is

$c_7$ = [0 3 0 7 3 4 4].

Taking a suitable control word as [0 0 3 5 3 6 5] from $G_{3,7}$, we obtain the final encoded word without the symbol 7 as [ 0 3 3 2 0 2 1 ]. It can also be verified that either [0 0 5 4 5 1 4] or [0 0 6 1 6 7 1] is a suitable control word.

## 7.2 Constrained Coded Modulation

Run Length Limited (RLL) codes are used in communication channels with power spectral limitations. The limitations (constraints) are translated into a longest (k-constraint) and a shortest run of d+1 transmitted symbols that are the same (d-constraint). The maximum run guarantees a variability in the transmitted symbols that improves the synchronization. In our case we assume $k = \infty$. The shortest run length constraint guarantees a minimum symbol duration $\tau$, which limits the bandwidth needed to transmit the constrained sequence. From a communication point of view, the bandwidth B needed to transmit the RLL sequence and the transition time $\tau$ are related as $2B = 1/\tau$. Thus, we use the same approach as for the coded modulation invented by Ungerböck [58]. For further information, we refer to the textbook by Schouhamer-Immink [32].

We describe the application of constrained coded binary modulation in the time domain. We combine error correcting codes and RLL codes before binary modulation. The RLL code is designed in such a way, that the minimum spacing between two transitions, $\tau$, in the binary RLL word is at least the same as for the uncoded symbol duration.

### 7.2.1 System design

In CD type of recording systems one often combines an RS code with an RLL constrained code [55]. The advantage is that errors in the constrained code can be seen as symbol errors for the RS code.

**Example** An example of a combination is given in Figure 7.3, where we encode 223 x 8 bits into RS code words of length 255 symbols of 8 bits and convert each RS symbol into a constrained RLL code word of length 14. The RLL code words are constructed in such a way that they can be concatenated without violating the d-constraint. For this, one can use stuff bits (3 additional bits for the CD length-14 RLL code) or the principle of RLL 1-symbol look-ahead encoding, see Hollmann [59]. The resulting bits are binary modulated, where the carrier period is a multiple of the RLL symbol duration.

The RS code needs an expansion of the time needed to transmit the information, whereas the constrained RLL code reduces the transmission time. The goal is not to change the total time to transmit the k information symbols. Since we aim at short code word lengths, we can apply soft-decision decoding on the constrained RLL code words if in addition to the d-constraint, we also have a minimum distance between any two RLL code words larger than one.

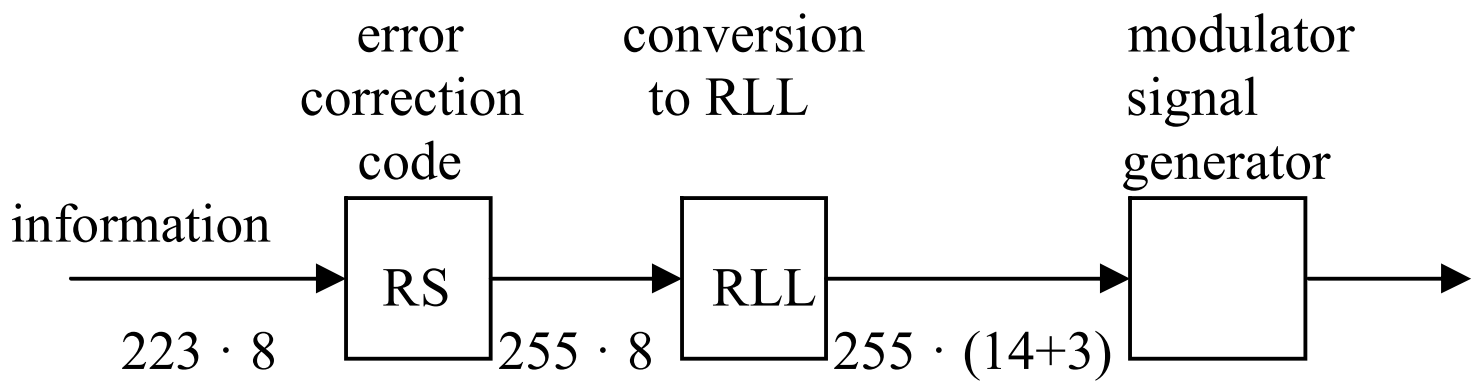

**Figure 7.3** Principle encoding/modulation scheme

### 7.2.2 Idea for the modulation

We use binary modulation to transmit the constrained words. The RLL constraint d and the RLL symbol duration τ' are chosen such that τ'(d+1) = τ. As a consequence, the RLL symbol is repeated at least d+1 times to give a new symbol of minimum duration τ. The minimum separation between two transitions in the binary RLL word is at least τ. In this way, the bandwidth B needed to transmit the RLL words is the same as for the uncoded transmission. In addition, the modulator output also has a constant envelope output.

### 7.2.3 Idea for the RLL-encoding

We use Figure 7.4 to explain the main idea in more detail.

1. we first convert binary words of length k information bits into code words of length n from an error correcting code and symbol duration τ. The efficiency of encoding is k/n;
2. the n symbols are converted into constrained binary words of length m with symbol duration τ' and a minimum run of d+1 symbols that are the same. The efficiency of the conversion is $R_{RLL} = n/m$.

Now, consider a time frame of length T with n = T/τ code word symbols. In the same time T, we can have m = T/τ' symbols from the constrained RLL code. Hence, in this case we have the relation T = nτ = mτ' and using the condition that τ'(d+1) = τ we have $R_{RLL}$ (d+1) = 1. If in addition we use an

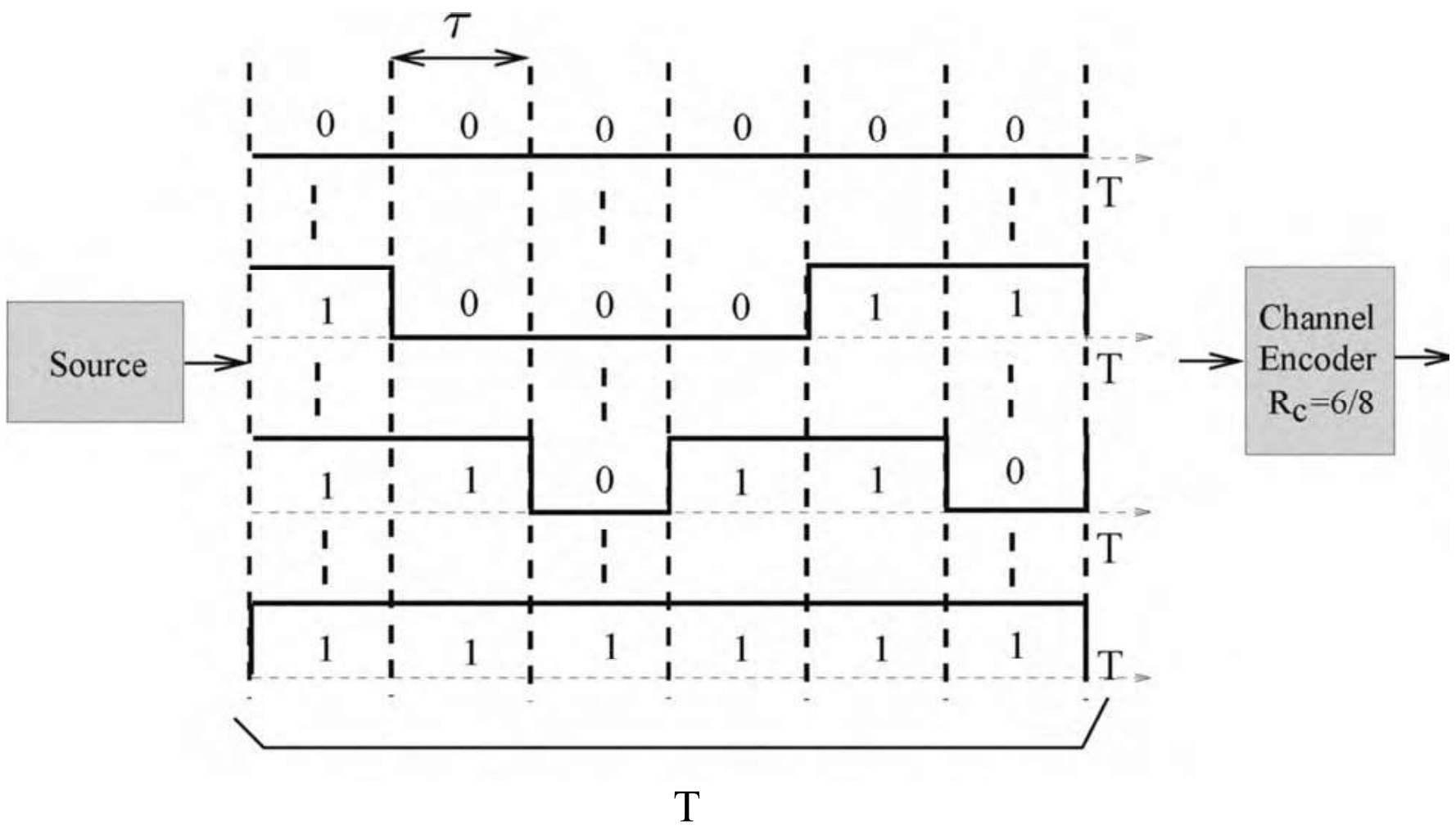

**Figure 7.4a** Message of length 6, encoded with rate 6/8

error correcting code with efficiency k/n, the final condition of "no bandwidth" expansion is

$$(k/n)\ R_{RLL}\ (d+1) = 1. \tag{7.1}$$

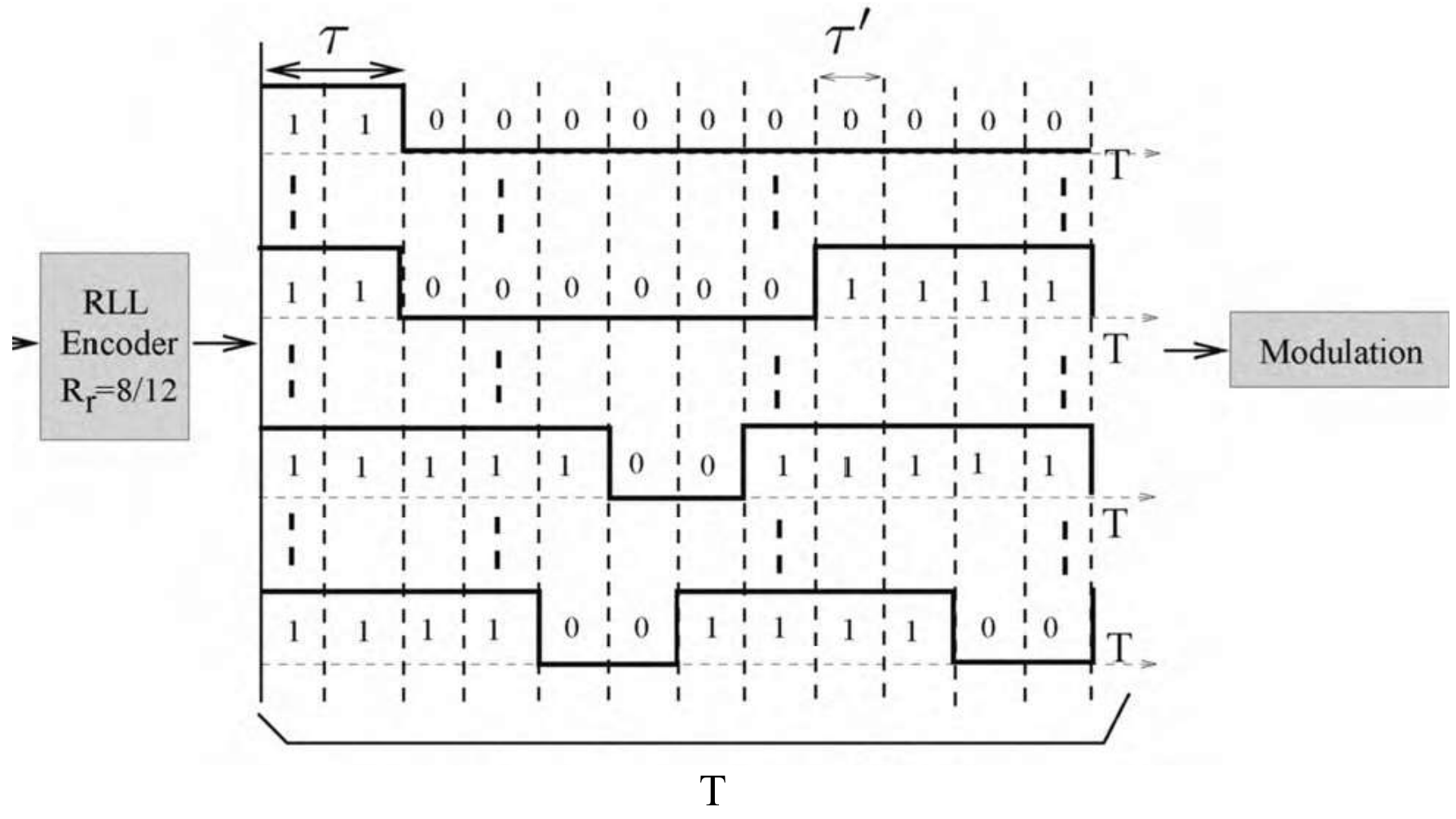

**Figure 7.4b** RLL code, d+1 = 2, with 64 code words of length 12

Figure 7.4 illustrates the idea in the time domain. Six information symbols are encoded into eight code symbols. The eight code symbols are translated into twelve RLL symbols. The overall rate satisfies (7.1). The research problem to solve, is the design of the combination of error correcting codes and constrained codes such that (7.1) is fulfilled (the d-constraint). The additional problem is the design of a demodulator. We will discuss a particular system design in the next section.

**Table 7.1**

Code Table: RLL(d+1 = 2), $R_{RLL} = 3/5$

| message | code word | message | code word |
|---|---|---|---|
| 0 | 00011 | 4 | 00001 or 00110 |
| 1 | 00111 | 5 | 11110 or 11001 |
| 2 | 11000 | 6 | 01111 or 10011 |
| 3 | 11100 | 7 | 10000 or 01100 |

**Example** In Table 7.1 we give the RLL words for (d+1) = 2 that can be implemented using a 1-symbol look-ahead method. For instance, if we store message 4 we can choose between 00001 and 00110. The code word is selected in such a way that the next code word does not violate the d-constraint. Thus, if the next message is 00011 we use 00110, whereas for message 2 we use 00001. In this way, we do not have to use extra bits between the RLL words to satisfy the d-constraint as is in recording systems like CD, Hollmann [59].

If we use an RS code over GF($2^3$), the binary code symbols are of length three. Every symbol can be converted into an RLL code word as shown in Table 7.1. For k = 5 and n = 6, $d_{min} = 2$ and the value of $R_{RS}$ $R_{RLL}$ (d+1) = 1. After demodulation, the RLL symbols can pass the likelihoods for every possible symbol to the RS decoder. Since $d_{min} = 2$, we expect a coding gain of 3 dB without any bandwidth expansion, see Ungerböck [58].

**Example** The RS code we focus on is of length n = 255 symbols of 8 bits. The code has minimum distance n – k + 1, where k is the number of information symbols. For the constrained code, we select an RLL code with parameter d = 1 or 2. The research problem, solved by Mengi [60,61], is the construction of RLL codes that satisfy the following constraints:

1. RLL code words can be concatenated without violating the d-constraint;
2. RLL code words must have a minimum distance larger than 1, to facilitate soft-decision RLL decoding;
3. the number of code words is at least 256;
4. the combination RS, RLL satisfies (7.1).

Ad 1). The first condition can be solved using the approach by Hollmann [59] for the construction of block decodable RLL codes.

Ad 2). The Hollmann construction is given for $d_{min} = 1$. In [60,61], an extension can be found for $d_{min} > 1$. The fact that the minimum distance is larger than one, gives the opportunity to do soft-decision on the RLL code words.

Ad 3). RLL codes are generated that satisfy the d-constraint. Then, we make a selection of the resulting code words that lead to an increase in minimum distance. The number of resulting code words should be larger than 255 to be able to make a concatenation with the 8-bit RS symbols.

Ad 4). If the product $R_{RLL}(d+1) < 1$, we can select the corresponding RS code to satisfy (7.1). In Table 7.2, we give some of the constructed RLL codes such that (7.1) is satisfied. More constructions can be found in [60,61].

**Table 7.2**

Example of RLL- RS combinations that satisfy $R_{RS}\, R_{RLL}\, (d+1) = 1$ and $d_{min} = d+1$

| d | $R_{RLL}$ | $d_{min}$ | $R_{RLL}\,(d+1)$ | $R_{RS}$ |
|---|---|---|---|---|
| 1 | 8/14 | 2 | 8/7 | 223/255 |
| 2 | 8/21 | 3 | 8/7 | 223/255 |

**Demodulation**
The last part of the system is the demodulator. We assume that the carrier frequencies are multiples of 1/τ' instead of 1/τ to obtain phase continuity. At the receiver the demodulator therefore changes the integrate-and-dump time interval to τ'. For non-coherent FSK we assume furthermore, that there is a constant phase within an RLL-block. After demodulation, we have 255 blocks of m soft values. For every block we find the RLL code word index i that maximizes

$$S_i = \sum_{j=1}^{m} c_j^i s_j, \qquad c_j^i = +/-1;\ i = 1,2,\ \cdots,2^8, \tag{7.2}$$

where $s_j$ is the soft output of the demodulator per interval τ' and $c_j^i$ the j-th symbol for block i. The complexity of this action is proportional to the number of code words times the code word length m. After this operation, we can apply hard-decision RS decoding.

**Performance** The performance depends on the soft-decision decoding of the RLL code words and the decoding of the RS code. We assume Gaussian noise with power spectral density or variance $N_0/2$. Since $\tau = (d+1)\tau'$, the RLL soft-decision error rate $P_{RLL}$ is given by

$$P_{RLL} \propto Q\left(\sqrt{d_{min}\frac{2E_s}{N_0}}\right) = Q\left(\sqrt{d_{min}\frac{2E_b}{(d+1)N_0}}\right) \quad . \tag{7.3}$$

For $d_{min}$ = d+1, the RLL symbol error rate is of the same order as for uncoded transmission. Therefore, we expect the coding gain to come from the RS code. In Figure 7.5, we give the resulting "waterfall" curves for the system. Three RS codes are simulated with $d_{min}$ = 1, 2, 3, respectively. The best performance is given by the (255,223) RS code combined with the RLL d = 2 and $d_{min}$ = 3 code. The RS code has minimum distance 255 - 223 = 32, and the decoder can correct up to 16 symbol errors. Different code constructions and combinations can be found in [60,61].

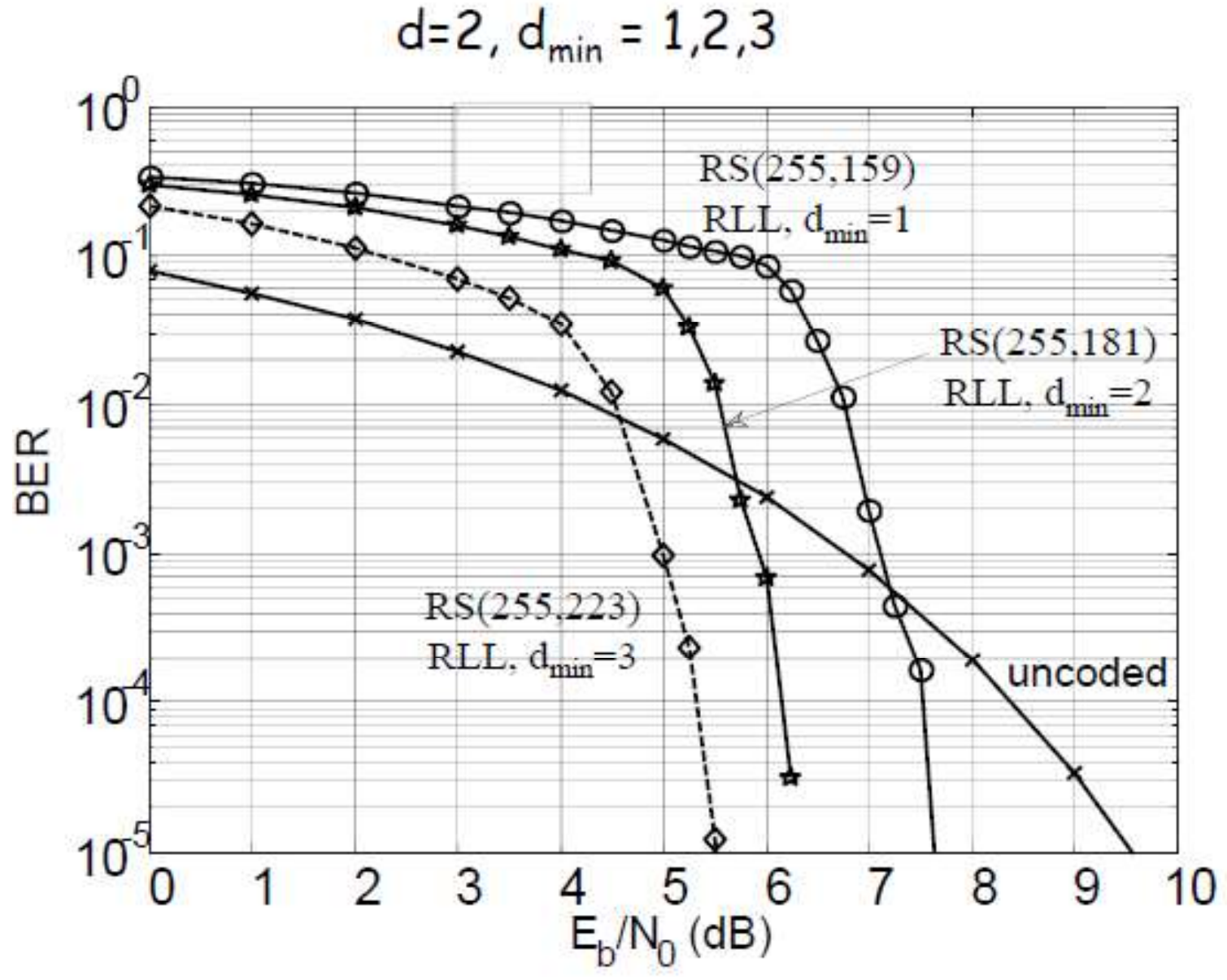

**Figure 7.5** Performance curves for several RLL-RS combinations

**Conclusion** We describe a coded modulation scheme where the redundancy for error correction is obtained from relaxation of the timing. The encoded symbols have the same minimum spacing between two transitions as for the uncoded case, but the symbols might be longer in discrete steps depending on the parameter d. For a particular example, we show a coding gain of about 4 dB. An additional advantage is that we have a constant envelope modulated signal. This can be of importance for systems where we have an absolute power constraint as for instance in the "Power-line Communications CENELEC band." Our system is considered for AWGN, but will also work for other types of noise, since we have a block wise detection and decoding without error propagation.

## 7.3 Optimum Distance Profile

**Motivation** Error control coding is very often a fixed part of a communication systems design. To change a code is impossible due to standards and already implemented decoding algorithms. In cognitive systems, we want to adapt the efficiency of a transmission scheme to the actual circumstances. For this we have to be able to change the modulation and also the error correcting or error detecting code.

**specify a code**
↓
**start with an encoding matrix**
↓
**generate the equivalent matrices**
↓
**calculate the distance profiles**
↓
**output the optimum distance profile**

**Figure 7.6** Concept of the optimum distance profile generation

The Transport Format Combination Indicators (TFCIs) are widely applied in CDMA systems, see [55,56]. In the realization of TFCI, input bits are used to combine some basis code words of a linear block code. When the number of the input bits increases or decreases, some basis code words will be included or excluded, respectively. In this process, a general consideration is how to realize large minimum distances of the generated sub-codes.

Some concepts are well known, such as puncturing, shortening or lengthening an encoding matrix of a linear code. In this way, the efficiency of the code is changed by changing the length n of the code words. We choose another method: row extension/deletion of the encoding matrix for an (n,k) code, and thus change the parameter k.

**Row deletion** means that we delete one row from the encoding matrix such that it generates a new (n,k-1) code. Here, the effect is reduction in the code efficiency from k/n to (k-1)/n but at the same time a possible improvement in the minimum distance [71].

We first consider one-by-one row deletion from a generator matrix $G_{k,n}$ that generates the starting "mother code". The mother code has minimum distance $d_k$. We delete a row from this matrix and obtain $G_{k-1,n}$ with minimum distance $d_{k-1}$. From $G_{k-1,n}$ we can again delete a row and obtain $G_{k-2,n}$ with minimum distance $d_{k-2}$. The process ends when we obtain a single row from $G_{1,n}$. This process generates a distance profile

$$DPB_{del}(G_{k,n}) = (d_k, d_{k-1}, \cdots, d_1).$$

A distance profile $DPB_{del}(G_{k,n})$ is optimum ($ODPB_{del}$ ) if

$$d_i = \tilde{d}_i \text{, for } k \geq i \geq t+1,\ t \geq 1,$$
$$d_t > \tilde{d}_t \text{ ,}$$

for any equivalent encoding matrix $\tilde{G}_{k,n}$ generating the same "mother code".

**Example** The (7,4) Hamming code can be generated by the encoding matrix

$$G_{4,7} = \begin{bmatrix} 1\,0\,0\,0\,1\,1\,1 \\ 0\,1\,0\,0\,1\,1\,0 \\ 0\,0\,1\,0\,1\,0\,1 \\ 0\,0\,0\,1\,0\,1\,1 \end{bmatrix}.$$

The minimum distance is three. Deleting row-by-row, starting from the bottom, we get the distance profile

$$DPB_{del}(G_{4,7}) = (3, 3, 3, 4). \qquad (7.4)$$

The same code can be generated by an equivalent matrix $\tilde{G}_{4,7}$, i.e.

$$\tilde{G}_{4,7} = \begin{bmatrix} 1\,0\,1\,1\,1\,0\,0 \\ 1\,1\,1\,0\,0\,1\,0 \\ 0\,1\,1\,1\,0\,0\,1 \\ 1\,1\,1\,1\,1\,1\,1 \end{bmatrix}.$$

Deleting the last row at first, gives the distance profile

$$DPB_{del}(\tilde{G}_{4,7}) = (3, 4, 4, 4), \qquad (7.5)$$

which is better than (7.4) and the best possible for this code. For any linear code, we can find the optimum distance profile by row deletion, $ODPB_{del}$, by taking the best possible distance profile over all possible equivalent encoder matrices, see Figure 7.6. Of course, this is a complicated task. In [54] Johannesson defines the optimum distance profile for convolutional codes.

**Row extension** means that we add a row to the encoding matrix such that it generates a new (n,k+1) code. The effect is an improvement in the efficiency from k/n to (k+1)/n, but at the same time a possible degradation in the minimum distance.

We start with a single particular row from a code generator matrix and extend the matrix row by row. Every matrix extension must generate code words from the desired "mother code". In every step, we select a row such that we maximize the minimum distance of the constructed code generated by code generator matrix. The starting code has minimum distance $\delta_1$, determined by the Hamming weight of the first selected row. We add a row to this matrix and obtain $G_{2,n}$ with minimum distance $\delta_2$. We can add a row to $G_{2,n}$ and obtain $G_{3,n}$ with minimum distance $\delta_3$. The process ends when we obtain the final code generator matrix $G_{k,n}$. We generate a distance profile

$$DPB_{ext}(G_{k,n}) = (\delta_1, \delta_2, \cdots, \delta_k).$$

A distance profile $DPB_{ext}(G_{k,n})$ is better than $DPB_{ext}(\tilde{G}_{k,n})$ when

$$\delta_i = \tilde{\delta}_i \quad \text{for } 1 \le i \le t-1, t \le k,$$

$\delta_t > \tilde{\delta}_t$ .

**Example** The (7,4) Hamming code can be generated by the encoding matrix

$$G_{4,7} = \begin{bmatrix} 1\,1\,1\,1\,1\,1\,1 \\ 1\,0\,1\,1\,1\,0\,0 \\ 1\,1\,1\,0\,0\,1\,0 \\ 0\,1\,1\,1\,0\,0\,1 \end{bmatrix} .$$

The matrix $G_{4,7}$ generates a distance profile $DPB_{ext}(G_{4,7})$ = (7, 3, 3, 3). Again, this is the best possible profile for the (7,4) Hamming code by row extension. Remark that $d_k = d_{min}$, whereas $\delta_1$ equals the maximum Hamming weight of a code word.

**Example** For RS codes, we take as a starting generator matrix

$$G_{RS} = \begin{bmatrix} 1 & 1 & 1 & \cdots & 1 \\ 1 & \alpha & \alpha^2 & \cdots & \alpha^{n-1} \\ 1 & \alpha^2 & \alpha^4 & \cdots & \alpha^{2(n-1)} \\ & \cdots & & & \\ 1 & \alpha^{k-1} & \alpha^{2(k-1)} & \cdots & \alpha^{(n-1)(k-1)} \end{bmatrix} .$$

Deleting row by row gives as a distance profile

$$DPB_{del}(G_{RS}) = (n-k+1, n-k+2, \cdots, n).$$

If we start with the first row and extend the matrix row by row, we get

$$DPB_{ext}(G_{RS}) = (n, n-1, \cdots, n-k+1).$$

Note that both these profiles are optimum.

**Remark** There is an extensive report on the optimum distance profile for block codes in [51,52,53,71].

## 7.4 Same-weight Code Construction

We give a particular code construction using an (n,k) RS code, such that the generated code words have the property that all code words have a maximum of k symbols that are the same.

**Motivation** In MFSK modulation, symbols from a code word are transmitted as one of the M possible frequencies. The transmission can be disturbed by permanent narrowband noise in such a way that the demodulator outputs a symbol that is always present [62]. An (n,k) RS encoding matrix containing the all-ones row produces code words that have n symbols that are the same. Hence, correct decoding will not be possible. We describe an encoding method that avoids this problem. An extensive description is given in [63].

Let an RS code be generated by the standard RS encoding matrix given by:

$$G_{k+1,n} = \begin{pmatrix} 1 & 1 & 1 & \dots & 1 \\ 1 & \alpha & \alpha^2 & \dots & \alpha^{n-1} \\ 1 & \alpha^2 & \alpha^4 & \dots & \alpha^{2(n-1)} \\ & & \dots & & \\ 1 & \alpha^{k-1} & \alpha^{2(k-1)} & \dots & \alpha^{(k-1)(n-1)} \\ 1 & \alpha^k & \alpha^{2k} & \dots & \alpha^{k(n-1)} \end{pmatrix} \quad \text{(first } k \text{ rows: } C_1\text{)} .$$

The matrix $G_{k+1,n}$ has the following properties:

- the minimum distance for this code is $d_{min} = n-k$;
- since the minimum distance is equal to the minimum number of differences between two code words, the code words agree in at most k positions;
- multiples of the first row give n symbols that are the same.

Given the above matrix, we construct a sub-code with the property that all code words of the sub-code have a maximum of k symbols that are the same.

**Construction:** Generate code words with the first k rows of $G_{k+1,n}$, which form the code matrix $C_1$. Add the last row from $G_{k+1,n}$ to every code word. After decoding, subtract the last row from the decoded code word.

As a consequence of the construction, code words with n symbols that are the same do not occur. Since we always add the last row of $G_{k+1,n}$, the all-0 code word will not occur. We therefore have the following properties:

**Properties**

- the "same-weight" code construction using by $C_1$ has $d_{min} = (n-k+1)$;
- all code words have a maximum of k symbols that are the same;
- code words contain at least n/k different symbols;
- the code generated by $C_1$ has efficiency k/n.

**Remark** Since any (n,k) linear code can be brought in systematic form, the number of the same symbols can be larger or equal to k. Using the constructed sub-code, we obtain k as the maximum number of the same symbols in a code word. Hence, RS codes are optimum.

We briefly summarize the operations at the receiver side that utilize the code properties. For simplicity, we suppose that the MFSK demodulator outputs the presence of a particular frequency. A narrowband disturbance may cause a particular frequency to be present during a long period of time. We can detect and delete this continuous presence. Using the same-symbol property, the positions where the code word symbols agree with the disturbance are considered as erasure positions. Since there is a maximum of k erasures per narrowband disturbance, the number of narrowband disturbances that can be corrected for this idealized situation is $NB < (n-k+1)/k$. The same situation occurs when the disturbance blocks the transmission of certain frequencies. The maximum number of "deletions" for our code is then equal to k and the same analysis can be made. Performance gets worse when in addition to narrowband noise we also have background noise. For the permutation codes in Chapter 4, every symbol occurs only once.

# 7.5 Concluding Remarks

This chapter contains several applications of RS codes. We start with the principle of symbol avoidance, where the output of an RS encoder does not contain a particular set of output symbols. The idea is from Solomon Golomb and was published in 1974. Golomb only considered one forbidden symbol. The principle is easy to extend to more than one forbidden symbol.

The second topic is coded modulation in the time domain. Ungerböck developed code modulation in the 80s for two-dimensional signal spaces, by expanding the number of symbols that can be used. In our coded modulation scheme for binary modulation, we only extend the length of symbols, while keeping the minimum time between two transitions constant. As a consequence, the bandwidth needed to transmit a message is not enlarged. We combine the modulation with look-ahead RLL coding as developed by Hollmann, and show that the combination RLL-RS gives a coding gain of about 4 dB for AWGN channels.

The third topic is concerned with a new property of block codes, the "optimum distance profile". This property can help to reduce the difficulty of changing the dimensions of an encoding matrix. We show that RS codes have the best possible profile.

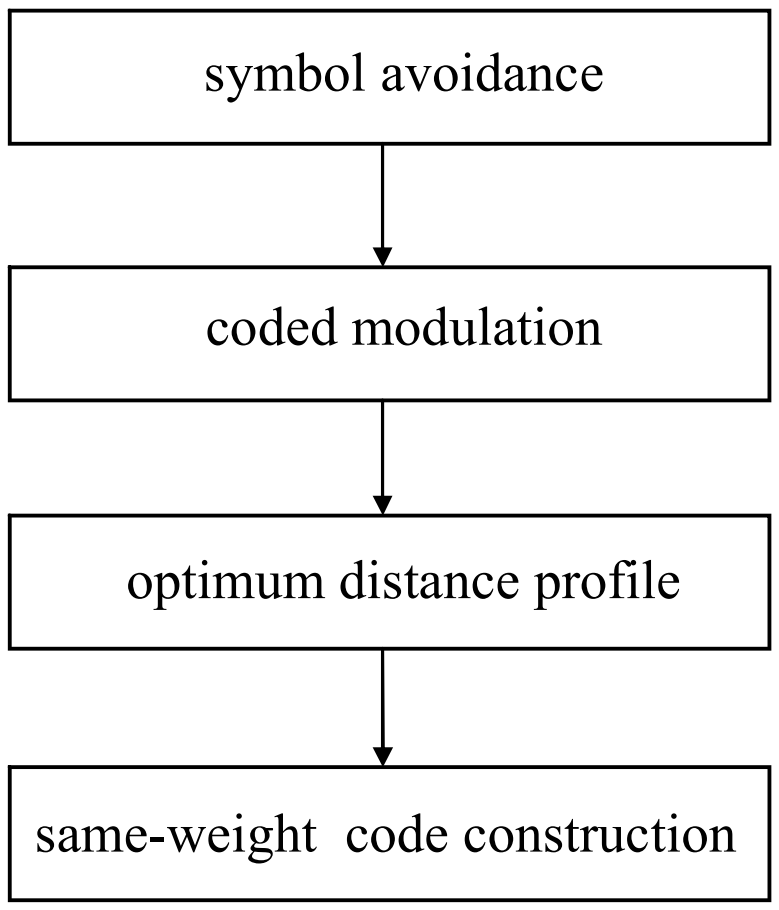

**Figure 7.7** Content of Chapter 7

The last topic describes a special application of RS encoding such that every code word contains a desired maximum number of the same type of symbols. This way of encoding can be beneficial when a channel is disturbed by narrowband noise.

The idea of constrained coding can be extended to convolutional codes, see [73], where we make sequences with an equal number of zeros and ones using the linearity property of convolutional codes.

The content of the chapter is summarized in Figure 7.7.

# Chapter 8

# Memories with Defects

## 8.1 Defect models

Improvements in process technology and clever circuit design make it possible to produce large memory systems on a chip with a high packing density. A high packing density has its limits and may cause errors in the memory cell structure. A popular error model is that of defects. A defect always produces the same output when being read, irrespective of the input. This might be “defect-0” or “defect-1”. The model for defective memory cells is given in Figure 8.1.

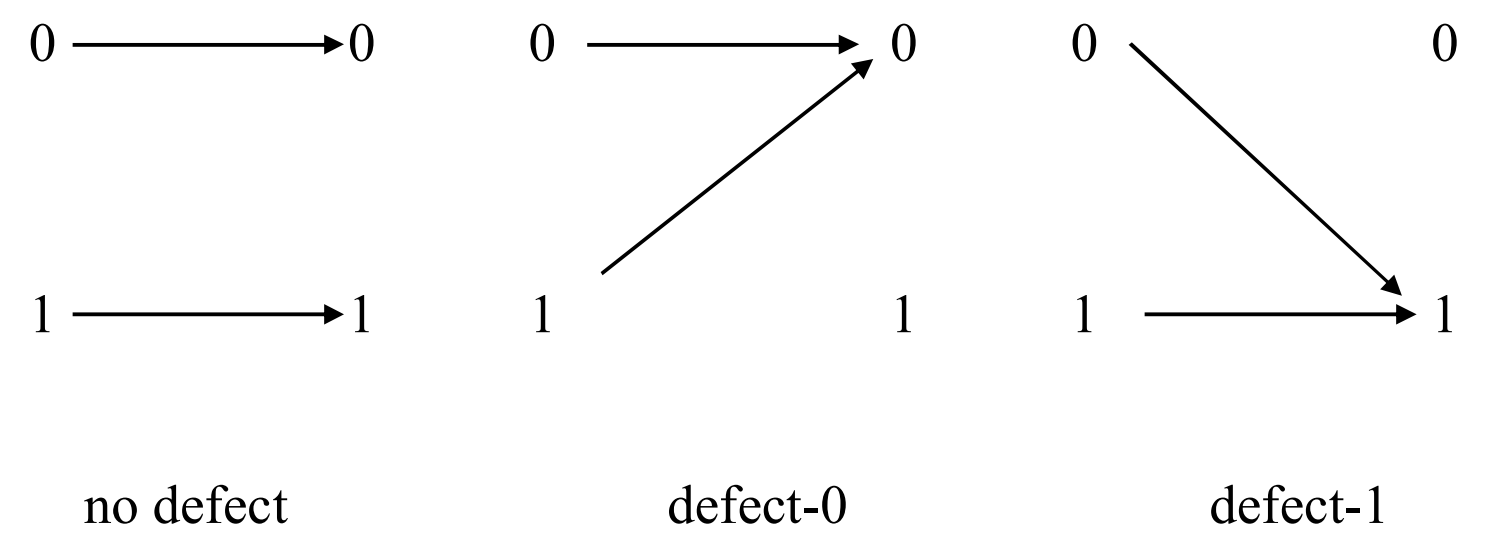

**Figure 8.1** Defect model for defect-0 and defect-1

In principle, we can write in the memory in coded or uncoded format. The reader may use a decoder to give an estimate for the encoded and stored information. We can distinguish between four situations, whether the writer/reader knows the value and position of the defects or not. This information may follow from inspection or by some additional side information.

The four situations give different results for the maximum storage efficiency or storage capacity. Assume that the fraction of defect-1 cells is equal to the fraction of defect-0 cells, and is equal to p/2. The situations can be described as:

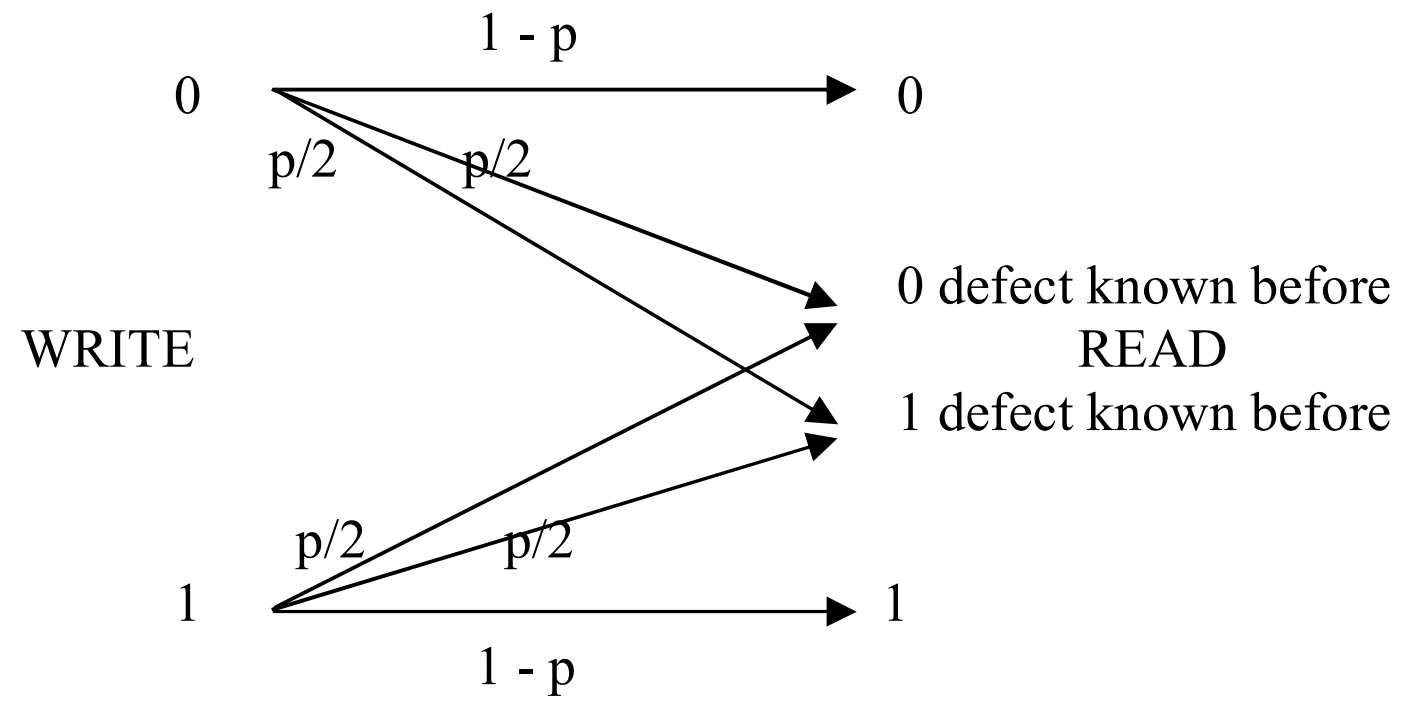

**Figure 8.2** Erasure channel model for known defects at receiver

1. writer and reader both know the positions and values of the defects. In this case, we can avoid the defective cells and thus the storage capacity is (1-p) bits per cell;
2. writer does not know, but the reader knows. The reader, knowing the position of a defect can treat this position as an erasure. Hence, the channel that he creates is an erasure channel with maximum storage capacity (1-p) bits/cell. The writer should use codes for the erasure channel to store his information, see the model in Figure 8.2;
3. the reader and writer do not know the defects. The defect can be considered as a random error and thus, on the average p/2 errors occur when reading. The channel that results is a BSC with storage capacity 1 – h(p/2), where h(*) is the binary entropy function, see the model in Figure 8.3;
4. the writer knows and the reader does not know the value and the position of the defect. This situation is the most interesting one, since we can design a special type of coding. Kuznetsov and Tsybakov [64,66] showed that in this case also the capacity is (1-p) bits/memory cell. In addition to that, they designed the "additive" coding method that asymptotically achieves this capacity, see Section 8.2.4.

We continue with situation 4 and we assume, that the writer is able to determine the location and value of a defect before using the memory.

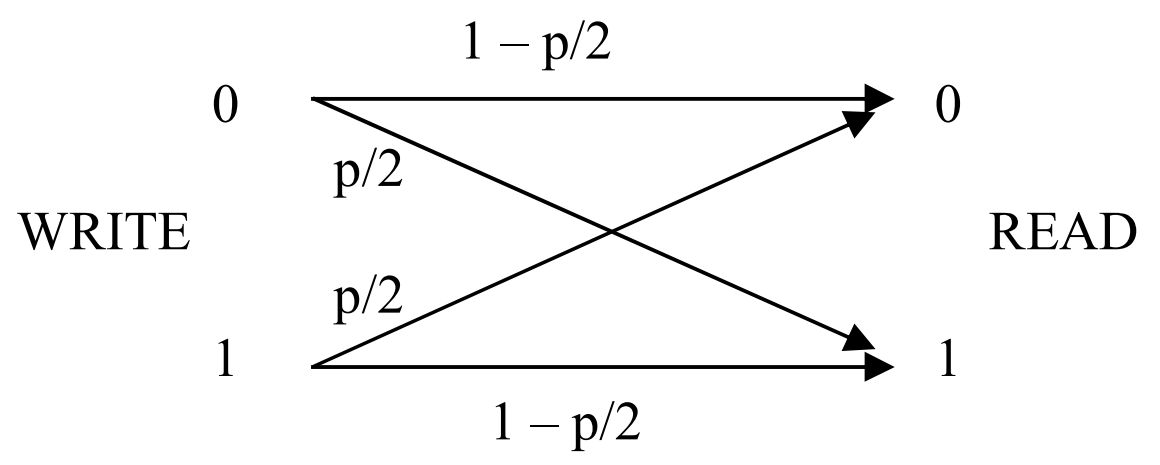

**Figure 8.3** BSC model for unknown defects at writer and reader

One of the possibilities is to use an error correcting code to store words in the memory. For a single defect per word, one could think of using the Hamming code. For instance, the (7,4) Hamming code can correct one error in 7 bits and stores 4 bits of information. Kutznetsov and Tsybakov initially worked on the (64,58) extended Hamming code that can correct one error or detect two errors (minimum distance 4), at the expense of 6 bits redundancy. Their task was to reduce the redundancy and their method of additive coding for defects will be explained in the sequel. The idea [64,66] finds wide spread application, for instance in coding for Write Once Memories (WOM) or flash memories.

## 8.2 Defect Matching Codes

### 8.2.1 One defect in n positions

Suppose that words of length n have a maximum of one defect per word. By inspection, the writer knows the value of the defect and encodes a word $x^{n-1}$ as $(0,x^{n-1})$ or $(1,\overline{x}^{n-1})$, respectively, where $\overline{x}^{n-1}$ is the complementary word for $x^{n-1}$. One of both words can always be stored error free and thus the redundancy is only one bit, irrespective of the value n. The reader can always find back the original message by inspecting the first bit of a stored word. This example shows that knowledge about the defect helps to reduce the redundancy needed for defect error correcting.

**Example** Let n = 7 and a defect-0 at position 3. The information to be stored is $x^{n-1}$ = (0 1 1 0 0 0). Since $(1,\overline{x}^{n-1})$ matches the defect-0 in the third position, we store (1 1 0 0 1 1 1). The information can be retrieved by inverting the last six bits.

### 8.2.2 n-1 defects in n positions

Another extreme example is the one that occurs when in n bits, n-1 bits are defective and only one bit is error free. In this case, since we can observe the value of the defects, we can always make the parity of the word even or odd with the remaining error free position. Hence, we store one bit of information in n positions.

**Example** Suppose that for the vector (1 0 1 1 0 - 1 1) only position six is free to chose. Then, a value 0 at that position makes the work parity equal to 1, whereas a value 1 makes the parity equal to 0.

**Note** Both strategies are optimal. In the first case, n-1 bits can be stored in n positions and the data storage efficiency R = 1 - 1/n. In the second situation, we can store only one bit in n positions and the storage efficiency R = 1 - (n-1)/n = 1/n.

### 8.2.3 Two-defects in n positions

We proceed by designing a method to handle a maximum of two defects in a word. A code word $c^n$ is called defect compatible or matching the defects, if it can be stored without any changes, i.e. the components of a code word agree with the values of the defects. The code word itself depends on the defects and the message $x^k$ to be stored.

We first construct a code matrix for which any pair of bits is present in some row. If this matrix exists and every row has a unique prefix, then this matrix can be used to match defects. The construction is as follows:

1. take all binary vectors of length (2α - 1) and weight α as columns of the code matrix;
2. add the all zero row to the matrix;

3. select $\lceil \log 2\alpha \rceil$ columns such that all rows of the $2\alpha \times \lceil \log 2\alpha \rceil$ sub matrix are different.

Ad 1) if we compare two arbitrary columns, then they contain at least the combinations (0,1), (1,0) and (1,1) in one of the rows. This follows from the fact that a column of length ($2\alpha$ - 1) has more than halve ones, and hence there must be some overlap between the two specific columns;

Ad 2) if we combine 1) with 2), we see that any pair of bits is present in some row and two arbitrary columns;

Ad 3) The construction for 3 can be done as follows. First take $2^{\lceil \log 2\alpha \rceil}$ different rows of length $\lceil \log 2\alpha \rceil$. The columns all have an equal number of zeros and ones. By deleting complementary pairs of rows, except for the all zero and all ones row, the equal weight property remains valid. We delete until $2\alpha$ rows are left, with the property that all columns have $\alpha$ ones. As this is part of the matrix construction, we place this matrix in front of the original matrix by column permutations. Note, that each row is uniquely specified by the first $\lceil \log 2\alpha \rceil$ bits.

These properties play an important role in the explanation of the coding method. To be more specific, we consider the case where $\alpha = 3$. The code matrix has 6 rows and 10 columns.

$$C = \begin{bmatrix} 0 & 0 & 0 & 0 & 0 & 0 & 0 & 0 & 0 & 0 \\ 0 & 0 & 1 & 0 & 1 & 1 & 0 & 1 & 1 & 1 \\ 0 & 1 & 0 & 1 & 0 & 1 & 1 & 0 & 1 & 1 \\ 1 & 1 & 0 & 0 & 1 & 0 & 1 & 1 & 0 & 1 \\ 1 & 0 & 1 & 1 & 0 & 0 & 1 & 1 & 1 & 0 \\ 1 & 1 & 1 & 1 & 1 & 1 & 0 & 0 & 0 & 0 \end{bmatrix} .$$

Any bit pattern of length 2 can be found in a row of length 10 and the first 3 digits uniquely specify each row of C.

**Encoding**

The message $x^7$ is represented by the vector

$$\tilde{x}^{10} = (0, 0, 0, x_1, x_2, \cdots, x_7).$$

The vector $\tilde{x}^{10}$ selects one out of 128 messages and the efficiency is 7/10.

Suppose that $\tilde{x}^{10}$ is not defect compatible in two of its components. The defects are specified in location and value by $d^{10} = (d_1, d_2, \cdots, d_{10})$, where we use a ? in case there is no defect at a particular location, otherwise we have the symbol 0 or 1. Then, from the code matrix we take a row vector $c^{10}(\tilde{x}^{10}, d^{10})$ that depends on $\tilde{x}^{10}$ and $d^{10}$ such that $\tilde{c}^{10} = \tilde{x}^{10} \oplus c^{10}(\tilde{x}^{10},d^{10})$ is defect compatible. The constructed vector $\tilde{c}^{10}$ is stored instead of $\tilde{x}^{10}$.

**Decoding**

The decoding is done as follows. The vector $\tilde{x}^{10}$ has 3 all zero initial components. Hence, the decoder (reader) knows which row of C is used in order to make $\tilde{x}^{10}$ defect compatible. This row is added modulo-2 to $\tilde{c}^{10}$ and the last 7 components of the result specify $x^7$ again, for

$$\tilde{c}^{10} \oplus c^{10}(\tilde{x}^{10},d^{10}) = \tilde{x}^{10} \Rightarrow x^7$$

**Example** Suppose that we want to store the message (1, 1, 0, 1, 0, 0, 0) as

$$\tilde{x}^{10} = (0, 0, 0, 1, 1, 0, 1, 0, 0, 0).$$

The memory has a defect-1 in the 3rd position and a defect-0 in the 4th position, i.e.

$$d^{10} = (?, ?, 1, 0, ?, ?, ?, ?, ?, ?).$$

The encoder adds modulo-2 the vector

$$c^{10}(\tilde{x}^{10},d^{10}) = (1, 0, 1, 1, 0, 0, 1, 1, 1, 0)$$

to

$$\tilde{x}^{10} = (0, 0, 0, 1, 1, 0, 1, 0, 0, 0)$$

and stores $\tilde{c}^{10} = (1, 0, 1, 0, 1, 0, 0, 1, 1, 0)$.

The decoder sees as three initial components (1, 0, 1) and adds the vector

$$(1, 0, 1, 1, 0, 0, 1, 1, 1, 0)$$

to $\tilde{c}^{10}$ which results in a decoded vector

$$\tilde{x}^{10} = (0, 0, 0, 1, 1, 0, 1, 0, 0, 0)$$

and message

$$x^7 = (1, 1, 0, 1, 0, 0, 0).$$

The overall, close to optimal, efficiency is 7/10 at a defect rate of 2/10. One problem is to show that this method is the best that we can have for 2 defects in a word of length n.

### 8.2.4 A general method for t = pn defects in n positions

The previously described coding methods are special cases of the general description by Kuznetsov and Tsybakov [64]. They show that there exists an encoding matrix C that facilitates utilization of the fraction (1-p) of non-defective memory cells, which is the best we can expect.

In the sequel, we assume that there are t defects in a length n binary vector and the defect fraction is $t/n = p$. The encoding matrix C is given in Figure 8.4, where the first (n-k) columns are used to uniquely identify a particular row (we have $2^{n-k}$ rows). A particular row is used to make the vector $\tilde{x}^n$ defect compatible.

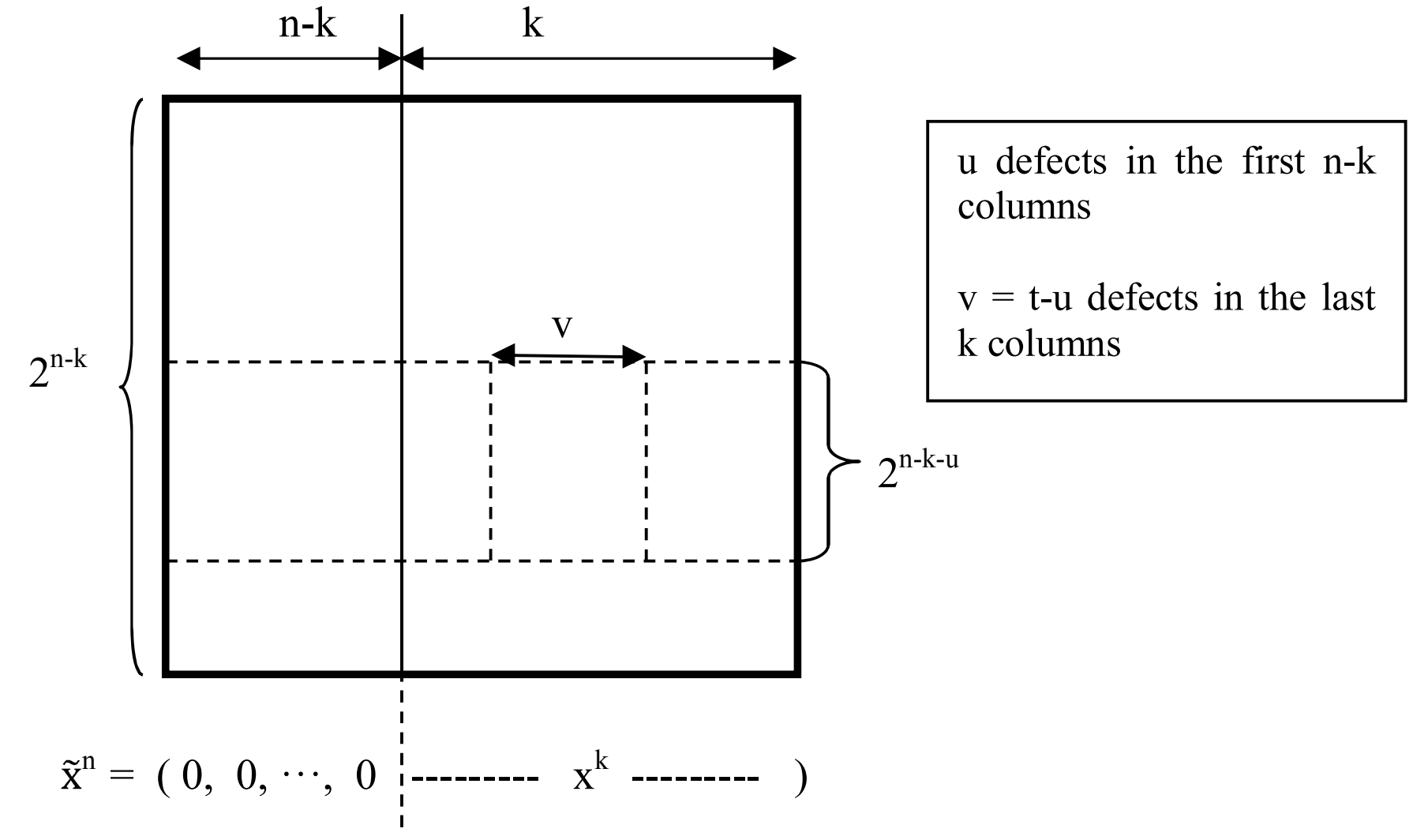

**Figure 8.4** Non-linear encoding (Kuznetsov-Tsybakov)

Suppose that for a particular defect vector $d^n$ we have u defects in the first (n-k) positions and v = t - u in the last k positions, respectively. Then, the maximum number of rows that can be used to match the defect vector is $2^{n-k-u}$, because u positions are equal to the defects and thus fixed. If the matrix is such that we have a pattern in the selected $2^{n-k-u}$ rows that can be used to match the defects in the last k positions, the matrix is called useful. We calculate the fraction of matrices that is not useful and when this fraction is smaller than 1, a useful matrix exists.

The v defects in the last k positions and the u defects in the first n-k positions specify a matrix of dimension $v \times 2^{n-k-u}$. The number of matrices where the given defect pattern does not occur is given by

$$f_1 = (2^v - 1)^{2^{n-k-u}}.$$

The remaining part of the $2^{n-k} \times k$ matrix can be chosen freely. The number of choices $f_2$ is

$$f_2 = 2^{2^{n-k}k - v2^{n-k-u}}.$$

For the particular defect vector, the fraction of useless matrices is thus given by

$$F = \frac{(2^v - 1)^{2^{n-k-u}} \times 2^{2^{n-k}k - v2^{n-k-u}}}{2^{2^{n-k}k}}$$

$$= \left(1 - 2^{-v}\right)^{2^{n-k-u}}$$

$$\leq e^{-2^{n-k-t}}.$$

Since we can have t defects with all possible combinations of defect-0 and defect-1, the total fraction of matrices that cannot be used is upper bounded by

$$F \leq \exp(-2^{n-k-t} + \ln\binom{n}{t}2^t).$$

If the efficiency is defined as R = k/n, then F is less than 1 for

$$R \leq 1 - \frac{t}{n} - \frac{1}{n}\log_2(\ln\binom{n}{t}2^t).$$

Hence, for large values of n, the storage efficiency approaches R → 1 - t/n = 1 - p. This is a surprising result with many applications, such as writing in WOM or flash memory [68]. Since the defect matching as designed by Kuznetsov-Tsybakov is a non-linear method, it might be useful to look at the potential application of linear error correcting block codes and analyze the performance.

## 8.3 Defects and Linear Codes

### 8.3.1 Defect matching

We start with the systematic binary encoding matrix $G_{k,n}$ for a code with minimum distance $d_{min}$ that has the form

$$G_{k,n} = [\ I_{k,k}\ \ H_{k,n-k}\ ],$$

where $I_{k,k}$ is the k × k identity matrix, and $H_{k,n-k}$ the binary check part. The parity check matrix for this encoder is given by

$$(H_{n-k,n})^T = \begin{bmatrix} H_{k,n-k} \\ I_{n-k,n-k} \end{bmatrix}.$$

Since the code has minimum distance $d_{min}$, any $d_{min}$ - 1 rows of $(H_{n-k,n})^T$ are linearly independent and thus in this case $d_{min}$ - 1 columns of

$$G_{n-k,n} = [\ I_{n-k,n-k}\ \ (H_{k,n-k})^T\ ].$$

We can use the systematic code $G_{n-k,n}$ to construct any combination of $d_{min}$ - 1 digits by some linear combination of rows. The matrix $G_{n-k,n}$ is taken as part of the bigger n × n matrix that has the structure as given in Figure 8.5, where $0_{k,n-k}$ is the all zero matrix with dimensions k and n-k. As before, the vector $x^k$ and the defect vector $d^n = (d_1, d_2, \cdots, d_n)$ determine the vector $c^{n-k}$ such that

$$\tilde{c}^n = (c^{n-k}, x^k)\, G_{n,n}$$

is defect compatible with $d^n$.

$$G_{n,n} = \begin{bmatrix} I_{n-k,n-k} & (H_{k,n-k})^T \\ 0_{k,n-k} & I_{k,k} \end{bmatrix}$$

**Figure 8.5** The ($d_{min}$-1)-defect matching code generator

**Example** For the encoder in Figure 8.6a, the first 4 rows are derived from a systematic encoder with a minimum distance of 4 and code word length 7. We can thus match 3 defects.

| | | | | | | |
|---|---|---|---|---|---|---|
| 1 | 0 | 0 | 0 | 1 | 1 | 1 |
| 0 | 1 | 0 | 0 | 1 | 1 | 0 |
| 0 | 0 | 1 | 0 | 1 | 0 | 1 |
| 0 | 0 | 0 | 1 | 0 | 1 | 1 |
| 0 | 0 | 0 | 0 | 1 | 0 | 0 |
| 0 | 0 | 0 | 0 | 0 | 1 | 0 |
| 0 | 0 | 0 | 0 | 0 | 0 | 1 |

**Figure 8.6a** A 3-defect matching code generator: n = 7, k = 3

## 8.3.2 Defect matching and random error correction

Extension to defect matching and random error correction is done as follows. We extend $G_{n,n}$ with some additional columns such that the resulting matrix is an encoding matrix for an error correcting code. At the same time the top rows are used for defect matching. Hence, at reading we first correct the random errors and then find back the information. The problem is to find the combination of defect matching and error correction with highest efficiency. This problem was tackled by Heegard [67]. We will illustrate the method by examples.

**Example** The encoding matrices in Figure 8.6b and Figure 8.6c follow from the (7,4) Hamming code with minimum distance 3. In both cases we have the same distance properties and thus the same error correcting capabilities (1-error correcting).

In Figure 8.6b we use the first row to make the vector $x^7$ defect compatible, and in Figure 8.6c we use the first 3 rows, respectively.

| | | | | | | |
|---|---|---|---|---|---|---|
| 1 | 1 | 1 | 1 | 1 | 1 | 1 |
| 0 | 1 | 0 | 0 | 1 | 1 | 0 |
| 0 | 0 | 1 | 0 | 1 | 0 | 1 |
| 0 | 0 | 0 | 1 | 0 | 1 | 1 |

**Figure 8.6b** 1-defect and 1-error correcting, n = 7, k = 3

| | | | | | | |
|---|---|---|---|---|---|---|
| 1 | 0 | 0 | 0 | 1 | 1 | 1 |
| 0 | 1 | 0 | 1 | 1 | 0 | 1 |
| 0 | 0 | 1 | 1 | 1 | 1 | 0 |
| 0 | 0 | 0 | 1 | 0 | 1 | 1 |

**Figure 8.6c** 2-defects and 1-error correcting, n = 7, k = 1.

Note, that the storage efficiency is the same for the situation 8.6a and 8.6b.

## 8.4 Defects as Symbols

Instead of using a binary code generator matrix, we can use a generator matrix of an RS code over GF($2^m$), where symbols represent m binary digits. We consider symbols as defects instead of binary digits.

In Figure 8.7 we give a particular partitioning for an (n–δ) × n RS encoding matrix $G_{n-\delta,n}$ with elements from GF($2^m$).

**Properties:**

- for the RS code generated by the encoding matrix $G_{n-\delta,n}$, the minimum distance is $(\delta + 1)$;
- any $(n-k) \times (n-k)$ sub-matrix of $G_{n-k,n}$ has rank n-k.

**Encoding**

For an information vector $x^{k-\delta}$, we can construct $c^{n-k}$ such that

$$\tilde{c}^n = (c^{n-k}, x^{k-\delta})\, G_{n-\delta,n}$$

is symbol defect matching.

We receive

$$r^n = \tilde{c}^n \oplus e^n.$$

**Decoding**

Using $G_{n-\delta,n}$ we can correct the random errors that might occur after storing the code vector $\tilde{c}^n$. After decoding the random errors, the vector $(c^{n-k}, x^{k-\delta})$ can be reconstructed and thus we can find back the original encoded information $x^{k-\delta}$.

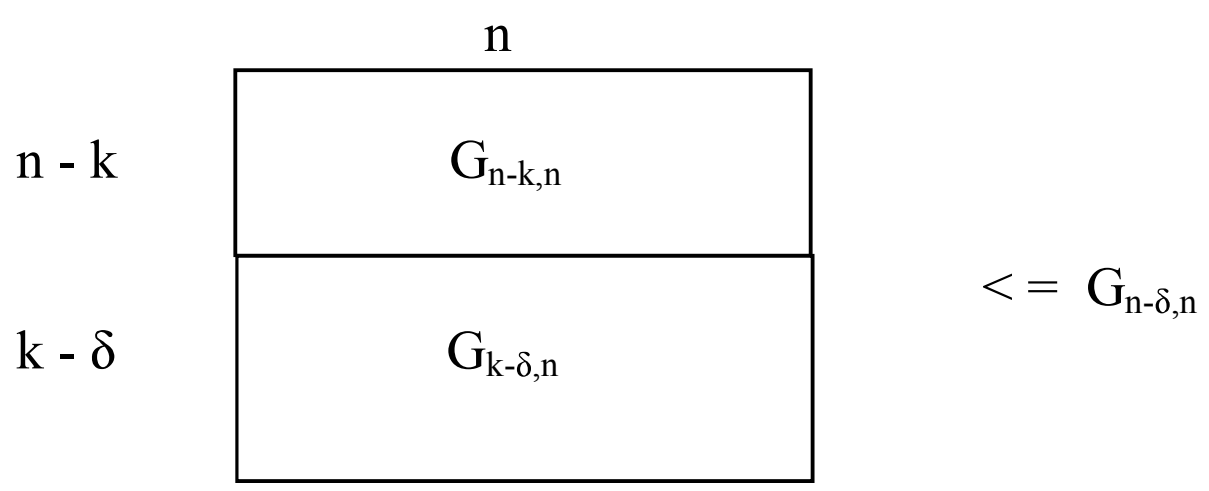

**Figure 8.7** Dimensions for the RS encoding matrix $G_{n-\delta,n}$

Note that the error correcting properties appear when $\delta > 0$. The defect matching properties and the efficiency depend on the values of $(n - k)$ and $(k - \delta)$. For $\delta = 0$, we store k symbols in n positions. The efficiency is

$$R = \frac{k}{n} = 1 - \frac{n-k}{n} = 1 - (\text{defect fraction}).$$

We observe, that the RS code is optimum!

**Remark.** Although not very practical, the same defect matching procedure can be designed for convolutional codes, see [65].

**Application to the Write Once Memory (WOM)**

Rivest and Shamir considered updating punch cards, punch tapes, and other storage media which in time [69]. In this section we show how to get the capacity of the punch card and other similar degrading memories by additive coding. In order to be able to use a punch card at least twice we must restrict the number of holes punched at the first use of a new punch card by some number p, 0 < pn < n. Under this restriction there are

$$M = \sum_{i=0}^{pn} \binom{n}{i}$$

different ways to punch a new card. Therefore, M messages can be represented in this way, and during the first use of the punch card we store $\log_2 M$ bits.

The configuration of holes on the punch card can be represented by the binary vector $d^n$, where $d_i = 1$ if the ith position of the card is punched (a hole), and $d_i = ?$ otherwise.

At the second use of the punch card the existing configuration of holes, represented by the vector $d^n$, can be used by a defect matching code. If we have a defect matching code for a fraction p of defects, we can store a maximum of (n - pn) bits in the second writing. Hence, the total normalized asymptotic number of bits stored in two writings is

$$T = \lim_{n\to\infty} \frac{1}{n}\left[\log_2 \sum_{i=0}^{pn} \binom{n}{i} + (n - pn)\right]$$
$$= h(p) + (1 - p) \quad \text{bit/memory cell.}$$

If the WOM is used three or more times, we can restrict the writing in every step and optimize the fraction of holes to be punched [66]. Note that the writing efficiency is different for subsequent writing. The writing with symmetric efficiencies is under investigation.

# 8.5 Concluding Remarks

This chapter describes the development of coding for memories with defects. The research was inspired by the researcher Alexander Kutznetsov, from the Russian Academy of Sciences, IPPI, in Moscow. The defect model is used today as a basic model in for instance coding for WOM and also flash memories, see also [78]. We describe the defect model and discuss the storage capacity for different modes of use. Defect matching, as developed by Kutznetsov and Tsybakov, uses non-linear codes for making information suitable to be stored in a memory where errors are permanent (defects). Their general method achieves the theoretical capacity. As a result, we can use a fraction (1-p) of a memory when a fraction p of the memory is defective. A surprising result! The matching is generalized to linear codes. When defects occur as symbols, RS codes obtain optimum performance.

The concept of the content of the chapter is given in Figure 8.8.

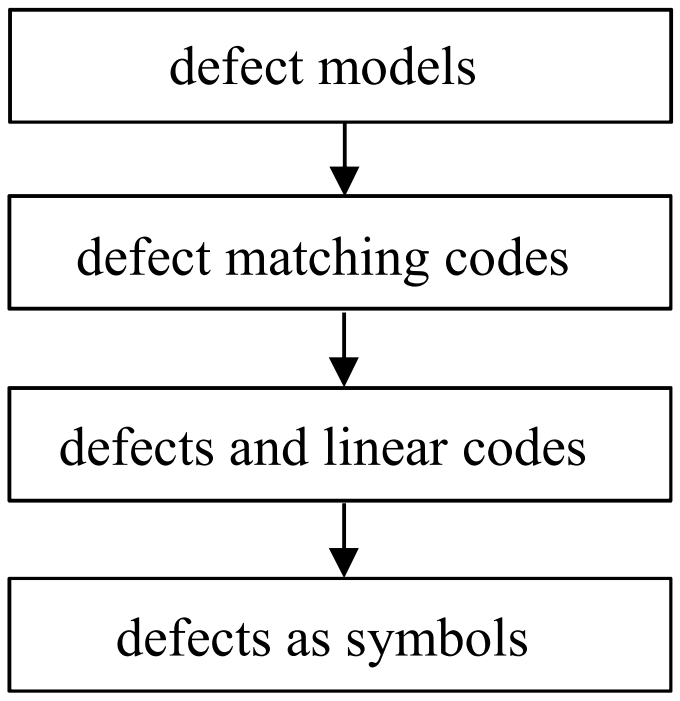

**Figure 8.8** Concept of the content of Chapter 8

# Appendix

## A.1 A Bit of Information Theory

Shannon entropy is a basic concept in information theory. For a source with output messages $S \in \{s_1, s_2, \cdots, s_M\}$, cardinality M and probabilities $(p_1, p_2, \cdots, p_M)$, the entropy is defined as

$$H(S) = -\sum_{i=1}^{M} p_i \log_2 p_i \quad \text{bit.} \tag{A.1.1a}$$

Shannon proved that H(S) is the minimum average number of bits needed to describe the source output. The binary entropy, h(p), is defined as

$$h(p) = -p\log_2 p - (1-p)\log_2(1-p) \text{ bit.} \tag{A.1.1b}$$

Figure A.1 gives the graphical representation of h(p).

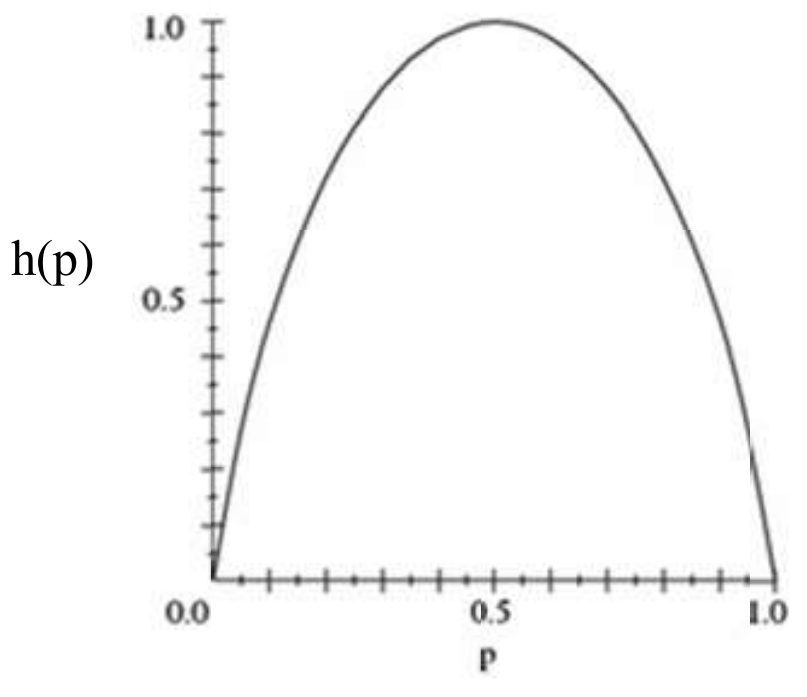

**Figure A.1** The graphical representation of h(p)

The conditional entropy H(S | T) is the minimum average description length of S given the side information T. It is defined as

$$H(S|T) = -\sum_{S,T} P(s,t)\log_2 P(s|t) . \quad \text{(A.1.1c)}$$

It follows from the Log sum inequality (Chapter 16, [22]) that $H(S|T) \leq H(S)$. The joint entropy,

$$\begin{aligned} H(S,T) &= H(T) + H(S|T) \\ &= H(S) + H(T|S). \end{aligned}$$

The difference

$$\begin{aligned} I(S;T) &= H(S) - H(S|T) \\ &= H(T) - H(T|S), \end{aligned}$$

is called the mutual information, and can be seen as the reduction in the average description length of S given the observation T (or the other way around). Conditional entropy, like H(S|T), is also called equivocation.

In data transmission we have a channel input ensemble X and a channel output ensemble Y connected via the mutual information I(X;Y). The maximum of I(X;Y) over the input probability distribution P(X) is called the channel capacity

$$\begin{aligned} C &= \max_{P(X)} I(X;Y) \\ &= \max_{P(X)} [H(X) - H(X|Y)]. \end{aligned} \quad \text{(A.1.2)}$$

Channel codes exist that can transmit at a rate as close to capacity as wanted, and at the same time achieve a decoding word error probability as close to zero as required. To achieve this, the word length has to go to infinity.

**Remark** Recommended text books:

- T. Cover and J. Thomas, Elements of Information Theory 2nd Edition, Wiley Series in Telecommunications and Signal Processing, 1991

- A classical book by R. Gallager is also highly recommended: R Gallager, Information Theory and Reliable Communication, John Wiley and Sons, 1968

# A.2 A Short Introduction to Galois fields

The purpose of this appendix is to give a brief introduction to Galois fields and it applications. We start with the system of calculations using the numbers modulo p, where p is a prime number. Then we continue to do the equivalent for symbols that can be considered as consisting of m binary digits.

## A.2.1 Calculations modulo a prime number, GF(p)

Suppose that we do calculation modulo a prime number p.

**Definition**: The Galois field GF(p) is the collection of p elements

$$\{0, 1, 2, \cdots, p-1\}.$$

Then, we use the following properties.

**Property** There exists a *primitive element*, called $\alpha$ such that all p-1 powers of $\alpha$ are <u>different</u> modulo p and $\alpha^{p-1} = 1$ modulo p.

**Proof** Suppose that $\alpha^i = 1$ modulo p, $i < p - 1$. Then $\alpha^{j-i}\,\alpha^{i} = \alpha^{j} = \alpha^{j-i}$ and thus not all elements are different, which contradicts the assumption.

**Example** let p = 7 and calculations modulo p. The 6 powers of the number 3 are: $3^1 = 3$, $3^2 = 2$, $3^3 = 6$, $3^4 = 4$, $3^5 = 5$, $3^6 = 1$. They are all different.

**Property** Every element has an inverse.

**Proof** Since $\alpha^{p-1-i}\,\alpha^i = 1$ modulo p, $\alpha^{p-1-i}$ is called the inverse of $\alpha^i$. Furthermore, $\alpha^j\,\alpha^i = \alpha^{j+i}$ modulo p.

**Example** Let p = 7 and calculations modulo p. The inverses are given by: $3^{-1} = 5$, $3^{-2} = 4$, $3^{-3} = 6$, $3^{-4} = 2$, $3^{-5} = 3$, and $3^{-6} = 1$.

We thus have defined multiplication, inverse and the unit element 1. The symbol 0 and all the nonzero elements together form the Galois field GF(p). We have an equivalent definition for a Galois field using polynomials.

## A.2.2 Calculations modulo a binary minimal polynomial

We consider polynomials of degree ≥ 0, i.e.

$$P(X) = p_0 + p_1 X + \cdots + p_{m-1} X^{m-1}, \qquad p_i \in \{0,1\}.$$

Polynomials can be added or multiplied, where the coefficients of equal exponents are assumed to be added modulo 2.

Calculations modulo a polynomial g(X) can be interpreted as follows.

P(X) modulo g(X) = r(X) means that P(X) = a(X)g(X) + r(X), where the degree{r(X)} < degree {g(X)}.

**Example** $(1+X^4)$ modulo$(1+X+X^2) \rightarrow (1+X^4) = (X+X^2)(1+X+X^2) + (1+X)$.

**Definition** A binary polynomial without divisors is called irreducible.

**Definition** An irreducible polynomial I(X) of degree m is called minimal if and only if (iff)

$$I(X) \text{ divides } (X^n + 1), \text{ where the smallest n is } n = (2^m - 1).$$

**Example** $(1 + X + X^2)$ divides $X^3 + 1$ and is minimal.

**Example** $(1 + X + X^3)$ divides $X^7 + 1$ and is minimal.

**Remark** At this position we want to remark that there are $2^m$ different polynomials of degree less than m. With every polynomial we can uniquely connect a binary m-tuple or binary symbol of length m.

**Example** The binary 4 tuple (0,1,0,1) can be represented as $X + X^3$.

We use the following important equivalence notation using the minimal polynomial I(X), i.e.

$$X^i \text{ modulo } I(X) \equiv \alpha^i.$$

**Definition** The Galois field $GF(2^m)$ is the collection of $2^m$ elements

$$\{0, 1, \alpha, \alpha^2, \cdots, \alpha^{p-2}\}.$$

As a consequence, every element corresponds to a binary polynomial of degree less than m and is equivalent to an m-tuple.

**Property** All the elements $\alpha^i$ are different.

**Proof** Suppose that two elements are the same, then for i < j,

$$(X^i + X^j) \text{ modulo } I(X) = X^i (X^{j-i} + 1) = 0 \text{ modulo } I(X).$$

Since $j - i < 2^m - 1$ and I(X) minimal, this is impossible.

**Property** The element $\alpha^{p-1} = 1$.

**Property** $\alpha^i \alpha^j = \alpha^{i+j}$.

**Property** Since $\alpha^{p-1} = 1$, we can reduce the exponents modulo p-1 as before.

**Property** Every nonzero element has an inverse.

This follows easily from $\alpha^i \alpha^{-i} = \alpha^i \alpha^{p-1-i} = \alpha^i \alpha^j$.

**Example** Take $I(X) = 1 + X + X^3$, which divides $X^7-1$. The following table lists the elements of the Galois field

**TABLE A.2.1**

| | | modulo $1 + X + X^3$ | binary 3-tuple | inverse |
|---|---|---|---|---|
| $\alpha$ | $X$ | $X$ | (010) | $\alpha^6$ |
| $\alpha^2$ | $X^2$ | $X^2$ | (001) | $\alpha^5$ |
| $\alpha^3$ | $X^3$ | $1 + X$ | (110) | $\alpha^4$ |
| $\alpha^4$ | $X^4$ | $X + X^2$ | (011) | $\alpha^3$ |
| $\alpha^5$ | $X^5$ | $1 + X + X^2$ | (111) | $\alpha^2$ |
| $\alpha^6$ | $X^6$ | $1 + X^2$ | (101) | $\alpha^1$ |
| $\alpha^0 = 1$ | $X^7$ | 1 | (100) | 1 |

Adding the element 0 gives the additional 3-tuple (000) and thus we have 8 elements. We can add, multiply and calculate an inverse with m-tuples represented as elements in a Galois field $GF(2^m)$.

# A.3 Vandermonde's Determinant

In the theory of RS codes, we need the property that any k × k submatrix from the generator matrix $G_{k,n}$ formed by k rows and any k columns has full rank. We prove this by showing that the corresponding determinant is unequal to zero.

Without loss of generality, we can consider the Vandermonde matrix

$$V_k = \begin{bmatrix} 1 & 1 & \cdots & 1 \\ \alpha_1 & \alpha_2 & \cdots & \alpha_k \\ \alpha_1^2 & \alpha_2^2 & \cdots & \alpha_k^2 \\ \cdots & & & \\ \alpha_1^{k-1} & \alpha_2^{k-1} & \cdots & \alpha_k^{k-1} \end{bmatrix}.$$

Let P(k) be the <u>proposition</u> that $D_k = \prod_{1 \le i < j \le k} (\alpha_i - \alpha_j)$ .

**Proof by <u>induction</u>**: P(1) is true, as this just says $D_1 = 1$. P(2) holds, as it is the case: $D_2 = (\alpha_1 - \alpha_2)$ . This is our <u>basis for the induction</u>.

Now we need to show that, if P(t) is true, where $t \ge 2$ , then it logically follows that P(t+1) is true. So this is our <u>induction hypothesis</u>:

$$D_t = \prod_{1 \le i < j \le t} (\alpha_i - \alpha_j).$$

Then we need to show: $D_{t+1} = \prod_{1 \le i < j \le t+1} (\alpha_i - \alpha_j)$.

This is our <u>induction step</u>: Consider the determinant:

$$D_{t+1} = \begin{vmatrix} 1 & 1 & 1 & \cdots & 1 \\ x & \alpha_2 & \alpha_3 & \cdots & \alpha_{t+1} \\ x^2 & \alpha_2^2 & \alpha_3^2 & \cdots & \alpha_{t+1}^2 \\ & & \cdots & & \\ x^t & \alpha_2^t & \alpha_3^t & \cdots & \alpha_{t+1}^t \end{vmatrix}.$$

If you use the "Expansion Theorem for Determinants" to expand it in terms of the first column, you can see it is a polynomial in x whose degree is no greater than t. Call that polynomial f(x).

If you substitute any $\alpha_i$ for x in the determinant, two of its columns will be the same. So the value of such a determinant will be 0. Such a substitution in the determinant is equivalent to substituting $\alpha_i$ for x in f(x). Thus it follows that $f(\alpha_2) = f(\alpha_3) = \ldots = f(\alpha_{t+1}) = 0$ as well. So f(x) is divisible by each of the factors $\alpha_2, \alpha_3, \ldots, \alpha_{t+1}$. All these factors are distinct otherwise the original determinant is zero. So:

$$f(x) = C(x - \alpha_2)(x - \alpha_3)\cdots(x - \alpha_t)(x - \alpha_{t+1}).$$

As the degree of f(x) is no greater than t , it follows that C is independent of x. From the Expansion Theorem for Determinants, we can see that the coefficient of $x^t$ is:

$$D_t = \begin{vmatrix} 1 & 1 & \cdots & 1 \\ \alpha_2 & \alpha_3 & \cdots & \alpha_{t+1} \\ \alpha_2^2 & \alpha_3^2 & \cdots & \alpha_{t+1}^2 \\ & \cdots & & \\ \alpha_2^{t-1} & \alpha_3^{t-1} & \cdots & \alpha_{t+1}^{t-1} \end{vmatrix},$$

which by the induction hypothesis is equal to $\prod_{2 \le i < j \le t+1} (\alpha_i - \alpha_j)$.
So this has to be our value of C. Thus, we have:

$$f(x) = (x - \alpha_2)(x - \alpha_3)\cdots(x - \alpha_t)(x - \alpha_{t+1}) \prod_{2 \le i < j \le t+1} (\alpha_i - \alpha_j).$$

Substituting $\alpha_1$ for x, we retrieve the proposition P(t+1). So P(t) $\Longrightarrow$ P(t+1) and the result follows by the Principle of Mathematical Induction. Therefore:

$$D_k = \prod_{1 \le i < j \le k} (\alpha_i - \alpha_j),$$

which is unequal to zero since all elements $\alpha_i$, $i = 1,2, \cdots, k$ are different. .

# A.4 RS Encoding and Syndrome Former

We first describe the general encoding matrix as it is convenient for our further considerations. We derive some of the RS code properties, like distance and encoding matrix rank. These are based on the Vandermonde matrix as explained in appendix A.3.

**The encoder**

The encoder performs the inner product of an information vector $a^k$ of length k with a k × n encoding matrix $G_{k,n}$. Both $a^k$ and $G_{k,n}$ have components from $GF(2^m)$. The k × n encoding matrix consists of k consecutive rows of the following matrix

$$G_{n,n} = \begin{bmatrix} 1 & 1 & 1 & \cdots & 1 \\ 1 & \alpha & \alpha^2 & \cdots & \alpha^{n-1} \\ 1 & \alpha^2 & \alpha^4 & \cdots & \alpha^{2(n-1)} \\ \cdots & & & & \\ 1 & \alpha^{n-1} & \alpha^{2(n-1)} & \cdots & \alpha^{(n-1)(n-1)} \end{bmatrix}, \qquad \text{(A.4.1)}$$

where $\alpha$ is a primitive element of $GF(2^m)$ and $n = 2^m - 1$.

**Example** The first k rows give as an encoding matrix

$$G_{k,n} = \begin{bmatrix} 1 & 1 & 1 & \cdots & 1 \\ 1 & \alpha & \alpha^2 & \cdots & \alpha^{n-1} \\ 1 & \alpha^2 & \alpha^4 & \cdots & \alpha^{2(n-1)} \\ \cdots & & & & \\ 1 & \alpha^{k-1} & \alpha^{2(k-1)} & \cdots & \alpha^{(k-1)(n-1)} \end{bmatrix}. \qquad \text{(A.4.2)}$$

The following two properties are basic for the applications of RS codes. Without loss of generality, we choose $G_{k,n}$ to be the first k rows from $G_{n,n}$ as in (A.4.2).

**Property 1** Any k columns of $G_{k,n}$ have rank k.

This follows easily from a particular selection of k columns

$$V(\alpha^p, \alpha^q, \cdots, \alpha^r) = \underbrace{\begin{bmatrix} 1 & 1 & \cdots & 1 \\ \alpha^p & \alpha^q & \cdots & \alpha^r \\ \alpha^{2p} & \alpha^{2q} & \cdots & \alpha^{2r} \\ \cdots & & & \\ \alpha^{p(k-1)} & \alpha^{q(k-1)} & \cdots & \alpha^{r(k-1)} \end{bmatrix}}_{k} \Big\updownarrow k$$

which has the form of a Vandermonde matrix (note that all second row elements are different).

**Property 2** The minimum distance $d_{min}$ of the code is (n-k+1).

For linear codes, the minimum distance is equal to the minimum Hamming weight of a nonzero code word. This property follows from the fact that any k columns are linearly independent and thus no linear combination of rows can give a code word with k zero symbols. Hence, the minimum nonzero Hamming weight of any code word is larger or equal to n-k+1. Since the Singleton upper bound gives $d_{min} \le n-k+1$, property 2 follows.

The same proof follows from the following observation (from Jack van Lint his lectures at Eindhoven University). Let A(X) be a polynomial over $GF(2^m)$ of degree k-1. We can evaluate this polynomial for all $n = 2^m-1$ possible different powers of the primitive element $\alpha$ of the Galois field $GF(2^m)$. This is equivalent to the inner product encoding operation as described before. Since A(X) can have only k-1 different roots, at least n-k+1 evaluations must be nonzero and thus any nonzero code word has Hamming weight larger or equal to n–k+1. Since the minimum distance for linear codes is less than or equal to n–k+1, $d_{min}$ = n-k+1.

**Example** The (7,3) non-systematic RS encoding matrix is given by

$$G = \begin{bmatrix} 1 & 1 & 1 & 1 & 1 & 1 & 1 \\ 1 & \alpha & \alpha^2 & \alpha^3 & \alpha^4 & \alpha^5 & \alpha^6 \\ 1 & \alpha^2 & \alpha^4 & \alpha^6 & \alpha & \alpha^3 & \alpha^5 \end{bmatrix}$$

In systematic form, we have

$$G = \begin{bmatrix} 1 & 0 & 0 & \alpha^3 & 1 & \alpha^3 & \alpha^2 \\ 0 & 1 & 0 & \alpha^5 & 1 & \alpha & \alpha \\ 0 & 0 & 1 & \alpha^4 & 1 & \alpha^6 & \alpha^4 \end{bmatrix} .$$

The matrix $G_{n,n}$ has an inverse, given by

$$H_{n,n} = \begin{bmatrix} 1 & 1 & \cdots & 1 & 1 \\ 1 & \alpha^{n-1} & \cdots & \alpha^2 & \alpha \\ 1 & \alpha^{2(n-1)} & \cdots & \alpha^4 & \alpha^2 \\ & \cdots & & & \\ 1 & \alpha^{(n-1)(n-1)} & \cdots & \alpha^{2(n-1)} & \alpha^{n-1} \end{bmatrix} . \quad \text{(A.4.3)}$$

The syndrome former $H^T$ that corresponds to the encoder (A.4.2), has dimension $n \times (n-k)$ and is given by the last n-k columns from $H_{n,n}$, i.e.

$$H^T = \begin{bmatrix} 1 & \cdots & 1 & 1 \\ \alpha^{n-k} & \cdots & \alpha^2 & \alpha \\ \alpha^{2(n-k)} & \cdots & \alpha^4 & \alpha^2 \\ & & \cdots & \\ \alpha^{(n-1)(n-k)} & \cdots & \alpha^{2(n-1)} & \alpha^{n-1} \end{bmatrix} . \quad \text{(A.4.4)}$$

**Property 3** The product $GH^T = 0$.

We recall that $n = 2^m-1$. For $GF(2^m)$, we have $\alpha^n = 1$ and thus for $\beta = \alpha^q$, $q \leq n-1$,

$$\beta^n - 1 = (\beta - 1)(\beta^{n-1} + \beta^{n-2} + \cdots + 1) = 0.$$

Any inner product P of a row of $G_{k,n}$ with a column of $H^T$ looks like

$$P = 1 + \alpha^{i+j} + \alpha^{2(i+j)} + \cdots + \alpha^{(n-1)(i+j)}$$

$$= 1 + \beta + \beta^2 + \cdots + \beta^{n-1}$$

$$= 0,$$

for $0 < i + j \leq n-1$, and thus $\beta \neq 1$.

The syndrome former $H^T$ can be used to decode a received code word at the receiver. We distinguish the following steps.

- Suppose that after transmission of $c^n = x^k G_{k,n}$, the receiver obtains a noisy version, $r^n = c^n + e^n$, where addition is in $GF(q = 2^m)$.

- The decoder calculates the product $r^n H^T = s^{n-k}$, which is called the syndrome. The syndrome only depends on the noise word $e^n$.

- For a code with minimum distance (n-k+1) all error words of Hamming weight less than $\lfloor (n-k)/2 \rfloor$ give a different syndrome. For, otherwise a vector of weight less than n-k+1 can be constructed that gives a syndrome $0^{n-k}$. According to the definition of the minimum distance, this is not possible.

- The number of different syndromes and thus correctable error words is $q^{n-k}$. The number of error words of weight $\lfloor (n-k)/2 \rfloor$ or less is given by

$$\sum_{i=0}^{\left\lfloor \frac{n-k}{2} \right\rfloor} \binom{n}{i} (q-1)^i < q^{\left\lfloor \frac{n-k}{2} \right\rfloor} \cdot \sum_{i=0}^{\left\lfloor \frac{n-k}{2} \right\rfloor} \binom{n}{i} \leq q^{n-k},$$

  where we assume that $n = q-1$. From this we can see, that for small k, almost all syndromes correspond to an error word of weight $\leq \lfloor (n-k)/2 \rfloor$. For larger k, the bound will not be tight.

**Example** We give two examples: the single parity check (n,n-1) RS code with efficiency k/n = (n-1)/n, minimum distance 2 and the repetition code with efficiency 1/n and minimum distance n.

For the (n,n-1) RS code, we start with the generator matrix

$$G_{n-1,n} = \begin{bmatrix} 1 & \alpha & \alpha^2 & \cdots & \alpha^{n-1} \\ 1 & \alpha^2 & \alpha^4 & \cdots & \alpha^{2(n-1)} \\ & \cdots & & & \\ 1 & \alpha^{n-1} & \alpha^{2(n-1)} & \cdots & \alpha^{(n-1)(n-1)} \end{bmatrix}.$$

The parity check matrix H is given by

$$H = [1, 1, \cdots, 1].$$

We give the equivalent (n-1) × n encoding matrix G in systematic form as

$$G_{n-1,n} = \begin{bmatrix} 1 & 0 & 0 & & \cdots & & 1 \\ 0 & 1 & 0 & & \cdots & & 1 \\ 0 & 0 & 1 & 0 & \cdots & & 1 \\ & \cdots & & & & & \\ 0 & 0 & \cdots & 0 & & 1 & 1 \end{bmatrix}. \qquad \text{(A.4.5)}$$

It is easy to see, that (A.4.5) generates all code words $c^n$ such the inner product $c^n H^T = x^{n-1} G_{n-1,n} H^T = 0^{n-k}$. This matrix $G_{n-1,n}$ is the simplest possible and simplifies the encoding procedure drastically.

Decoding for this code is in principle the same as for the SPC. After taking hard decision, the syndrome gives the error value. The position can be determined by finding the position that minimizes the Euclidean distance.

**Example** For the R = 1/n repetition code, the encoding matrix is

$$G_{1,n} = [1, 1, , \cdots, 1]. \qquad \text{(A.4.6)}$$

The corresponding parity check matrix $H^T$ is given by

$$H^T = \begin{bmatrix} 1 & \cdots & 1 & 1 \\ \alpha^{n-1} & \cdots & \alpha^2 & \alpha \\ \alpha^{2(n-1)} & \cdots & \alpha^4 & \alpha^2 \\ \cdots & & & \\ \alpha^{(n-1)(n-1)} & \cdots & \alpha^{2(n-1)} & \alpha^{n-1} \end{bmatrix}. \qquad \text{(A.4.7)}$$

After hard decision, the decoding procedure is a simple majority vote. In this case, more than n/2 errors lead to decoding decision errors.

**Property 4** Any (n-k) rows of $H^T$ as given in (A.4.4) have rank n-k.

This follows directly from the minimum distance of the code generated by $G_{k,n}$, which is n-k+1. Thus, all n-k or less rows of $H^T$ are linearly independent. Furthermore, $H_{n-k,n}$ is an RS encoding matrix with minimum distance (k + 1).

**Remark** Suppose that we represent a code word $c^n$ of length n as a degree (n-1) polynomial C(X). Then, $c^n H^T = x^k G H^T = 0^{n-k}$ is equivalent to the n-k evaluations of C(X) for $X = \alpha, \alpha^2, \cdots, \alpha^{n-k}$, where $H^T$ is given in (A.4.4) We can say that

$$C(X) = A(X)g(X) = A(X)(X-\alpha)(X-\alpha^2)\cdots(X-\alpha^{n-k}),$$

where g(X) is called the generator of the RS code and A(X) the information polynomial of degree (k-1).

The calculation $X^{n-k}A(X)$ modulo g(x) gives a rest of degree less than n-k, also called check part. Hence, we can have a systematic code word containing the original information by writing $C(X) = (X^{n-k}A(X) - X^{n-k}A(X)$ modulo g(x)). Note that C(X) is still a multiple of g(X), since C(X) modulo g(X) = 0 and the – sign can be replace for + when calculations are done in the Galois field $GF(2^m)$.

**Remark** We developed the theory of syndrome decoding for convolutional codes in [92,93,94]. This approach leads to lower complexity decoding when statistical properties of the noise are considered. Furthermore, structure in the encoding procedure can give rise to state-space reductions.

# A.5 The Decoding Algorithm

Recall that the syndrome former has the following form

$$H^T = \begin{bmatrix} 1 & \cdots & 1 & 1 \\ \alpha^{n-k} & \cdots & \alpha^2 & \alpha \\ \alpha^{2(n-k)} & \cdots & \alpha^4 & \alpha^2 \\ \cdots & & & \\ \alpha^{(n-1)(n-k)} & \cdots & \alpha^{2(n-1)} & \alpha^{(n-1)} \end{bmatrix}.$$

If we write the code words in the form of a polynomial, then the code word $c^n$ is written as C(X). In the same way, we write the received word as

$$R(X) = C(X) + E(X),$$

where E(X) represents the error symbols. Using the properties of the code word C(X), we calculate the syndromes

$$S_j = R(X = \alpha^{j+1}) = C(X = \alpha^{j+1}) + E(X = \alpha^{j+1}), \quad 0 \le j \le 2t-1$$

$$= E(X = \alpha^{j+1})$$

$$= \sum_{i=0}^{n-1} E_i (\alpha^i)^{j+1}$$

$$= \sum_{k=1}^{t} E_{i_k} (\alpha^{i_k})^{j+1},$$

where we have assumed t errors at position $i_1, i_2, \cdots, i_t$, $0 \le i_k \le n-1$. The syndrome values $S_j$, $0 \le j \le 2t-1$ define the syndrome polynomial

$$S(X) = \sum_{j=0}^{\infty} S_j X^j$$

$$= \sum_{j=0}^{\infty} X^j \sum_{k=1}^{t} E_{i_k} (\alpha^{i_k})^{j+1}$$

$$= \sum_{k=1}^{t} E_{i_k}(\alpha^{i_k})/(1 - X\alpha^{i_k}).$$

The error locator polynomial for t errors at position $i_k$, k = 1,2, ⋯, t is defined as

$$L(X) = \prod_{k=1}^{t}(1 - X\alpha^{i_k}).$$

The product L(X)S(X) can be written as

$$L(X)S(X) = \sum_{k=1}^{t} E_{i_k}(\alpha^{i_k}) \prod_{\ell \neq k}^{t} (1 - X\alpha^{i_\ell}) = W(X).$$

We remark that the degree of W(X) less than t. Actually, the polynomial S(X) is known only through the coefficient of $X^{2t-1}$ so that in terms of the known syndrome components we can write the key equation

$$L(X)S(X) = W(X) \text{ modulo } X^{2t}.$$

The polynomials L(X) and W(X) are found according to Euclid's algorithm. This algorithm gives a unique valid solution in a finite number of steps (an alternative is the Berlekamp-Massey algorithm). We therefore reformulate the key equation as

$$L(X)S(X) + F(X)\, X^{2t} = W(X).$$

We have to find the solution L(X) under the condition that the greatest common divisor of S(X) and $X^{2t}$ is given by

$$\gcd(S(X), X^{2t}) = W(X).$$

For t errors, the polynomials L(X) and W(X) have degree t and degree < t, respectively and are the only solution to the key equation and hence to the problem.

We use the polynomial L(X) for finding the error locations. For an error at position $i_m$, according to the definition,

$$L(\alpha^{-i_m}) = \prod_{k=1}^{t}(1-\alpha^{-i_m}\alpha^{i_k}) = 0.$$

The search for the positions where L(*) = 0, is called the Chien search. Knowing the incorrect positions, also gives at least k correct positions and thus, this allows us to reconstruct the encoded information. An interesting topic that remains, is the complexity of this procedure.

**Example** We use the $GF(2^3)$ from Table A.2.1 and a (7,5) RS code defined as

$$C(X) = A(X)g(X) = A(X)(X\text{-}\alpha)(X\text{-}\alpha^2).$$

For an information polynomial

$$A(X) = \alpha + \alpha^3X^4,$$

The code word polynomial becomes

$$C(X) = \alpha^4 + \alpha^5X + \alpha\, X^2 + \alpha^6X^4 + X^5 + \alpha^3X^6.$$

We receive

$$R(X) = C(X) + E(X) = \alpha^4 + \alpha^5X + \alpha\, X^2 + \alpha^6X^4 + \alpha^6X^5 + \alpha^3X^6.$$

The syndrome

$$S_0 = R(\alpha) = \alpha^4 + \alpha^6 + \alpha^3 + \alpha^3 + \alpha^4 + \alpha^2 = 1 = E_i\alpha^i\,,$$

$$S_1 = R(\alpha^2) = \alpha^4 + \alpha^7 + \alpha^5 + 1 + \alpha^2 + \alpha = \alpha^5.$$

Thus, $S(X) = 1 + \alpha^5X$. For t = 1, we have to find the $\gcd(1 + \alpha^5X, X^2)$. From this, we find

$$(1 + \alpha^5X)(1 + \alpha^5X) + \alpha^3X^2 = 1.$$

The error location is at position 5. The error value is $S_0\,\alpha^{-5} = \alpha^2$.

# A.6 Middleton Class-A Impulse Noise Model

The Class-A noise model is a model for noise that is impulsive in combination with Gaussian noise. A Gaussian noise component is added to model the (almost) always present thermal receiver and background noise.

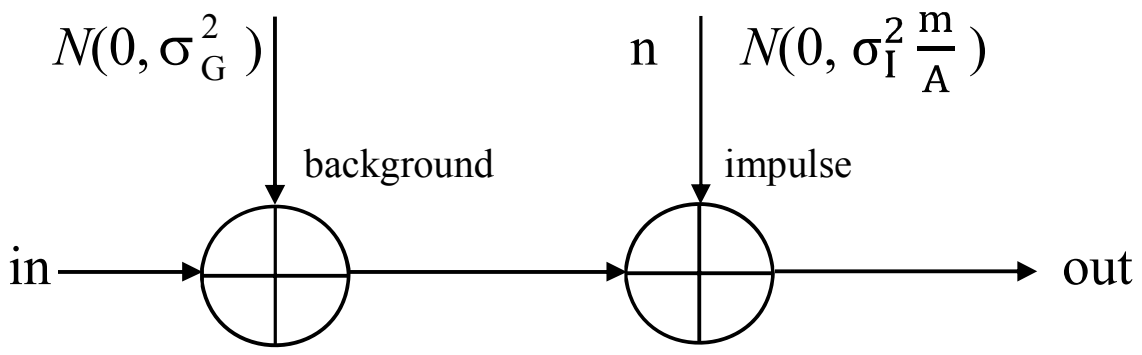

**Figure A2** Representation of the Middleton model

The model, Figure A2, is characterized by:

- $\sigma_G^2$, the variance of the Gaussian distributed background noise;
- $\sigma_I^2$ is the average variance of the Gaussian distributed impulse noise;
- the parameter A is called impulse index and is given by the product of the average number of impulses per unit time and the mean duration of the emitted impulses entering the receiver. For small A the noise has a structured/impulsive character;
- the parameter T gives the impulse strength or ratio between the mean power of the Gaussian and the mean power of the impulsive noise component, i.e. T:= $\sigma_G^2/\sigma_I^2$;
- the state m of the model is selected according to the Class-A noise probability density function [25],

$$P_m = e^{-A}\frac{A^m}{m!};m \ge 0.$$

The impulse noise probability distribution is given by

$$p(n\,|\,m) = \frac{1}{\sqrt{2\pi\sigma_m^2}} e^{-\frac{n^2}{2\sigma_m^2}}, \qquad p(n) = \sum_{m=0}^{\infty} P_m p(n\,|\,m),$$

where $\sigma_m^2 = \sigma_I^2 \frac{m}{A}, m \geq 0.$ The average noise variance $\overline{\sigma^2} = \sigma_I^2 + \sigma_G^2$. Note that for m = 0, we only have background noise.

**Example** A simple two-state model can be defined as follows:

$P(m = 0) = 1 - P(m = 1) = 1 - A;$

$\overline{\sigma^2} = \sigma_I^2 + \sigma_G^2$ ; $T := \sigma_G^2/\sigma_I^2$ .

The impulse noise channel can be seen as an infinite number of parallel channels each with specific probability density p(n|m), where before transmission, one of them is selected with probability $P_m$. The Class-A Middleton channel model is depicted in Figure A3. Note that the two models from Figure A2 and Figure A3 are fully equivalent.

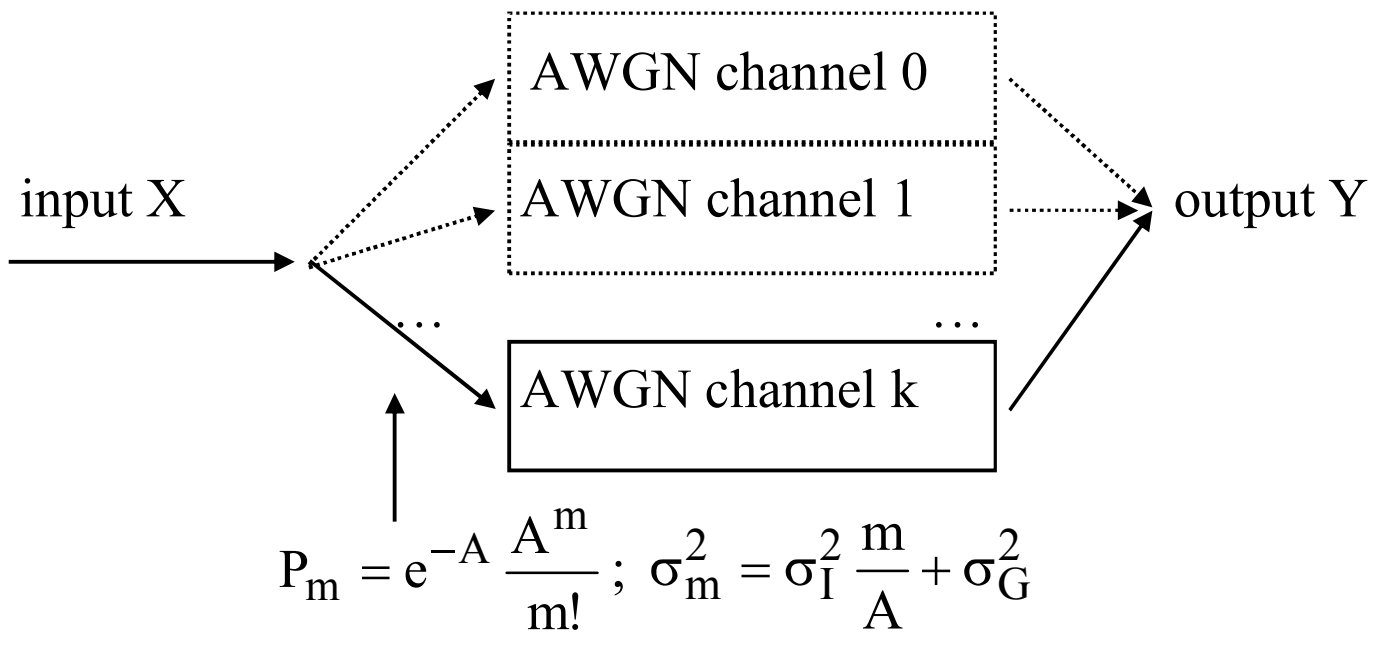

**Figure A3** Alternative Class-A Middleton channel model

**Example** As an example, for an impulse duration of $10^{-4}$ seconds and on the average 100 impulses per second, we have $A = 10^{-2}$. For a symbol transmission rate of $10^4$ symbols/second, the probability that a symbol is hit by impulse noise is $10^{-2}$. It can be seen that the impulse duration plays an important role in the transmission scheme and it is therefore very important to have reliable measurements. For instance, for a transmission rate of $10^5$ symbols/second and an impulse width of about $10^{-4}$ seconds, about 10 symbols are hit by the impulse noise.

**Example** In Table 4 we give an example of the noise power spectral density for A = 0.01 and impulse noise variance $\sigma_I^2 = 100\sigma_G^2$. From a communication point of view, we may say that the impulse noise is 20 dB above the background noise. Note that the probability of selecting a particular channel depends on A. Channel m = 0 is the most likely one. For larger A, channels with a higher number become more likely. The Class-A Middleton channel model is in principle a memoryless channel model.

**Table 4**

Example of the noise power spectral density for
A = 0.01 and T = $\sigma_G^2/\sigma_I^2 = 0.01$

| channel state | $P_m = e^{-A}\frac{A^m}{m!}$ | $\sigma_m^2 = \sigma_I^2\frac{m}{A} + \sigma_G^2$ |
|---|---|---|
| m = 0 | 0.99 | $\sigma_G^2$ |
| m = 1 | 0.01 | $10^4 \cdot \sigma_G^2 + \sigma_G^2$ |
| m = 2 | $5.0 \cdot 10^{-5}$ | $2 \cdot 10^4 \cdot \sigma_G^2 + \sigma_G^2$ |

# A.7 Impulse Noise Channel Capacity

Suppose that we have the two-state impulse noise channel model as given in Figure A4. The channel has two states with additive Gaussian noise:

Gaussian: probability 1-A, variance $\sigma_G^2$ ;
Impulsive: probability A, variance $\sigma_G^2 + \sigma_I^2 / A$ .

The average variance is $\overline{\sigma^2} = \sigma_I^2 + \sigma_G^2$. The average input power spectral density is P/2B.

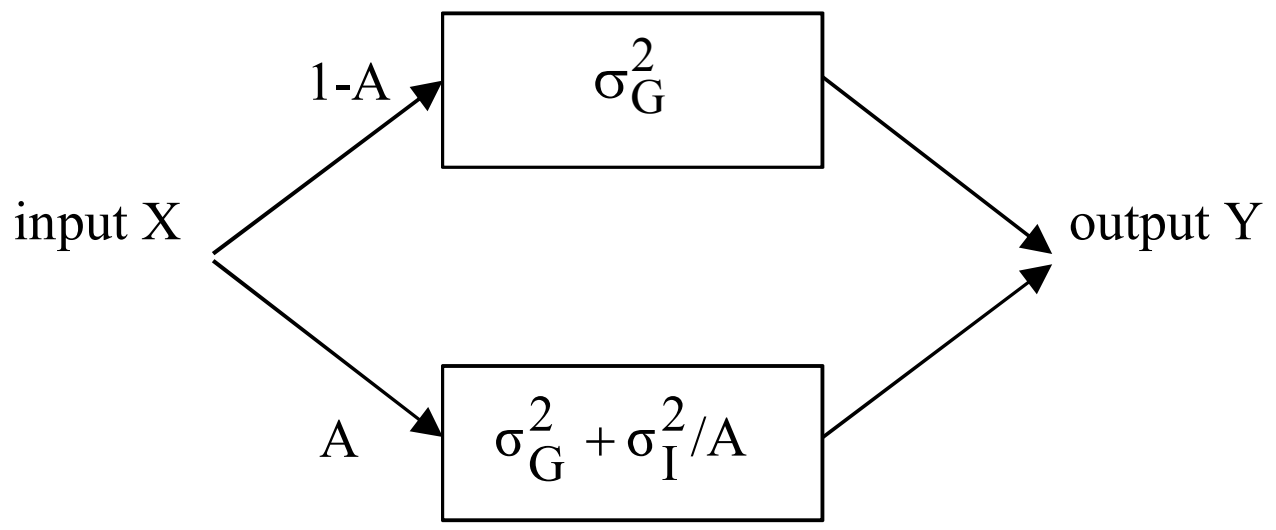

**Figure A4** Simple impulse noise model

We can distinguish four different situations for the communication, depending on whether the transmitter and the receiver are informed about the state of the channel:

**1. State known at transmitter and receiver**

In this case, the transmitter can use a water-filling argument to optimize his transmission. As a result, the input power spectral density is

$$P_A = P/2B + \sigma_I^2 - \sigma_I^2 / A ,$$

$P_{1-A} = P/2B + \sigma_I^2$,

for $P/2B \geq \frac{1-A}{A}\sigma_I^2$.

The channel capacity is given by

$$C_{+,+} = H_G(\frac{P}{2B} + \sigma_I^2 + \sigma_G^2) - (1\text{-}A)H_G(\sigma_G^2) - AH_G(\sigma_G^2 + \sigma_I^2/A) \text{ bits}$$

$$= B\log_2\left(\frac{\frac{P}{2B} + \sigma_G^2 + \sigma_I^2}{\sigma_G^2}\right) + AB\log_2\left(\frac{\sigma_G^2}{\sigma_G^2 + \frac{\sigma_I^2}{A}}\right) \text{bit/second,} \quad \text{(A7.1)}$$

where we define $H_G(s) = ½\log_2(2\pi es)$. For $P/2B < \frac{1-A}{A}\sigma_I^2$, only the Gaussian channel is considered and thus,

$$C_{+,+} = (1\text{-}A)\, H_G\left(\frac{\frac{P}{2B(1-A)} + \sigma_G^2}{\sigma_G^2}\right) \text{ bit/transmission}$$

$$= (1\text{-}A)\, B\log_2\left(\frac{\frac{P}{2B(1-A)} + \sigma_G^2}{\sigma_G^2}\right) \text{bit/second} \quad \text{(A7.2)}$$

**Remark** We see that in the high power region, water-filling is the same as fully randomizing the channel for the output entropy H(Y). For low power, we only use the Gaussian channel and thus no power is wasted on the bad channel.

**2. State not known at transmitter**

In this case, the transmitter always uses the same average input power since it has no information on the channel state. The receiver has the information, and thus

$$C_{-,+} = (1\text{-}A)\{H_G(P/2B + \sigma_G^2) - H_G(\sigma_G^2)\} +$$

$$+ A\{H_G\left(\frac{P}{2B} + \sigma_G^2 + \frac{\sigma_I^2}{A}\right) - H_G\left(\sigma_G^2 + \frac{\sigma_I^2}{A}\right)\}\text{bit/transmission}$$

(A7.3)

$$= (1-A)B\log_2\left(\frac{\frac{P}{2B} + \sigma_G^2}{\sigma_G^2}\right) + AB\log_2\left(\frac{\frac{P}{2B} + \sigma_G^2 + \frac{\sigma_I^2}{A}}{\sigma_G^2 + \frac{\sigma_I^2}{A}}\right)\text{bit/second.}$$

**3. State not known at transmitter and receiver**

For this situation, we use a randomization or orthonormal transform (FFT) at the transmitter and the receiver, see Figure A5. The output Y is assumed to

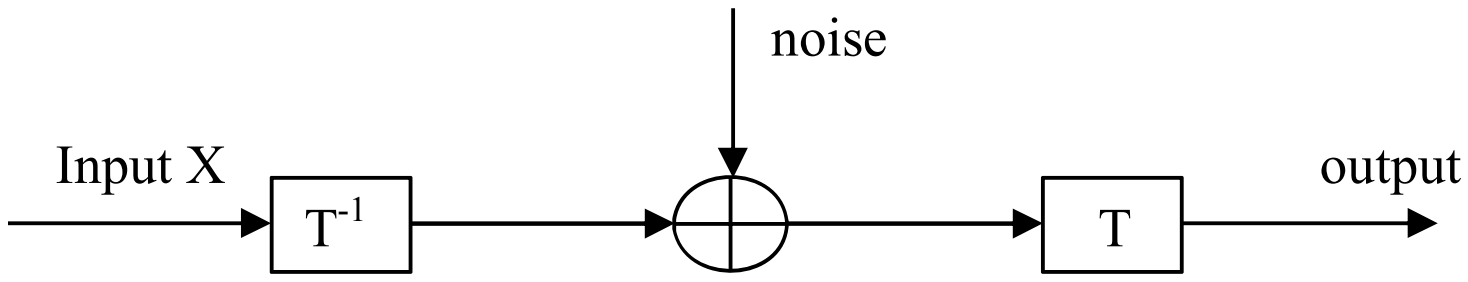

**Figure A5** Randomized transmission scheme with transform T

be Gaussian with variance P/2B + $\sigma_I^2 + \sigma_G^2$. The capacity can then be estimated as

$$C_{-,-} = H_G(\frac{P}{2B} + \sigma_I^2 + \sigma_G^2) - H_G(\sigma_I^2 + \sigma_G^2)\text{ bit/transmission}$$

$$= B\log_2\left(\frac{\frac{P}{2B} + \sigma_G^2 + \sigma_I^2}{\sigma_G^2 + \sigma_I^2}\right)\text{bit/second.} \qquad \text{(A7.4)}$$

**4. State known at transmitter, but not at receiver**

We use the water-filling argument for the power when the channel state is known. Since H(Y|X) ≥ H(Y|X, state), we can estimate the channel capacity as

$$C_{+,-} \leq C_{+,+}$$

A better approximation of the $C_{+,-}$ has to be found.

**Conclusion** It can be seen from (A7.2) and (A7.3), that $C_{-,+} \approx C_{+,+}$. Hence, this shows that an informed receiver can be as effective as an informed transmitter. In principle, no feedback is needed.

The difference Δ between informed and un-informed transmitter and receiver is defined as

$$\Delta = 10\log_{10}\left(\frac{\sigma_G^2}{\sigma_G^2 + \sigma_I^2}\right) \text{dB}.$$

For $\sigma_I^2 = 100\,\sigma_G^2$, the gain in capacity by the channel state information is about 20 dB.

For more information on impulsive noise mitigation and capacity bounds we refer to the work by Jürgen Häring[1] in [75,76,77,86,87].

[1]Jürgen Häring, PhD 2001, University Duisburg-Essen, Germany, "Error Tolerant Communication over the Compound Channel"

# A.8 Channel Capacity and Water-filling

## A.8.1 Gaussian channel capacity

Before we consider the impulse noise channel, we repeat the channel capacity for the additive white Gaussian noise channel with input X and output Y, under the following conditions:

- channel bandwidth: 2B (double sided);
- input power spectral density: $\sigma_X^2 = P/2B$;
- noise power spectral density: $\sigma_G^2$;
- entropy of the Gaussian noise: $H_G(\sigma_G^2) = ½ \log_2(2\pi e \sigma_G^2)$bits.

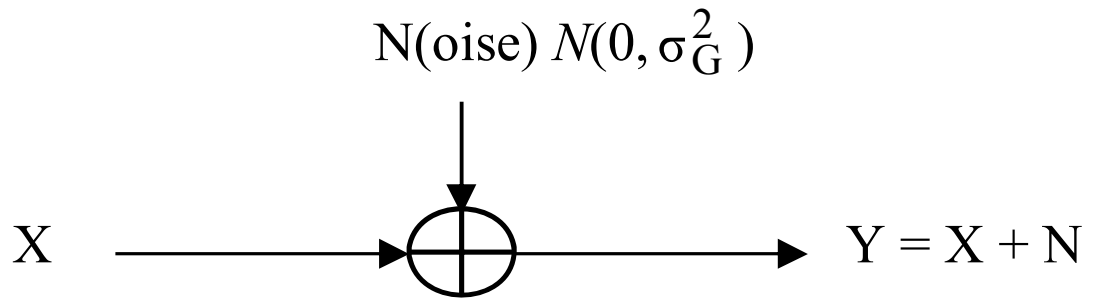

**Figure A6** Additive white Gaussian noise channel model

The capacity for the AWGN channel is

$$C_{AWGN} = H_G(Y) - H_G(Y \mid X)$$

$$= H_G(Y) - H_G(\sigma_G^2)$$

$$= \frac{1}{2}\log_2\left(\frac{\sigma_Y^2}{\sigma_G^2}\right) = \frac{1}{2}\log_2\left(\frac{\sigma_X^2 + \sigma_G^2}{\sigma_G^2}\right) \text{bit/transmission.}$$

Assuming that the channel has bandwidth B, we can represent both the input and the output by samples taken 1/2B seconds apart. If the channel noise is Gaussian with power spectral density $\sigma_G^2$, each of the 2B noise samples per second has variance $\sigma_G^2$. The power spectral density per sample is P/2B and thus, the channel capacity in bit/second is

$$C = B\log_2\left(1 + \frac{P}{2B\sigma_G^2}\right) \text{bit/second.}$$

## A.8.2 Capacity of the degraded broadcast channel

The broadcast channel ($\gamma^2 > \sigma^2$ and $\overline{X^2} \le P$) is depicted in Figure A7.

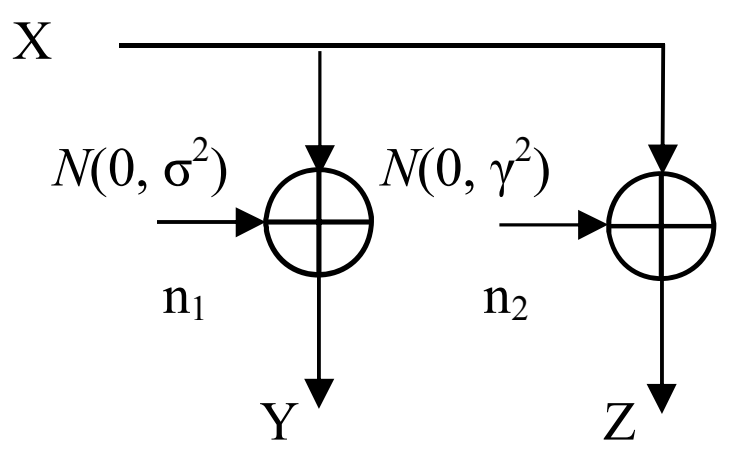

**Figure A7** The degraded broadcast channel

We can represent the transmission from X to Y and X to Z in Figure A8, where X = U + V, U and V Gaussian input variables with average power $\alpha$P and (1-$\alpha$)P, respectively. The message U is for receiver Y and the message V is for user Z. Since $\gamma^2 > \sigma^2$, the first user is more powerful and can thus also decode V and subtract its influence from X. A simple AWGN channel U→Y remains. The second user experiences the signal U and $n_2$ as additive white Gaussian noise and thus the achievable rate from transmitter to both the receivers is

$$C_1 = \frac{1}{2}\log_2\left(1 + \frac{\alpha P}{\sigma^2}\right) \text{ bit,}$$

$$C_2 = \frac{1}{2}\log_2\left(1 + \frac{\bar{\alpha} P}{\alpha P + \gamma^2}\right) \text{ bit.} \qquad \text{(A8.1)}$$

What remains is to prove that this is indeed the capacity region for the degraded broadcast channel. For this, we refer to [20].

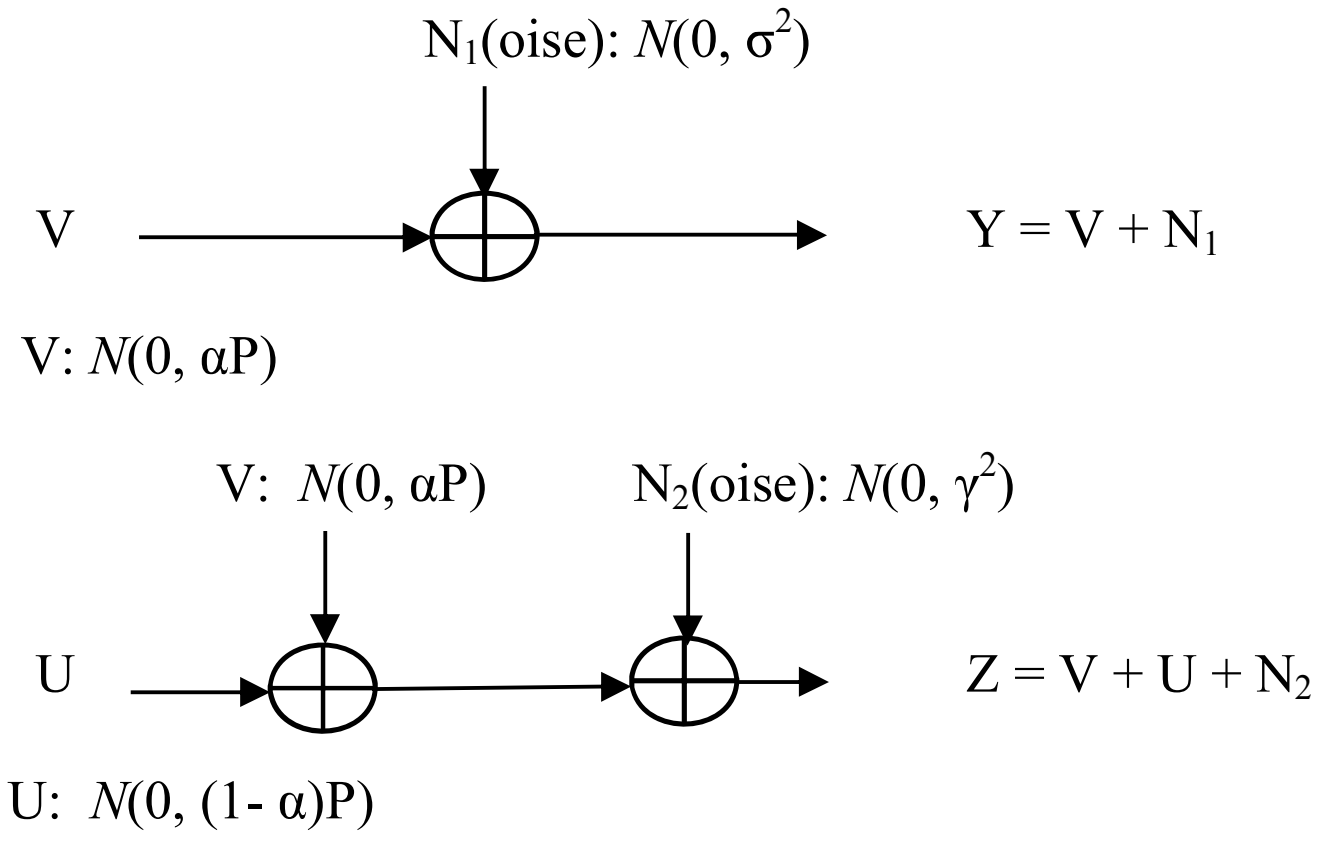

**Figure A8** The degraded broadcast channel where X = U + V

### A.8.3 Water-filling for two parallel channels

Suppose that we have an AWGN channel with bandwidth B, divided into two channels in portions αB and (1-α)B, respectively. The maximum average power $P = P_1 + P_2$, where $P_1$ and $P_2$ are the average power for channel 1 and channel 2, respectively. In addition, one channel has noise $N_1$ with variance $\gamma^2$ and the other channel has noise $N_2$ with variance $\sigma^2$, $\gamma^2 > \sigma^2$. The capacities in bit/second are

$$C_1 = \alpha\, B \log_2\left(1 + \frac{P_1}{2\alpha\, B\, \gamma^2}\right) \quad \text{bit/second} \tag{A8.2a}$$

$$C_2 = (1-\alpha) B \log_2\left(1 + \frac{P - P_1}{2(1-\alpha)\, B\, \sigma^2}\right) \text{bit/second} \tag{A8.2b}$$

We maximize $C_1 + C_2$ as a function of $P_1$, and find that for the maximum,

$$P_1 = \alpha(P - 2B(\gamma^2 - \sigma^2)\,\bar{\alpha}), \qquad \text{(A8.3a)}$$

$$P_2 = \bar{\alpha}(P + 2B(\gamma^2 - \sigma^2)\,\alpha), \qquad \text{(A8.3b)}$$

for $P \geq 2B(\gamma^2 - \sigma^2)\bar{\alpha}$, otherwise $P_1 = 0$. Figure A9 gives a graphical interpretation.

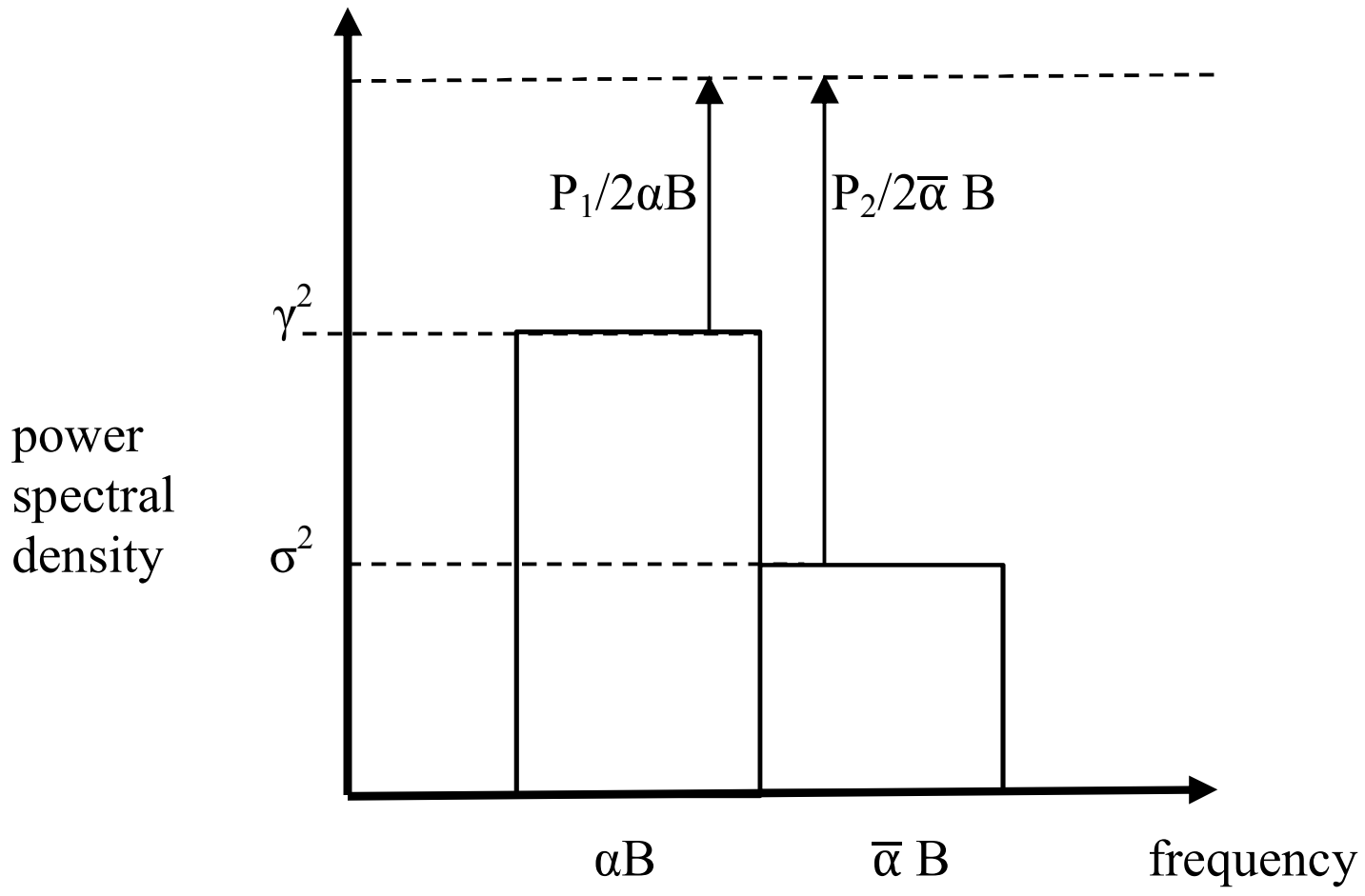

**Figure A9** Graphical presentation of the water-filling argument

Note that $P_1/2\alpha B + \gamma^2 = P_2/2\bar{\alpha} B + \sigma^2$. If the frequency band is divided into more parts, a complicated optimization algorithm is necessary, see [22].

### A.8.4 Water-filling for n parallel channels

For n parallel channels, each with the same bandwidth B, the sum capacity in bit/second is given by

$$C \leq B\sum_{i=1}^{n} \log_2\left(1 + \frac{P_i}{2B\sigma_i^2}\right) \text{ bit/second,} \qquad \text{(A8.4.1)}$$

$$\sum_{i=1}^{n} P_i \leq P \quad ,$$

where for every sub-channel the capacity is optimized by having a Gaussian input distribution with power constraint $P_i \geq 0$ subject to the sum power constraint P. To maximize (A8.4.1) one uses the technique of Lagrange multipliers. The Lagrange multiplier is given by

$$J(P_1, P_2, \cdots, P_n) = \sum_{i=1}^{n} \log_2\left(1 + \frac{P_i}{2B\sigma_i^2}\right) + \lambda(P - \sum_{i=1}^{n} P_i)\,. \qquad \text{(A8.4.2)}$$

Differentiating with respect to $P_i$ gives

$$\frac{1}{2\lambda \ln 2} = \frac{P_i}{2B} + \sigma_i^2\,.$$

The value of $\lambda$ can be solved by using the sum power constraint P, where for physical reasons $P_i \geq 0$. See Figure A10 for a graphical interpretation. For the left situation, enough power is available for all channels to have a positive capacity.

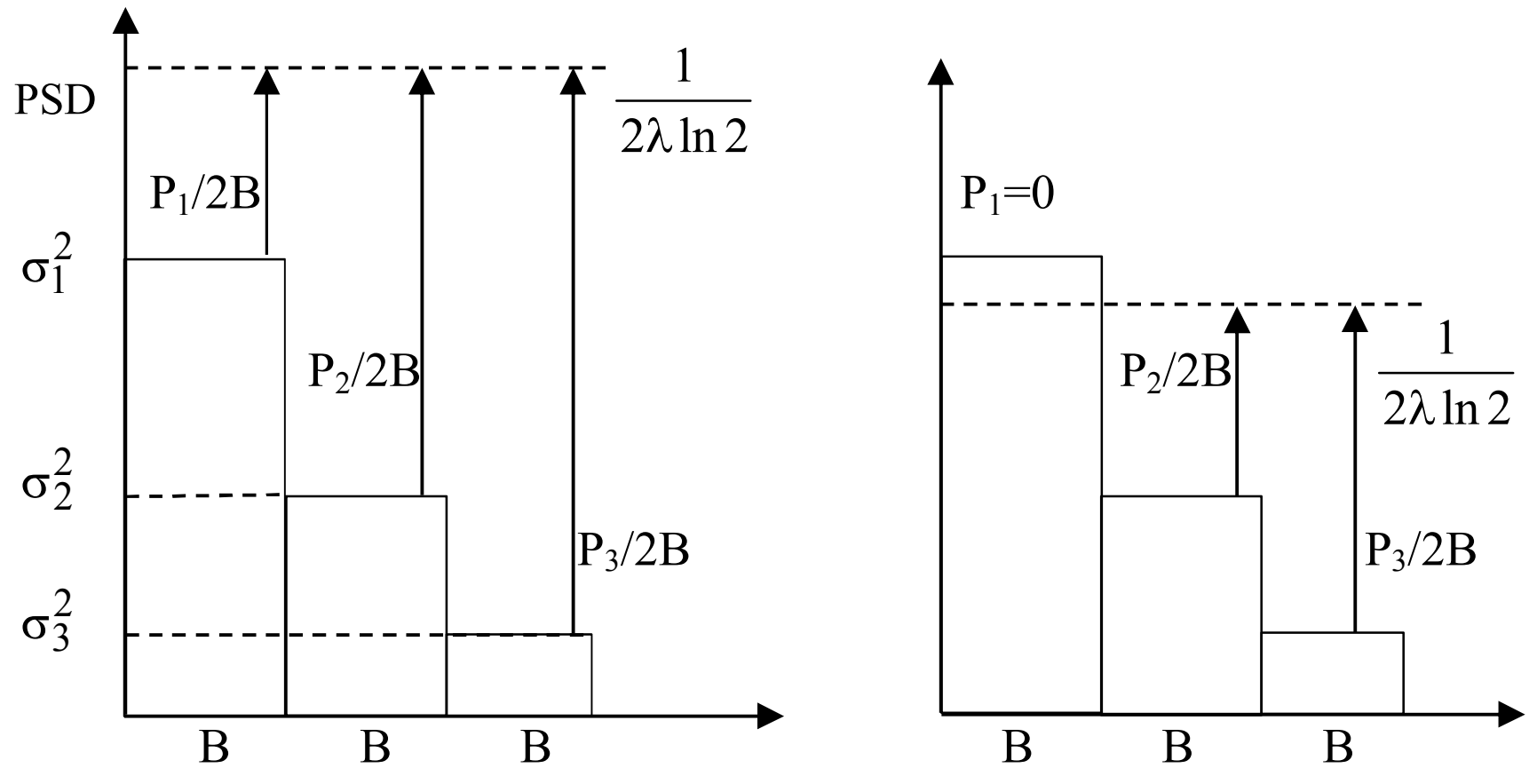

**Figure A10** Graphical interpretation of the water-filling argument for n parallel channels. Left: all channels are used

# INDEX

# My co-authors for the referenced papers

| | |
|---|---|
| 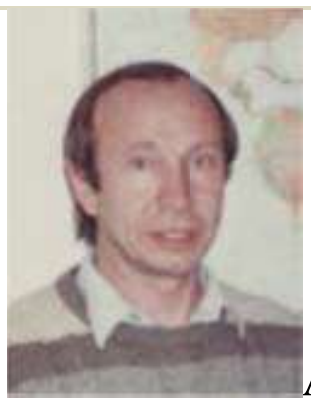 A.V. Kuznetzov,<br>member of IPPI, Moscow, Russia<br>Guest IEM | 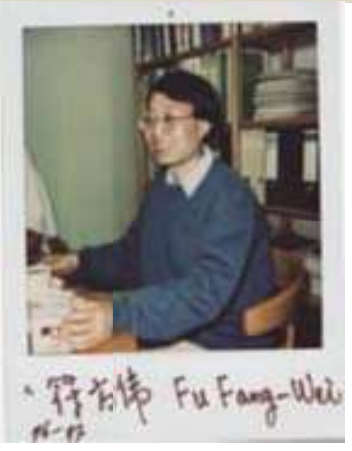 Fangwei Fu<br>Nankei University, China, Guest IEM |
| 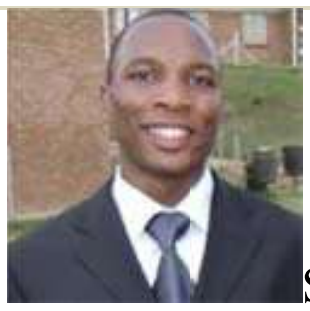 S. Thokozani,<br>PhD Univ. of Johannesburg, South<br>Africa, visiting scientist IEM | 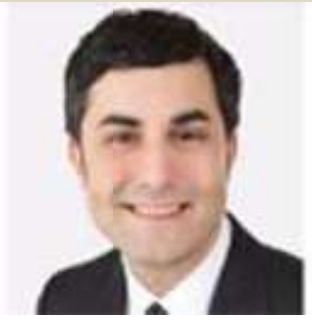 Anil Mengi,<br>PhD, IEM |
| 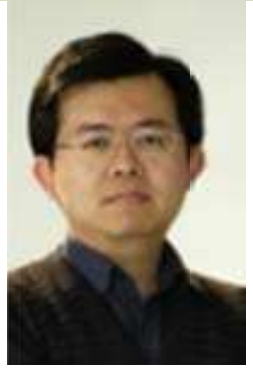 Yuan Luo,<br>Tjiao-Tong University Shanghai<br>Guest researcher IEM, | 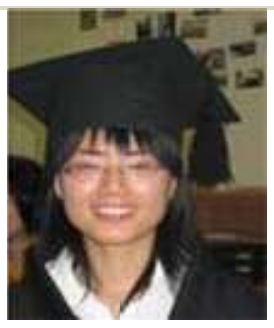 Yanling Chen<br>PhD, IEM |
| 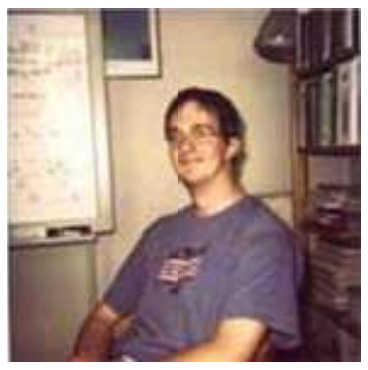 Christof Haslach,<br>PhD, IEM | 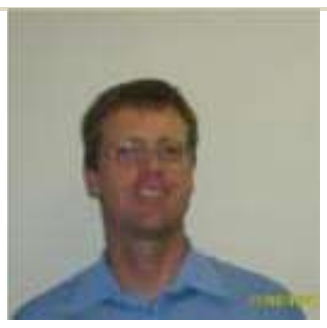 Jaco Versveld<br>Guest IEM, Univ. Witwatersrand,<br>South Africa |
| 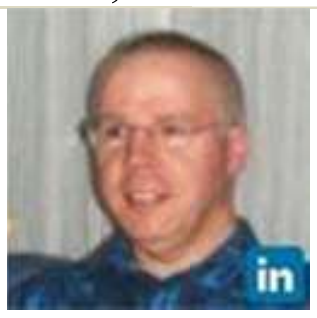 Wim Hoeks<br>University of Eindhoven<br>Master Student, 1983 | 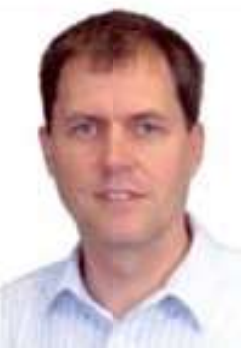 A. van Wijngaarden<br>PhD, IEM |
| 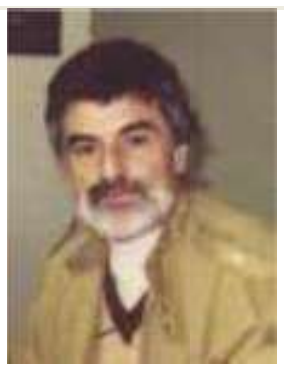 Samwel Martirossian<br>Guest IEM, Univ. Duisburg-Essen<br>Armenia Academy of Sciences | 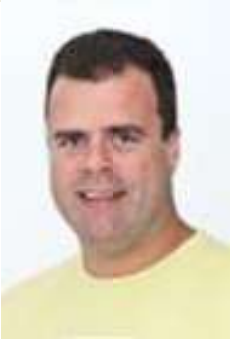 Jeroen Keunig,<br>Guest IEM<br>University Eindhoven, The Netherlands |
| 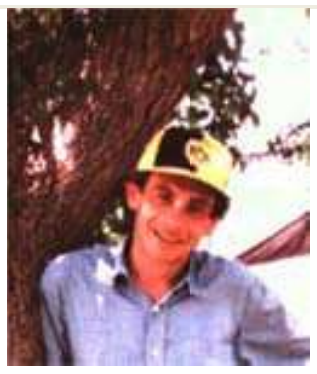 Vladimir Balakirsky,<br>Guest IEM, Russia | 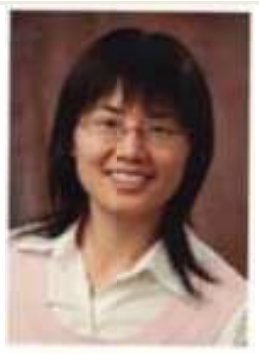 Yanling Chen,<br>PhD University Duisburg-Essen |
| 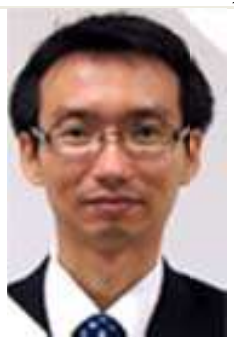 Chai Mitrpant, PhD University Duisburg-Essen, 2004 | 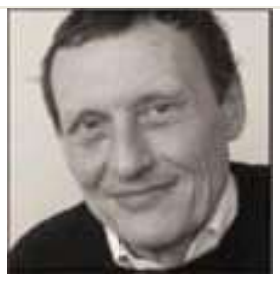 Robert Schweikert,<br>Audens, former DLR Oberpfaffenhofen, Germany |
| 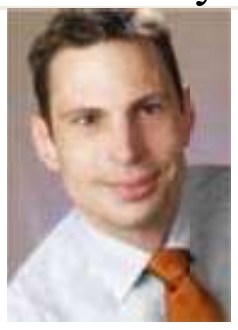 Peter Gober<br>PhD, IEM, Univ. Duisburg Essen | 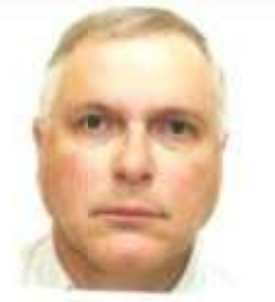 Hendrik Ferreira<br>Univ. of Johannesburg, South Africa<br>Guest IEM, Univ. Duisburg-Essen |
| 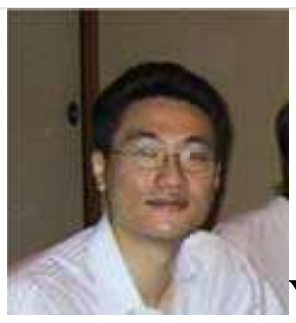 Young Gil Kim<br>Guest IEM,<br>Seoul University, South Korea | 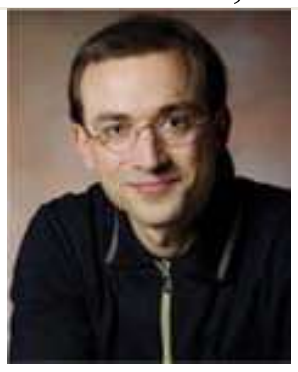 Lutz Lampe,<br>Prof. UBC, Canada, Humboldt Fellow, Guest IEM |
| 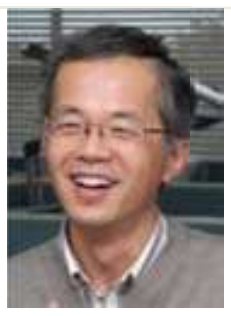 Hiroyoshi Morita<br>Chofu Univ. of Electro-Comm, Japan,<br>Guest IEM, Univ. Duisburg-Essen. | 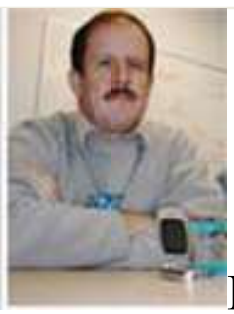 Francis Swart<br>Univ. of Johannesburg, South Africa<br>Guest IEM, Univ. Duisburg--Essen |
| 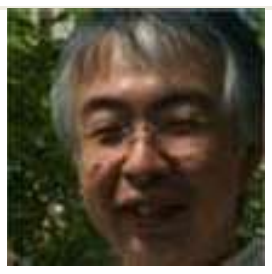 Tadashi Wadayama<br>Guest IEM, Nagoya University, Japan | 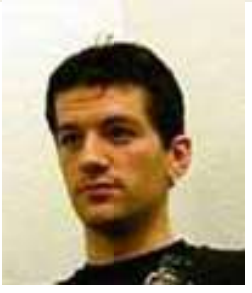 Jürgen Häring<br>2001, PhD, Univ. Duisburg Essen |
| 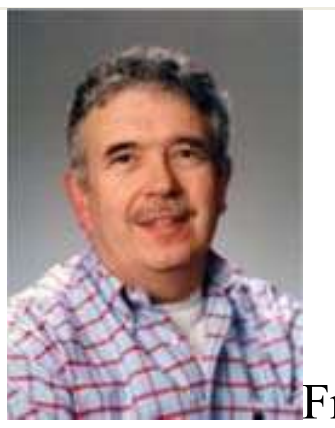Frans Willems,<br>Univ. of Eindhoven, the Netherlands | 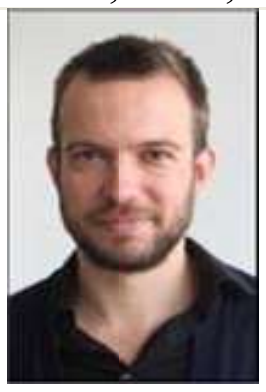 Martin Vinck, PhD Univ. of Amsterdam, the Netherlands |
| 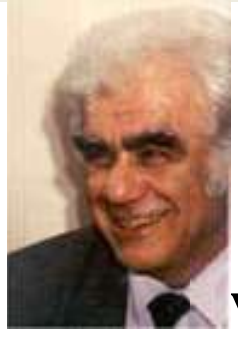Vladimir Levenshtein<br>Keldish Inst., Moscow, RU, visitor IEM | 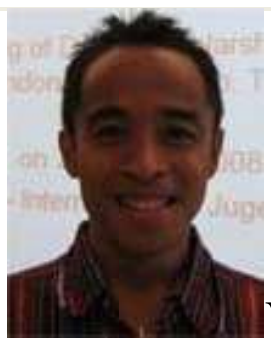Victor Papilaya, PhD<br>Student IEM, Univ. Duisburg-Essen |
| 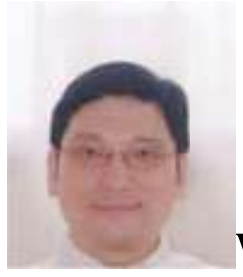Wai Ho Mow,<br>Chinese University of HongKong | 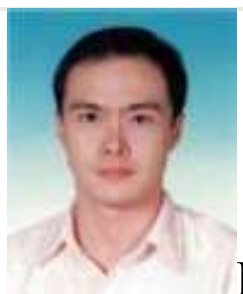Der Feng Tseng<br>National Taiwan University of Science and Technology, Taipei |
| 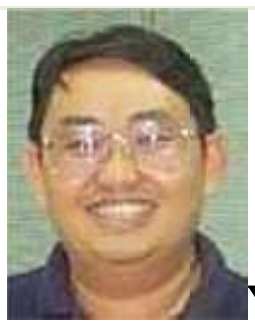Yunghsiang Sam Han<br>National Taiwan University of Science and Technology, Taipei | 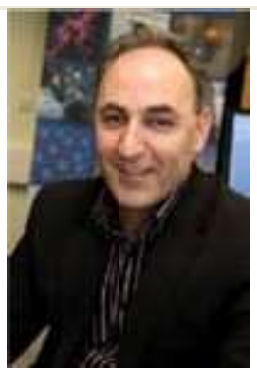 Javad Yazdani, University of Central Lancashire, England |
| 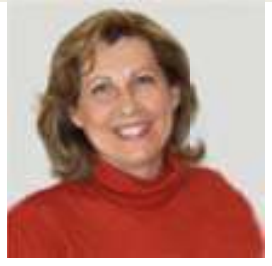 Niovi Pavlidou, University of Thessaloniki, Greece | 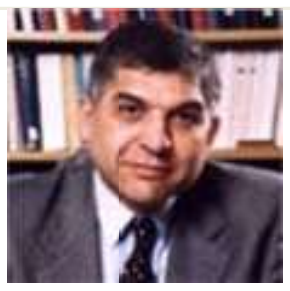 Bahram Honary, University of Lancaster, England |
| 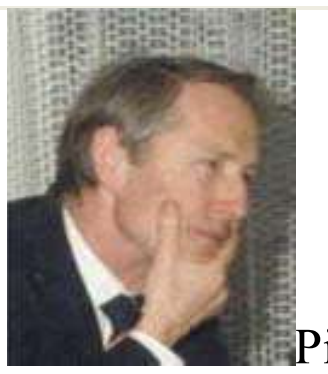Piet Schalkwijk<br>University of Eindhoven | 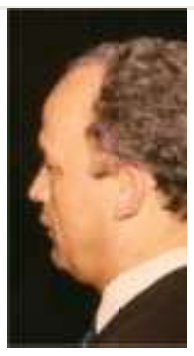 Dr. Karel Post,<br>University of Eindhoven |
| 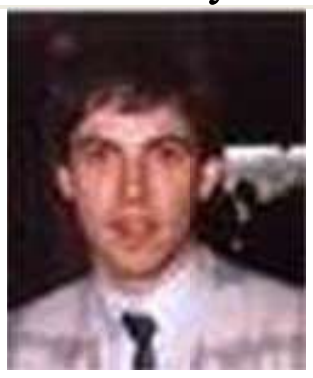 Hans Peter Foerster<br>DLR Oberpfaffenhofen, Germany | 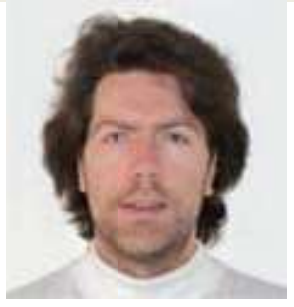 Theo Swarts<br>University of Johannesburg<br>South Africa |
| 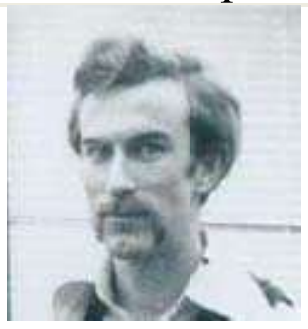 Andre de Paepe<br>University of Eindhoven | 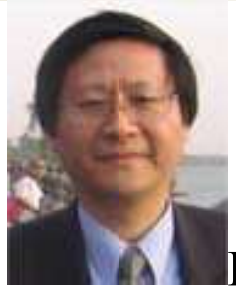Hsiao-Hwa Chen<br>National Cheng Kung University<br>Tainan City , 70101, Taiwan |
| 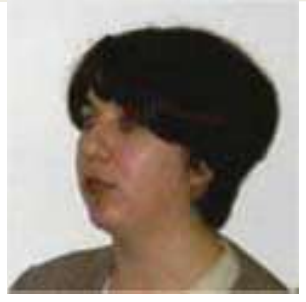 Anahit Ghazaryan<br>Guest IEM | |

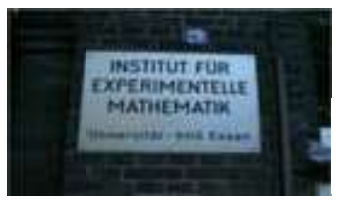

## Institute for Experimental Mathematics

To enable mathematicians, computer experts and telecommunications engineers to engage in uncomplicated and trans-disciplinary collaboration under one roof, the Institute for Experimental Mathematics (IEM) was founded, with the support of the Volkswagen Foundation, as a central scientific facility of the former University of Essen in 1989. With the addition of the Alfried Krupp von Bohlen und Halbach Foundation Chair on 1 January 1999, the Institute was expanded in the area of "Computer Networking Technology". The areas of finite mathematics, digital communications, computer networking technology and theory of numbers are all represented at the IEM.

The primary objective of the Institute is to foster interactions between the fields of mathematics, computer science and the engineering sciences. Several of the activities carried out by scientists at IEM in pursuit of this objective are listed below:

- basic research in algebra, theory of numbers, and algebraic and technical coding theory;
- improvement of possibilities for using computers in mathematic research by development algorithms and more efficient software;
- development of methods for digital communication data backup for theoretical and practical applications.

The Working Group on Digital Communications focuses on problems in the areas of information theory, communication theory and data security.

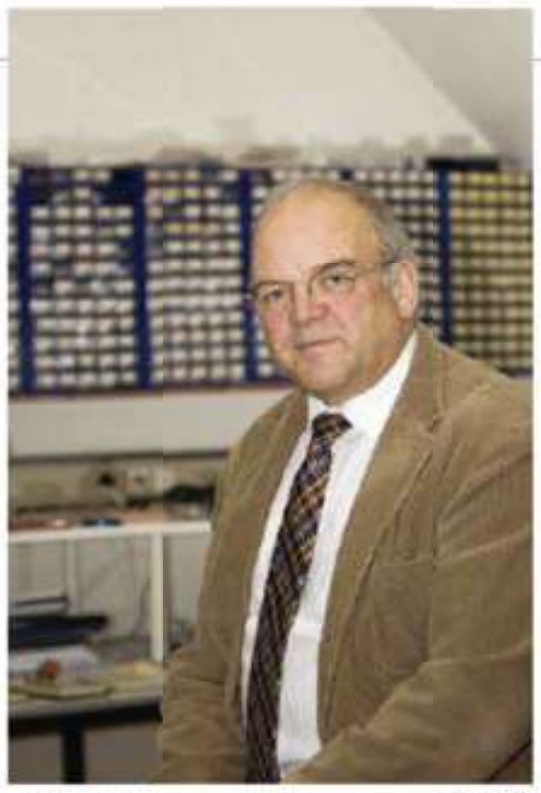

Geschäftsführender Direktor / Managing Director: Prof. Dr. A. J. Han Vinck

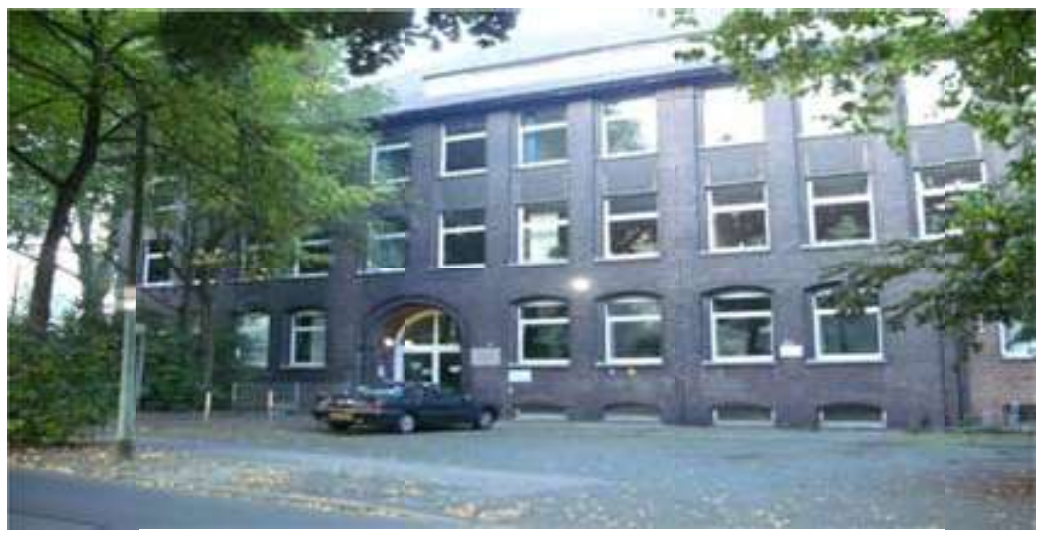